\documentclass[fleqn,usenatbib]{mnras}
\usepackage{newtxtext,newtxmath}
\usepackage[T1]{fontenc}
\usepackage{ae,aecompl}
\usepackage [english]{babel}
\usepackage{epsfig}
\usepackage{graphicx}
\usepackage{amsmath}
\usepackage{amsfonts}
\usepackage{amssymb}
\usepackage{braket}
\usepackage{placeins}
\usepackage{xcolor}
\usepackage{soul}
\usepackage{multirow}

\def\zs{\mbox{{$z_{\rm spec}$}}}
\def\zp{\mbox{{$z_{\rm phot}$}}}

\title[Statistical analysis of PDFs for the KiDS-DR3]{Statistical analysis of  probability density functions for photometric redshifts through the KiDS-ESO-DR3 galaxies}
\author[Amaro et al. 2017]{V.~Amaro$^{1}$\thanks{E-mail: amaro@na.infn.it}, S.~Cavuoti$^{1,2,3}$, M.~Brescia$^{2}$, C.~Vellucci$^{6}$, G.~Longo$^{1}$, M.~Bilicki$^{4,5}$, \and J.~T.~A.~de Jong$^{7}$,  C.~Tortora$^{7}$, M.~Radovich$^{8}$, N.~R.~Napolitano$^{2}$, H. Buddelmeijer$^{4}$ \\ 
 $^{1}$Department of Physical Sciences, University of Napoli Federico II, via Cinthia 9, 80126 Napoli, Italy\\
 $^{2}$INAF - Astronomical Observatory of Capodimonte, via Moiariello 16, 80131 Napoli, Italy\\
 $^{3}$INFN section of Naples, via Cinthia 6, I-80126, Napoli, Italy\\
 $^{4}$Leiden Observatory, Leiden University, P.O. Box 9513, 2300 RA Leiden, the Netherlands\\
 $^{5}$National Centre for Nuclear Research, Astrophysics Division, P.O. Box 447, PL-90-950 \L{}\'{o}d\'{z}, Poland\\
 $^{6}$DIETI, University of Naples Federico II, Via Claudio, 21, Napoli, Italy\\
 $^{7}$Kapteyn Astronomical Institute, University of Groningen, P.O. Box 800, 9700 AV Groningen, the Netherlands\\
 $^{8}$INAF - Osservatorio Astronomico di Padova, via dell'Osservatorio 5, 35122 Padova, Italy\\
}

\date{Accepted . Received ; in original form }

\pagerange{\pageref{firstpage}--\pageref{lastpage}} \pubyear{2018}
\begin{document}
\label{firstpage}
\maketitle
\begin{abstract}
Despite the  high accuracy of photometric redshifts (zphot) derived using Machine Learning (ML) methods, the quantification  of errors through reliable and accurate Probability Density Functions (PDFs) is still an open problem. First, because it is difficult to accurately assess the contribution from different sources of errors, namely internal to the method itself and from the photometric features defining the available parameter space. Second, because the problem of defining a robust statistical method, always able to quantify and qualify the PDF estimation validity, is still an open issue. We present a comparison among PDFs obtained using three different methods on the same data set: two ML techniques, METAPHOR (Machine-learning Estimation Tool for Accurate PHOtometric Redshifts) and ANNz2, plus the spectral energy distribution template fitting method, BPZ. The photometric data were extracted from the KiDS (Kilo Degree Survey) ESO Data Release 3, while the spectroscopy was obtained from the GAMA (Galaxy and Mass Assembly) Data Release $2$. \\
The statistical evaluation of both individual and \textit{stacked} PDFs was done through quantitative and qualitative estimators, including a \textit{dummy} PDF, useful to verify whether different statistical estimators can correctly assess PDF quality. We conclude that, in order to quantify the reliability and accuracy of any zphot PDF method, a combined set of statistical estimators is required.
\end{abstract}

\begin{keywords}
galaxies: photometry - galaxies: distances and redshifts - methods: data analysis -
methods: statistical
\end{keywords}

\section{Introduction}\label{SEC:Introduction}
Redshifts, by allowing the calculation of distances for large samples of galaxies, are at the core of most extragalactic and cosmological studies and are needed for many purposes, such as, to quote just a few, to constrain the dark matter and dark energy contents of the Universe through weak gravitational lensing \citep{Serjeant2014,Hild2017,Fu2018}, to reconstruct the cosmic Large Scale Structure \citep{Aragon2015}, to identify galaxy clusters and groups \citep{Capozzi2009,Annunziatella2016,Rad2017}, to disentangle the nature of astronomical sources \citep{Brescia2012,Tortora2016}; to map the galaxy colour-redshift relationships \citep{Masters2015} and to measure the baryonic acoustic oscillations spectrum \citep{gorecki2014,Ross2017}.\\
\indent The last few years have seen a proliferation of multi-band photometric galaxy surveys, either ongoing  (see KiDS - Kilo-Degree Survey, \citealt{dJ2015, dJ2017}; DES - Dark Energy Survey, \citealt{Annis2013}), planned (LSST, \citealt{Ivezic2009}, \citealt{LSSTB2009} and Euclid, \citealt{Laureijs2014}, \citealt{EucRedB2011}). All these surveys require redshift estimates for hundreds of millions or billions of galaxies that cannot be observed spectroscopically and therefore must be obtained via multi-band photometry (photometric redshifts or zphot).
This is possible due to the existence of a highly non-linear correlation between photometry and redshift, caused by the fact that the stretching introduced by the redshift induces the main spectral features to move through the different filters of a photometric system (\citealt{Baum1962, Connolly1995}). \\
\indent In a broad but widespread oversimplification, there are two main classes of methods commonly used to derive zphot: the Spectral Energy Distribution (SED) template fitting methods (e.g., \citealt{Bolzonella2000, Arnouts1999, Ilbert2006, Tanaka2015}) and the empirical (or interpolative) methods  (e.g., \citealt{Firth2003,Ball2008,CeB2013,Brescia2014b,graff2014,Cavuoti2015a,Cavuoti2015, Masters2015,Sadeh2016,Soo2017,disanto}), both characterized by their advantages and shortcomings. There are also recent experiments that try to combine these two zphot estimation classes, in order to merge their respective capabilities (e.g., \citealt{Cavuoti2017b, duncan2017, Hoyle2018}).\\
\indent SED template fitting methods are based on a fit (generally a $\chi^{2}$ minimization) to the multi-band photometric observations of the objects. The starting point is a set of template (either synthetic or observed) spectra covering different morphological types and physical properties. Each of these template SEDs is convolved with the transmission functions of any given filters, in order to create synthetic magnitudes as a function of the redshift.\\
\indent SED fitting methods are capable to derive all at once the zphot, the spectral type and the Probability Density Function (or PDF) of the error distribution of each source. However, these methods suffer from several  shortcomings: the potential mismatch between the templates used for the fitting, the properties of the selected sample of galaxies \citep{Abdalla2011}, colour/redshift degeneracies  and template incompleteness. 
Such issues are stronger at high redshift, where galaxies are fainter and photometric errors larger. Furthermore, at high redshifts there are fewer or no empirical spectra available to build a reliable template library.\\ 
Among empirical methods, those based on various Machine Learning (ML) algorithms are the most frequently used. They infer (not analytically) the complex relation existing between the input, mainly multi-band photometry (i.e. fluxes, magnitudes and/or derived colours) and the desired output (the spectroscopic redshift, hereafter zspec). In supervised ML the learning process is regulated by the spectroscopic information (i.e. redshift) available for a subsample of the objects, whereas in the unsupervised approach, the spectroscopic information is not used in the training phase, but only during the validation phase. 
There are many ML algorithms that have been used for zphot estimation. To quote just a few: neural networks (\citealt{Taglia2002,CollandLahav2004,Brescia2013,Sadeh2015}), boosted decision trees \citep{Gerdes2010}, random forests \citep{CeB2013a}, self organized maps \citep{carrasco2014a, Masters2015}. 
ML techniques are endowed with several advantages: \textit{(i)} high accuracy of predicted zphot within the limits imposed by the spectroscopic Knowledge Base (hereafter KB); \textit{(ii)} ability to easily incorporate external information in the training, such as surface brightness, angular sizes or galaxy profiles \citep{Taglia2002, cavuoti2012, Soo2017, bilicki2017}. \\ 
\indent On the other hand,  ML methods have a very poor capability to extrapolate information outside the regions of the parameter space properly sampled by the training data that, for instance, implies that they cannot be used to estimate redshifts for objects fainter than those present in the spectroscopic sample. 
Furthermore, supervised methods are viable only if accurate photometry and spectroscopy are available for a quite large (few thousands of objects at least) number of objects. See \cite{Hild2010}, \cite{Abdalla2011} and \cite{Sanchez2014} for reviews about the zphot estimation techniques.\\
\indent Finally, due to their intrinsic nature of self-adaptive learning models,  the ML based methods do not naturally provide a PDF estimate of the predicted zphot, unless special procedures are implemented.\\
\indent In recent years it has been demonstrated in several studies that PDFs can increase the accuracy of cosmological parameter measurements. For example, \cite{Mand2008} have shown that most common statistics (bias, outlier rate, standard deviation etc.) are not sufficient to evaluate the accuracy of zphot required by weak lensing studies. In particular the measurement of the critical mass surface density requires a reliable PDF estimation to remove any calibration bias effect.\\
\indent Over the last few years, particular attention has been paid to develop techniques and procedures able to compute a full zphot PDF for an astronomical source as well as for an entire galaxy sample \citep{Bonnet2013,CeB2013a,carrasco2014a,carrasco2014b}.
The PDF contains more information than the single redshift estimate, as it is also confirmed by the  improvement in the accuracy of cosmological and weak lensing measurements \citep{Mand2008, Viola2015}, when PDFs are used rather than zphot point estimates. However, to the best of our knowledge, the positive role played by the PDFs has been demonstrated only for zphot obtained with SED fitting methods.\\
\indent In this paper we perform a comparative analysis of zphot and associated PDF performance among different methods. The data used for this analysis were extracted from the KiDS ESO (European Southern Observatory) Data Release 3 (hereafter, KiDS-ESO-DR3), described in \cite{dJ2017}.  In that work, three different methods for photometric redshifts were used and the  corresponding catalogues made publicly available\footnote{Available at \url{http://kids.strw.leidenuniv.nl/DR3/ml-photoz.php}.}: two ML methods, respectively, METAPHOR (Machine-learning Estimation Tool for Accurate PHOtometric Redshifts, \citealt{Cavuoti2017}) and ANNz2 (\citealt{Sadeh2016,bilicki2017}), plus one template fitting method, the Bayesian Photometric Redshifts (hereafter, BPZ, \citealt{Benitez2000}). For the purpose of the present paper, we also build a \textit{dummy} PDF, independent from method errors and photometric uncertainties, useful to compare and assess the statistical estimators used to evaluate the reliability of PDFs.\\
\indent The paper is structured as follows: in Sec.~\ref{SEC:data} we present the KiDS-ESO-DR3 data used for the analysis. In Sec.~\ref{SEC:PDF} we give a general overview about the calculation of PDFs and we describe the methods as well as the statistical estimators involved in our analysis. In Sec.~\ref{SEC:thecomparison} we perform the comparison among the PDF methods and a critical discussion about the statistical estimators. Finally, in Sec.~\ref{SEC:discussion} we draw our conclusions.

\section{THE DATA}\label{SEC:data}
The sample of galaxies used to estimate zphot and their individual and stacked PDFs was extracted from the third data release of the ESO Public Kilo-Degree Survey (KiDS-ESO-DR3, \citealt{dJ2017}).
When completed, the KiDS survey will cover $1500$ $deg^{2}$ \citep{dJ2017}, distributed over two survey fields, in four broad-band filters (\textit{u, g, r, i}). Compared to the previous data releases \citep{dJ2015}, the DR3 not only covers a larger area of the sky, but it also relies on an improved photometric calibration and provides photometric redshifts along with shear catalogues and lensing-optimized image data. The total DR3 data set consists of $440$ tiles for a total area covering approximately $450$ $deg^{2}$, with respect to the $160$ $deg^{2}$ of the previous releases.\\
The DR3 provides also an aperture-matched multi-band catalogue for more than $48$ million sources, including homogenized photometry based on Gaussian Aperture and PSF (hereafter GAaP) magnitudes \citep{Kuij2008}. All the measurements (star/galaxy separation, source position, shape parameters) are  based on the \textit{r}-band images, due to their better quality (see Table A.2 of \citealt{dJ2017}).\\
\indent KiDS was primarily designed for Weak Lensing (WL) studies, in order to reconstruct the Large Scale Structure (LSS) of the Universe. Indeed, the first 148 tiles of the first two data releases produced their first scientific results on weak lensing for galaxies and groups of galaxies in the Galaxy And Mass Assembly (GAMA, \citealt{driver2011}) fields \citep{dJ2015}, as the reader can find in \cite{Viola2015}.\\
The photometry used in this work consists of the \textit{ugri} GAaP magnitudes, two aperture magnitudes, measured within circular apertures of 4$\arcsec$  and 6$\arcsec$ diameter ($20$ and $30$ pixels, referred in Table~\ref{tab:maglimtot} as $MAG\_APER\_20\_X$ and $MAG\_APER\_30\_X$), respectively, corrected for extinction and zeropoint offsets and the derived colours, for a total of $21$ photometric parameters for each object. \\
\indent The original data set was cleaned by removing objects affected by missing information (the performance of ML methods may degrade if data is missing) and by clipping the tails of the magnitude distributions in order to ensure a proper density of training points in the sampled regions of the parameter space. The lower and upper cuts, applied to exclude the tails of the distributions, are reported in Table~\ref{tab:maglimtot}. \\
\indent Furthermore, as we shall specify in Sec.~\ref{SEC:METAPHOR}, the fundamental concept of the PDF estimation in METAPHOR is the perturbation of the data photometry, based on a proper fitting function of the  flux errors in specifically defined bins of flux. Therefore, in the preparation phase, we excluded from the KB all entries with a photometric error higher than a given threshold (e.g. $1$ magnitude) in order to provide a dataset used for the polynomial fitting of the errors, as prescribed by the \textit{mixture} perturbation law (see Sec.~\ref{SEC:METAPHOR}).\\
\indent In order to perform a zphot comparison through a common spectroscopic base in the work of \cite{dJ2017}, each of the three zphot catalogues (obtained, respectively, by METAPHOR\footnote{In \cite{dJ2017} it is referred to as MLPQNA, the internal zphot estimation engine of METAPHOR},  ANNz2 and BPZ), has been cross-matched in coordinates with the spectroscopic information extracted from the second data release (DR2) of GAMA \citep{Liske2015}, containing spectroscopy in the KiDS-North field (composed of $77\%$ objects from GAMA, $18\%$ from SDSS/BOSS DR10 \citealt{Ahn2014} and $5\%$ from 2dFGRS \citealt{Colless2001}). 
For what concerns this paper, since the ANNz2 catalogue released with DR3 does not include individual PDFs \citep{bilicki2017}, these have been derived for the purposes of the present work, by uniforming the training and test sets with those used by METAPHOR in \cite{dJ2017}. Finally, since in the present work we were interested in performing the zphot PDF comparison among the two mentioned ML methods and BPZ using a uniform data sample, we followed the same approach as \cite{dJ2017}, based on the cross-matching between KiDS-DR3 photometry and SDSS DR9 + GAMA DR2 + 2dFGRS spectroscopy. As described in Sec.~$4.2$ of \cite{dJ2017}, we performed a random shuffling and split procedure, obtaining a training set of $\sim 71,000$ and a blind validation set of $\sim 18,000$ objects. The final comparison test among methods has been done on a supplementary blind set of $\sim 64,000$ galaxies, as detailed in Sec.~$4.4$ of \cite{dJ2017}.

\begin{table}
\centering
\caption{ Brighter and fainter limits imposed on the magnitudes and defining the region of the parameter space used for training and test experiments.}\label{tab:maglimtot}
 \begin{tabular}{lcc}
{\bf Input magnitudes}              & {\bf brighter limit} & {\bf fainter limit} \\ \hline \hline
MAG\_APER\_20\_U    & $16.84$   & $28.55$\\
MAG\_APER\_30\_U    & $16.81$   & $28.14$\\
MAG\_GAAP\_U        & $16.85$   & $28.81$\\
MAG\_APER\_20\_G    & $16.18$   & $24.45$\\
MAG\_APER\_30\_G    & $15.86$   & $24.59$\\
MAG\_GAAP\_G        & $16.02$   & $24.49$\\
MAG\_APER\_20\_R    & $15.28$   & $23.24$\\
MAG\_APER\_30\_R    & $14.98$   & $23.30$\\
MAG\_GAAP\_R        & $15.15$   & $23.29$\\
MAG\_APER\_20\_I    & $14.90$   & $22.84$\\
MAG\_APER\_30\_I    & $14.56$   & $23.07$\\  
MAG\_GAAP\_I        & $14.75$   & $22.96$\\
 \end{tabular}
\end{table}

\section{The methods}\label{SEC:PDF}

In general terms, a PDF is a way to parametrize the uncertainty on the zphot solution. 
In the context of  zphot estimation, a PDF is strictly dependent both on the measurement methods and on the physical assumptions. This simple statistical consideration renders the real meaning of PDFs quite complex to grasp in the case of zphot error evaluation. \\
\indent Furthermore, a PDF should provide a robust estimate of the reliability of an individual redshift. The factors affecting such reliability are: photometric errors, intrinsic errors of the methods and statistical biases. In fact, under the hypothesis that a perfect reconstruction of the redshift is possible, the PDF would consist of a single Dirac delta. However, since the zphot cannot be perfectly mapped to the true redshift, the corresponding PDF represents the intrinsic uncertainties of the estimate.
In other words, as anticipated in Sec.~\ref{SEC:Introduction}, PDFs are useful to characterize zphot estimates by providing more information than the simple estimation of the error on the individual measurements.\\  
\indent In the following paragraphs we shortly summarize the characteristics of the methods used in our study. Besides the already mentioned ML methods (METAPHOR, ANNz2) and BPZ, we introduce also a special way to assess the validity of the statistical estimators used to measure the PDF reliability, called \textit{dummy} PDF.

\subsection{METAPHOR}\label{SEC:METAPHOR}

The METAPHOR (Machine-learning Estimation Tool for Accurate PHOtometric Redshifts; \citealt{Cavuoti2017}) method is a modular workflow, designed to produce both zphot and related PDFs. The internal zphot estimation engine is our model MLPQNA (Multi Layer Perceptron trained with Quasi Newton Algorithm; \citealt{Brescia2013,Brescia2014a}). METAPHOR makes available a series of functional modules: 
\begin{itemize}
	\item[-]{Data Pre-processing: data preparation, photometric evaluation of the Knowledge Base (KB), followed by its perturbation based on the given magnitude error distributions;}
	\item[-]{zphot prediction: single photometric redshift estimation, based on training/test of the KB with the MLPQNA model;}
	\item[-]{PDF estimation: production of the individual zphot PDFs and evaluation of their cumulative statistical properties.}
\end{itemize} 
 
\begin{figure*}
 \centering
  {\includegraphics[width=0.49 \textwidth]{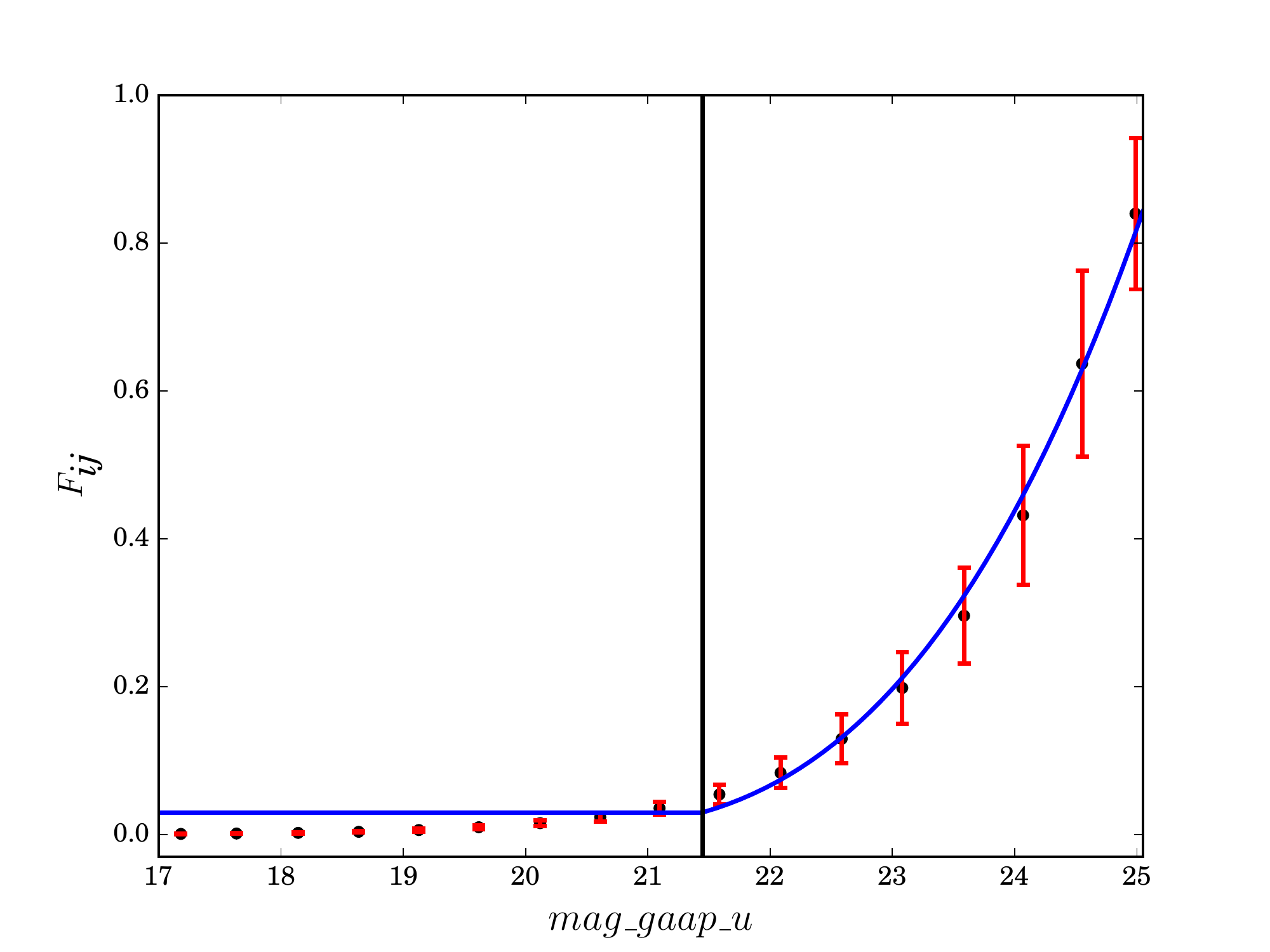}}
  {\includegraphics[width=0.49 \textwidth]{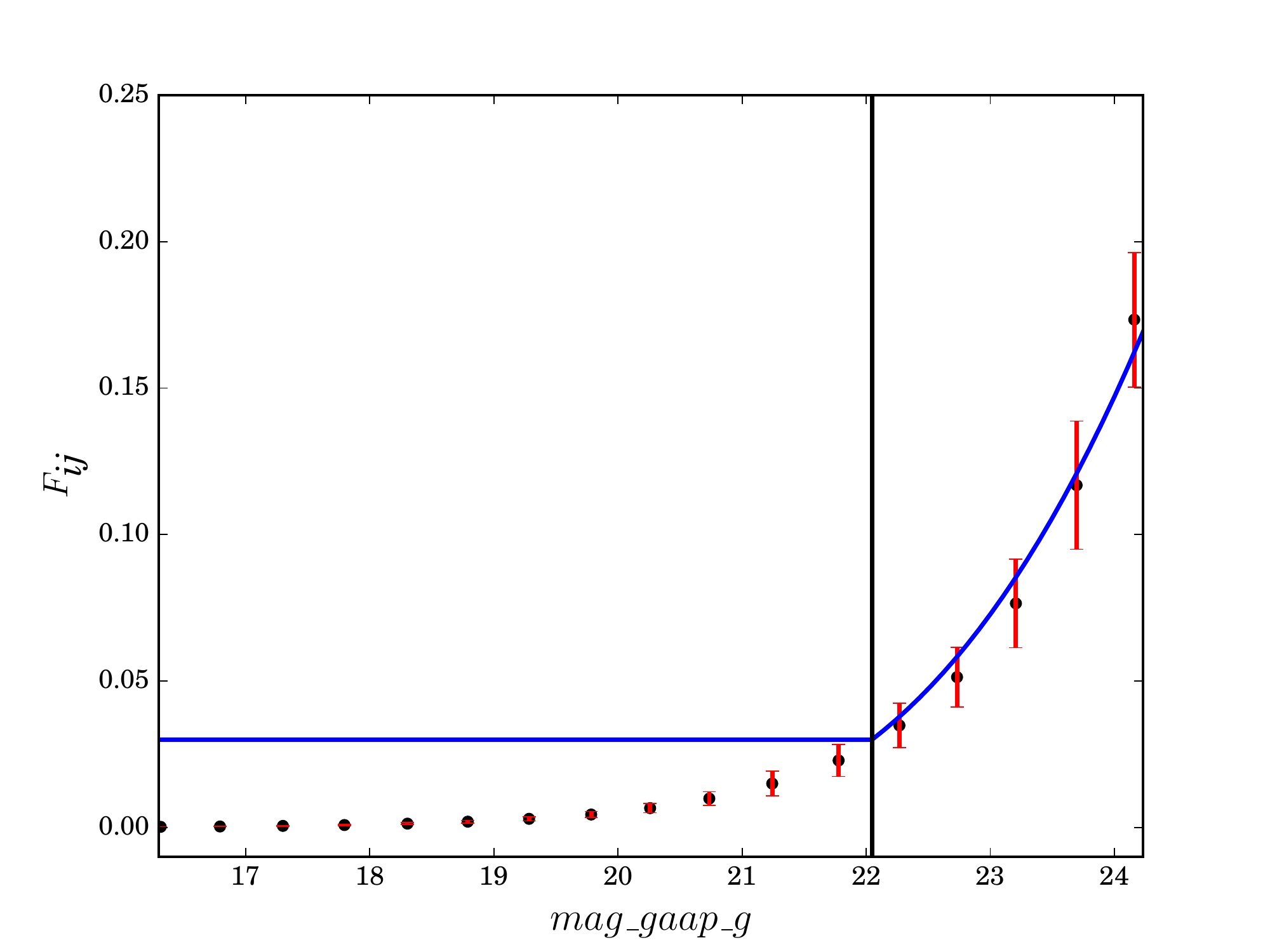}}
  {\includegraphics[width=0.49 \textwidth]{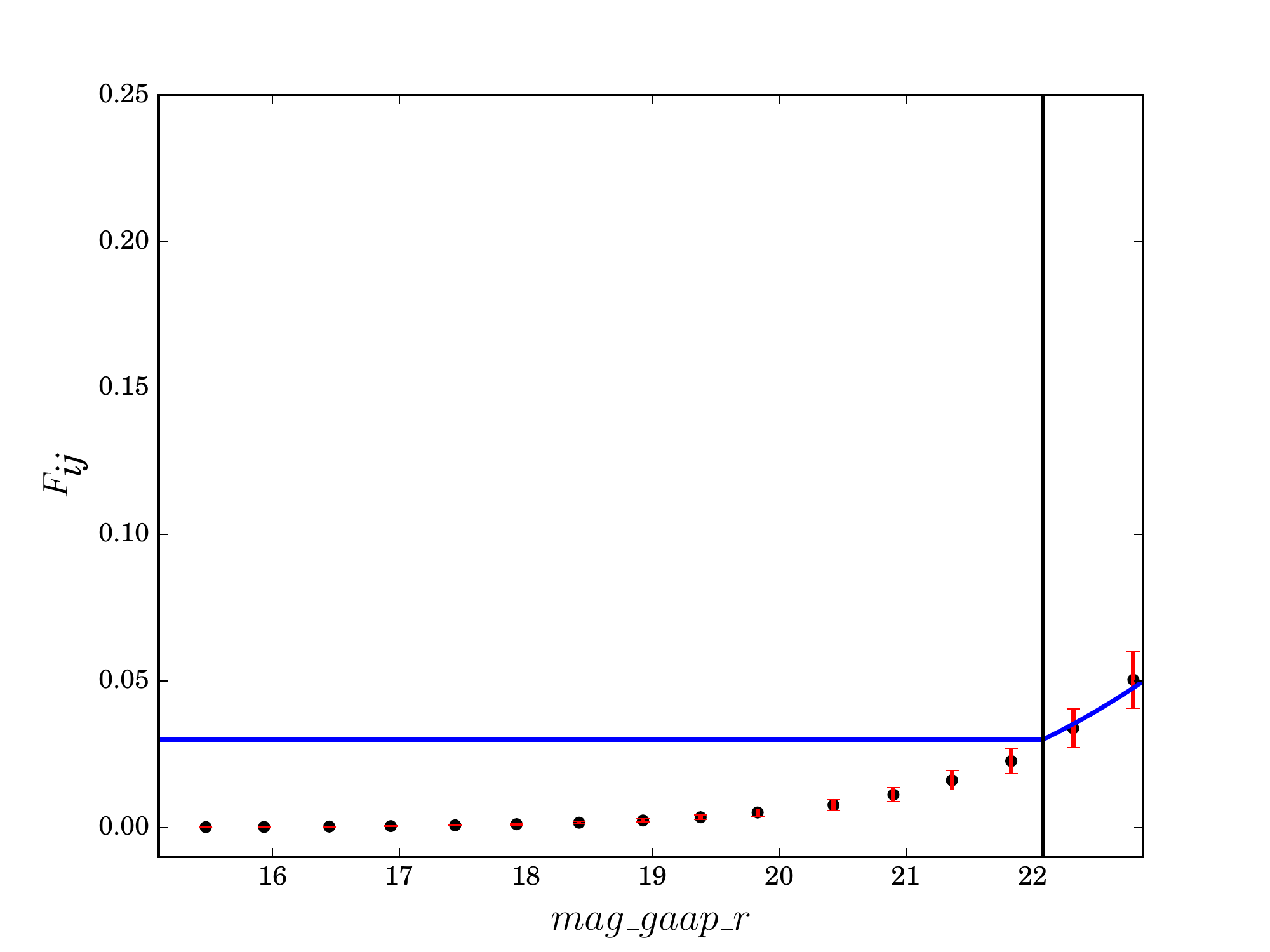}}
  {\includegraphics[width=0.49 \textwidth]{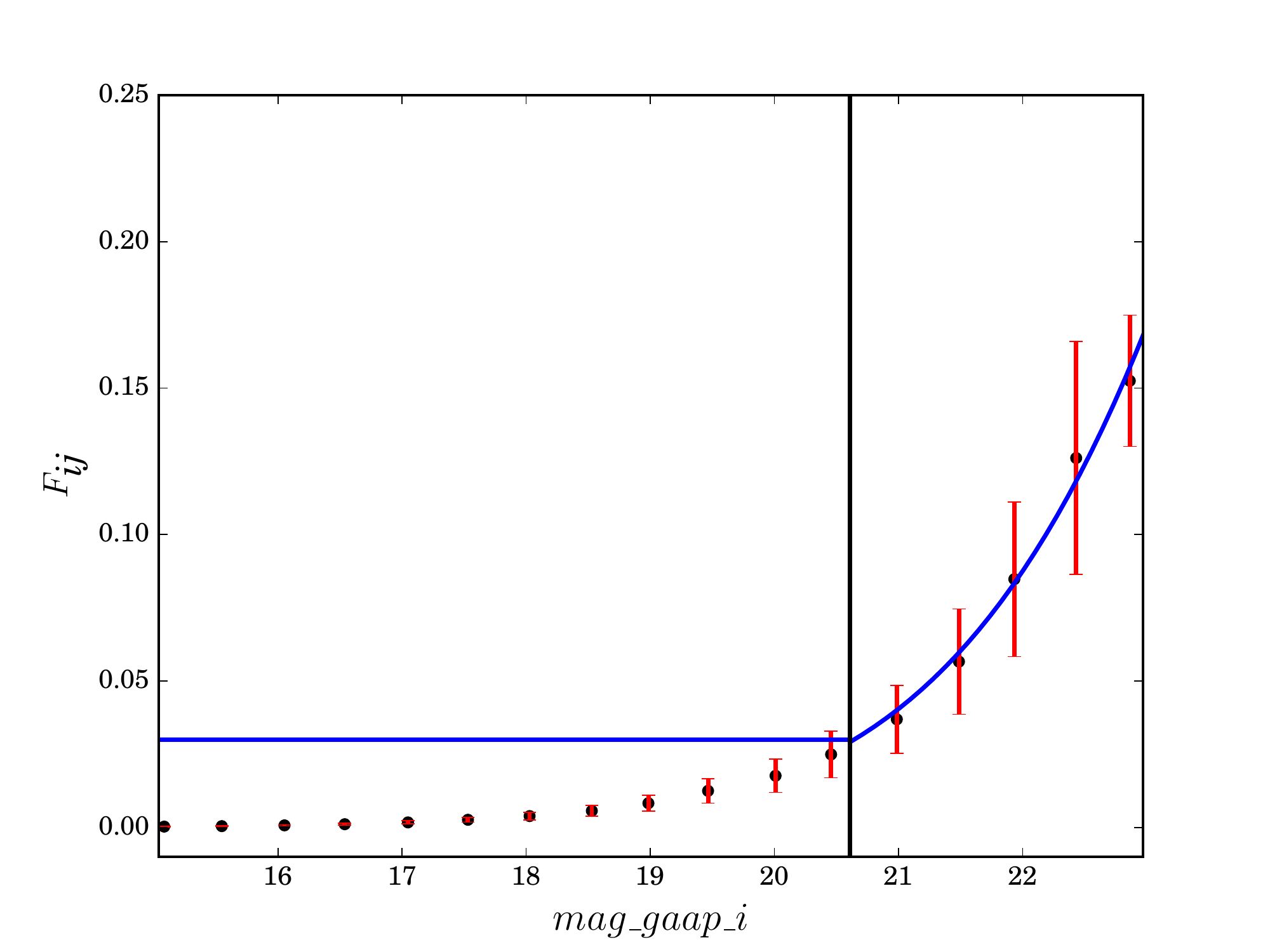}}
\caption{\textit{Mixture} perturbation function $F_{\textit{ij}}$ in  Eq.~\ref{equationPert} for the KiDS GAaP magnitudes, composed of a flat perturbation for magnitudes lower than a selected threshold (black solid lines) and a polynomial perturbation $p_i(m_{ij})$ for higher magnitude values (see Sec.~\ref{SEC:METAPHOR}). The  switching thresholds between the two functions are, respectively, $21.45$ in \textit{u} band, $22.05$ in \textit{g}, $22.08$ in \textit{r} and $20.61$ in \textit{i} band. The black points are the average of the magnitude errors for each magnitude bin. The red lines report the corresponding standard deviation.}

\label{fig:allpolynomfitting}
\end{figure*}

\indent As anticipated in the introduction, in the context of ML techniques, the determination of individual PDFs is a challenging task. This is because we would like to determine a PDF starting from several estimates of zphot,  embedding the information on the photometric uncertainties on those estimates. Therefore, we derived an analytical law to perturb the photometry by taking into account the magnitude errors provided in the catalogues. \\
\indent Indeed, the procedure followed to determine individual source PDFs consists of a single training of the MLPQNA model and the perturbation of  the photometry of the given blind test set in order to obtain an arbitrary number $N$ of test sets, each characterized by a variable photometric noise contamination. The decision to perform a single training is mainly motivated by the idea of excluding the contribution of the intrinsic error of the method itself from the PDF calculation. Appendix $A$ is dedicated to the analysis of error contributions.\\
\indent With this goal in mind, we use in this work the following \textit{perturbation} law: \begin{equation}\label{equationPert}
\tilde{m}_{\textit{ij}}= m_{\textit{ij}} +\alpha_{\textit{i}} {F_\textit{ij}} u_{(\mu=0,\sigma=1)}
\end{equation}
where \textit{j} denotes the j-th object's magnitude and \textit{i} the reference band; $\alpha_{\textit{i}}$ is a multiplicative constant, heuristically chosen by the user (generally useful to take into account cases of heterogeneous photometry, i.e. derived from different surveys and in this particular case fixed to $0.9$ for all the bands); the term $u_{(\mu=0,\sigma=1)}$ is a random value from a standard normal distribution; finally,  $F_{\textit{ij}}$ is the function used to perturb the magnitudes.\\
\indent In this work, the selected perturbation function ({$F_{\textit{ij}}$}) is the \textit{mixture}, i.e. a function composed of a constant threshold (in this case heuristically fixed to $0.03$) and a polynomial fitting of the average magnitude errors computed in several magnitude bins for each given photometric band. The role of the constant function is to act as a threshold under which the polynomial term is too low to provide a significant noise contribution to the perturbation (see \citealt{Cavuoti2017} for further details; in that paper the \textit{mixture} function was called \textit{bimodal}). This choice was made in order to take into account that there are very low  average errors for the brighter objects within the catalogues. These perturbations were applied to both GAaP and aperture magnitude types.\\
For the calculation of the individual PDFs, we submit the $N + 1$ test sets (i.e. $N$ perturbed sets plus the original one) to the trained model, thus obtaining $N + 1$ zphot estimates. Then we perform a binning in zphot, thus calculating the probability that a given zphot value belongs to each bin. We selected a binning step of $0.01$ for the described experiments and a value of N equal to $1,000$. The same binning step has been adopted by all three methods compared in this work.\\
\indent In Fig.~\ref{fig:allpolynomfitting} we can see the \textit{mixture} functions $F_{\textit{ij}}$ for the homogenized magnitudes  $mag\_gaap\_x$ (with x=\textit{u,g,r,i}).

Concerning the zphot production,  the \textit{best-estimate} zphot values are not always corresponding to the given unperturbed catalogue estimate of zphot (hereafter photo-$z_0$), as calculated by MLPQNA.  In particular it coincides with  photo-$z_0$ if this measurement falls into the interval (or \textit{bin}) representing the \textit{peak} (maximum) of the PDF; otherwise, it coincides with the zphot estimate (among the $N+1$ zphot estimates mentioned above) closest to photo-$z_0$ and falling in the bin to which corresponds the PDF \textit{peak}.

\subsection{ANNz2}\label{SEC:ANNz2}
ANNz2 \citep{Sadeh2016} is a versatile ML package\footnote{Available from \url{https://github.com/IftachSadeh/ANNZ}.}, designed primarily for deriving zphot, but appropriate also for other ML applications such as automated classification. The main ML method used by ANNz2 is based on artificial neural networks (ANNs), but it is also possible to employ boosted decision and regression trees; here we use the ANNs only. We work in the randomized regression mode of ANNz2, in which a (preferably) large number of randomly-designed ANNs (100 in our case) are trained on the input spectroscopic calibration data. This ensemble of trained ANNs is used for deriving both zphot point estimates and their PDFs. Here we provide a brief overview of the PDF generation procedure in the software version employed for this work, referring the reader to \cite{Sadeh2016} and to the online documentation of ANNz2 for more details\footnote{Here we used version $2.2.2$ of the ANNz2 software, while some significant changes in the PDF estimation have been introduced since version $2.3.0$.}.
Once the desired number of ANNs have been trained, then in the validation phase (called `optimization' in ANNz2) each source from the spectroscopic validation set\footnote{We used the ANNz2 option to randomly split the spectroscopic calibration sample into disjoint training and validation sets in proportion 1:1.} is assigned to a distribution of zphot solutions from the individual ANNs. These solutions are then ranked by their performance, and the top one is used to derive the individual zphot estimate, \texttt{Z\_BEST}, which we use in this paper as the zphot point estimation from ANNz2. In order to derive PDFs, the various ANNs are first folded with their respective single-value uncertainty estimates, derived via the k-Nearest Neighbour method \citep{Oyaizu2008}. A subset of ranked solutions is combined in different random ways to obtain a set of candidate PDFs. In order to select  the final PDF, these candidates are compared using their Cumulative Distribution Functions (CDFs), defined as the integrated PDF for redshifts smaller than the reference value of the true redshift, $z_{spec}$:
\begin{equation}
\mathcal{C}(z_{spec})= \int_{z_0}^{z_{spec}} p_{reg}(z) dz \; .
\end{equation}
The function $p_{reg}(z)$ is the differential PDF for a given redshift and $z_0$ is the lower bound of the PDF ($z_0=0$ in our case). The final PDF is chosen as the candidate for which the distribution of $\mathcal{C}$ is the closest to uniform \citep{Bordoloi2010}.\\
\indent ANNz2 may generate two types of PDFs, depending on how the $\mathcal{C}$ function is chosen. In the first case, denoted as \texttt{PDF\_0}, the CDF is based on $z_{spec}$ from the validation sample; in the second option, \texttt{PDF\_1}, the results of the best ML solution are used as reference. In this work we use the \texttt{PDF\_1} option as we found it to perform generally better than the other one.

\subsection{BPZ}\label{SEC:BPZpdf}

The BPZ method \citep{Benitez2000}, as usual for SED fitting techniques, is able to provide a PDF estimation law, based on the equation:
\begin{equation}\label{equation1}
\chi^{2}(z,T,A) = \sum_{i=1}^{N_{f}} { \left( \frac{F_{\rm obs}^{f}-A\times F_{\rm pred}^{f}(z,T)}{\sigma_{obs}^{f}} \right)^2}
\end{equation}

\noindent where $F_{\rm pred}^{f}(z,T)$ is the flux predicted for a template T at redshift z. $F_{\rm obs}^{f}$ is the observed flux, $\sigma_{obs}^{f}$ the associated error, while $A$ is a normalization factor. From Eq. \ref{equation1} it is clear that the spectroscopic information is not needed, thus implying the possibility to estimate the zphot for all sources.\\
\indent Individual PDFs are a natural by-product of every SED fitting method. In the case of BPZ for the KiDS-DR3 data, the PDFs are obtained by multiplying the probability by the used priors and then performing a summation  over all the templates, in order to obtain the full posterior probability. The theory, implemented in the BPZ code, is expressed by equations $6$ to $12$ in the paper of \cite{Benitez2000}. This method has been used to obtain BPZ KiDS-DR3 zphot and PDFs, by utilizing the priors specified in \cite{Hild2012}. \\
\indent Finally, the reference to the selected re-calibrated template set \citep{Capak2004}, as well as more details about the use of BPZ, are provided in \cite{dJ2017}.
 

\subsection{Dummy PDF}\label{SEC:thedummypdf}

In order to have a benchmark tool useful to analyze and compare the statistical validity of previous methods, we set to zero the multiplicative constant parameter $\alpha_{\textit{i}}$ of  Eq. \ref{equationPert} for all bands in order to produce a \textit{dummy} perturbation law.

The relative \textit{dummy} PDF obtained by METAPHOR is made by individual source PDFs, for which the one hundred per cent of the zphot estimates (coincident with  photo-$z_0$, i.e. the unperturbed estimate of zphot) fall in the same redshift interval (by fixing the binning step at $0.01$, as described in Sec.~\ref{SEC:METAPHOR}).\\ 
The main goal in determining the \textit{dummy} PDFs is to assess the reliability of several statistical estimators used to evaluate an ensemble of PDFs. 

\subsection{Statistical Estimators}\label{SEC:statindicators}

This section is dedicated to describing the set of statistical estimators adopted to evaluate zphot estimates and relative PDFs performance.\\ 

\indent The basic statistics are calculated on the residuals:
\begin{equation}\label{equationDelta}
\Delta z = (\zs-\zp)/(1+\zs)
\end{equation}
As the individual zphot estimates, in all the presented statistics the following quantities have been considered:  the zphot \textit{best-estimates} for METAPHOR and ANNz2 (see, respectively, Sec.~\ref{SEC:METAPHOR} and Sec.~\ref{SEC:ANNz2}); the zphot values $Z\_B$ provided in the KiDS-DR3 catalogue for BPZ \citep{dJ2017}; and the photo-$z_0$ estimates for the \textit{dummy} PDF calculated via METAPHOR (see Sec.~\ref{SEC:thedummypdf}).\\
\indent The most common estimators of the zphot accuracy, which we use here, are the standard first four central moments of the residual distribution, respectively, the mean (or bias), standard deviation $\sigma$, skewness and kurtosis, the fraction of catastrophic outliers, defined as $|\Delta z| > 0.15$, plus the normalized median absolute deviation (NMAD), defined as:
\begin{equation}\label{equationNMAD}
NMAD = 1.4826 \times median (|\Delta z - median (\Delta z)|)
\end{equation}

The cumulative performance of the stacked PDF on the entire sample is evaluated by means of the following three estimators:
\begin{itemize}
\item $f_{0.05}$: the percentage of residuals $\Delta$z within $\pm 0.05$;
\item $f_{0.15}$: the percentage of residuals $\Delta$z within $\pm 0.15$;
\item $\Braket{\Delta z}$: the average of all the residuals $\Delta$z of the stacked PDFs.
\end{itemize}

Here, by stacked PDFs we mean the individual zphot PDFs transformed into the PDFs of scaled residuals $\Delta z$ defined in equation~\ref{equationDelta}, and then stacked for the entire sample.\\
\indent Furthermore, the quality of the individual PDFs is evaluated against the single corresponding zspec from the test set, by defining five categories of occurrences:
\begin{itemize}
\item \textit{zspecClass} = 0: the zspec is within the \textit{bin} (see Sec.~\ref{SEC:METAPHOR}) containing the peak of the PDF;
\item \textit{zspecClass} = 1: the zspec falls in one bin from the peak of the PDF;
\item \textit{zspecClass} = 2: the zspec falls into the PDF, e.g. in a bin in which the PDF is different from zero;
\item \textit{zspecClass} = 3: the zspec falls in the first bin outside the limits of the PDF;
\item \textit{zspecClass} = 4: the zspec falls out of the first bin outside the limits of the PDF.
\end{itemize}

By definition, the \textit{zspecClass} term depends on the chosen bin amplitude  (see Sec.~\ref{SEC:METAPHOR}), which also determines the accuracy level of PDFs. The quality evaluation of the entire PDF can be hence measured in terms of fractions of occurrences of these five categories within the test data set. In particular, these quantities should be regarded as complementary statistical information, useful to complete the PDF reliability analysis. For example, classes 3 and 4 could quantify the amount of objects falling outside the PDF. The distinction between the two classes gives the supplementary information about how far from the PDFs is their zspec, thus contributing to evaluate their  reliability.

Finally, we use two additional diagnostics to analyze the \textit{cumulative} performance of the PDFs:  the credibility analysis presented in \cite{Witt2016} and the Probability Integral Transform (hereafter PIT), described in \cite{Gneiting2007}.

The credibility test should assess if PDFs have the correct \textit{width} or, in other words, it is a test of the \textit{confidence} of any method used to calculate the PDFs. In particular, the method is considered overconfident if the produced PDFs are too narrow, i.e. too sharply peaked; underconfident otherwise. In order to measure the credibility, rather than the Confidence Intervals (hereafter CI), the Highest Probability Density Confidence Intervals (hereafter HPDCI) are used  \citep{Witt2016}.

The implementation of the credibility method is very straightforward, and it involves the computation of the threshold credibility $c_{i}$ for the \textit{i}th galaxy with 
\begin{equation}\label{eq:eqWitt}
c_{i} = \sum_{z\in p_{i} \geq p_{i} (z_{spec,i}) } { p_{i}(z)}
\end{equation}
where $p_{i}$ is the normalized PDF for the \textit{i}-th galaxy. \\
\indent The credibility is then tested by calculating the cumulative distribution \textit{F(c)}, which should be equal to \textit{c}. \textit{F(c)} resembles a q-q plot, (a typical quantile-quantile plot used for comparing two distributions), in which \textit{F} is expected to match \textit{c}, i.e. it follows the bisector in the \textit{F} and \textit{c} ranges equal to [0,1]. Therefore, the \textit{overconfidence} corresponds to \textit{F(c)} falling below the bisector, otherwise the \textit{underconfidence} occurs. In both cases this method indicates the inaccuracy of the error budget \citep{Witt2016}.\\
\indent The PIT histogram measures the predictive capability of a forecast, which is generally probabilistic for continuous or mixed discrete-continuous random variables \citep{Gneiting2007} and that has been already used to assess the reliability of PDFs in the case of photometric redshifts (see for instance \citealt{disanto}). We can define the PIT as the histogram of the various $p_{i}$:
\begin{equation}\label{eq:PIT}
p_{i}=F_{i}(x_{i})
\end{equation}
where in our case $F_{i}$ is the CDF of the i-th object and $x_{i}=z_{spec_{i}}$.
Ideal forecasts produce continuous $F_{i}$ and PIT with a uniform distribution on the interval $(0,1)$. In other words, we can check the forecast by investigating the uniformity of the PIT: the closer the histogram to the uniform distribution, the better the calibration, i.e. the statistical consistency between the predictive distributions and the validating observations \citep{Baran2016}.
Nevertheless, it is possible to show that the uniformity of a PIT is a necessary but not sufficient condition for having an ideal forecast \citep{Gneiting2007}.\\
A strongly U-shaped PIT histogram indicates a highly \textit{underdispersive} character of the predictive distribution \citep{Baran2016}.

\section{Comparison among methods}\label{SEC:thecomparison}

A preliminary comparison among the three methods METAPHOR, ANNz2 and BPZ, only in terms of zphot prediction performance, has been already given in \cite{dJ2017}. That comparison was based on statistics applied to the residuals defined by the Eq.~\ref{equationDelta}, reported in Table 8 and Fig. 11 of \cite{dJ2017}. In that figure the upper panel shows the plots of zphot vs GAMA-DR2 spectroscopy, while in the bottom panel residuals vs \textit{r}-magnitude are shown for the three methods.\\
\indent More recently, in \cite{bilicki2017} a comparison among the three methods has also been presented on KiDS-DR3 data, more in terms of zphot estimation quality at the full spectroscopic depth available, confirming the better behavior of ML methods at bright end of KiDS data sample ($z<0.5$), as well as comparable quality of ML methods and BPZ at higher redshift ($z\sim1$).

\subsection{Statistics on zphot and stacked PDFs}

The statistical comparison among the three methods on the dataset obtained by cross-matching KiDS-DR3 and GAMA data (see Sec.~\ref{SEC:data}), is summarized in Table~\ref{tab:photozstat}. It shows a better performance in terms of bias and fraction of outliers for METAPHOR, while BPZ and ANNz2 obtain, respectively, a lower $\sigma$ and $NMAD$ of the errors. 

\begin{table}
\centering
\caption{Statistics of zphot estimation obtained with MLPQNA (zphot estimation engine of METAPHOR), ANNz2, BPZ, on the GAMA DR2: respectively, the bias, the standard deviation, the Normalized Median Absolute Deviation, the fraction of outliers outside the 0.15 range, kurtosis and skewness.}\label{tab:photozstat}
 \begin{tabular}{|c|c|c|c}
 {\bf Estimator}		& {\bf MLPQNA}    & {\bf ANNz2}        & {\bf BPZ} \\ \hline
 $bias$  	& $-0.004$   & $-0.008$     & $-0.020$ \\
 $\sigma$		& $0.065$    & $0.078$      & $0.048$ \\
 $NMAD$         & $0.023$   & $0.019$      & $0.028$ \\
 $outliers$     & $0.98\%$  & $1.60\%$     & $1.13\%$ \\
$Kurtosis$ & $774.1$ & $356.0$  & $52.2$  \\                                   
$Skewness$ & {$-21.8$}  & $-15.9$  & $-2.9$  \\ \hline
 \end{tabular}
\end{table}

\begin{figure*}
\centering
 {\includegraphics[width=0.49 \textwidth]{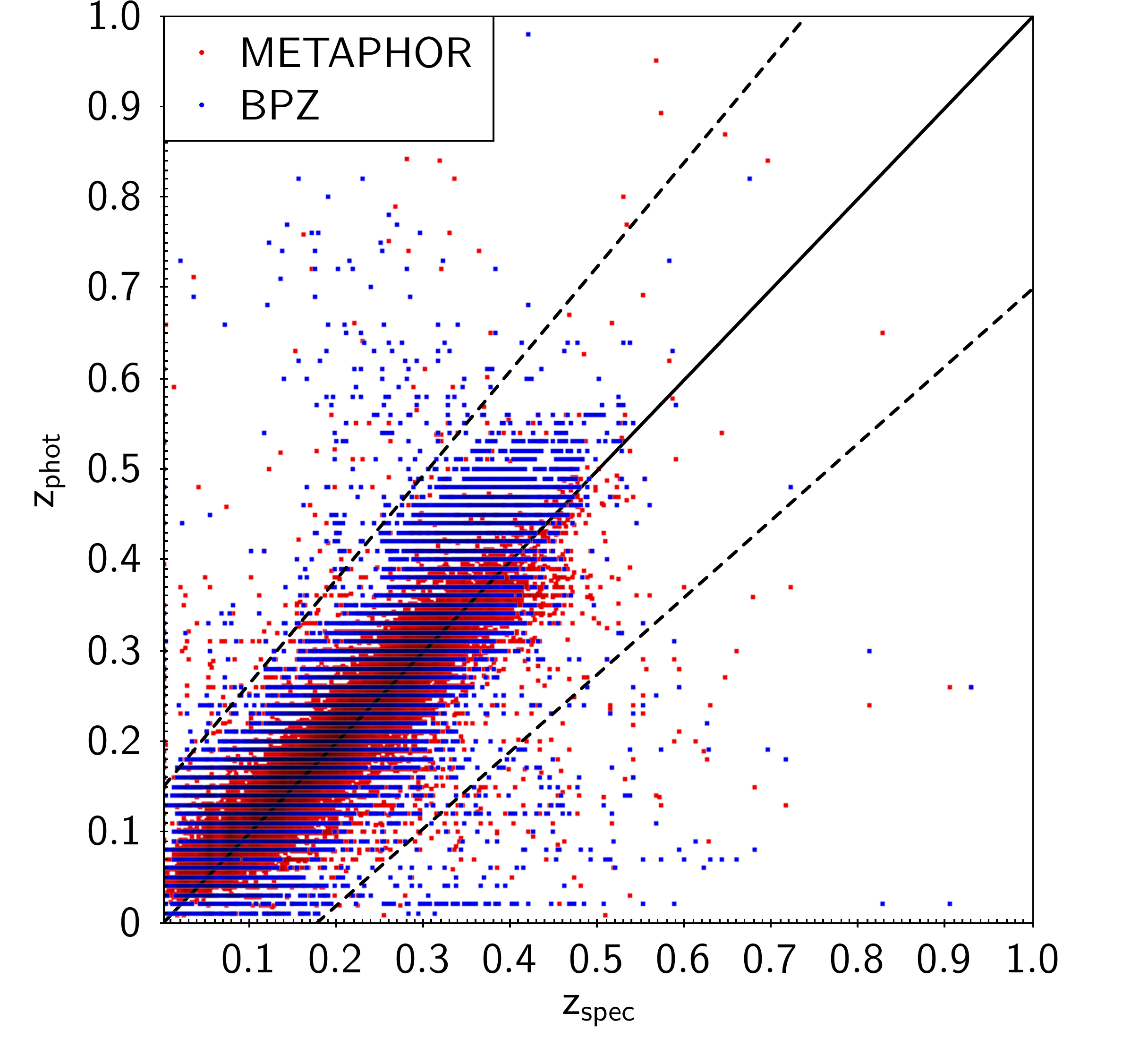}}
 {\includegraphics[width=0.49 \textwidth]{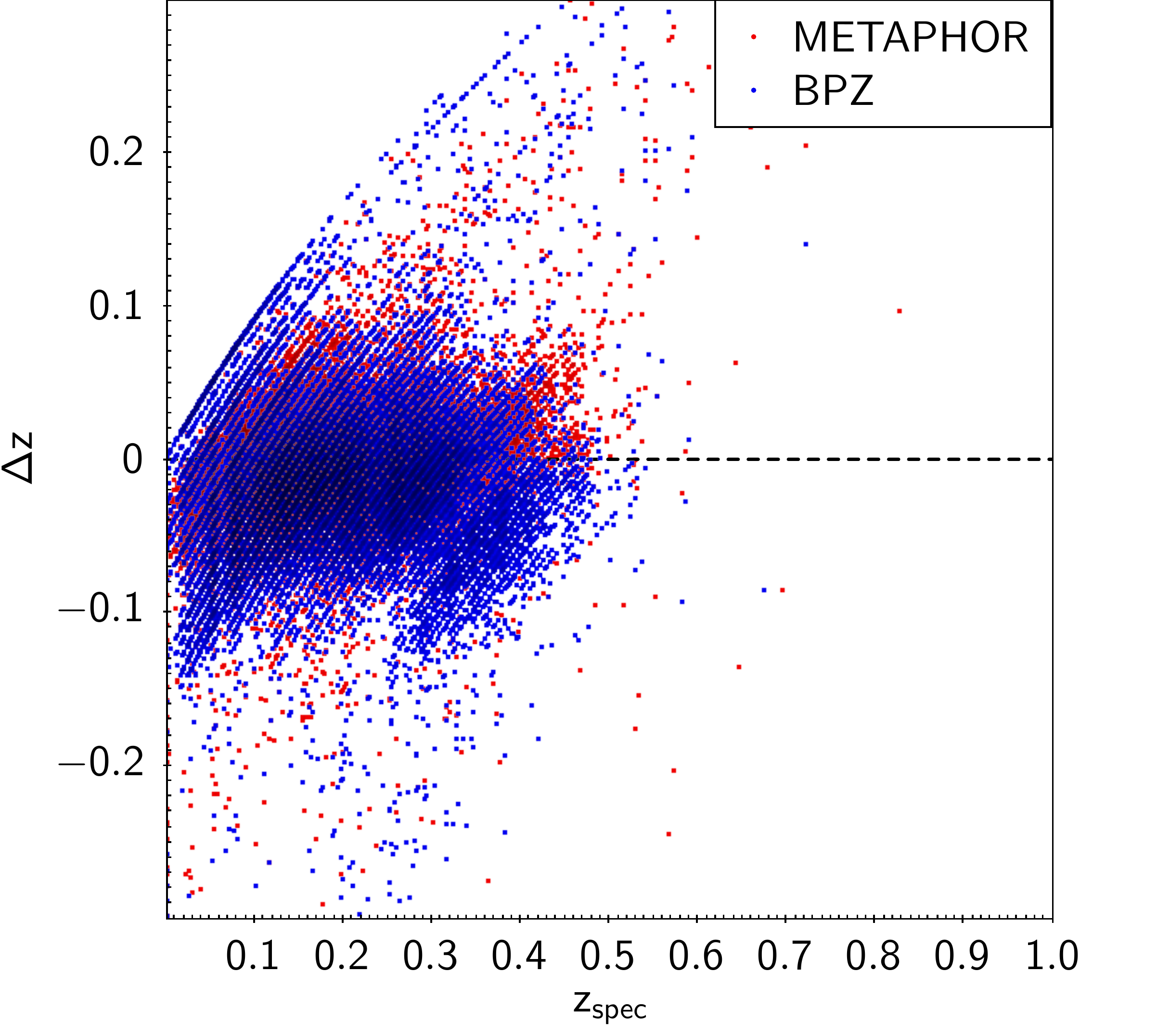}}
 {\includegraphics[width=0.49 \textwidth]{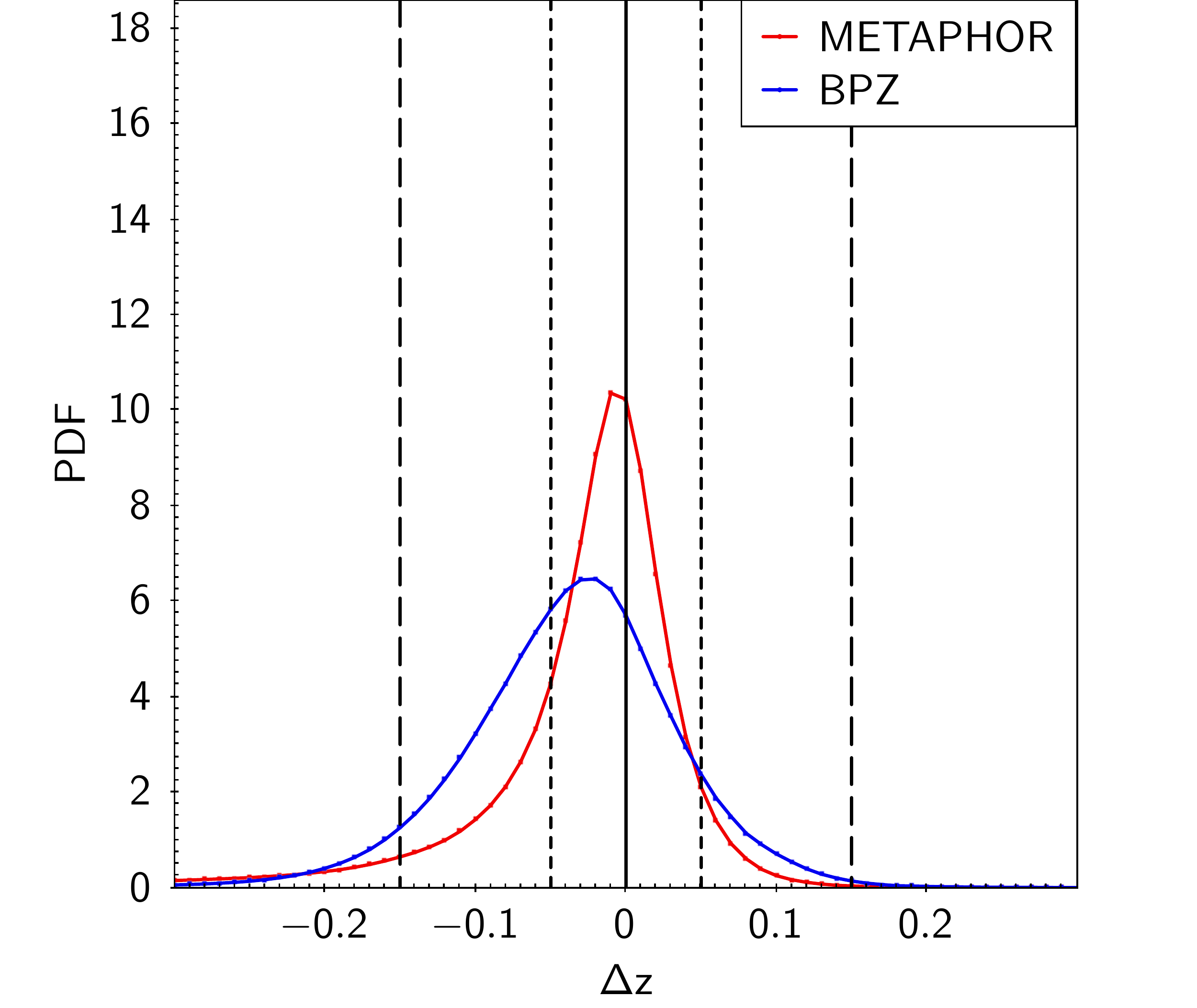}}
\caption{Comparison between METAPHOR (red) and BPZ (blue). Upper row: scatter plot of photometric redshifts as function of spectroscopic redshifts (left-hand panel) and scatter plot of the residuals as function of the spectroscopic redshifts (right-hand panel). Lower row: \textit{stacked} representation of the residuals of PDFs (with redshift bin equal to 0.01).}
\label{fig:BPZ_vs_METAPHOR}
\end{figure*}

\begin{figure*}
 \centering
  {\includegraphics[width=0.49 \textwidth]{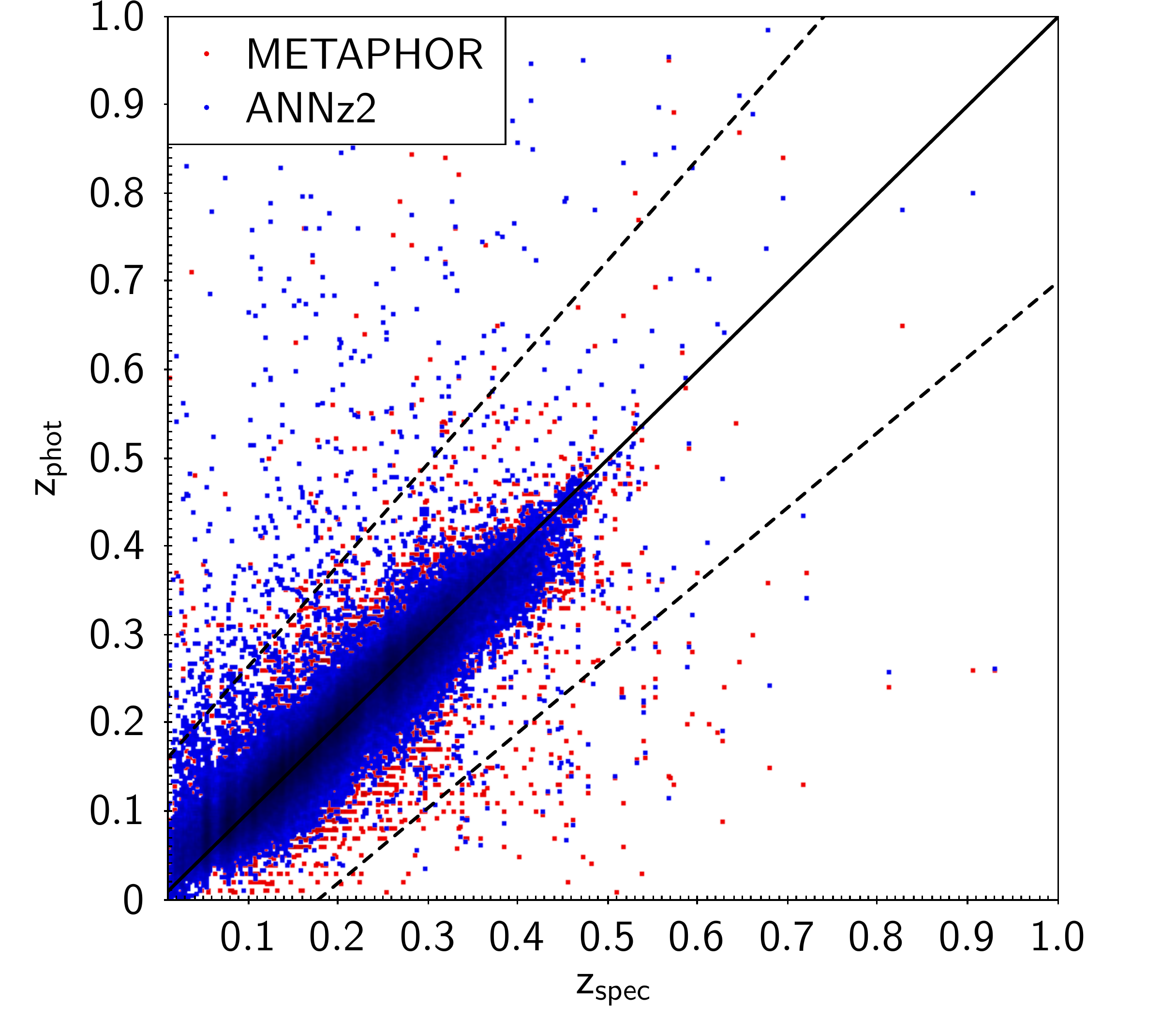}}
  {\includegraphics[width=0.49 \textwidth]{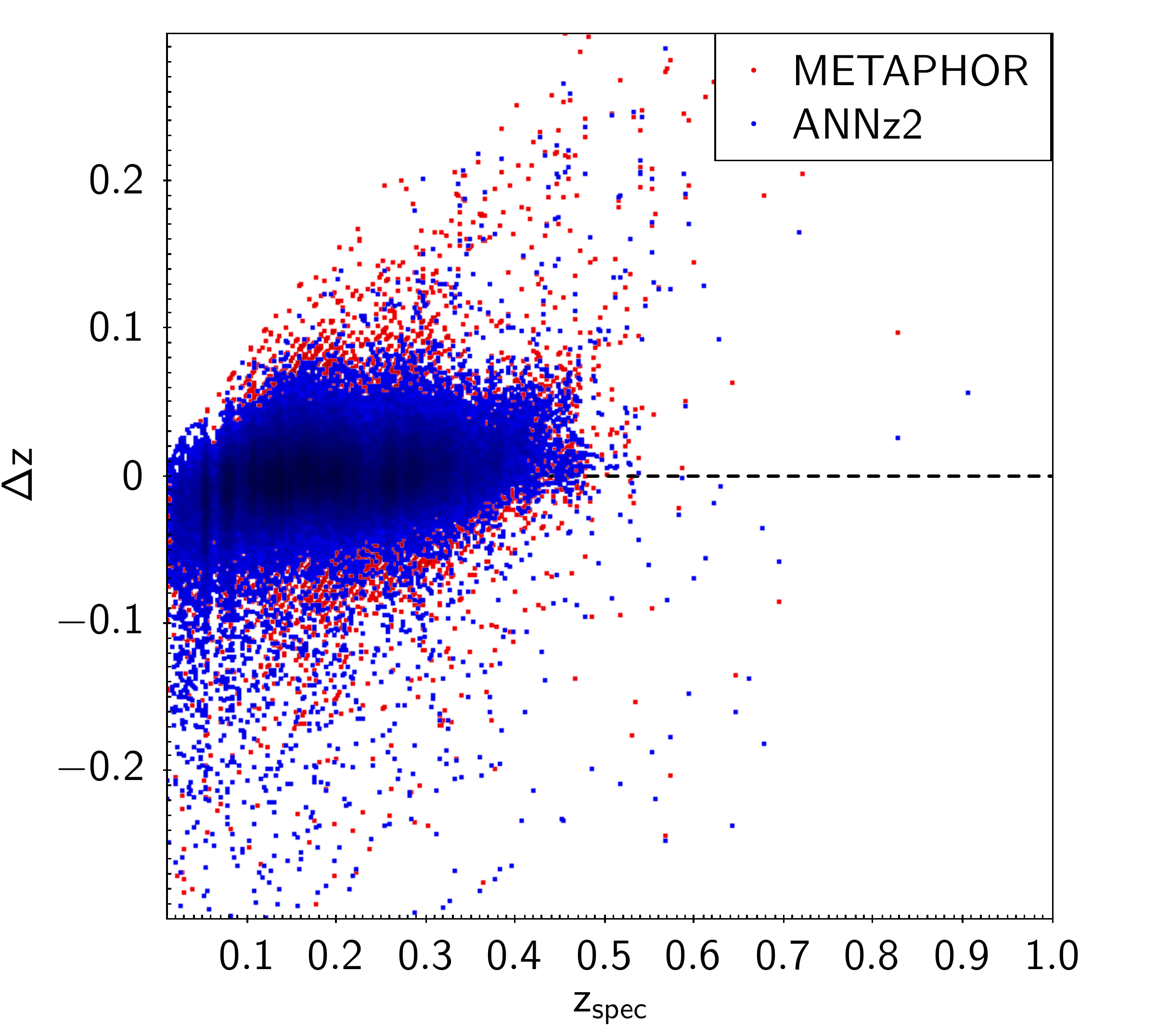}}
  {\includegraphics[width=0.49 \textwidth]{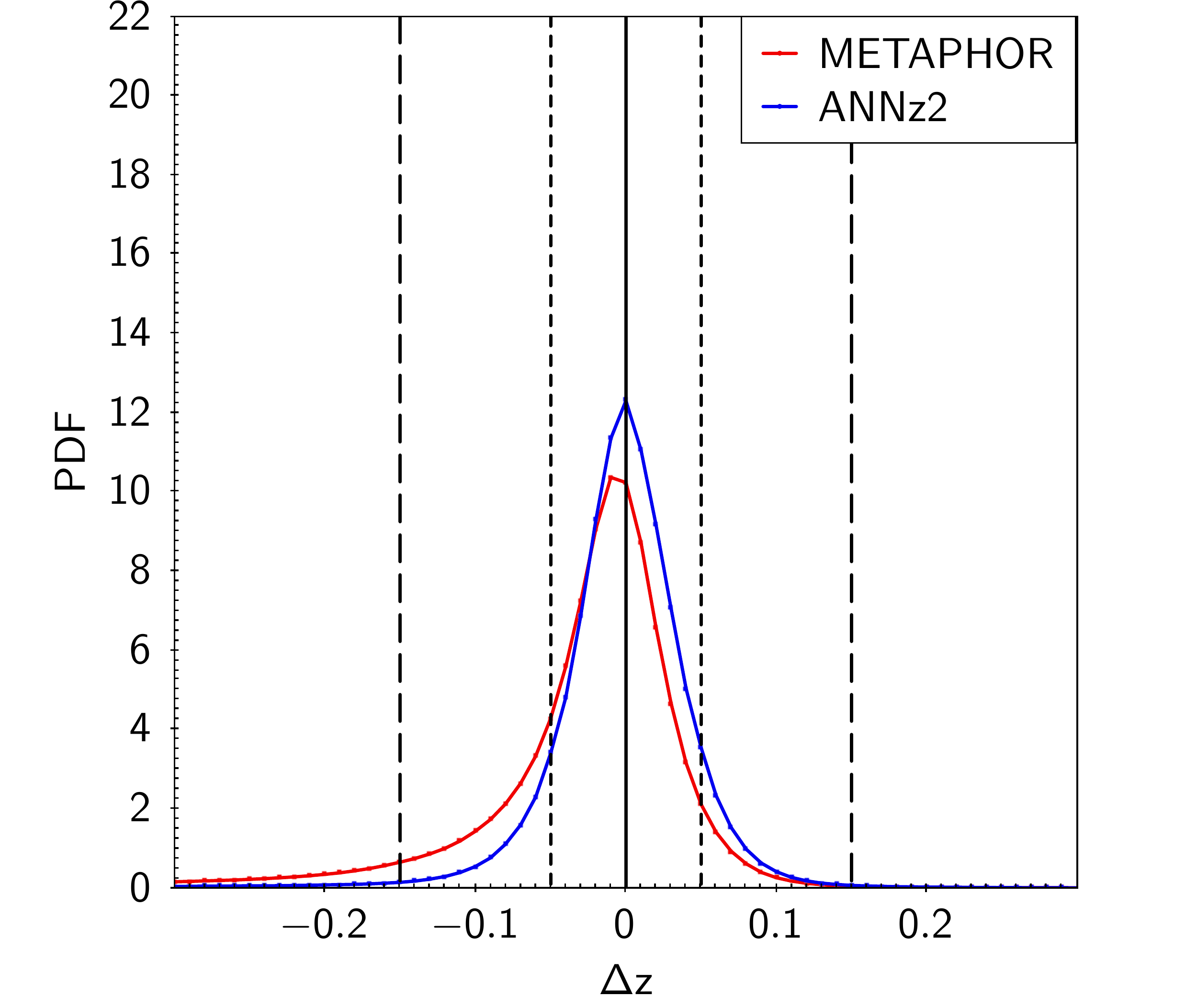}}
\caption{Comparison between METAPHOR (red) and ANNz2 (blue). Upper row: scatter plot of photometric redshifts as function of spectroscopic redshifts (left-hand panel) and scatter plot of the residuals as function of the spectroscopic redshifts (right-hand panel). Lower row: \textit{stacked} representation of the residuals of PDFs (with redshift bin equal to 0.01).}
\label{fig:METAP_vs_ANNz2}
\end{figure*}

In Figures \ref{fig:BPZ_vs_METAPHOR} and \ref{fig:METAP_vs_ANNz2} we show the comparison on the GAMA field between METAPHOR and respectively, BPZ and ANNz2, in terms of  graphical distributions of predicted zphot and stacked PDFs of the residuals.\\
\indent From Fig.~\ref{fig:BPZ_vs_METAPHOR} it is apparent that the correlation between zphot and zspec is tighter for METAPHOR than for BPZ. In terms of stacked PDF, the distributions are in agreement with statistics of Table~\ref{tab:stackedstat}, since the BPZ PDF is more enclosed within the $\pm 0.15$ residual range.

\begin{figure*}
 \centering
  {\includegraphics[width=0.49 \textwidth]{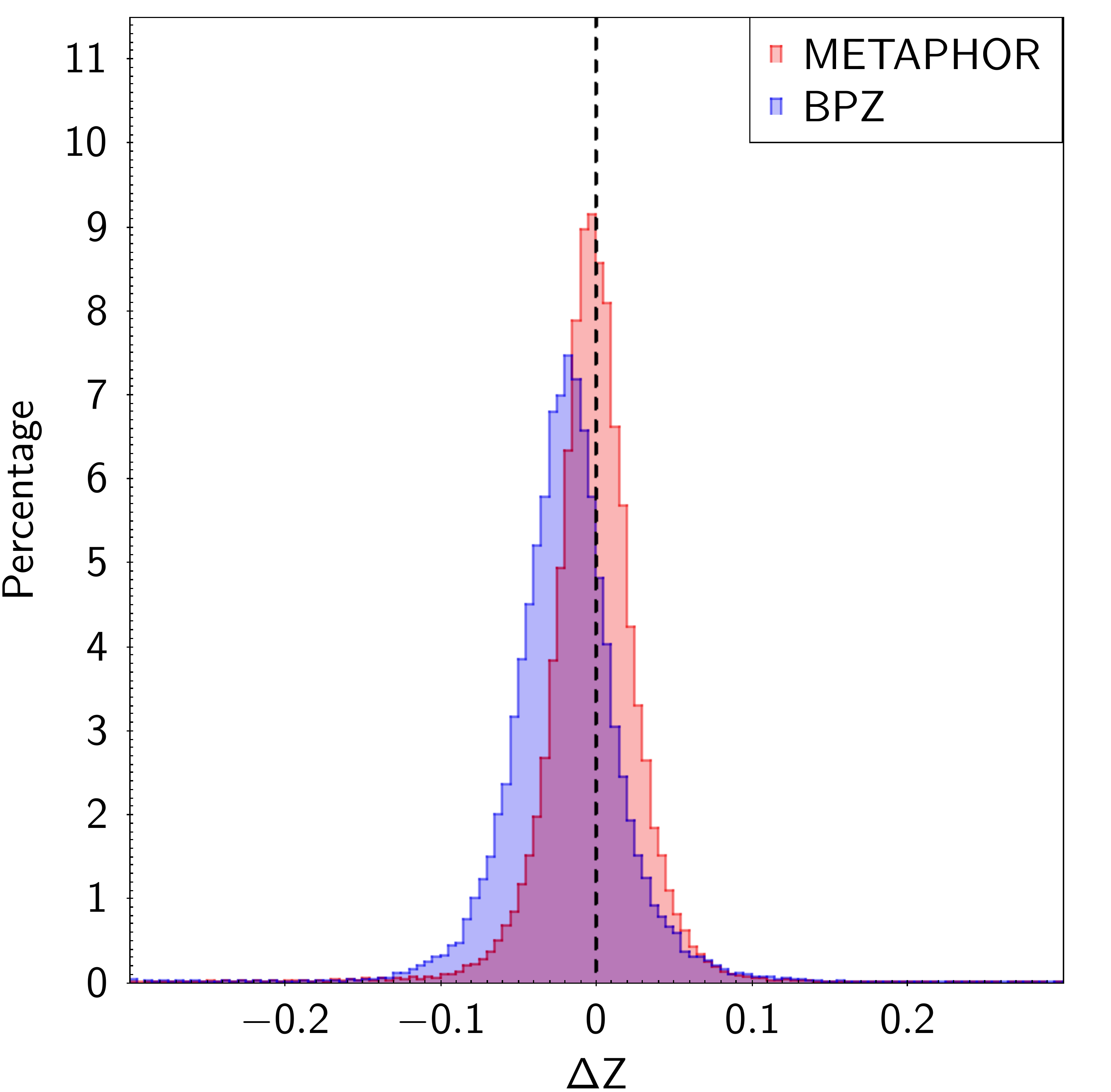}}
  {\includegraphics[width=0.49 \textwidth]{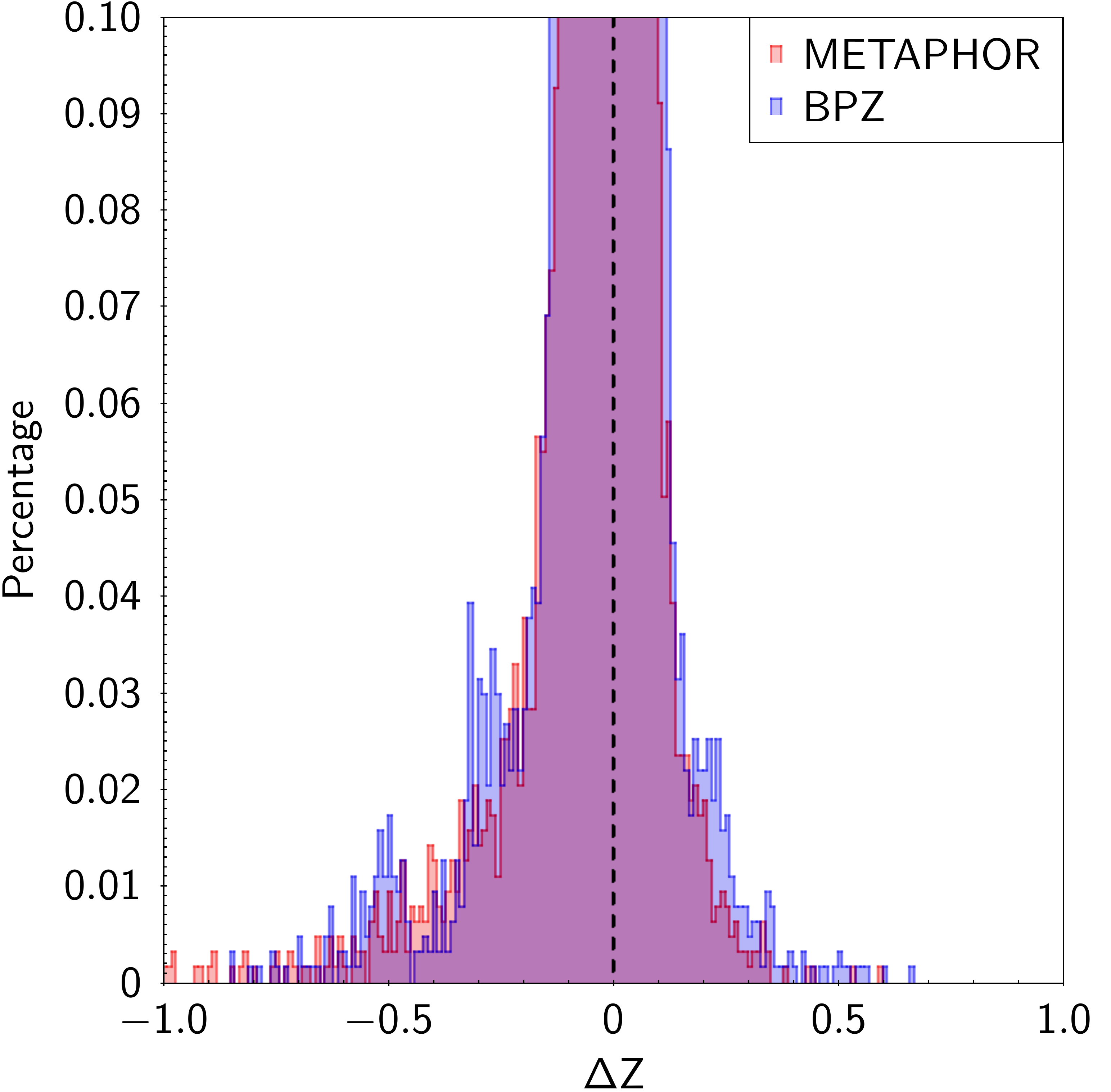}}
  {\includegraphics[width=0.49 \textwidth]{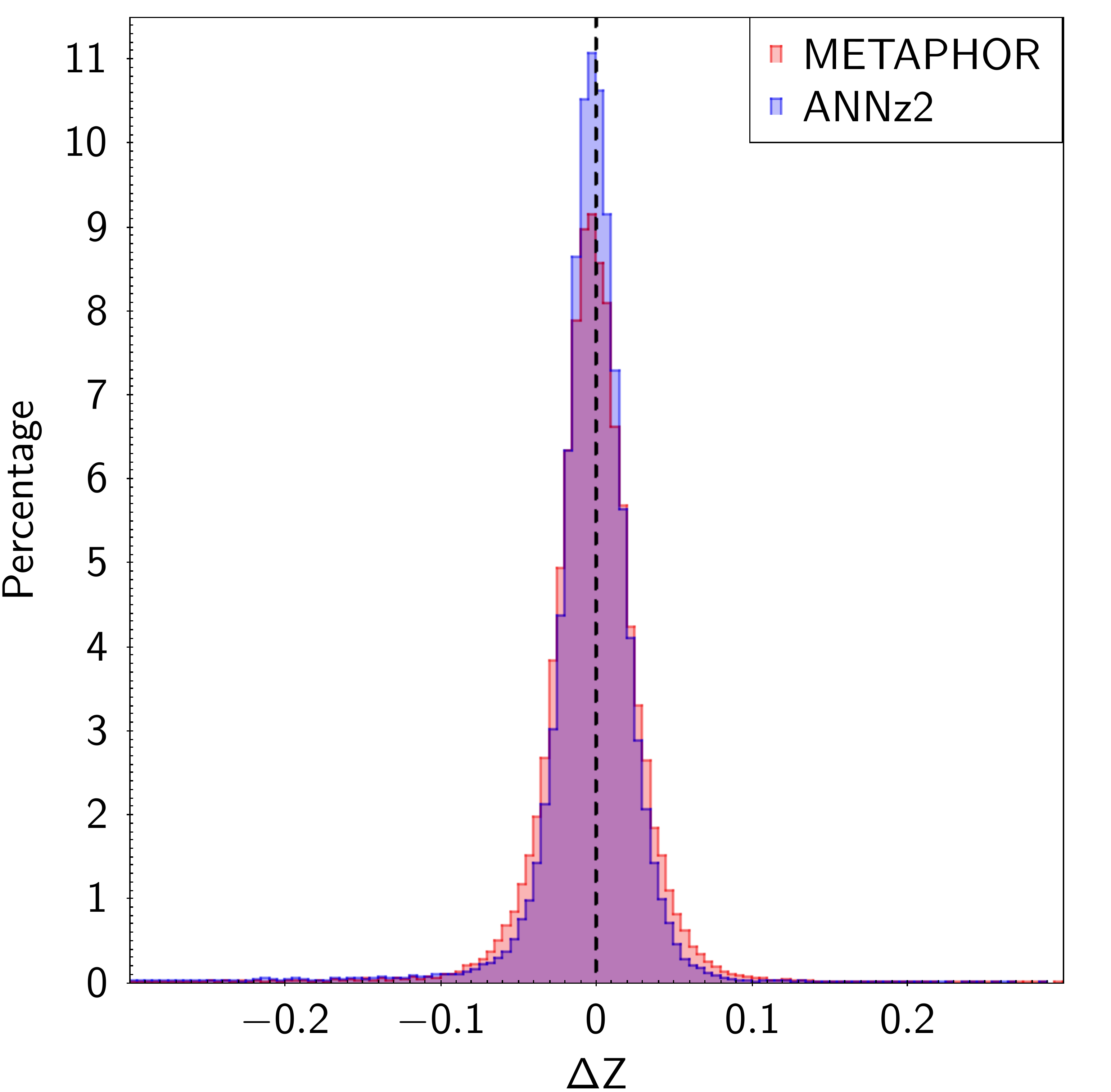}}
  {\includegraphics[width=0.49 \textwidth]{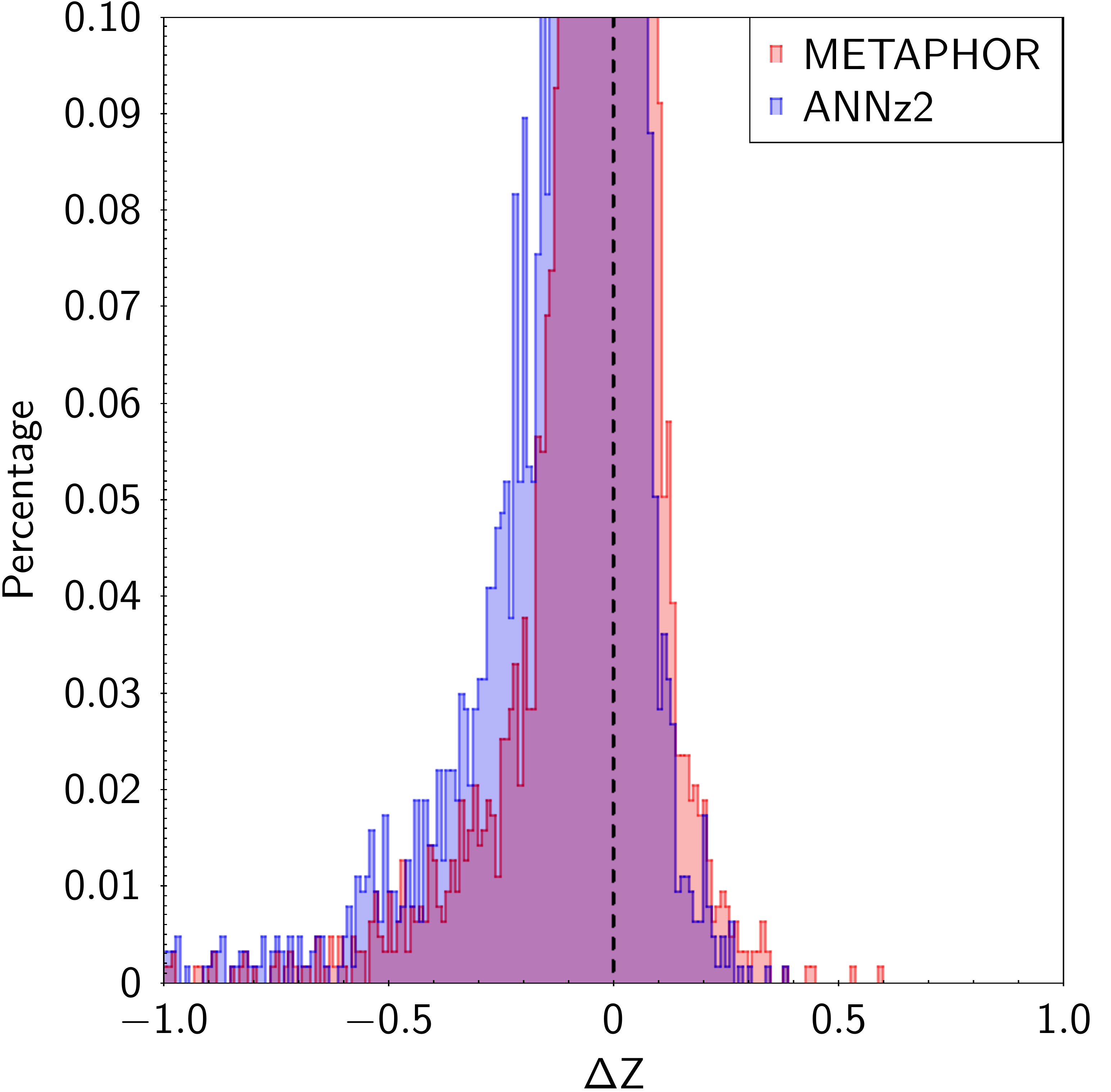}}
\caption{Top panels: comparison between METAPHOR (red) and BPZ (blue); bottom panels: comparison between METAPHOR (red) and ANNz2 (blue). Left-hand panels show the histograms of residual distributions, zoomed in the right-hand panels in order to make more visible the skewness effect. The values are expressed in percentage, after normalizing the distributions to the total number of objects of the blind test set (see Sec.~\ref{SEC:data}).}
\label{fig:residualhisto}
\end{figure*}

Fig.~\ref{fig:METAP_vs_ANNz2} shows a tighter photo-spectro redshift correlation for ANNz2 as well as a better symmetry of the stacked PDF. \\
\indent The effects of kurtosis and skewness are evident from Fig.~\ref{fig:residualhisto}. The kurtosis is a measure of the shape of the residual distribution, particularly suitable for characterizing its tails. From Fig.~\ref{fig:residualhisto} and Table~\ref{tab:photozstat}, all three methods show a leptokurtic behavior. This means that the distributions asymptotically approach zero faster than the Gaussian distribution, therefore indicating a small amount of outliers with respect to the Gaussian limit at $2\sigma$ ($\sim 0.2\%$ for METAPHOR, $\sim 0.0005\%$ for BPZ and $\sim 1.5\%$ in the case of ANNz2). This also implies that in this case the standard deviation could be considered a poor estimator for the zphot prediction performance. \\
\indent The skewness is a measure of the symmetry around zero of the $\Delta z$ distribution. All the three compared methods show a negative value (see Table~\ref{tab:photozstat}), mostly due to a longer tail towards negative than to positive $\Delta z$. This is more pronounced in the case of METAPHOR and ANNz2 (right panels of Fig.~\ref{fig:residualhisto}), but a negative skewness is expected in zphot residual distributions, because of an inherent tendency to overestimate the redshift. By calculating the residuals through eq.~\ref{equationDelta}, all methods naturally tend towards negative $zspec-zphot$ in the low redshift regime, because negative photometric redshifts are removed (meaningless), introducing the above negative bias in $zspec-zphot$. \\
\indent In Table~\ref{tab:stackedstat} we report the fraction of residuals in the two ranges $[-0.05, 0.05]$ and $[-0.15, 0.15]$  and the average of residuals for all the probed methods. Last column shows such statistics also for the \textit{dummy} PDF. Table~\ref{tab:zspecclass} summarizes the distribution of fractions of samples among the five categories of individual PDFs, obtained by the evaluation of their spectroscopic redshift position with respect to the PDF. 

\begin{table}
\centering
\caption{Statistics of the zphot error stacked PDFs for METAPHOR, ANNz2, BPZ and  \textit{dummy} obtained by METAPHOR, for the sources cross-matched between KiDS-DR3 photometry and GAMA spectroscopy.}
 \begin{tabular}{|c|c|c|c|c}
 {\bf Estimator}		  & {\bf METAPHOR}    & {\bf ANNz2}           & {\bf BPZ}   & {\bf dummy}\\ \hline
 $f_{0.05}$ & $65.6\%$\phantom{AA}	& $76.9\%$  & $46.9\%$  & $93.1\%$\phantom{A}\\
 $f_{0.15}$ & $91.0\%$\phantom{AA}	& $97.7\%$  & $92.6\%$  & $99.0\%$\phantom{A}\\
 $\Braket{\Delta z}$ & $-0.057$\phantom{AA}	& $0.009$  & $-0.038$     & $-0.006$\phantom{A}\\ \hline
 \end{tabular}
\label{tab:stackedstat}
\end{table}
\begin{table}
\centering
\caption{\textit{zspecClass} fractions for METAPHOR, ANNz2 and BPZ on the GAMA field.} \label{tab:zspecclass}
 \begin{tabular}{|c|rrrrrr}
 {\bf zspecClass}  & \multicolumn{2}{c}{\bf METAPHOR}  & \multicolumn{2}{c}{\bf ANNz2}   & \multicolumn{2}{c}{\bf BPZ} \\ \hline
 0           & $9042$ & $(14.2\%)$           & $12426$& $(19.4\%)$         & $4889$ &$(7.7\%)$ \\
 1           & $16758$ &$(26.3\%)$           & $19040$ &$(29.9\%)$         & $9650$ &$(15.1\%)$ \\
 2           & $37233$ & $(58.4\%)$          & $31927$ &$(50.1\%)$         & $49170$& $(77.15\%)$\\
 3           & $200$ &$(0.3\%)$              & $8$ &$(0.01\%)$ & $0$       & $(0\%)$ \\
 4           & $516$& $(0.8\%)$              & $324$& $(0.5\%) $           & $31$ &$(0.05\%)$\\ \hline
\end{tabular}
\end{table}

\begin{figure*}
 \centering
  {\includegraphics[width=0.49 \textwidth]{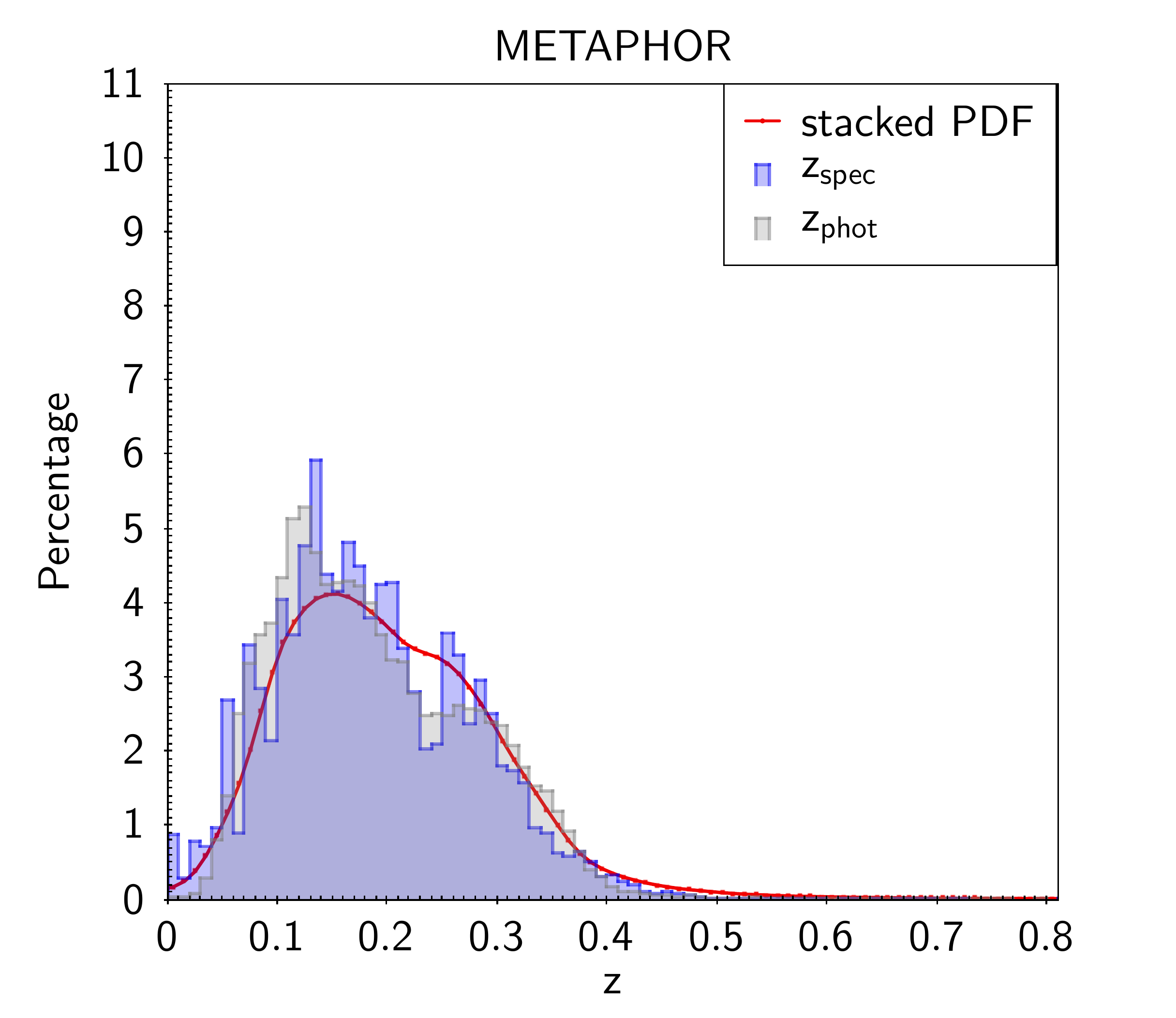}}
  {\includegraphics[width=0.49 \textwidth]{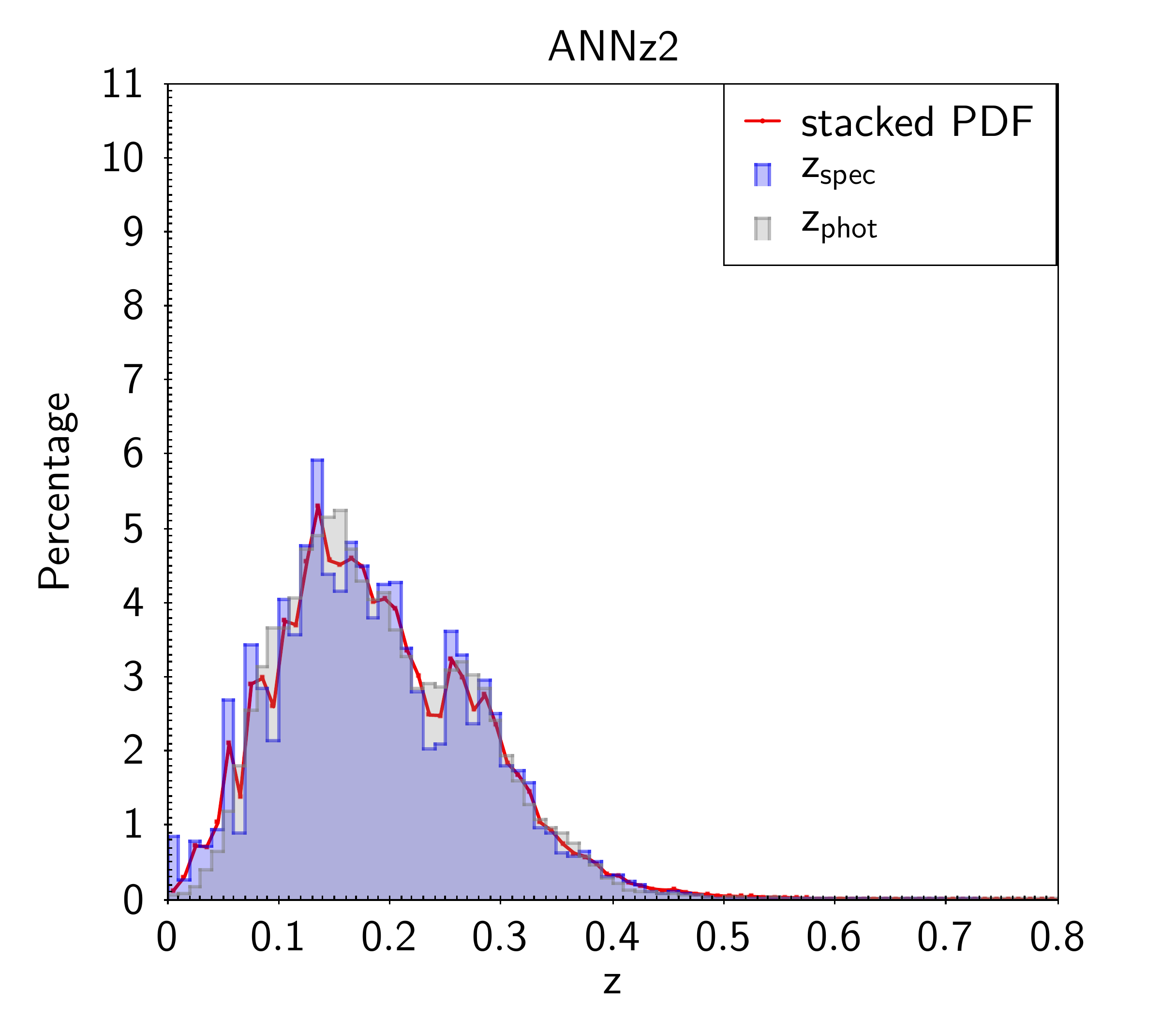}}
  {\includegraphics[width=0.49 \textwidth]{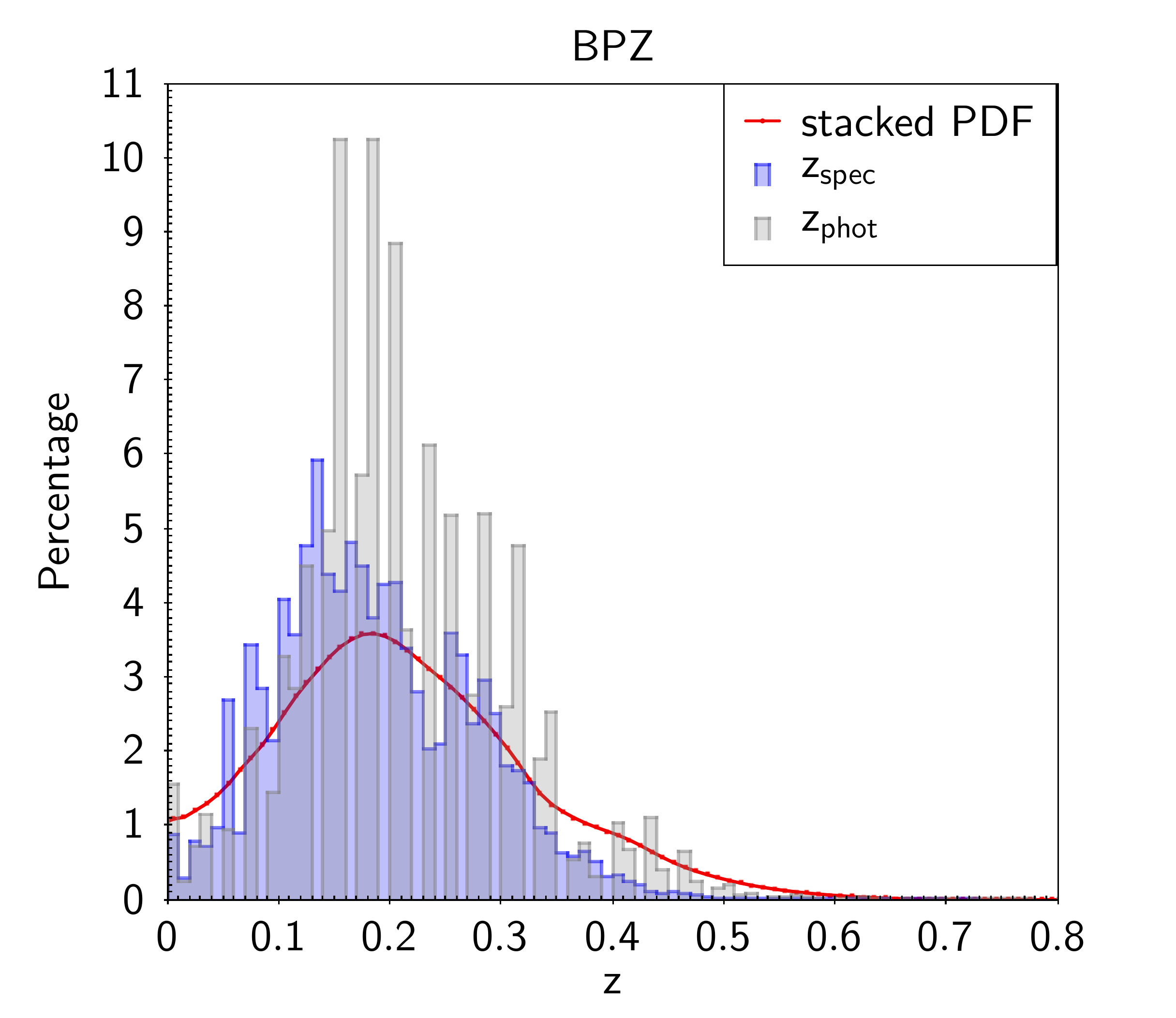}}
  {\includegraphics[width=0.49 \textwidth]{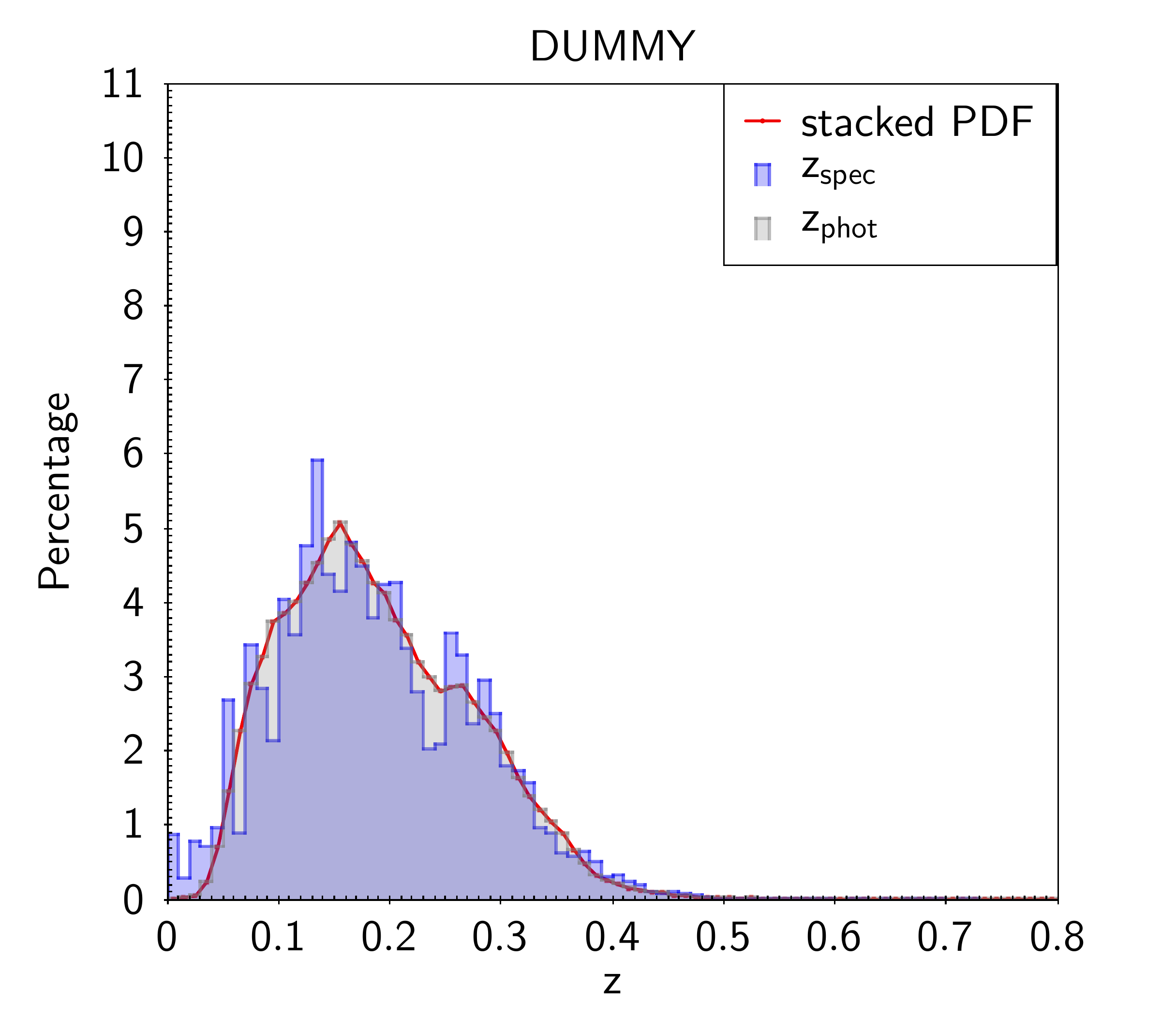}}
\caption{Superposition of the stacked PDF (red) and estimated zphot (gray) distributions obtained by METAPHOR, ANNz2, BPZ and for the \textit{dummy} (in this last case the zphot distribution corresponds to that of the photo-$z_{0}$ estimates, Sec.~\ref{SEC:thedummypdf}) to the z-spec distribution (in blue) of the GAMA field.}
\label{fig:stackedALL}
\end{figure*}

From Table~\ref{tab:stackedstat} it appears evident that in terms of PDFs, ANNz2 performs quantitatively better than the other two methods, while the \textit{dummy} PDF, derived from METAPHOR, obtains the best estimates. This demonstrates that the statistical estimators adopted for the stacked PDF show low robustness in terms of quality assessment of zphot errors and that there is a need for a deeper understanding of the real meaning of a PDF in the context of zphot quality estimation as well as a careful investigation of the statistical evaluation criteria.\\
\indent The former statement about ANNz2 performance is also supported by Table~\ref{tab:zspecclass}, where ANNz2 shows a percentage of  49.4$\%$ of samples falling within one bin from the PDF peak (the sum of fractions for \textit{zspecClass} 0 and 1) against, respectively, the 40.5$\%$ and 22.8$\%$, of the other two methods.\\ 
However, for all the \textit{stacked} PDF estimators, the \textit{dummy} PDF obtains better statistical results than all other methods.
By construction, the \textit{dummy} PDFs are non-zero only at a single value; therefore it is not worth to report its statistics regarding the \textit{zspecClass} estimator (see Sec.~\ref{SEC:statindicators}), since, as expected, most of the spectroscopic redshifts fall outside the PDF. Furthermore, the \textit{zspecClass} estimator for the \textit{dummy} PDF is equal to $0$ and $4$, i.e. the  zspec falls either in the bin to which corresponds the PDF peak or outside the PDF. The \textit{dummy} PDF method is then particularly suitable to verify that the residual fractions reported in Table~\ref{tab:stackedstat} are not sufficient to quantify the performance of a PDF.
In Fig.~\ref{fig:stackedALL}, we superimpose the stacked distribution of PDFs, derived by the three methods plus the \textit{dummy} PDF, on the photometric and spectroscopic redshift distributions.  
The stacked trend of the \textit{dummy} PDF method reproduces the photometric distribution, since it does not take into account the redshift error contribution arising from the photometric uncertainties introduced through the perturbation law in Eq.~\ref{equationPert}.  
Very close to the spectroscopic redshift distribution is the stacked PDF of \textit{dummy} and ANNz2, while BPZ and METAPHOR, although still able to follow the spectroscopic distribution, differ from the first two methods. Nevertheless, METAPHOR and ANNz2 PDFs show a better agreement with the individual photometric redshift distributions.

\subsection{Credibility analysis and PIT}

\begin{figure*}
 \centering
  {\includegraphics[width=0.49 \textwidth]{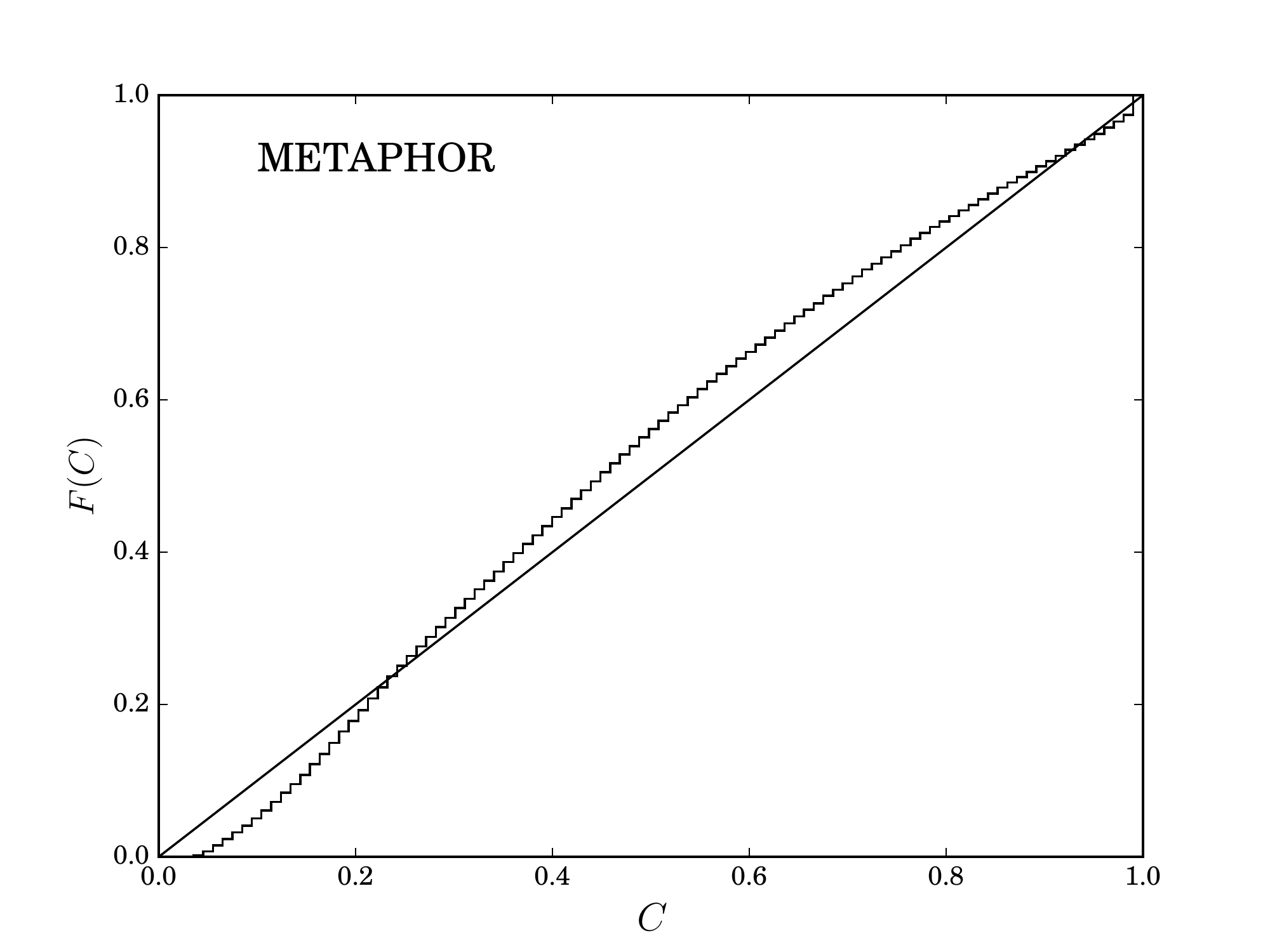}}
  {\includegraphics[width=0.49 \textwidth]{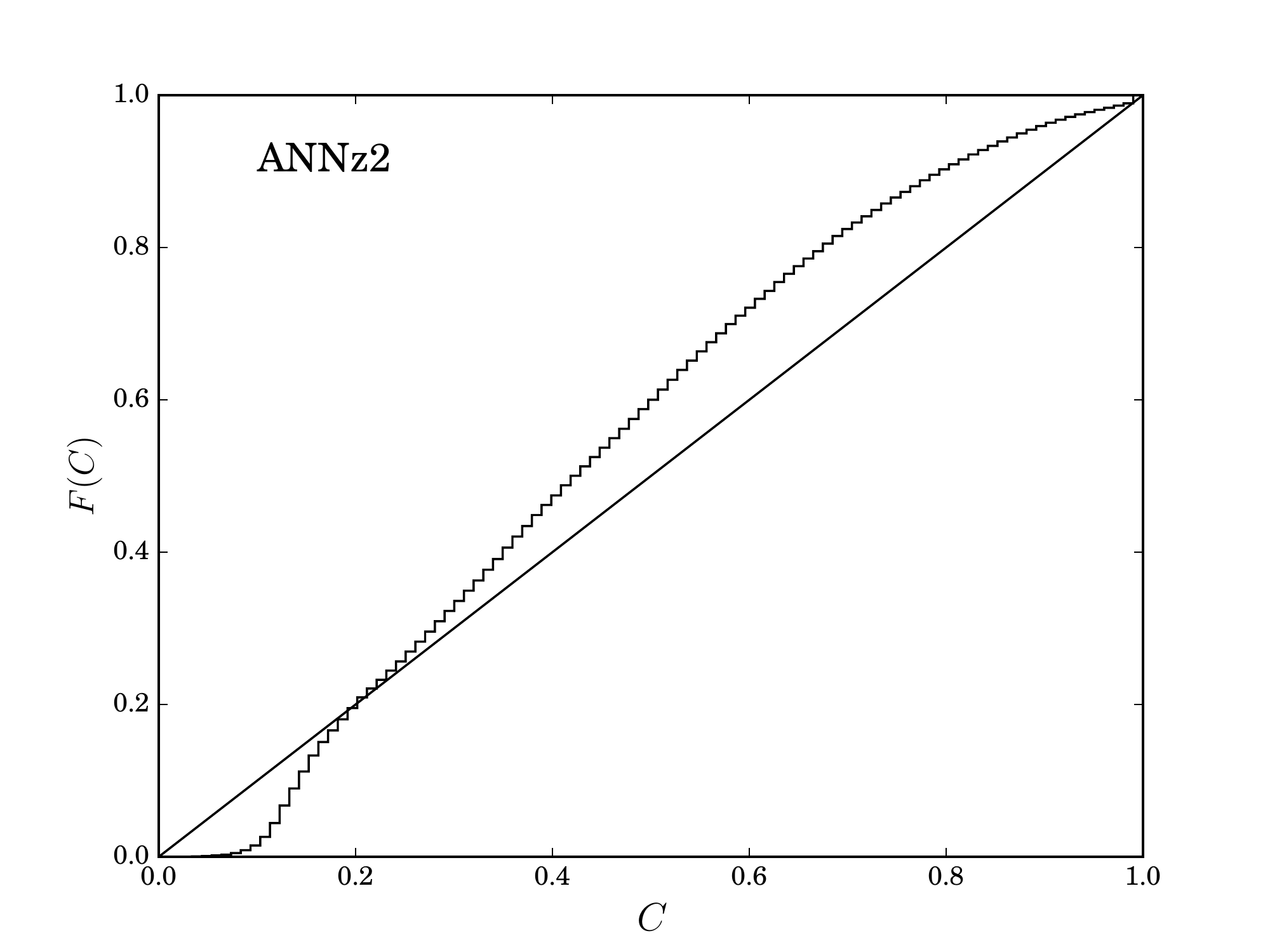}}
  {\includegraphics[width=0.49 \textwidth]{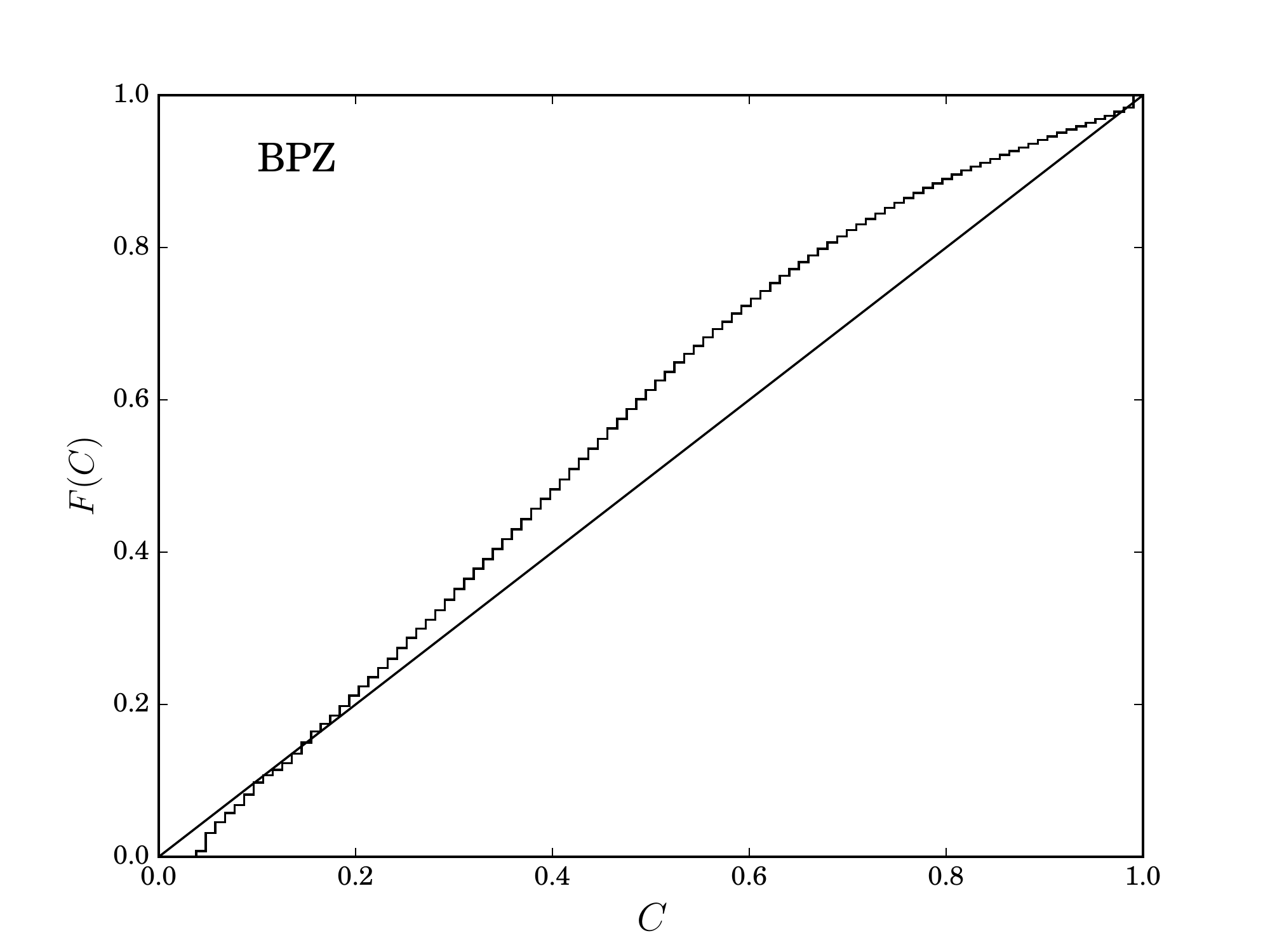}}
  {\includegraphics[width=0.49 \textwidth]{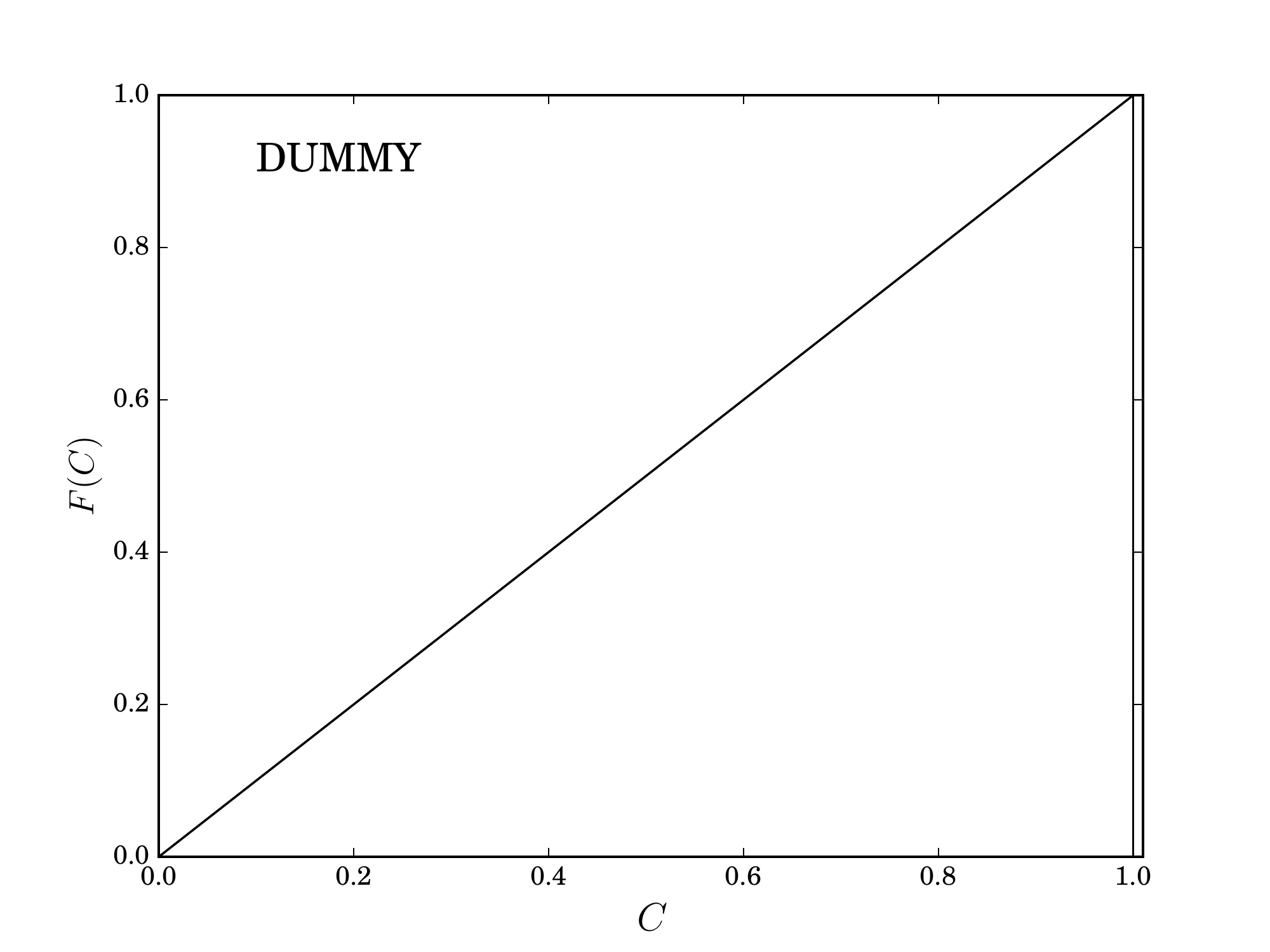}}
\caption{ Credibility analysis (see Sec.~\ref{SEC:statindicators}) obtained for METAPHOR, ANNz2, BPZ and the \textit{dummy} PDF.} 
\label{fig:WittmanALL}
\end{figure*}

\begin{figure*}
 \centering
  {\includegraphics[width=0.49 \textwidth]{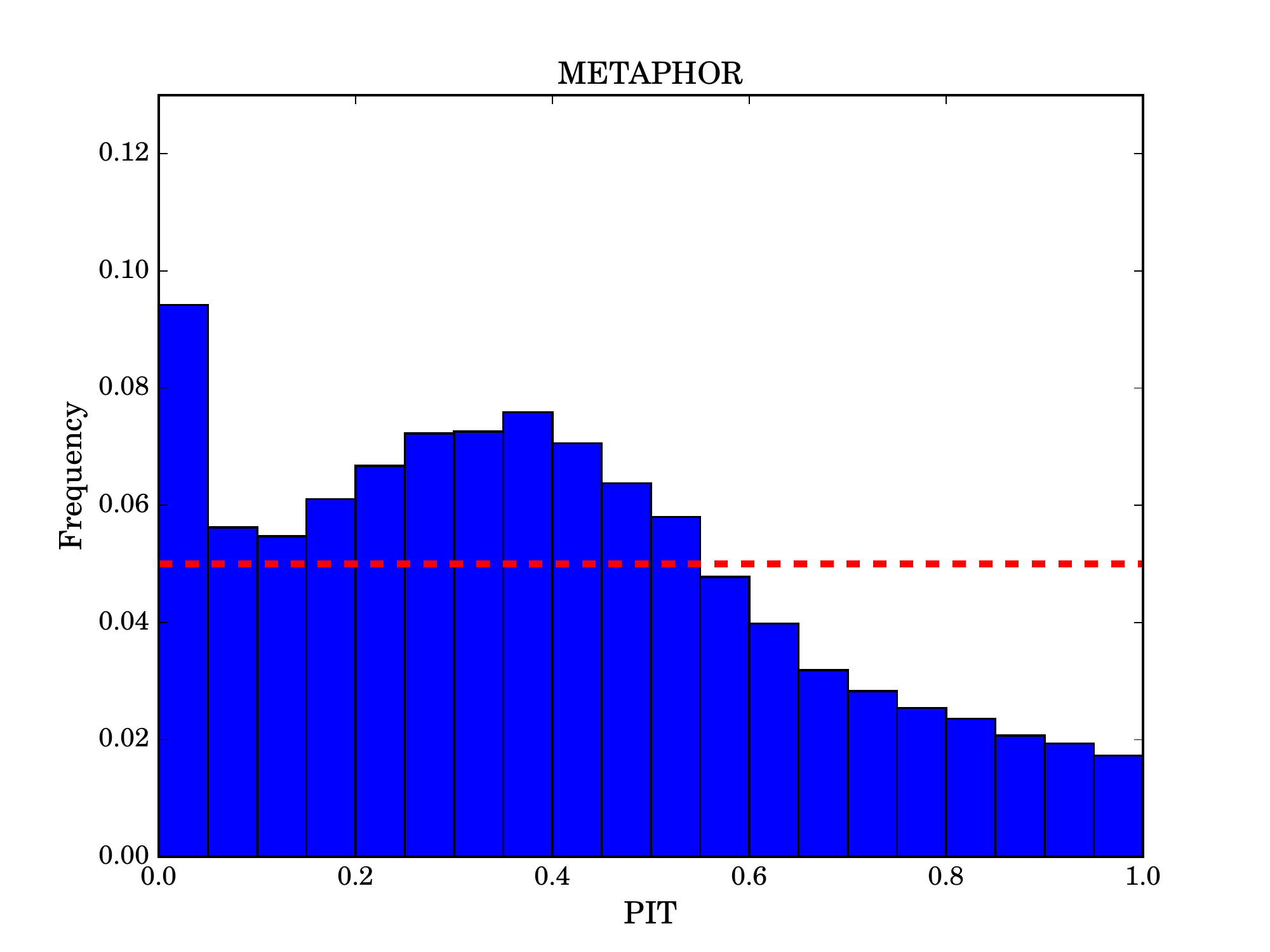}}
  {\includegraphics[width=0.49 \textwidth]{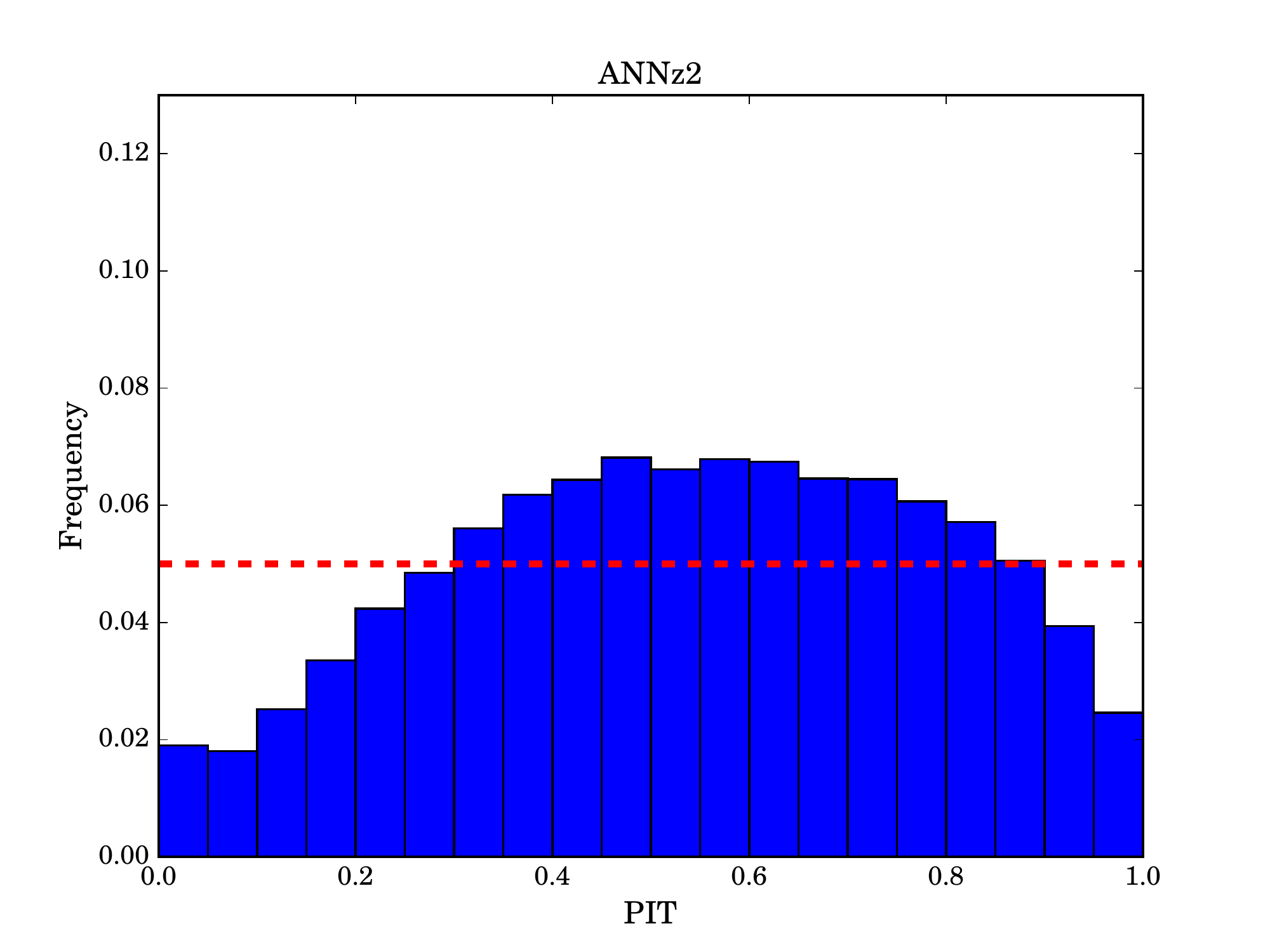}}
  {\includegraphics[width=0.49 \textwidth]{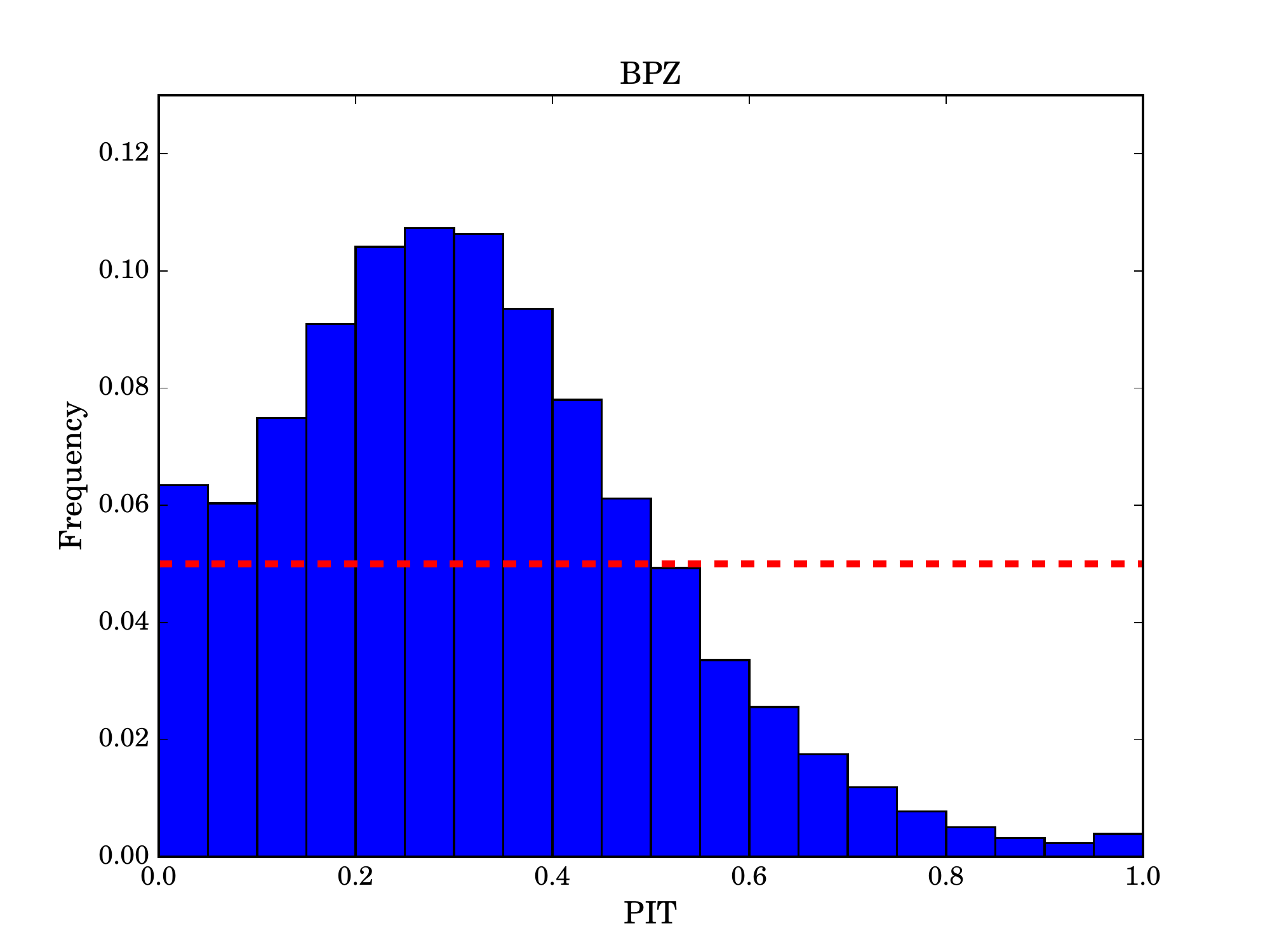}}
  {\includegraphics[width=0.49 \textwidth]{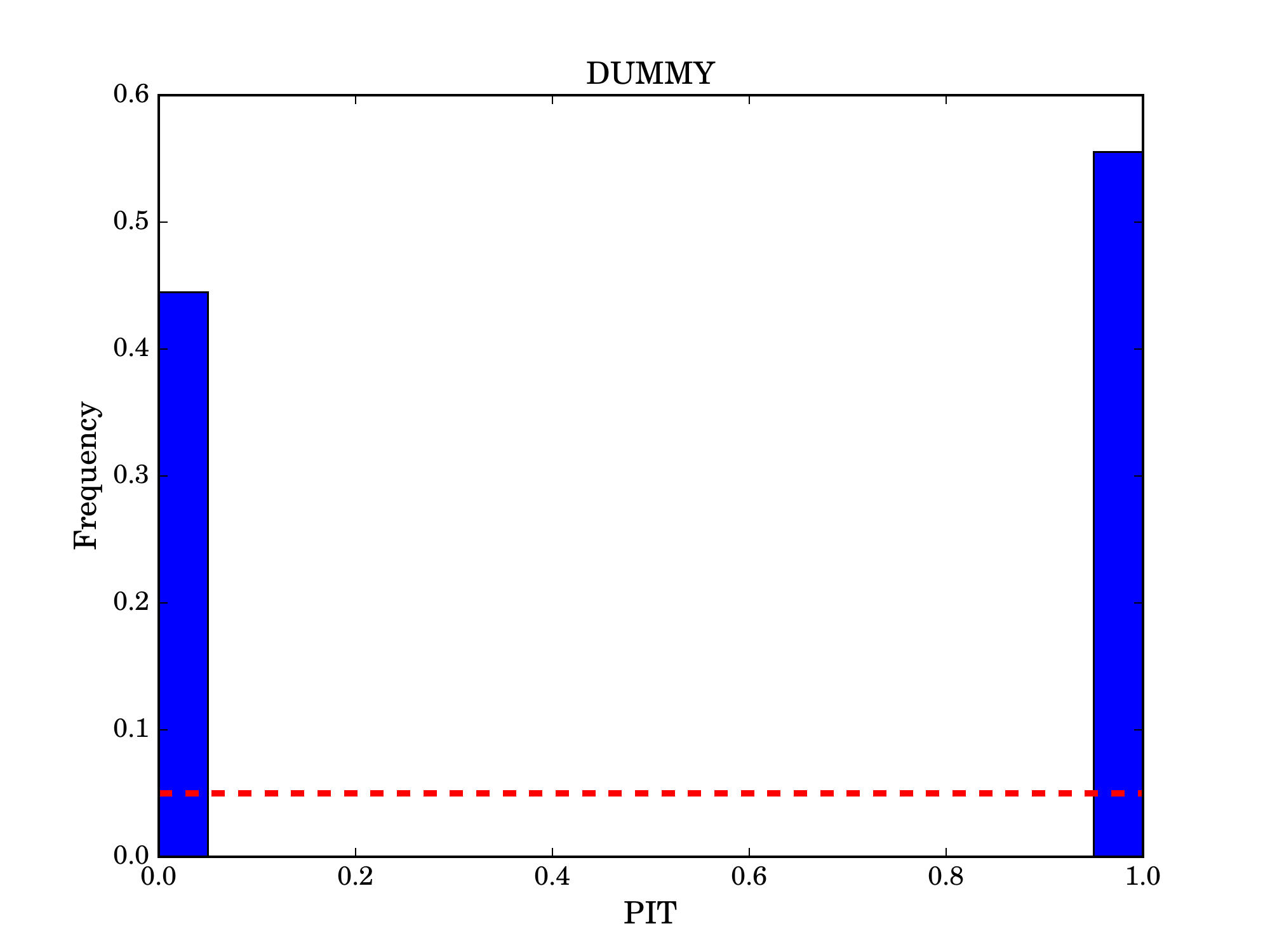}}
\caption{ Probability Integral Transform (PIT) obtained for METAPHOR (top left panel), ANNz2 (top right panel), BPZ (bottom left panel), and for the \textit{dummy} PDF (bottom right panel).}
\label{fig:pitALL}
\end{figure*}

We also show in a graphical form the two estimators introduced in Sec.~\ref{SEC:statindicators}, namely the credibility analysis on the cumulative PDFs and the PIT. Figures~\ref{fig:WittmanALL}  and \ref{fig:pitALL} show these two respective diagrams for the three methods and the \textit{dummy} PDF. 
The credibility analysis trend of METAPHOR (top left panel of Fig.~\ref{fig:WittmanALL}) reveals a higher degree of credibility with respect to ANNz2 and BPZ (respectively, top right and bottom left panels of Fig.~\ref{fig:WittmanALL}), the latter being characterized by a higher \textit{underconfidence}. However, the credibility diagram of the \textit{dummy} PDF (bottom right panel of Fig.~\ref{fig:WittmanALL})  is identically unitary for each galaxy of the data set. This is evidence of the inability to evaluate the credibility of a zphot error PDF in an objective way. 
In other words, according to the construction of the HPDCI for the credibility analysis (see Sec.~\ref{SEC:statindicators}), the \textit{dummy} PDF method shows that the 100$\%$ of the photo-$z_0$'s fall in the 100$\%$ of the HPDCI, thus the predictions are entirely \textit{overconfident}.\\
\indent The statistical evaluation of the three methods and the \textit{dummy} PDF based on the PIT diagram is shown in Fig.~\ref{fig:pitALL}. We observe a better behavior of ANNz2 (top right panel) than the two other methods, METAPHOR (top left panel) and BPZ (bottom left panel). For ANNz2, the \textit{overdispersive} and \textit{underdispersive} trends appear less pronounced than for the other cases, especially BPZ. However, the PIT histogram for \textit{dummy} PDFs shows an entirely degraded (i.e.  \textit{underdispersive}) behavior of the zphot distribution (bottom right panel of Fig.~\ref{fig:pitALL}). This result was expected, since by definition its CDF is a step function, thus allowing only values $0$ or $1$, corresponding to the two bars in Fig.~\ref{fig:pitALL}. This is in some contradiction to the previous statistics, shown for the quantitative estimators for the \textit{dummy} PDFs (Tables \ref{tab:stackedstat} and \ref{tab:TOMOG}), which were indicating the best behavior for the dummy stacked PDF.

\begin{figure}
 \centering
   {\includegraphics[width=0.49 \textwidth]{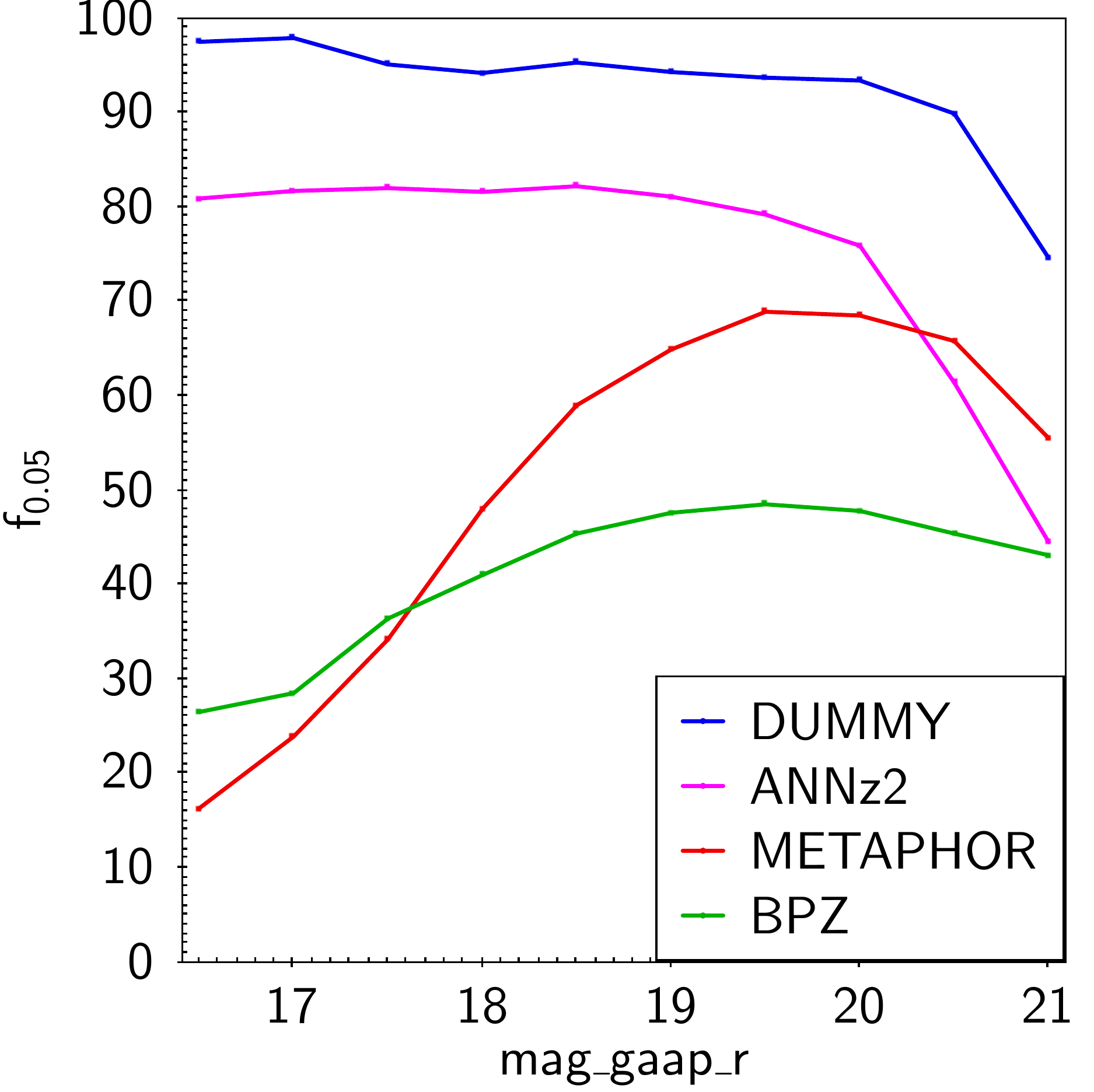}}
\caption{ Residuals fraction in the range [-0.05, 0.05] of the PDFs versus magnitude $mag\_gaap\_r$ in the range [16.0, 21.0], used for the tomographic analysis shown in Tab.~\ref{tab:TOMOG}. From top to bottom, \textit{dummy} (blue), ANNz2 (violet), METAPHOR (red) and BPZ (green).}
\label{fig:f005ALL1}
\end{figure}

\begin{figure*}
 \centering
  {\includegraphics[width=0.245 \textwidth]{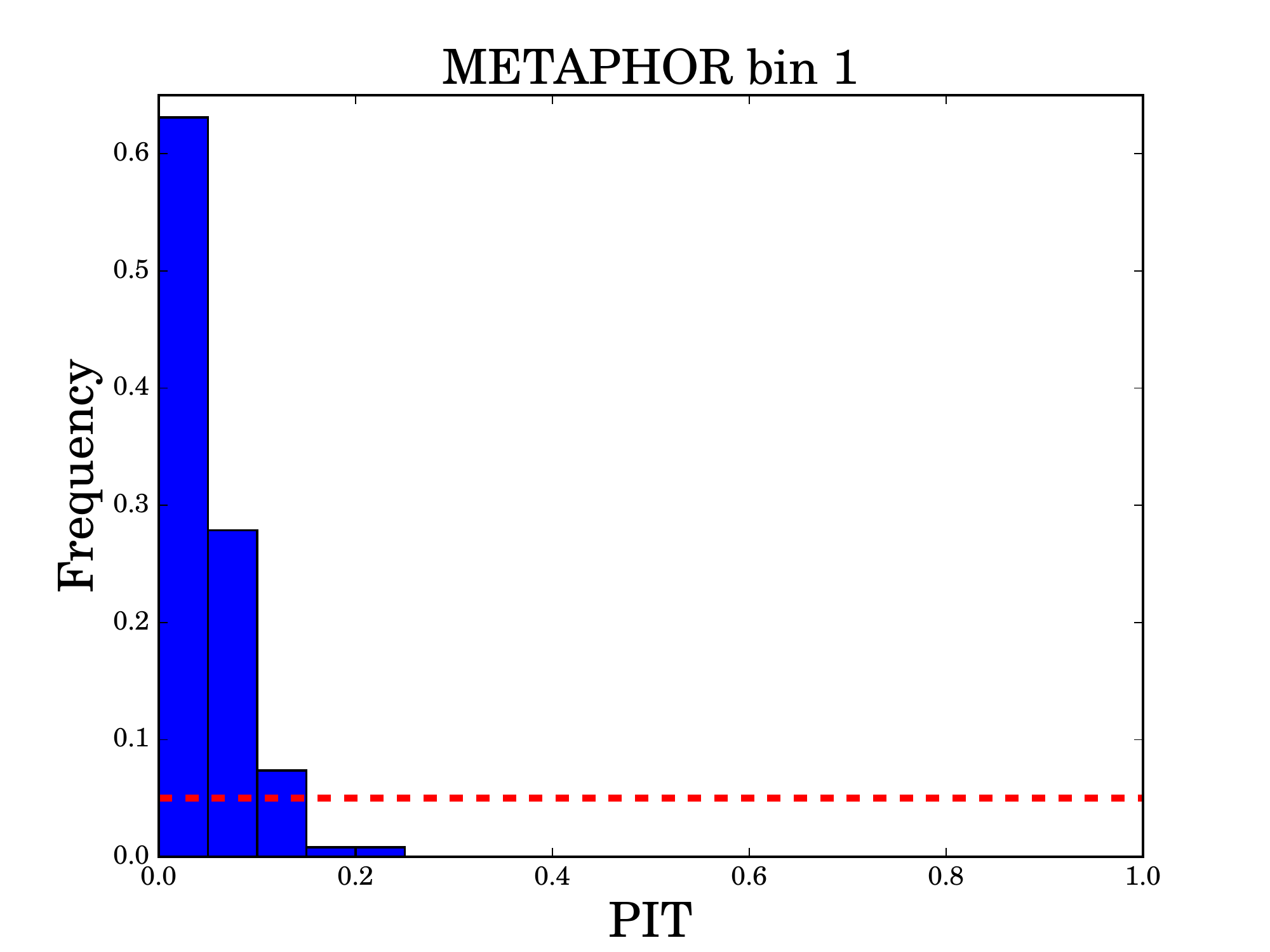}}
  {\includegraphics[width=0.245 \textwidth]{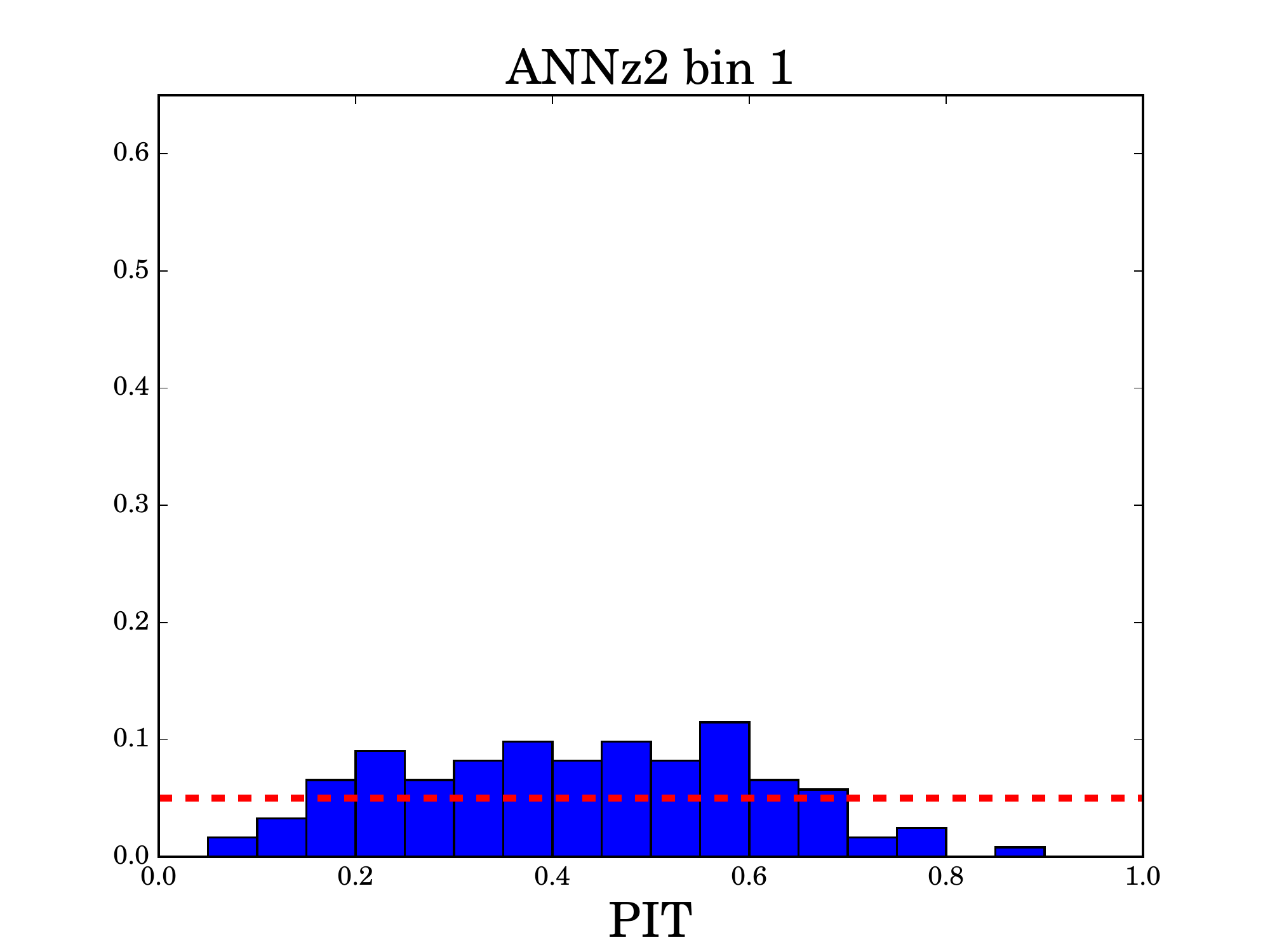}}
  {\includegraphics[width=0.245 \textwidth]{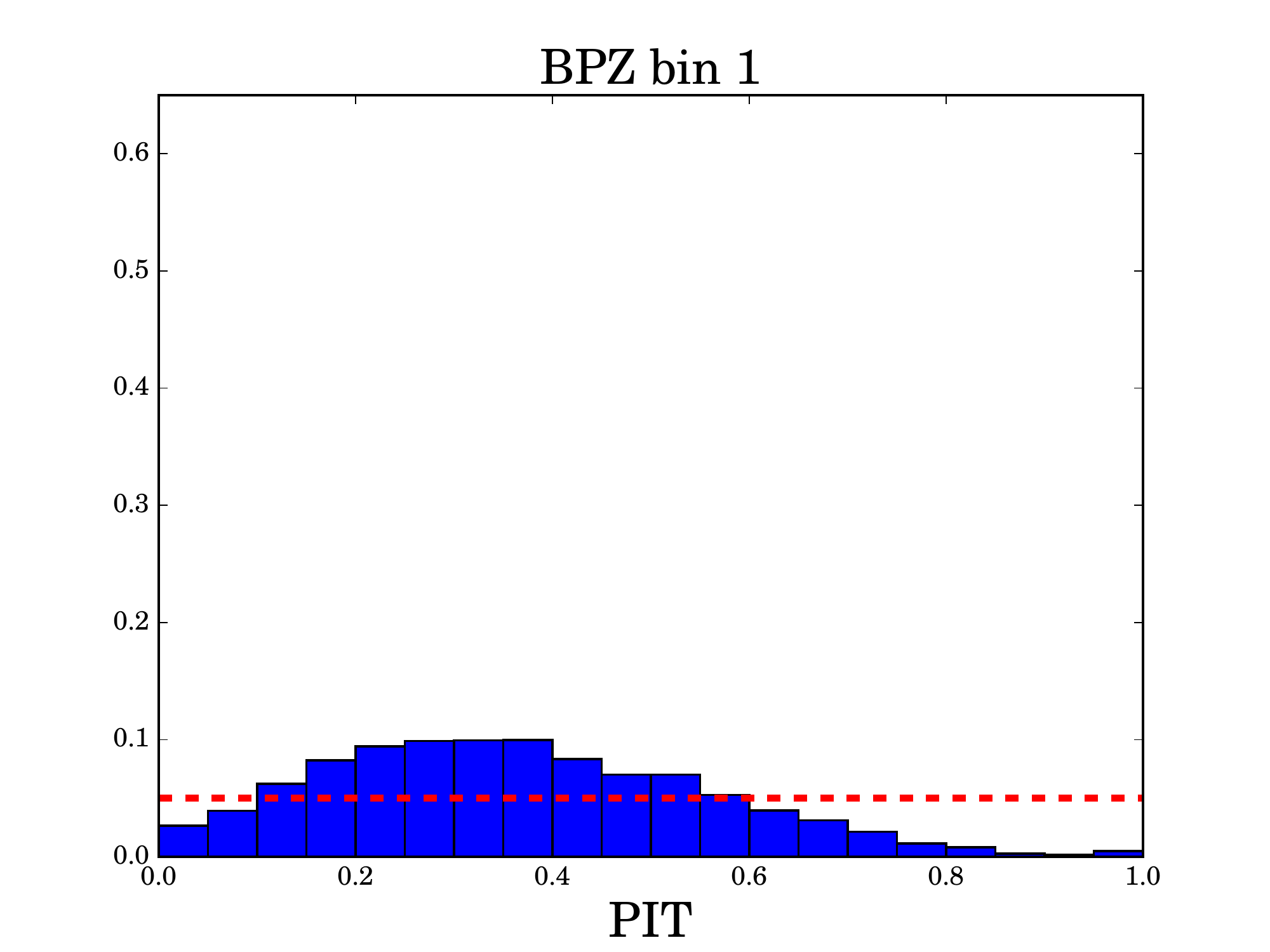}}
  {\includegraphics[width=0.245\textwidth]{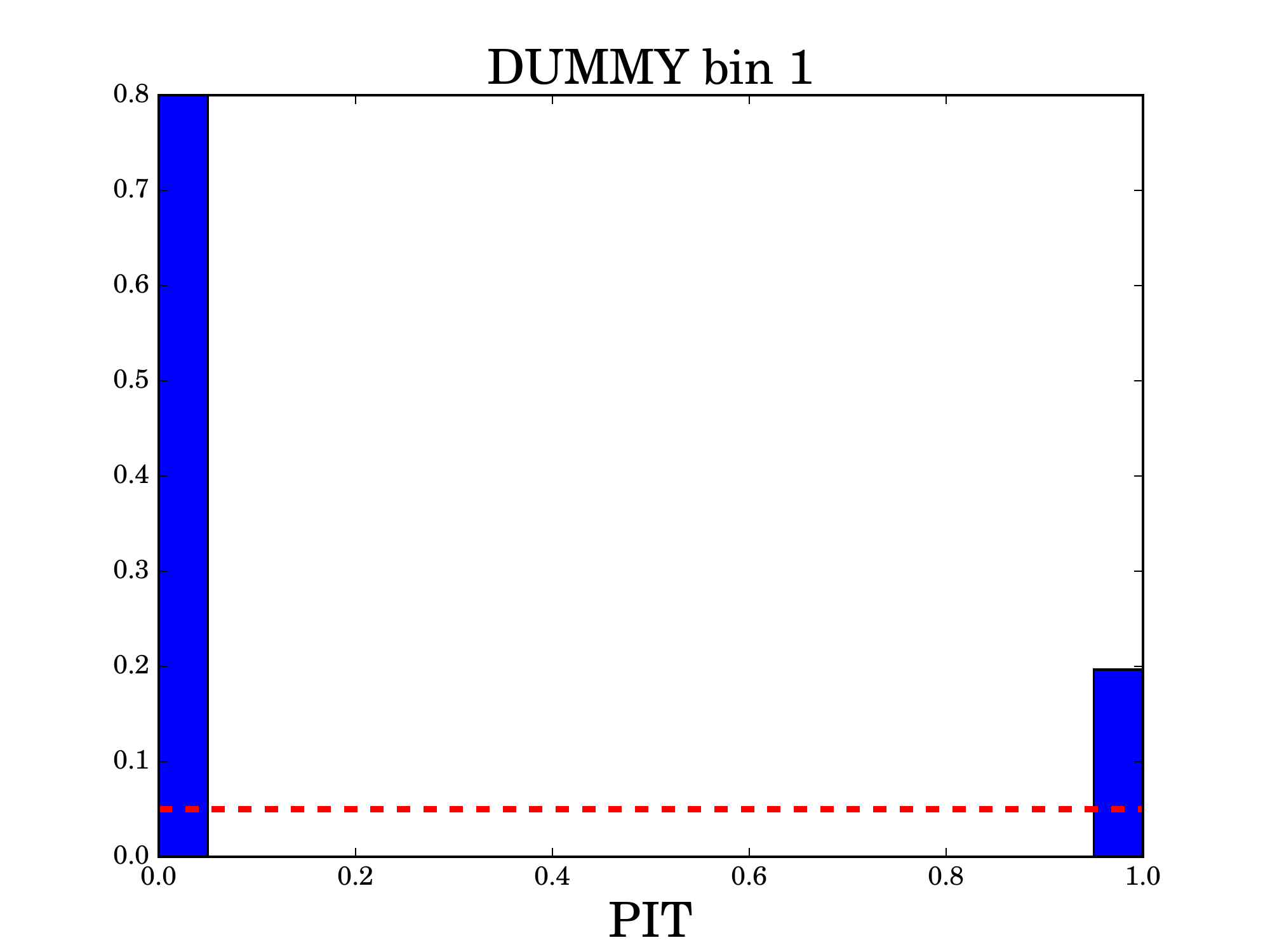}}
   {\includegraphics[width=0.245 \textwidth]{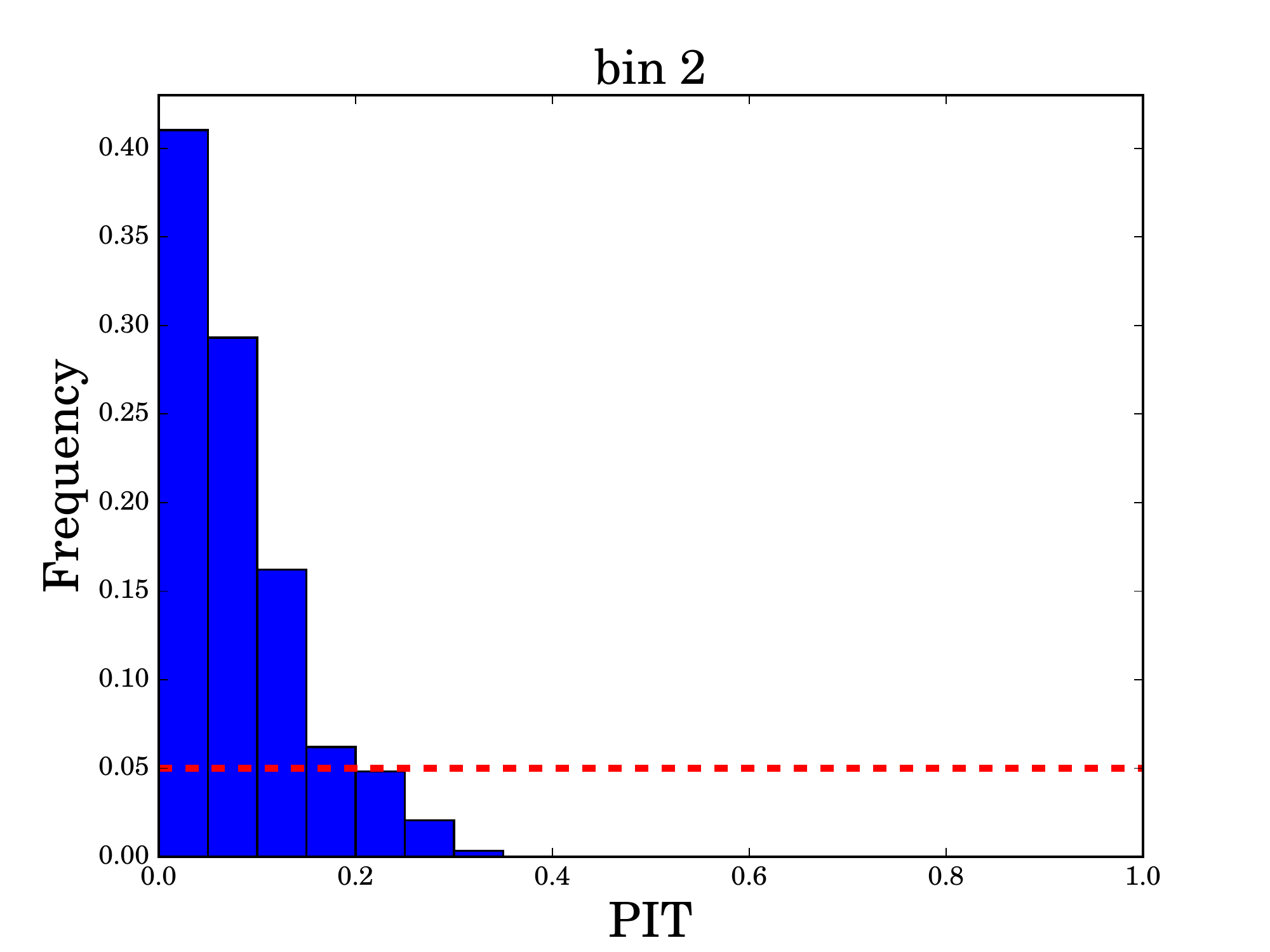}}
  {\includegraphics[width=0.245 \textwidth]{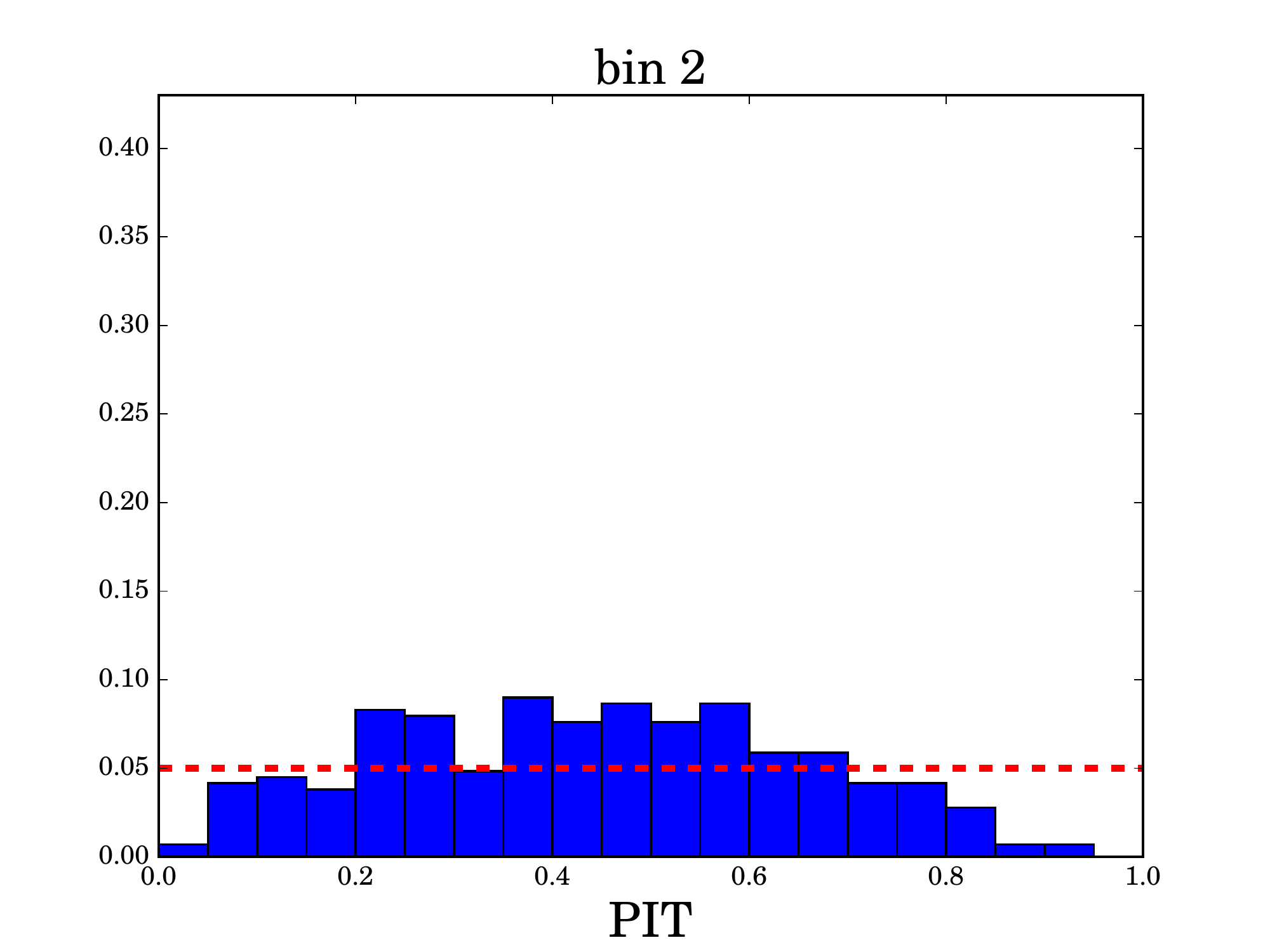}}
  {\includegraphics[width=0.245 \textwidth]{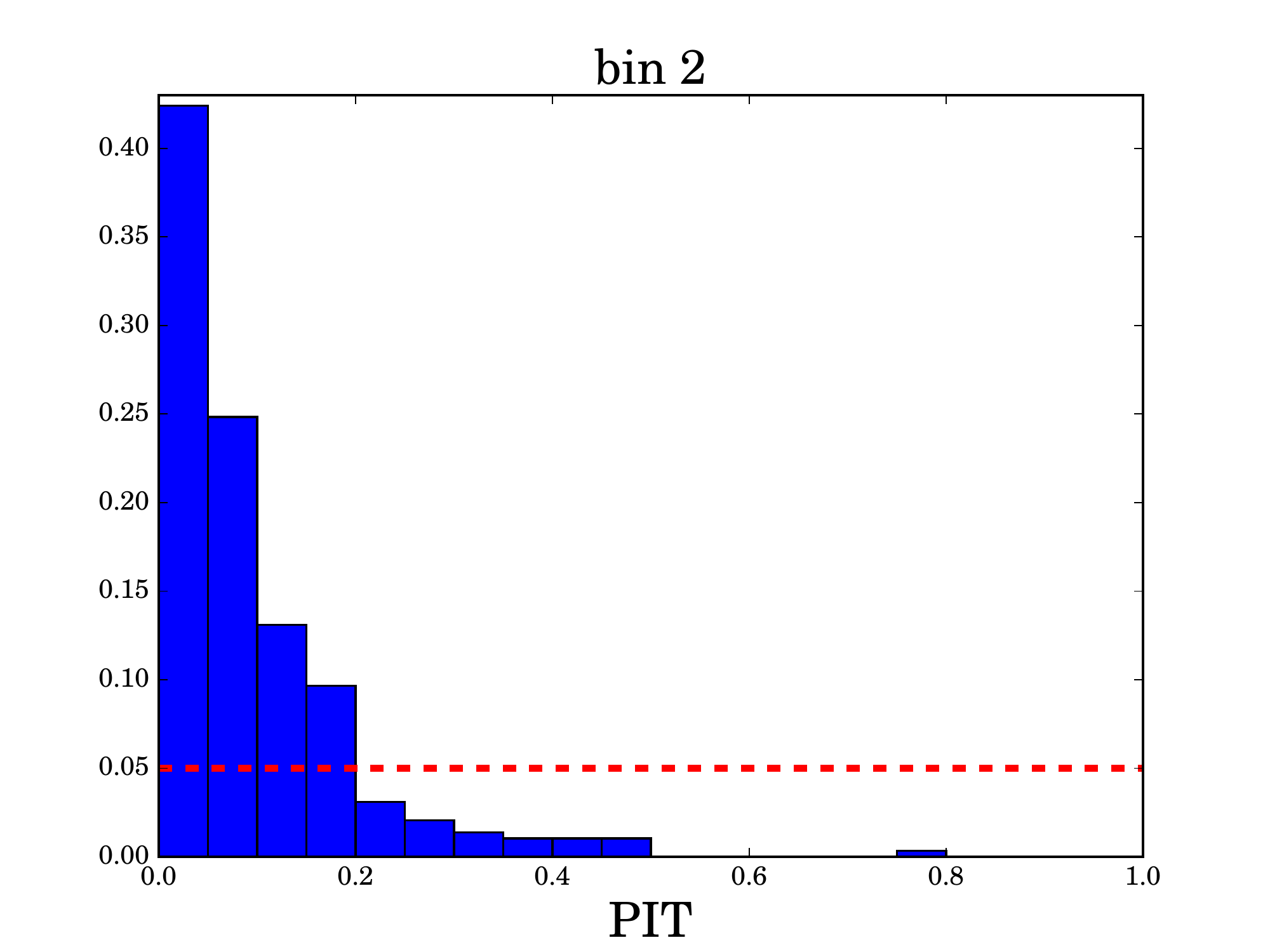}}
  {\includegraphics[width=0.245\textwidth]{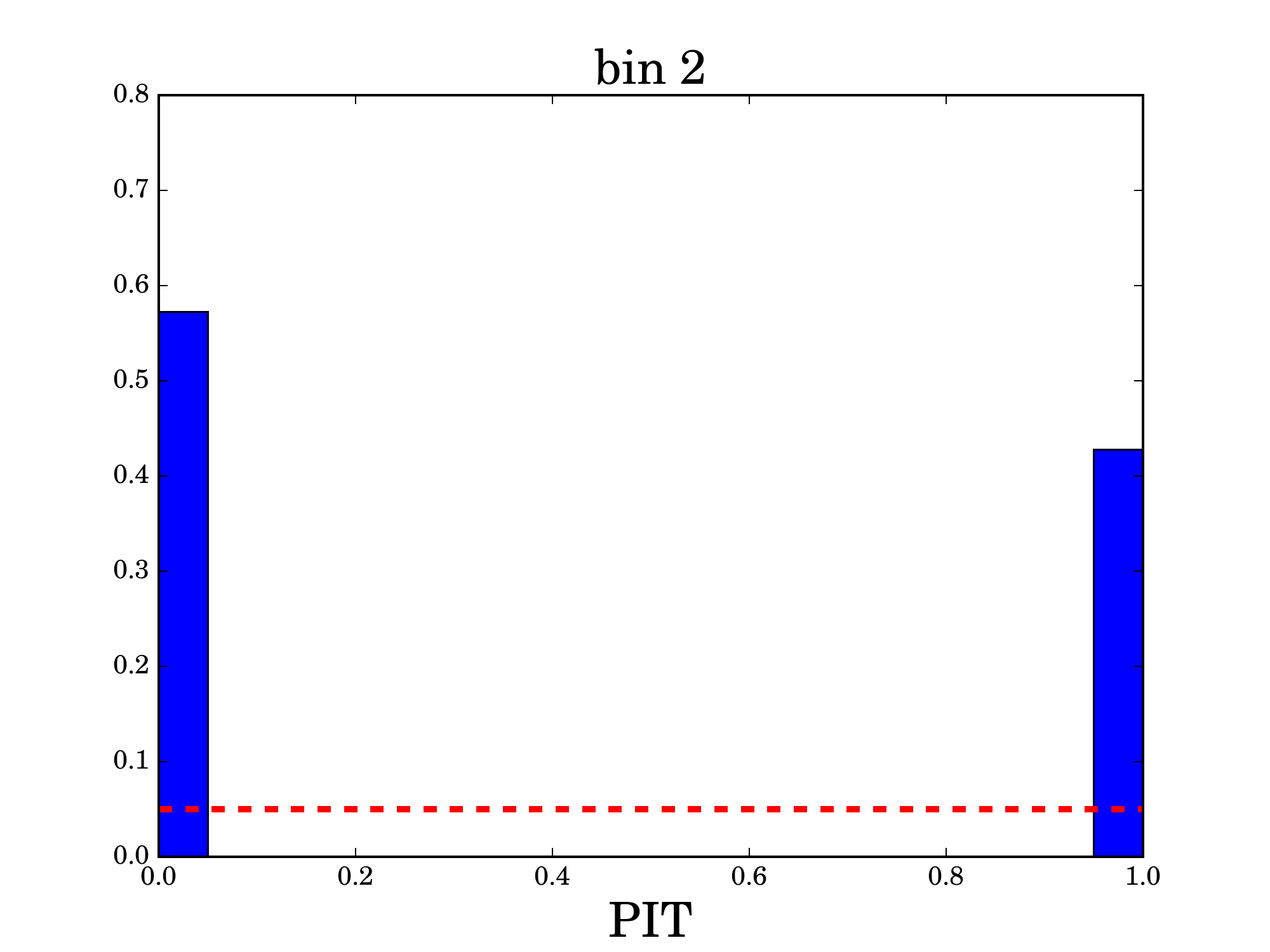}}
   {\includegraphics[width=0.245 \textwidth]{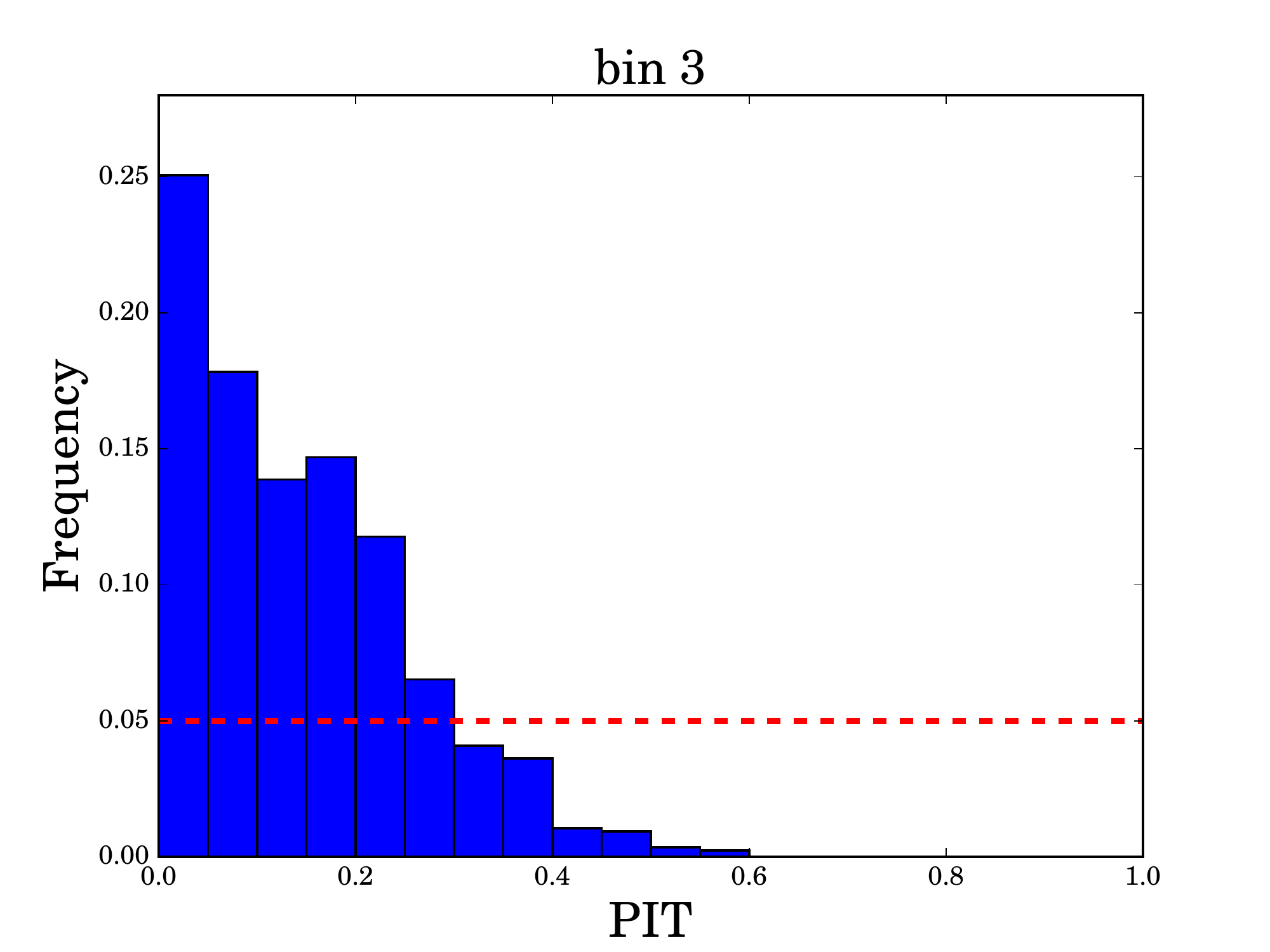}}
  {\includegraphics[width=0.245 \textwidth]{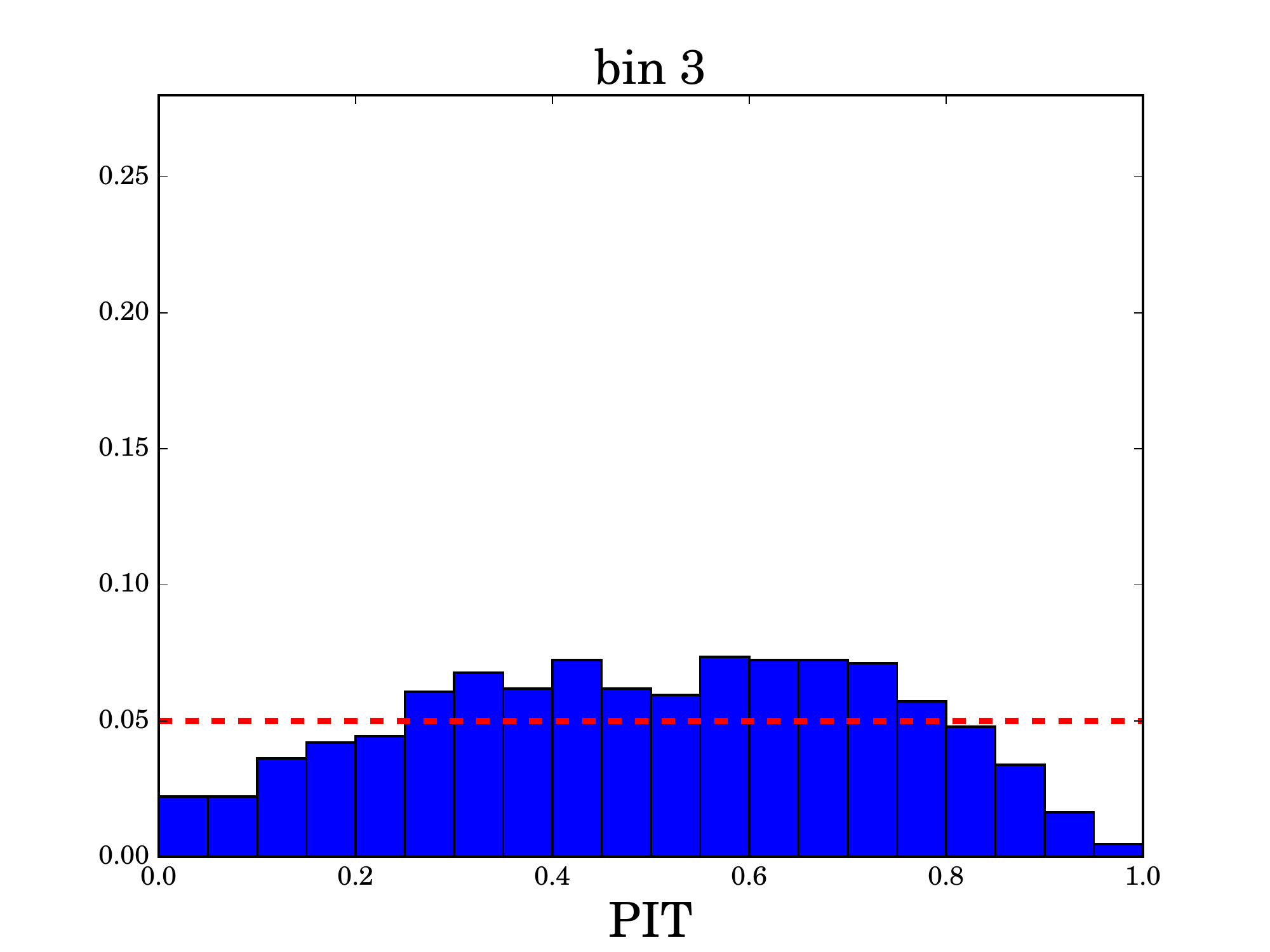}}
  {\includegraphics[width=0.245 \textwidth]{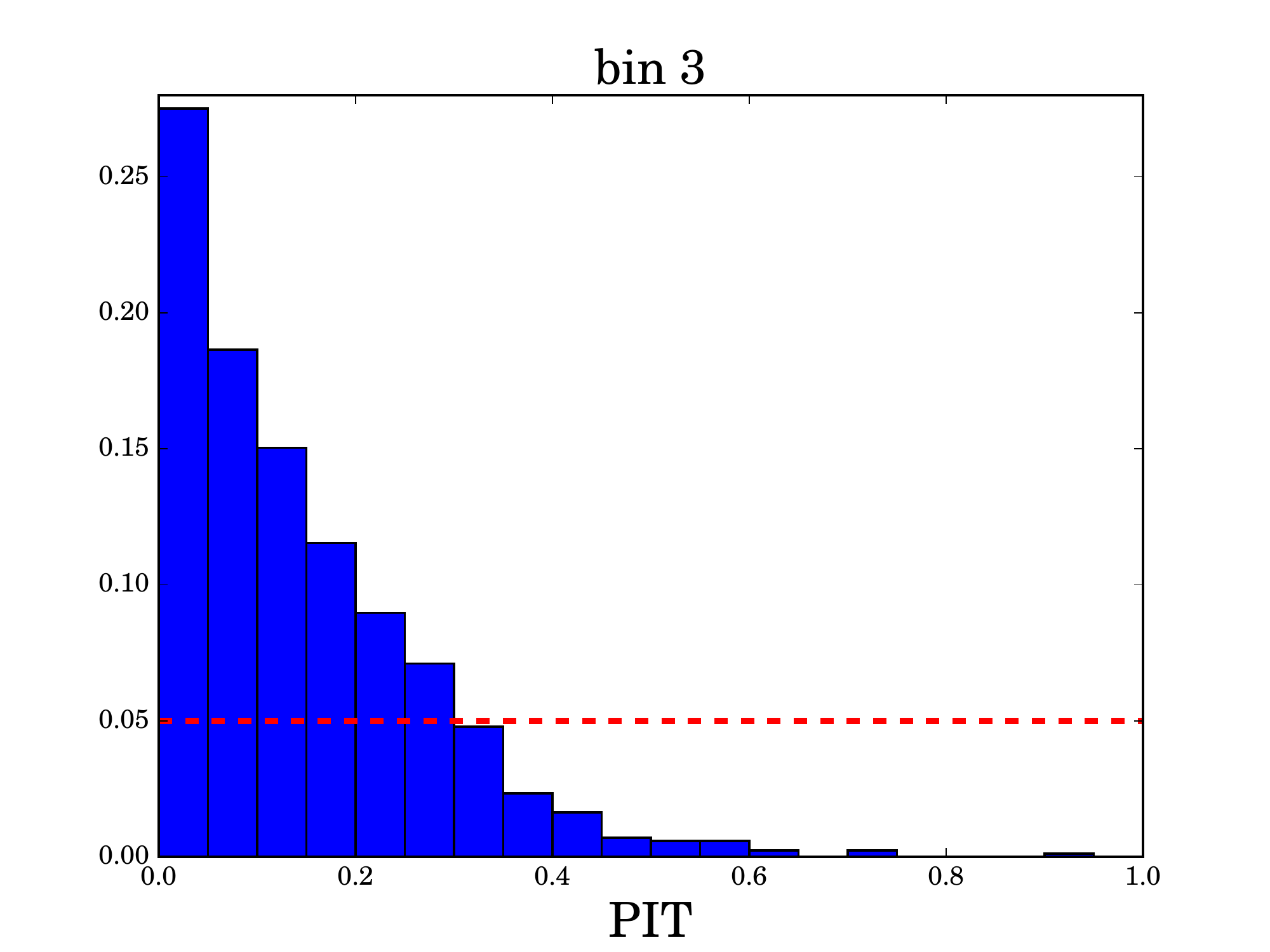}}
  {\includegraphics[width=0.245\textwidth]{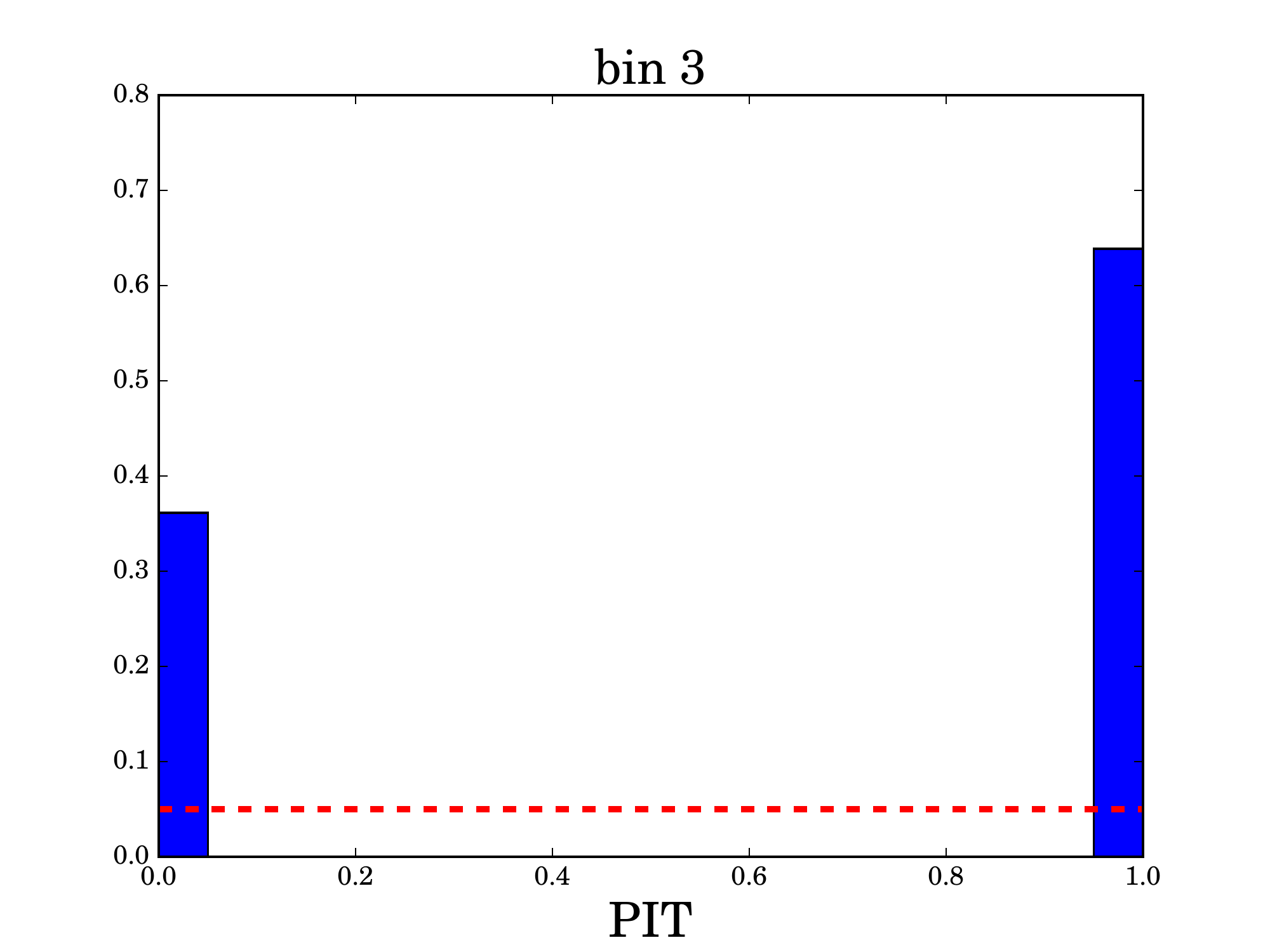}}
   {\includegraphics[width=0.245 \textwidth]{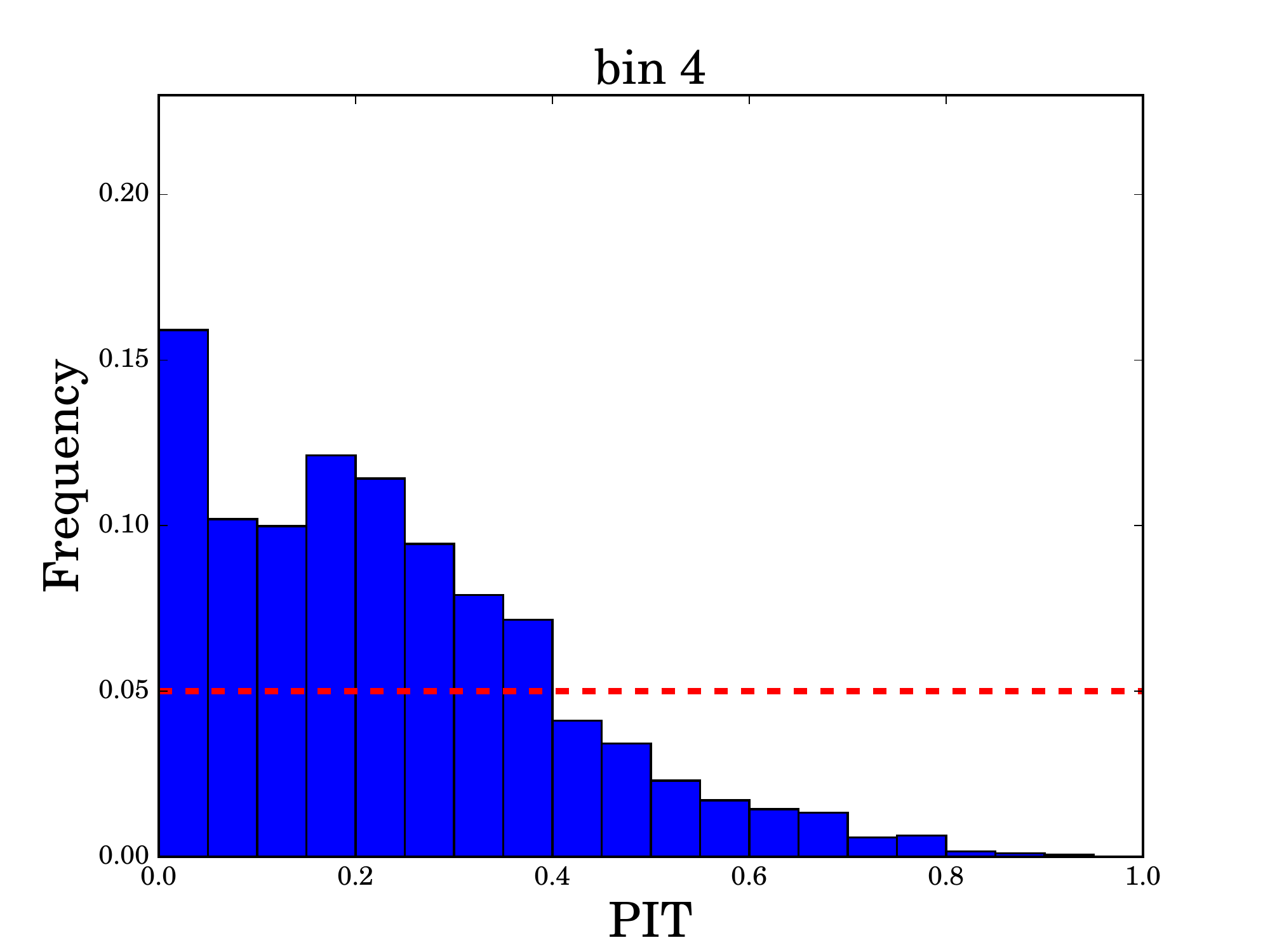}}
  {\includegraphics[width=0.245 \textwidth]{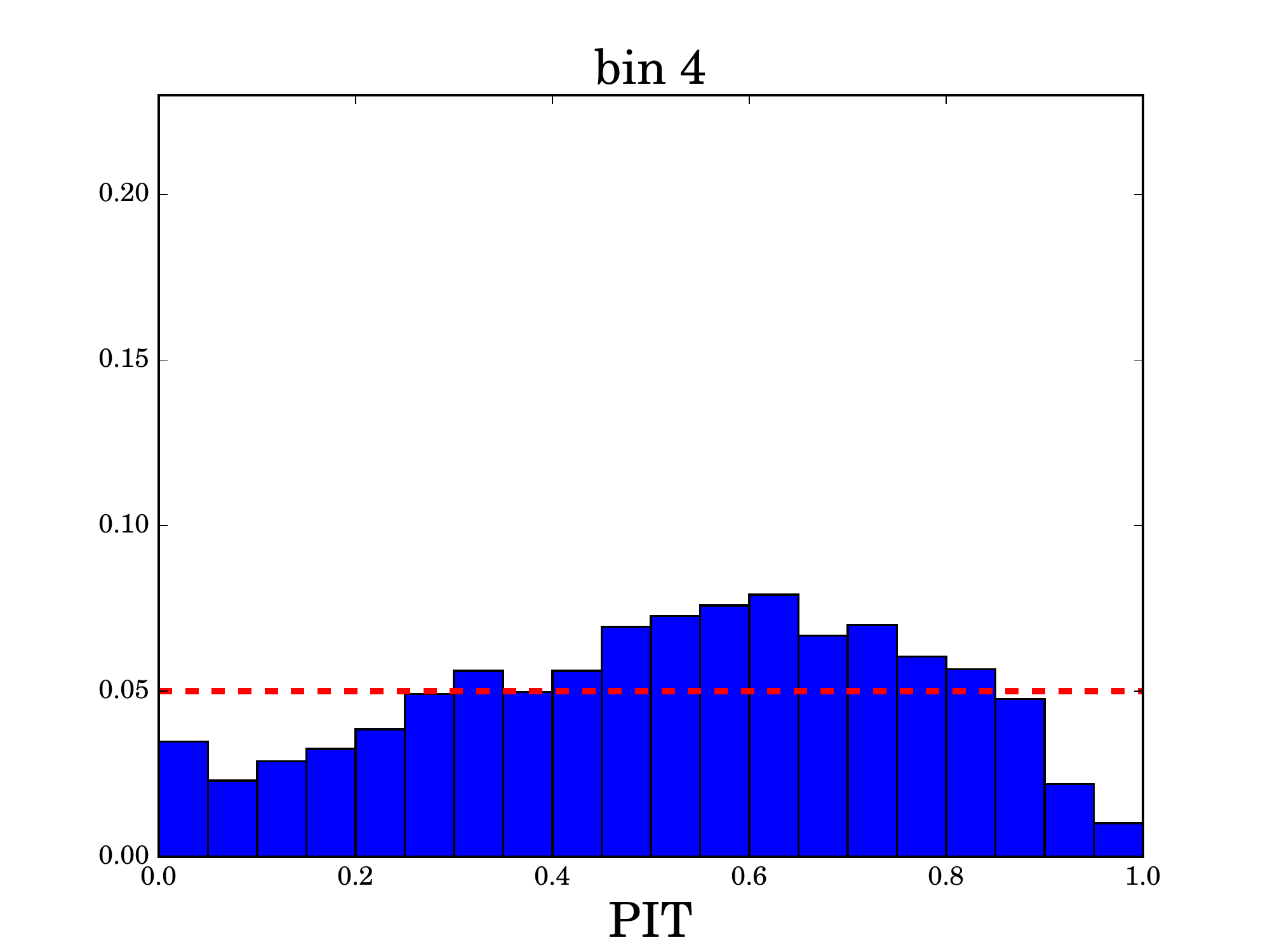}}
  {\includegraphics[width=0.245 \textwidth]{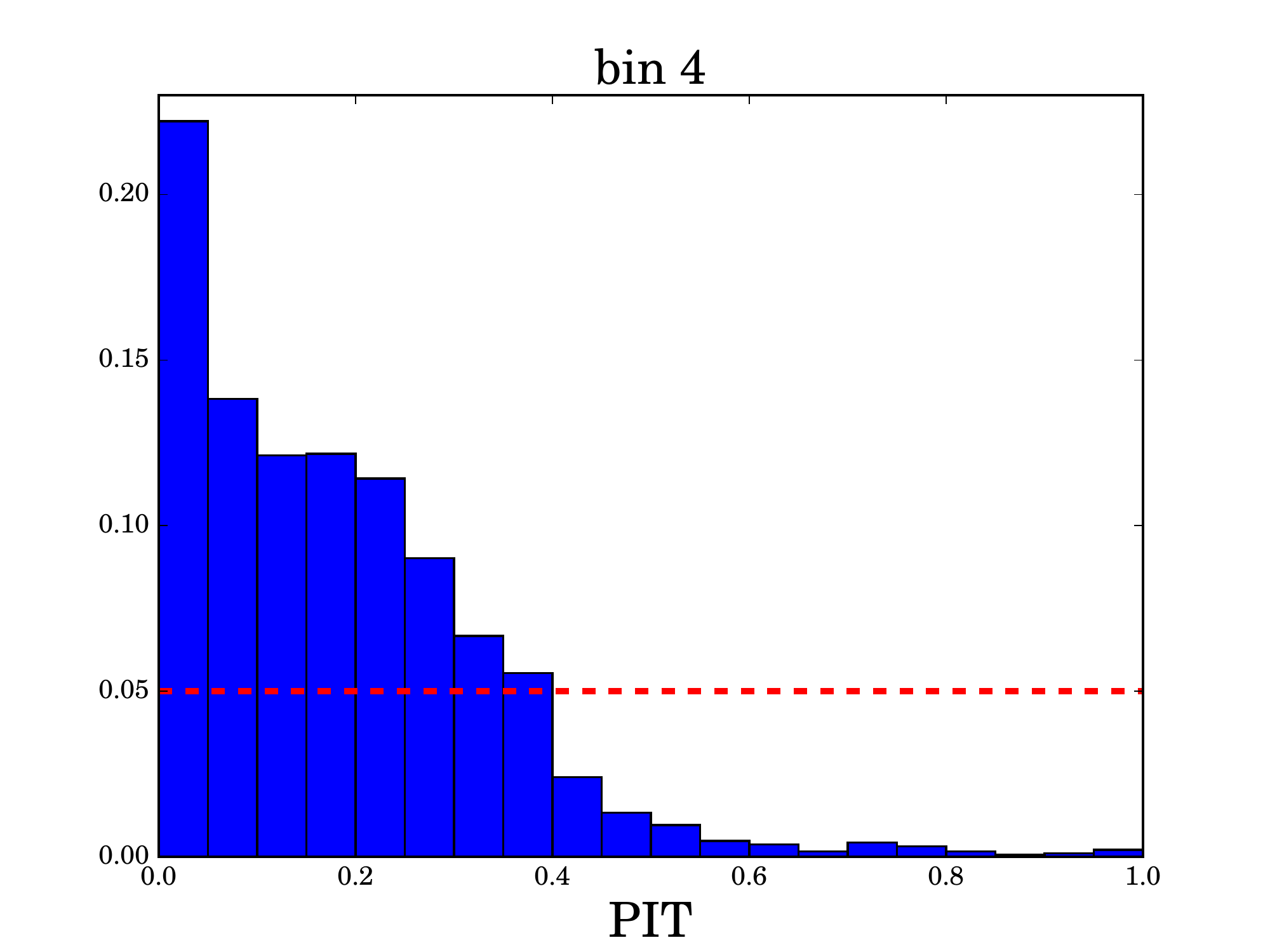}}
  {\includegraphics[width=0.245\textwidth]{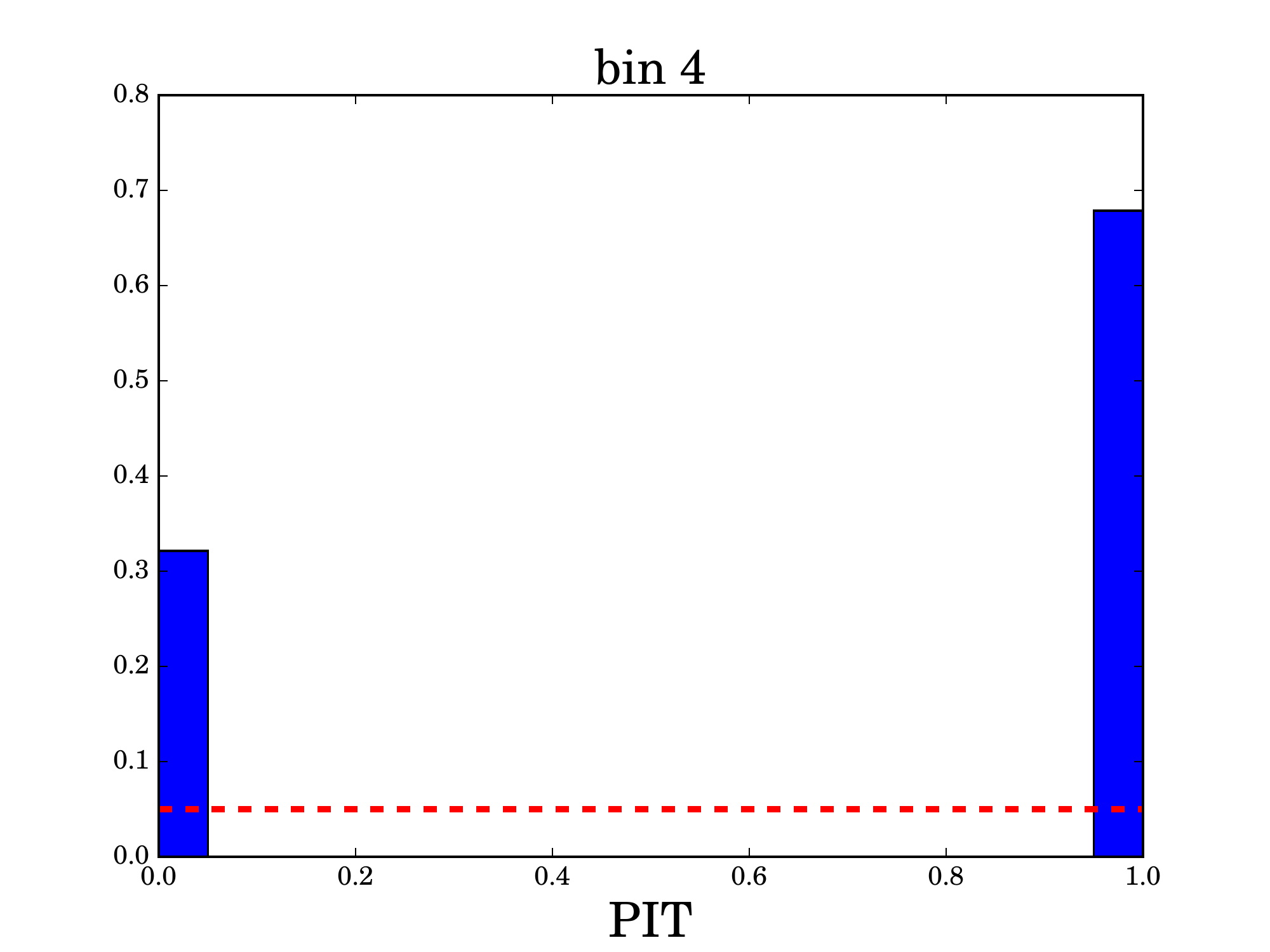}}
   {\includegraphics[width=0.245 \textwidth]{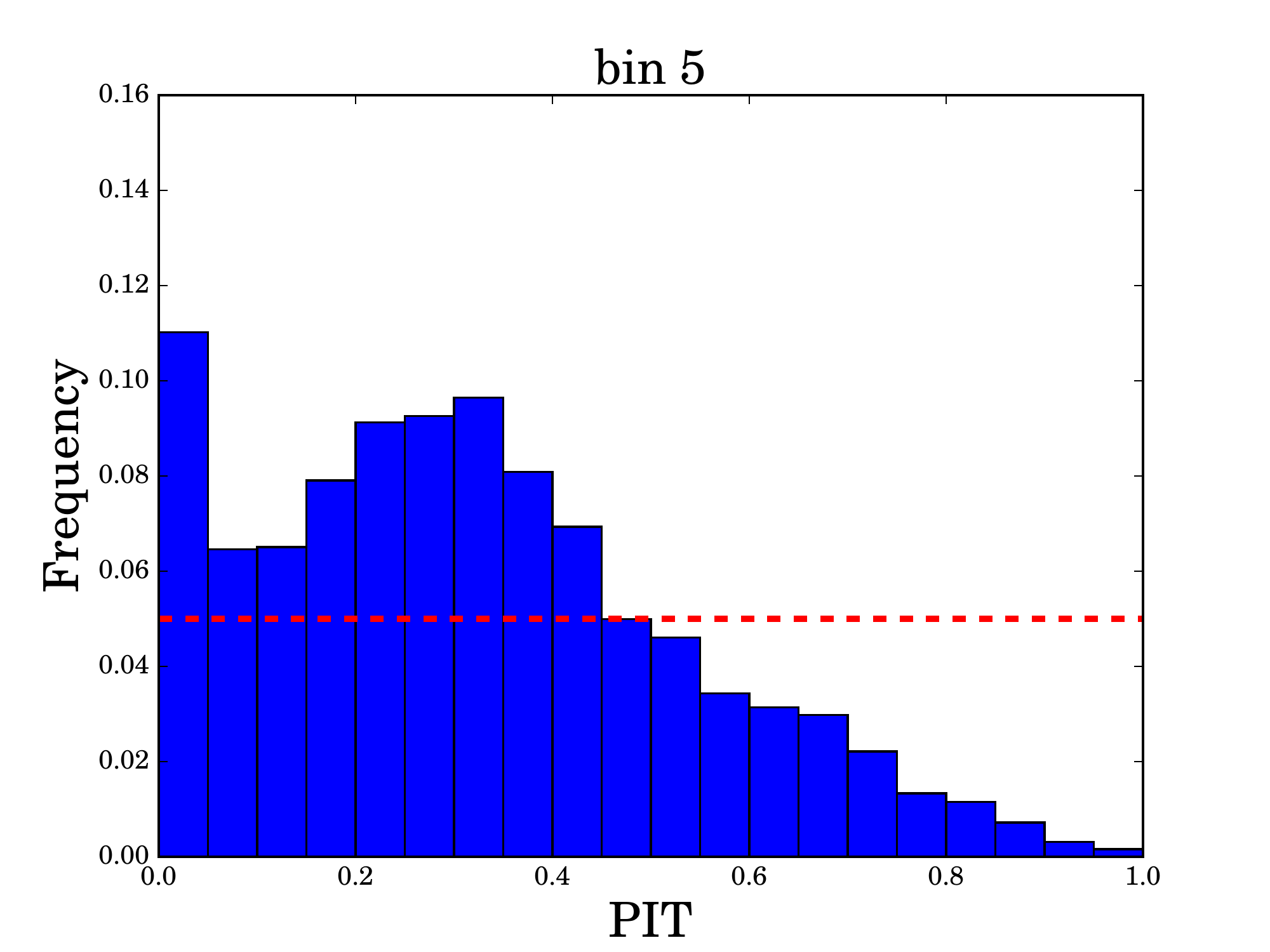}}
  {\includegraphics[width=0.245 \textwidth]{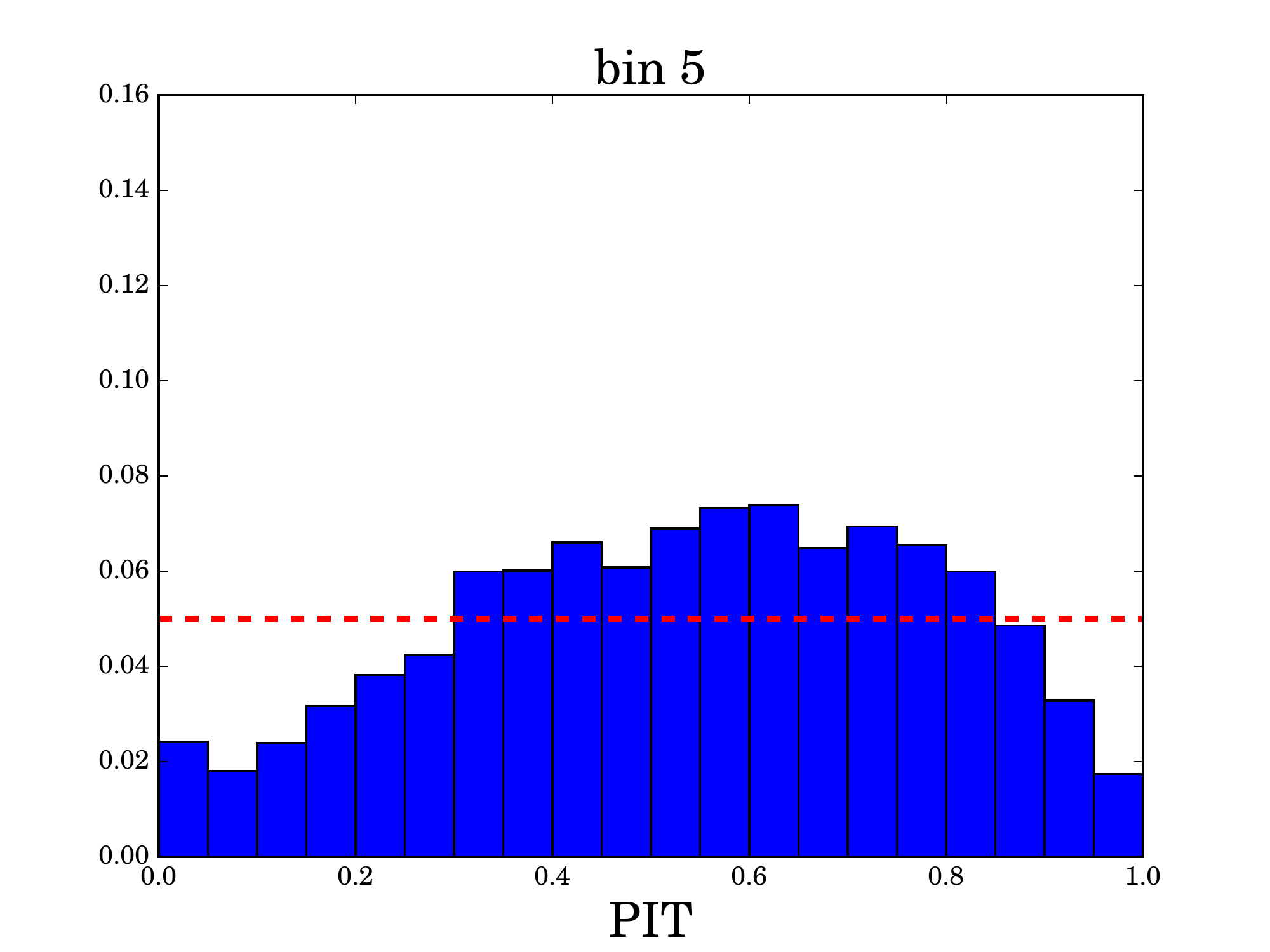}}
  {\includegraphics[width=0.245 \textwidth]{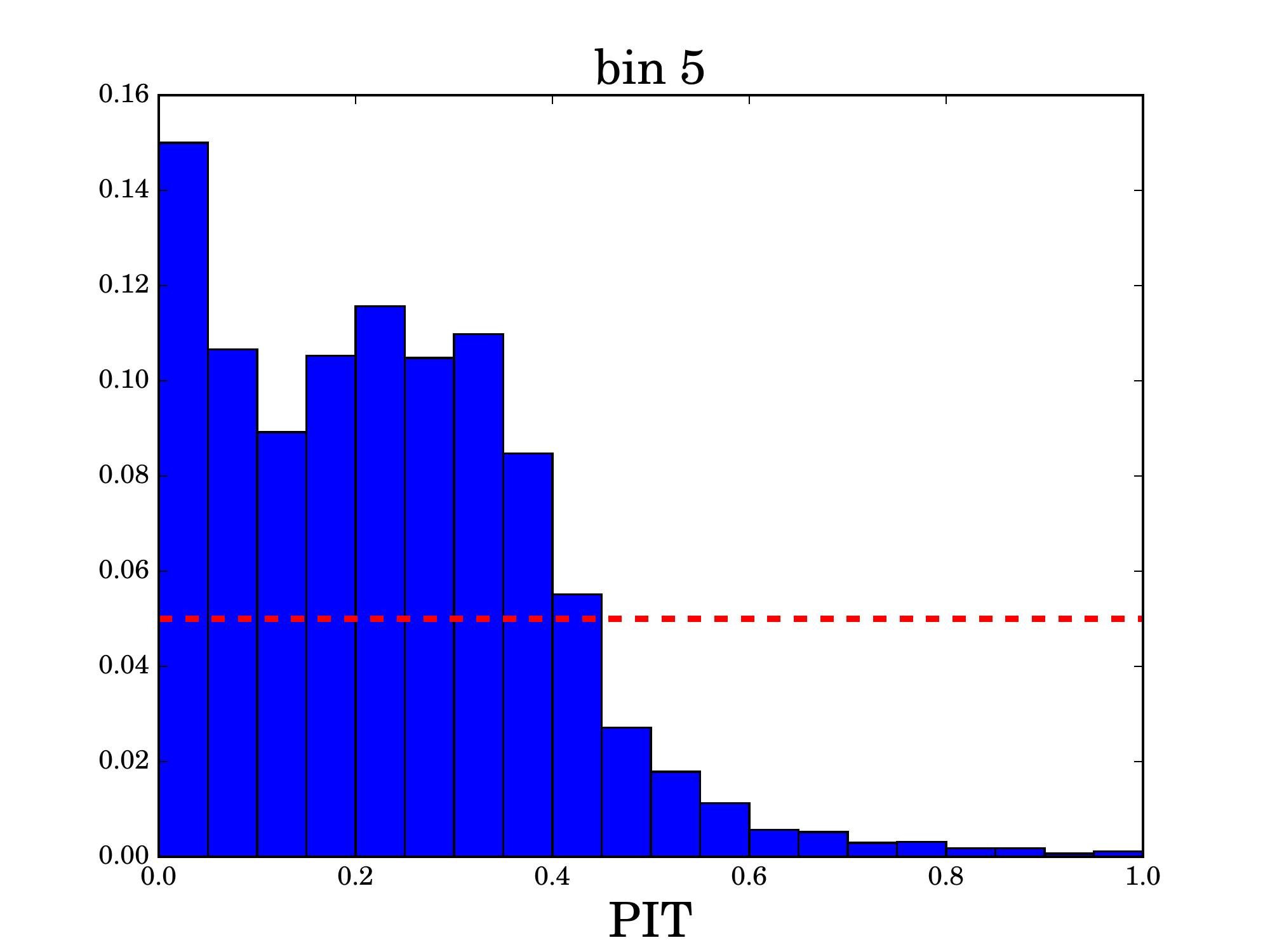}}
  {\includegraphics[width=0.245\textwidth]{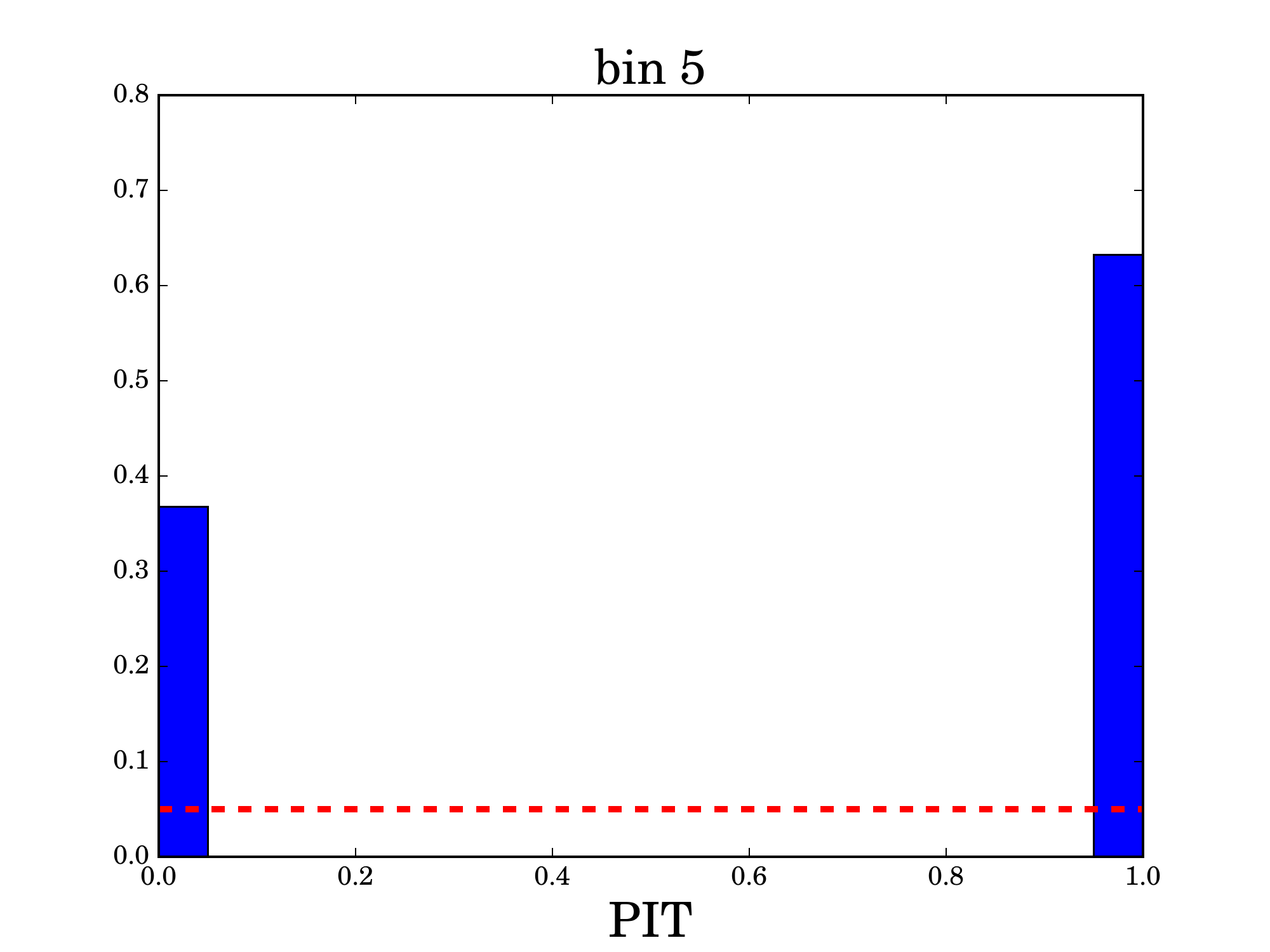}}
\caption{ Probability Integral Transform (PIT) obtained for METAPHOR ( first  column panels), ANNz2 (second column panels), BPZ (third column panels), and for the \textit{dummy} PDF (fourth column panels) in the first five magnitude tomographic bins from Table~\ref{tab:TOMOG}.}
\label{fig:pitALLtom1}
\end{figure*}
\begin{figure*}
 \centering
   {\includegraphics[width=0.245 \textwidth]{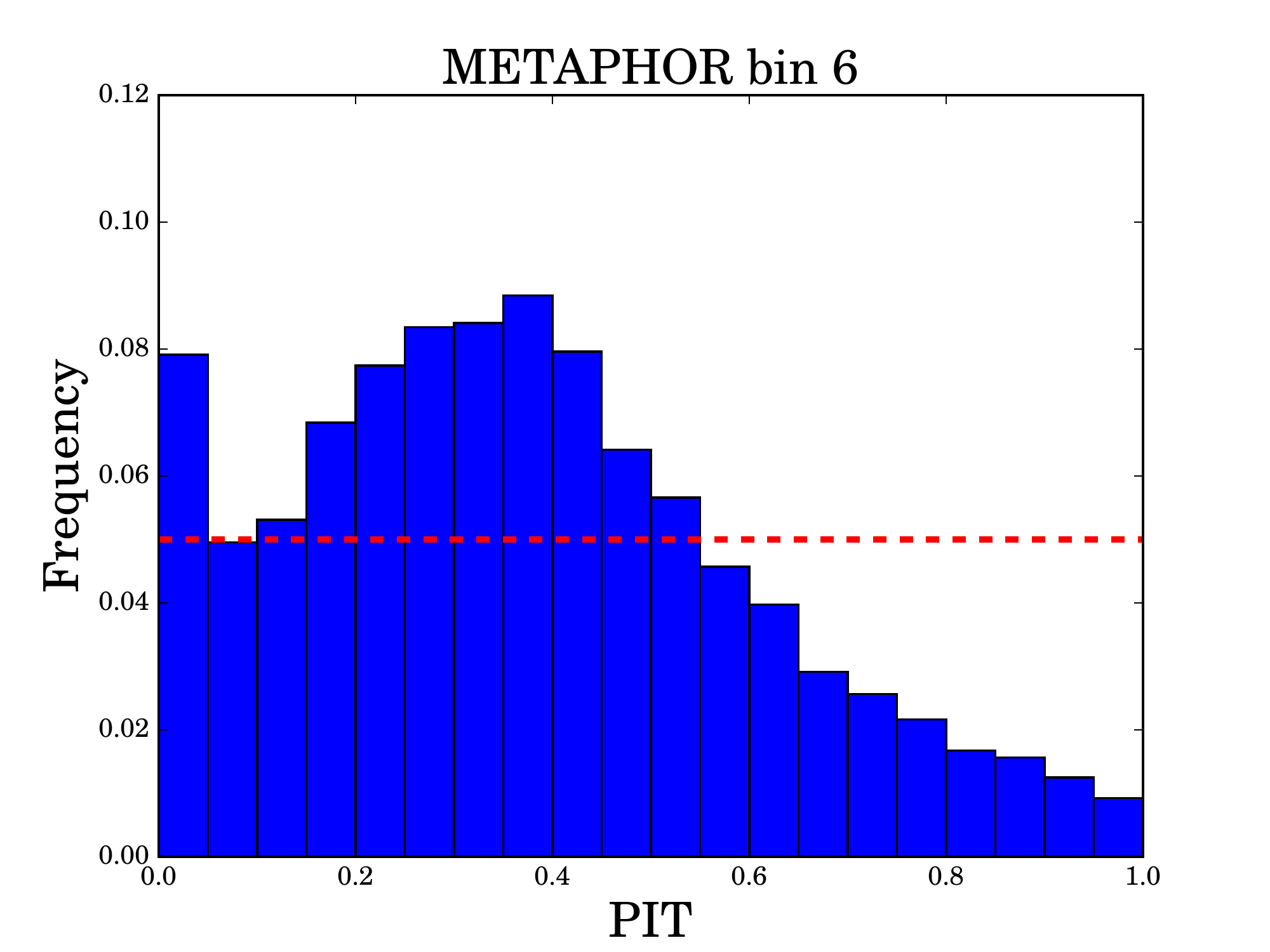}}
  {\includegraphics[width=0.245 \textwidth]{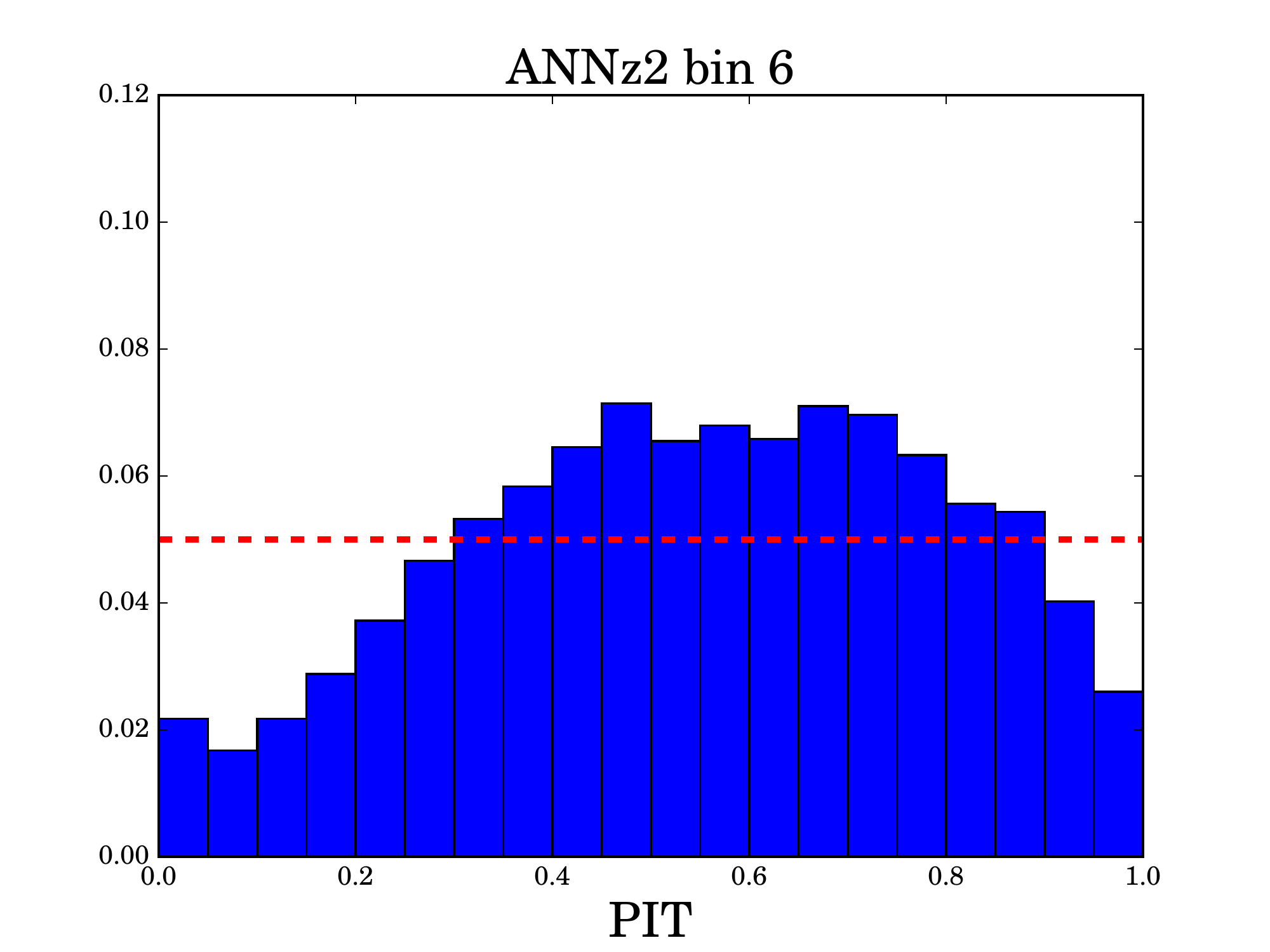}}
  {\includegraphics[width=0.245 \textwidth]{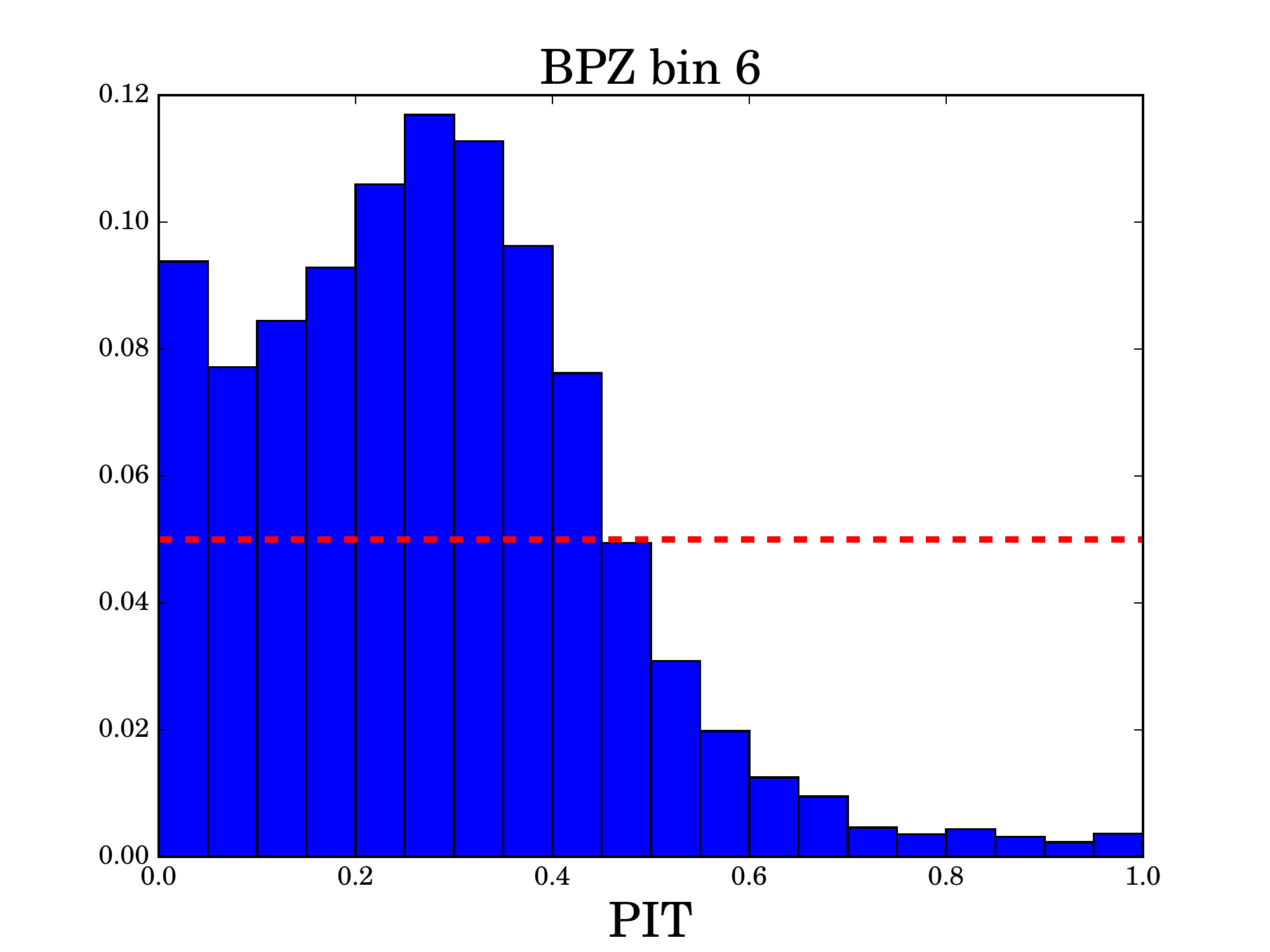}}
  {\includegraphics[width=0.245\textwidth]{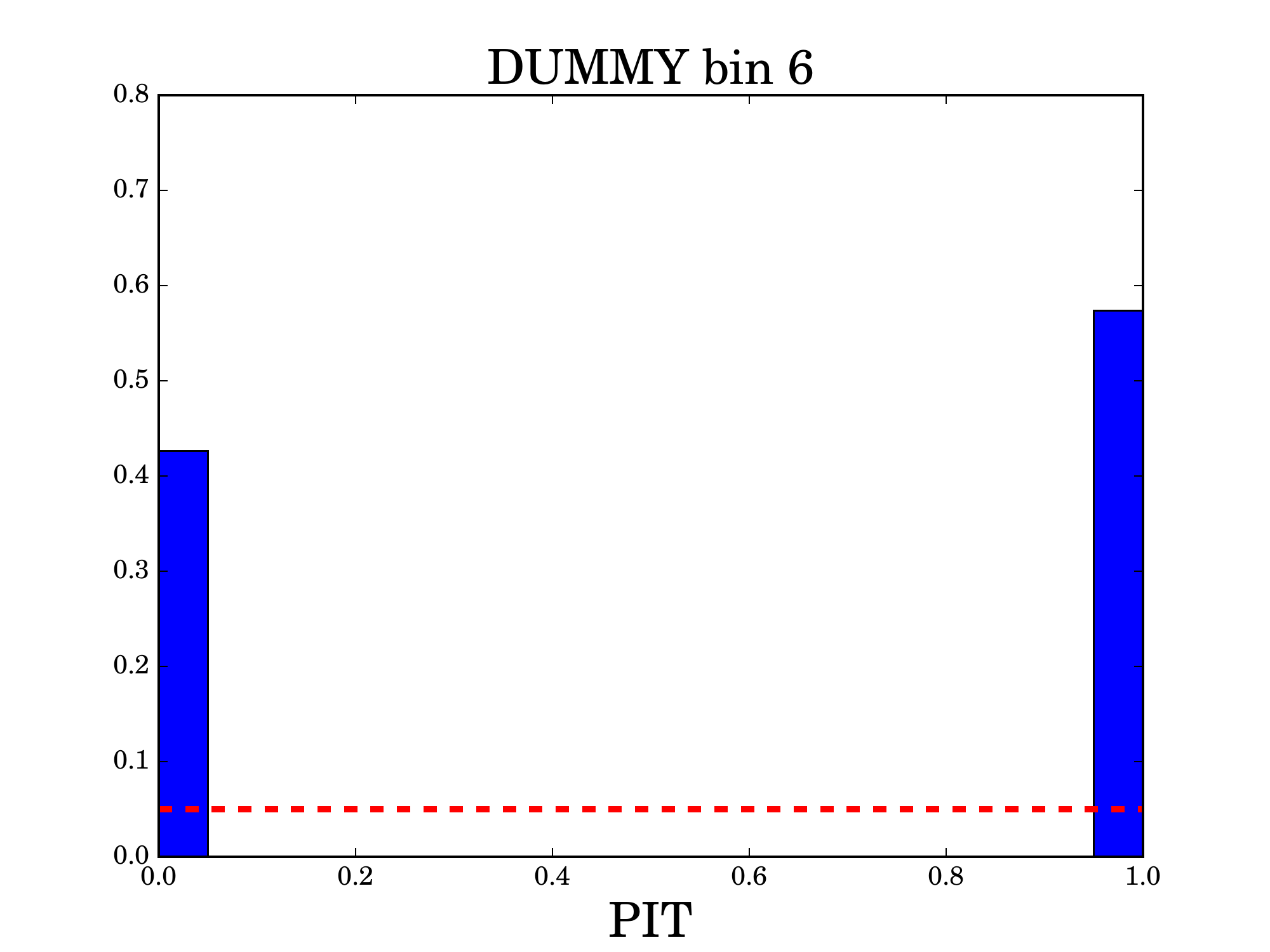}}
   {\includegraphics[width=0.245 \textwidth]{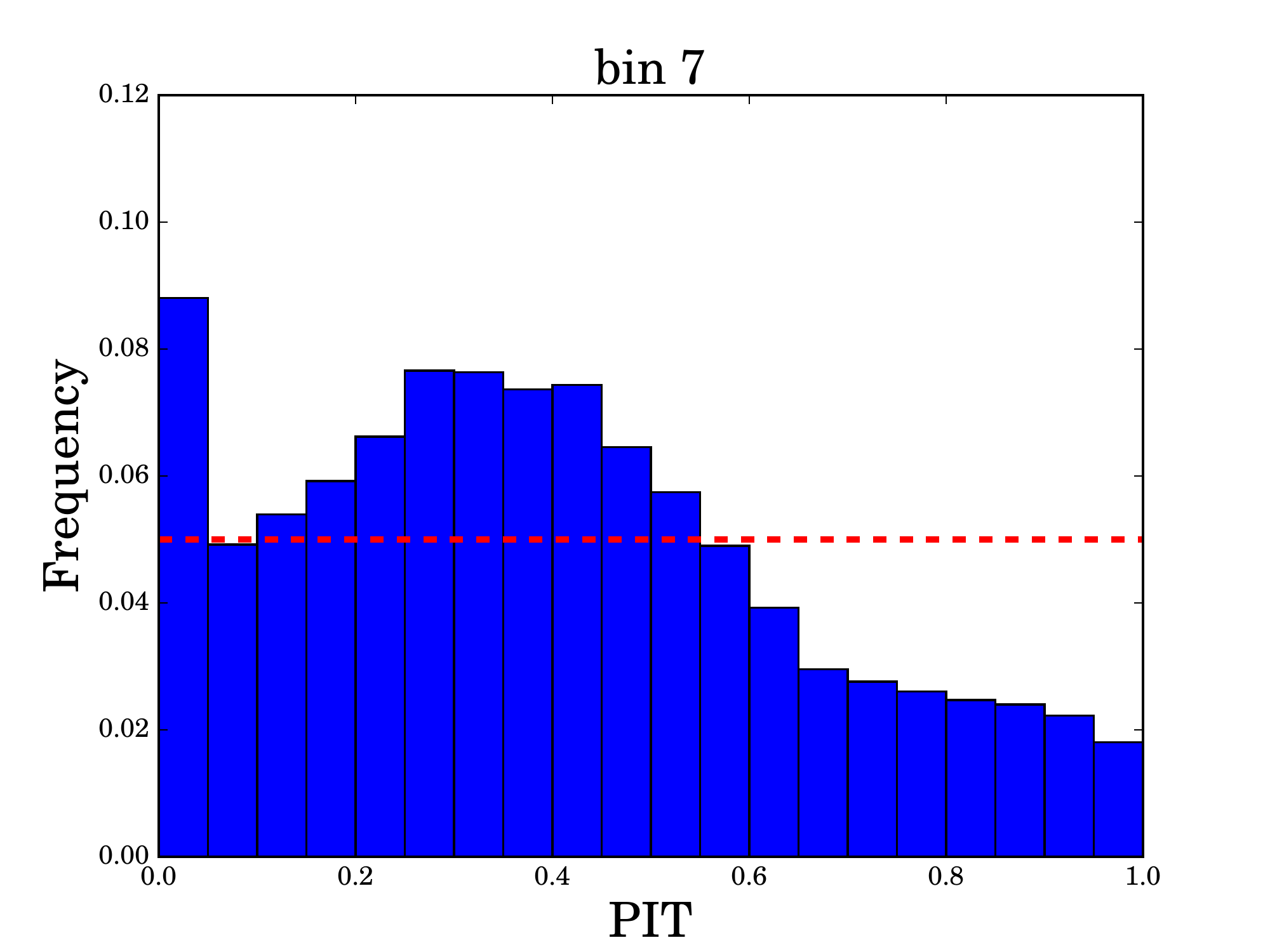}}
  {\includegraphics[width=0.245 \textwidth]{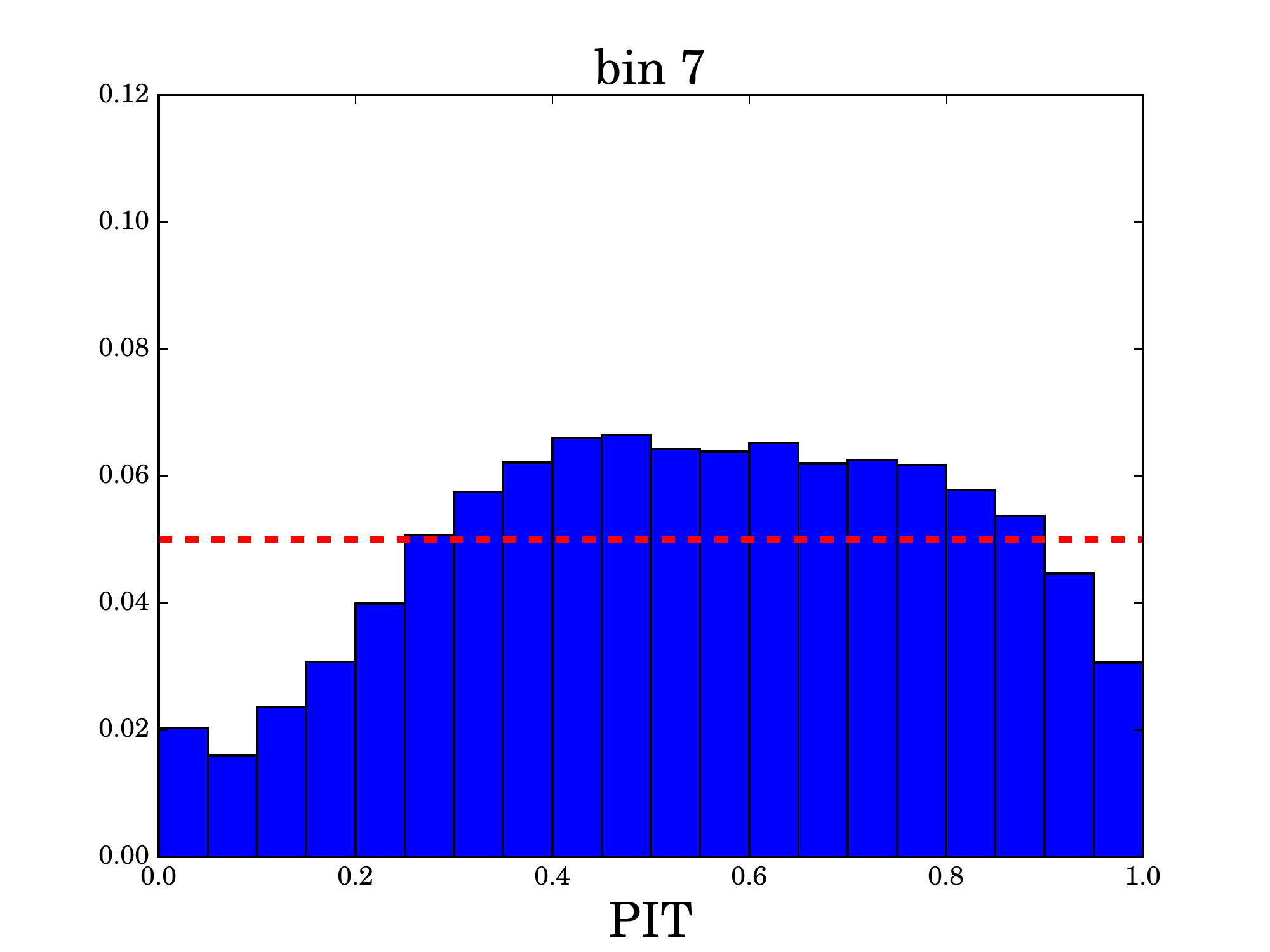}}
  {\includegraphics[width=0.245 \textwidth]{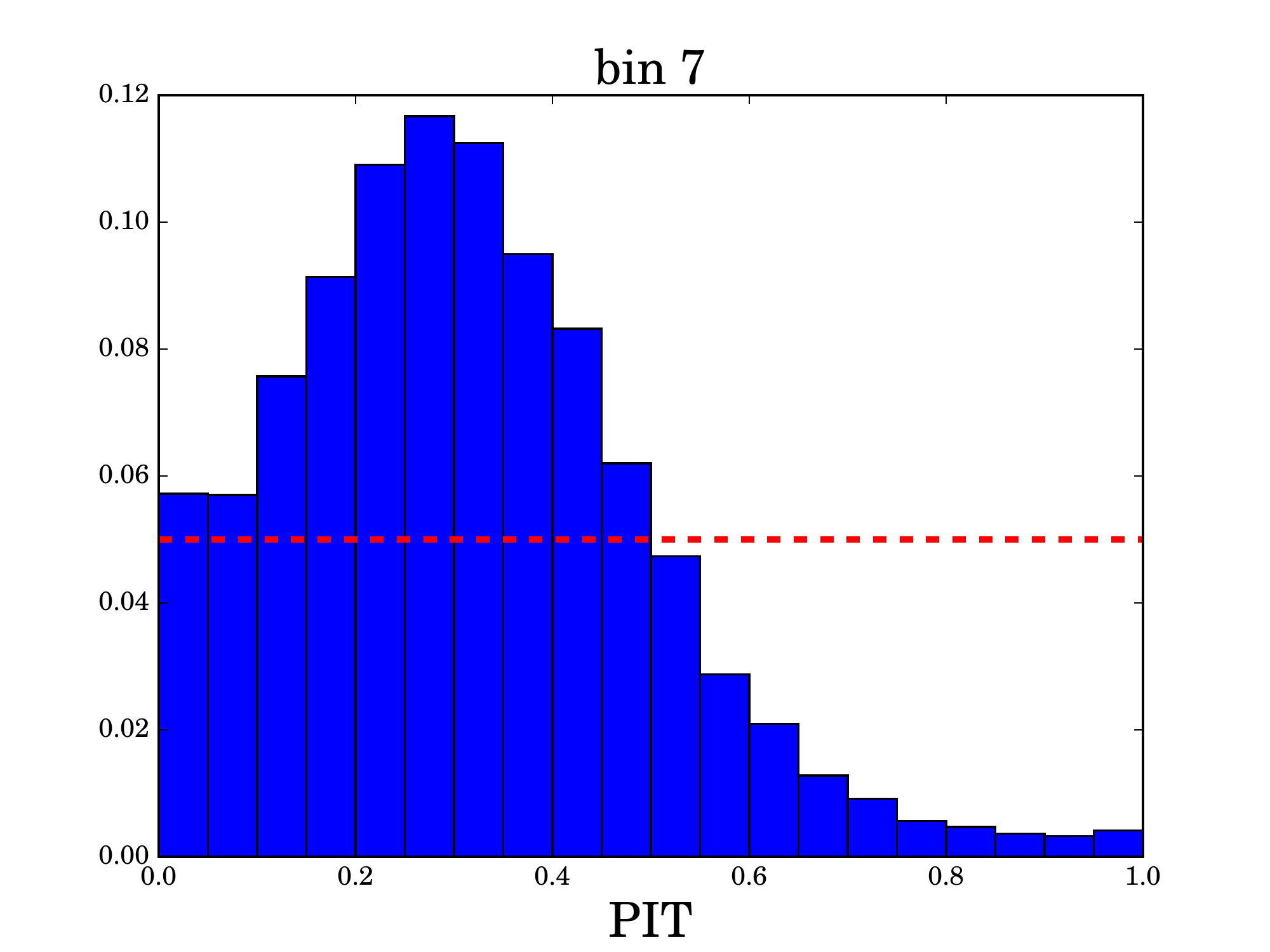}}
  {\includegraphics[width=0.245\textwidth]{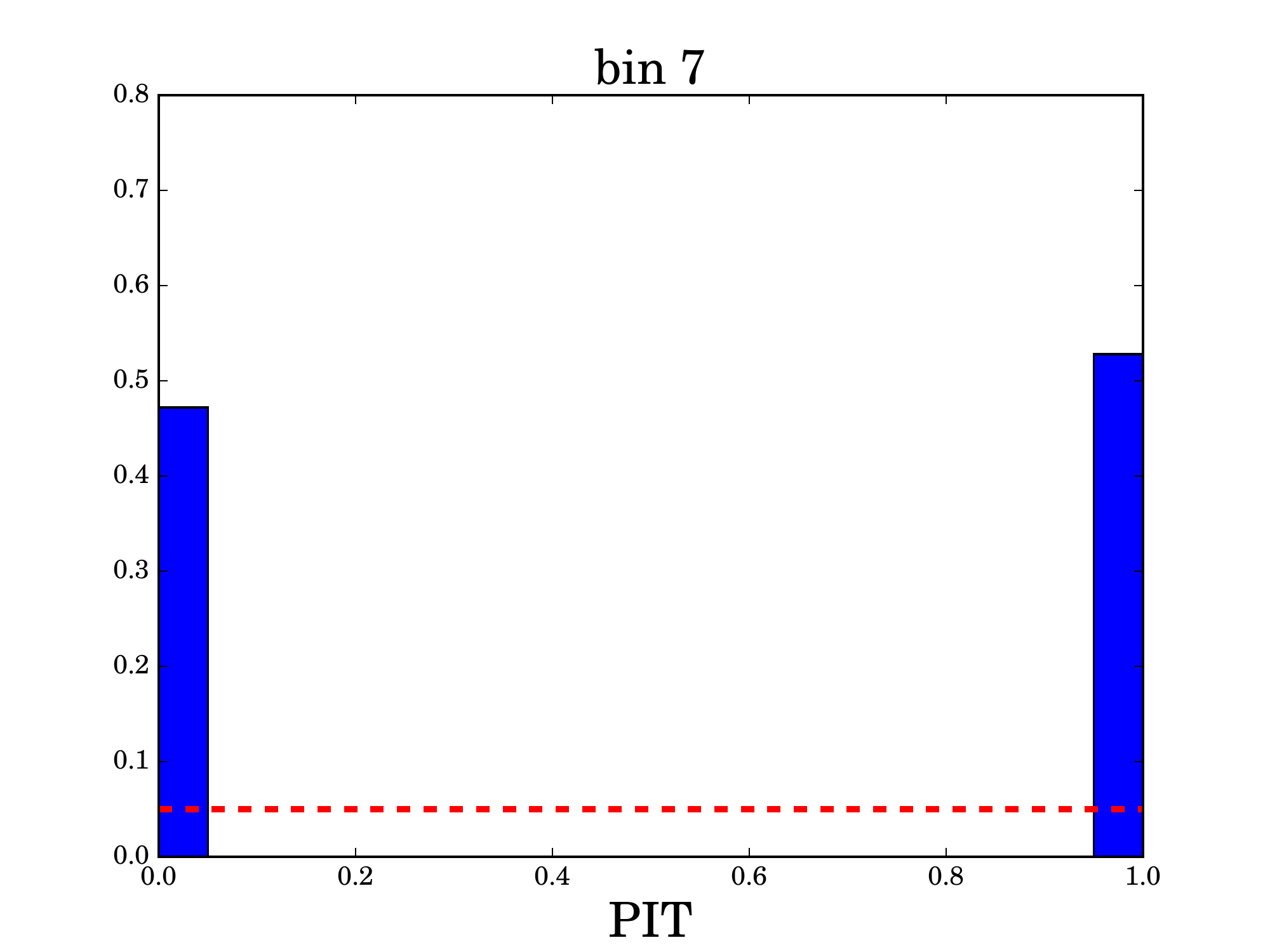}}
   {\includegraphics[width=0.245 \textwidth]{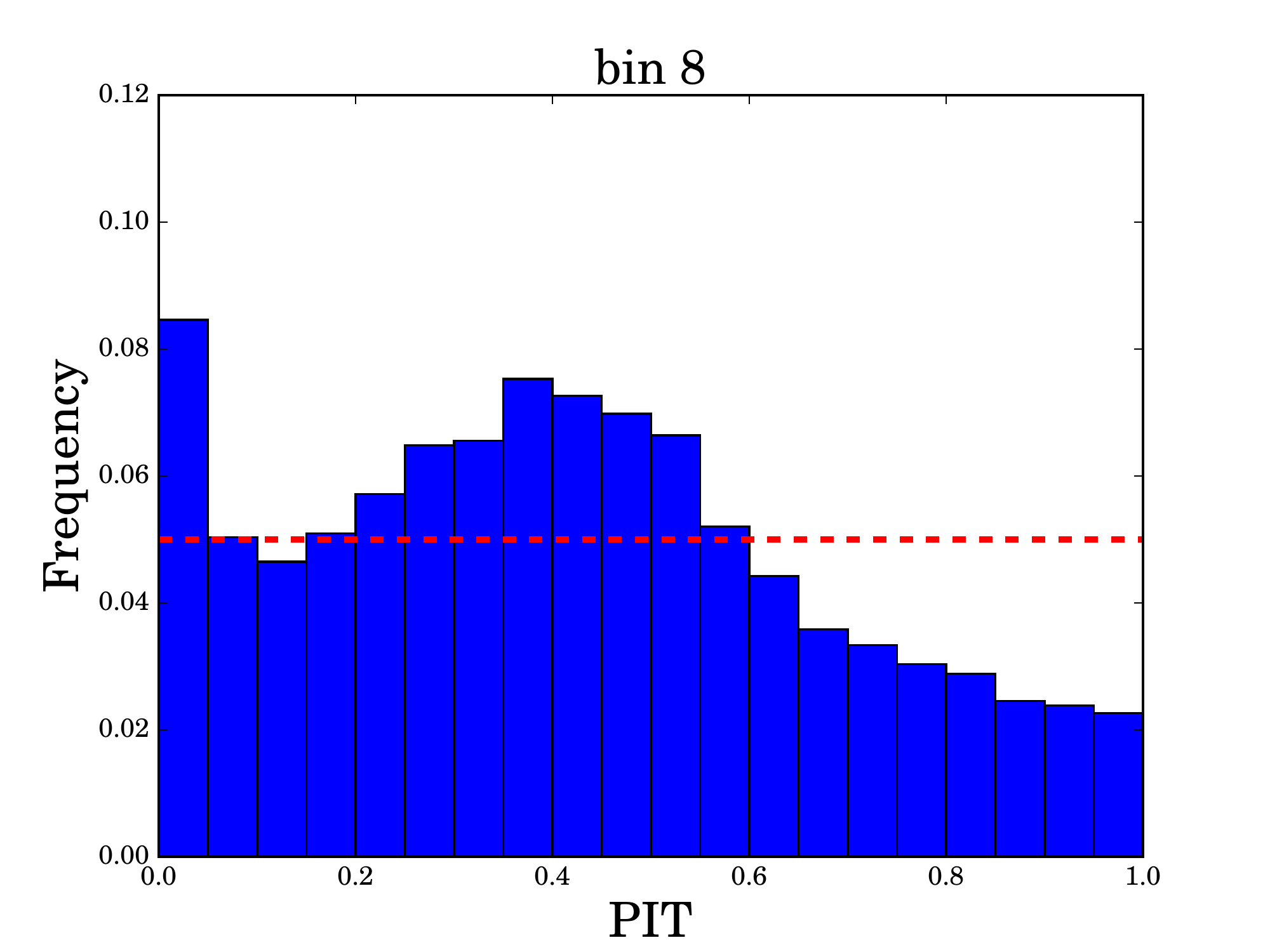}}
  {\includegraphics[width=0.245 \textwidth]{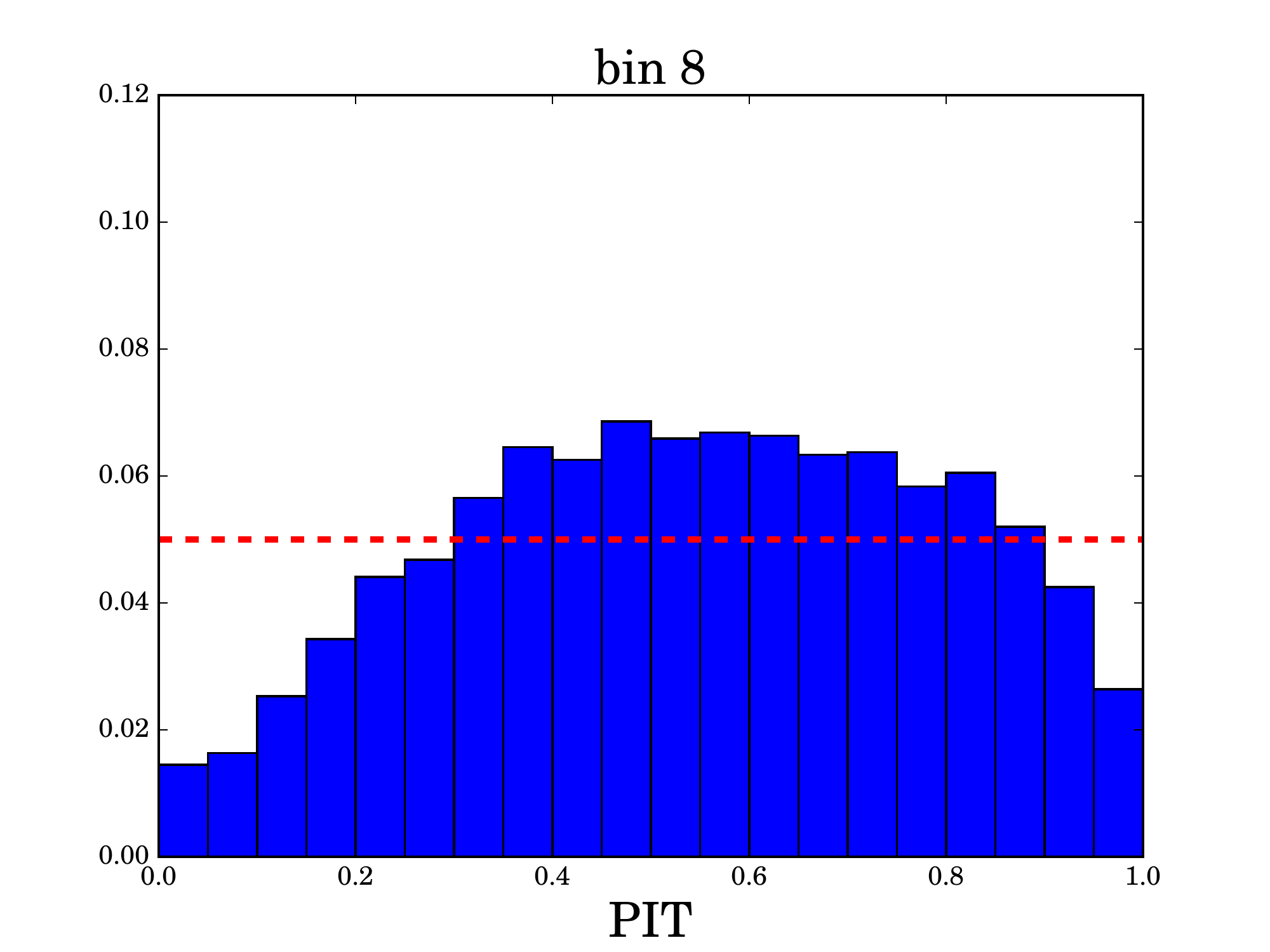}}
  {\includegraphics[width=0.245 \textwidth]{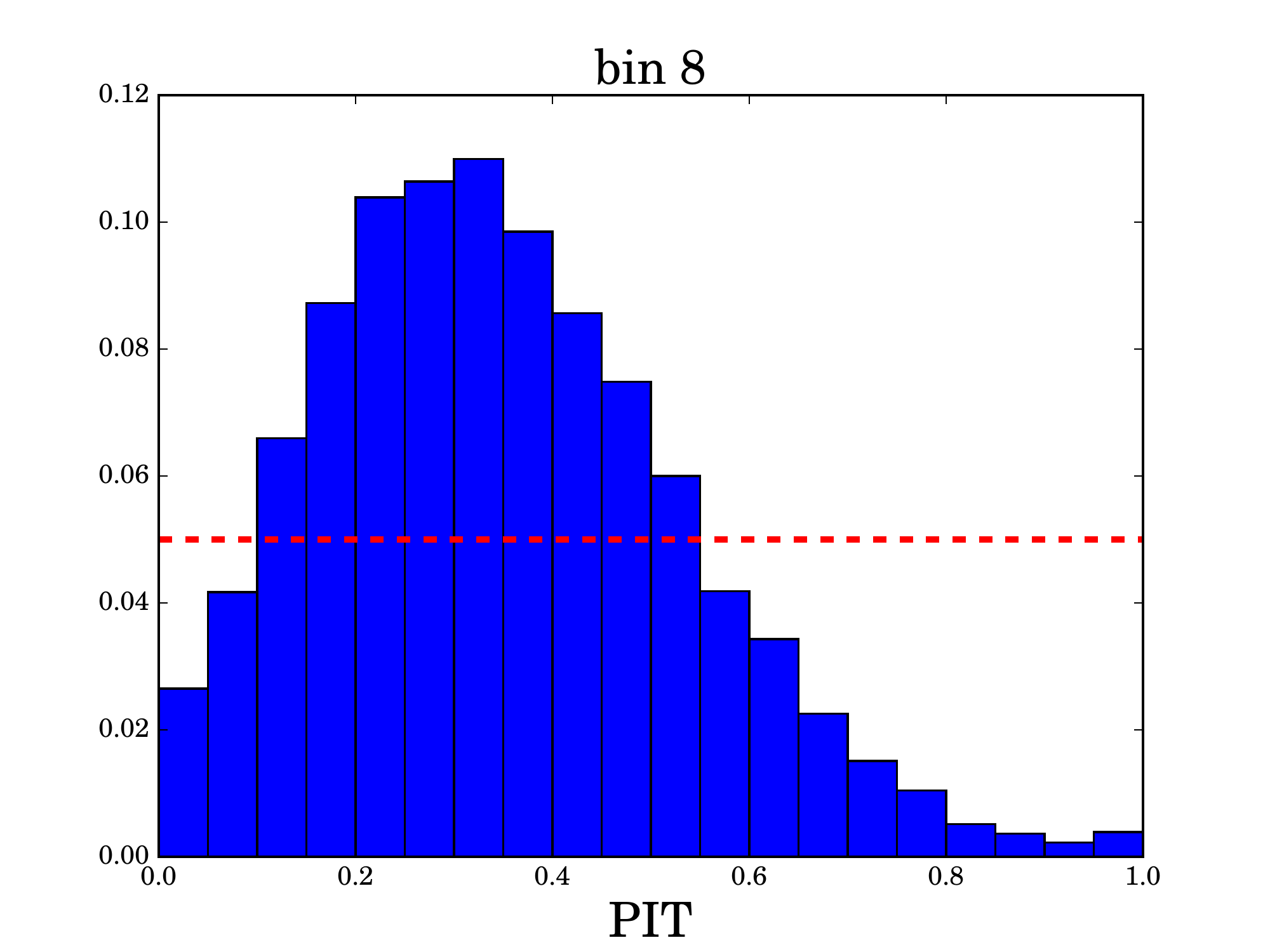}}
  {\includegraphics[width=0.245\textwidth]{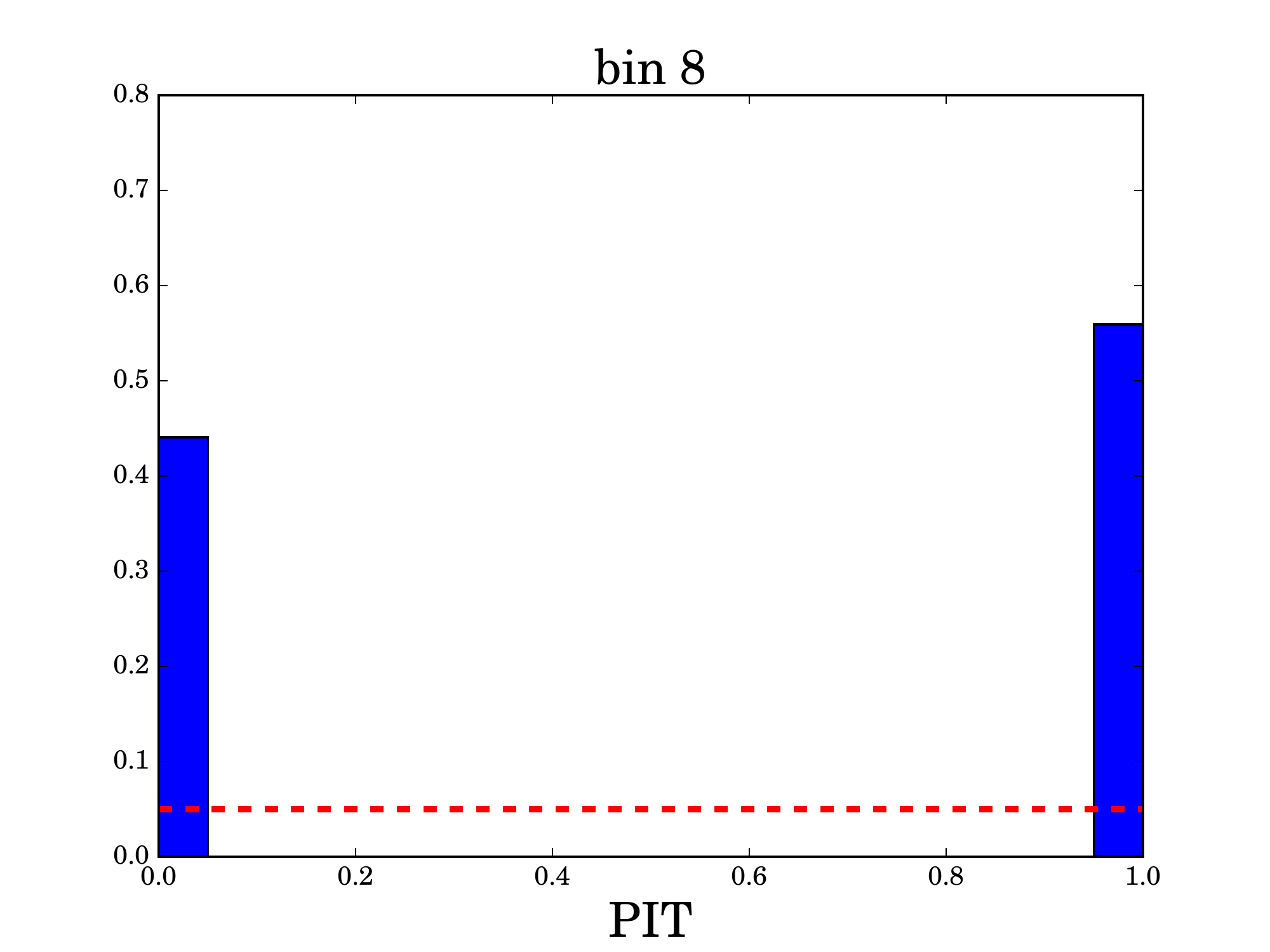}}
   {\includegraphics[width=0.245 \textwidth]{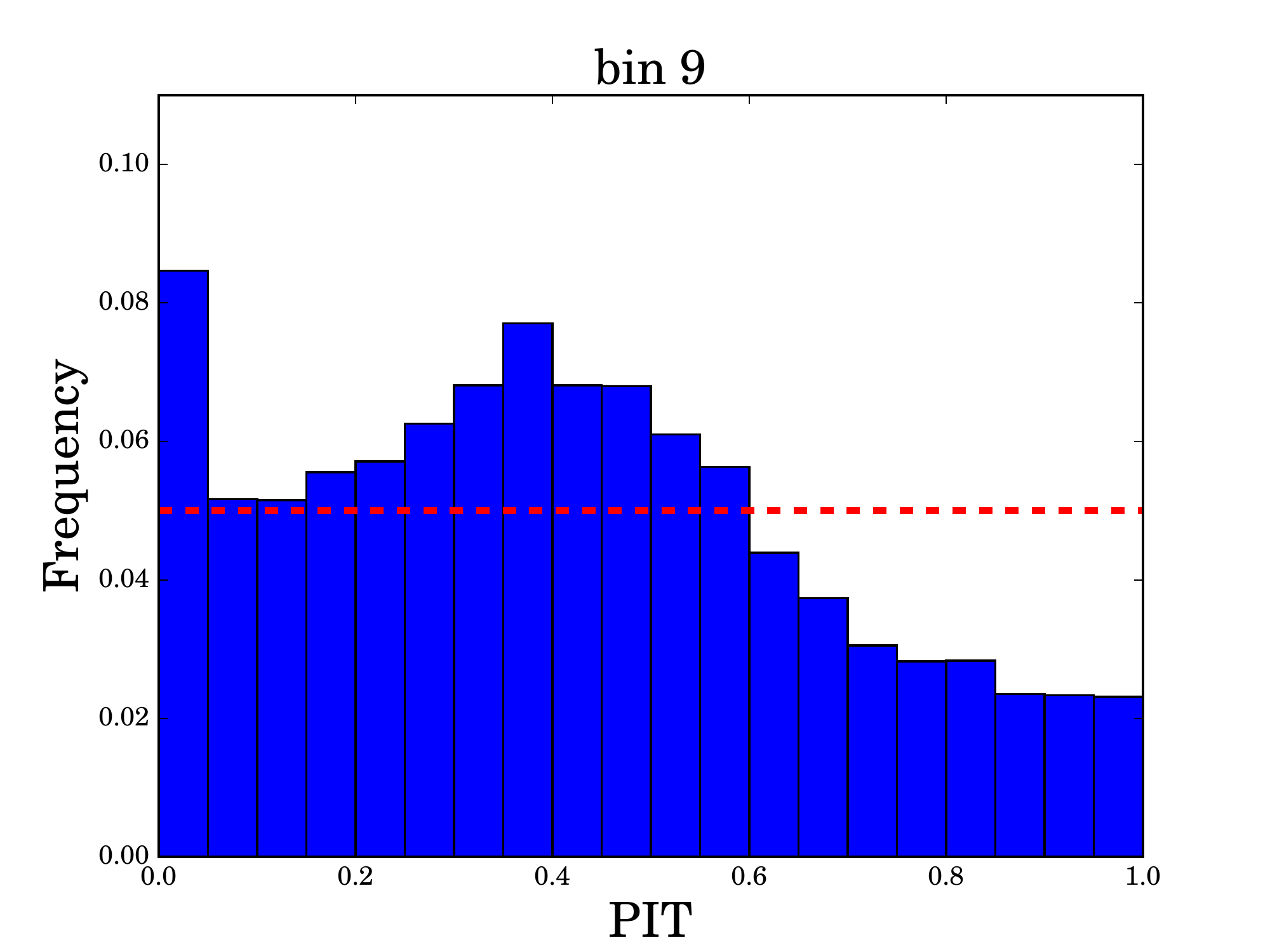}}
  {\includegraphics[width=0.245 \textwidth]{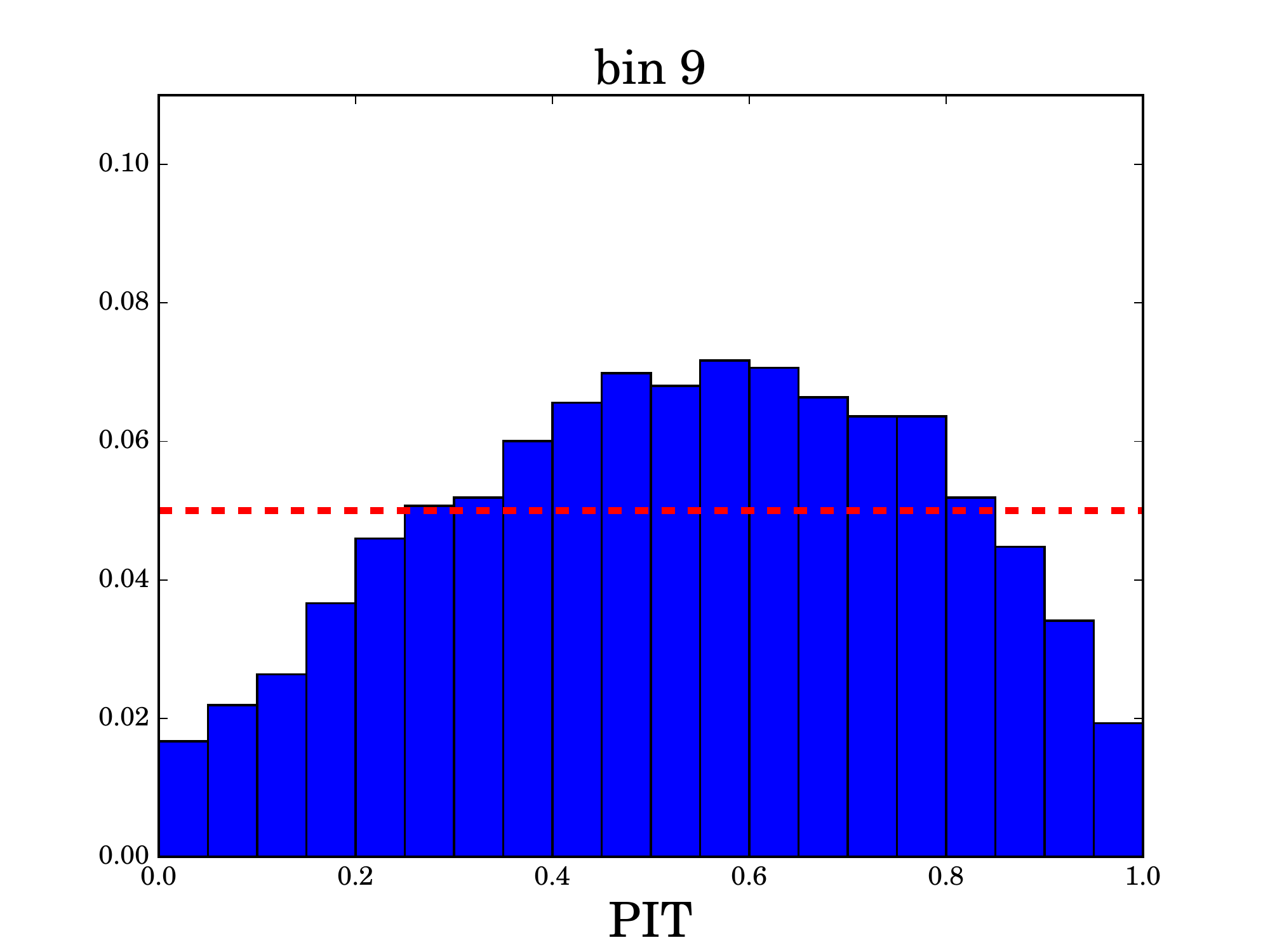}}
  {\includegraphics[width=0.245 \textwidth]{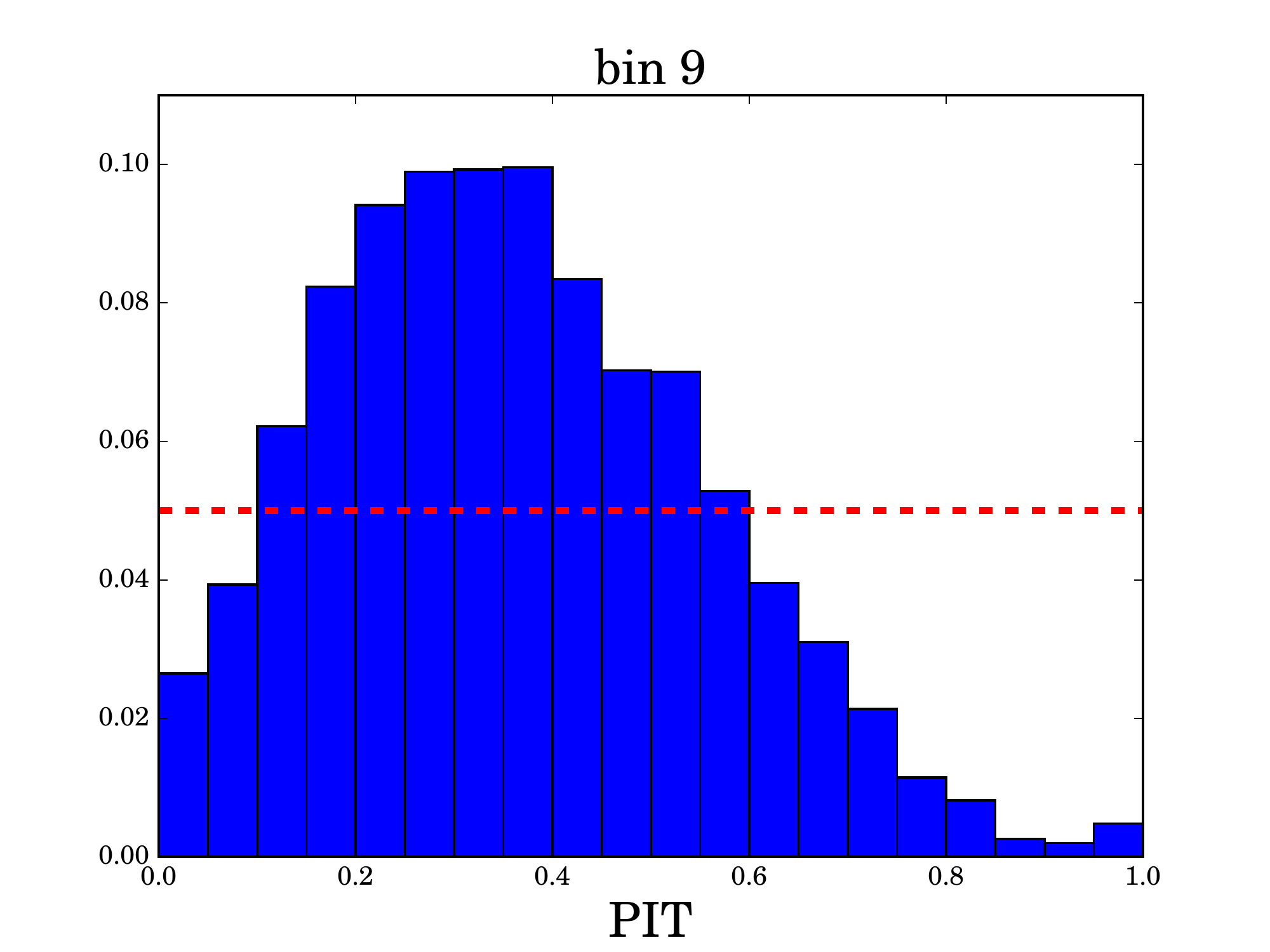}}
  {\includegraphics[width=0.245\textwidth]{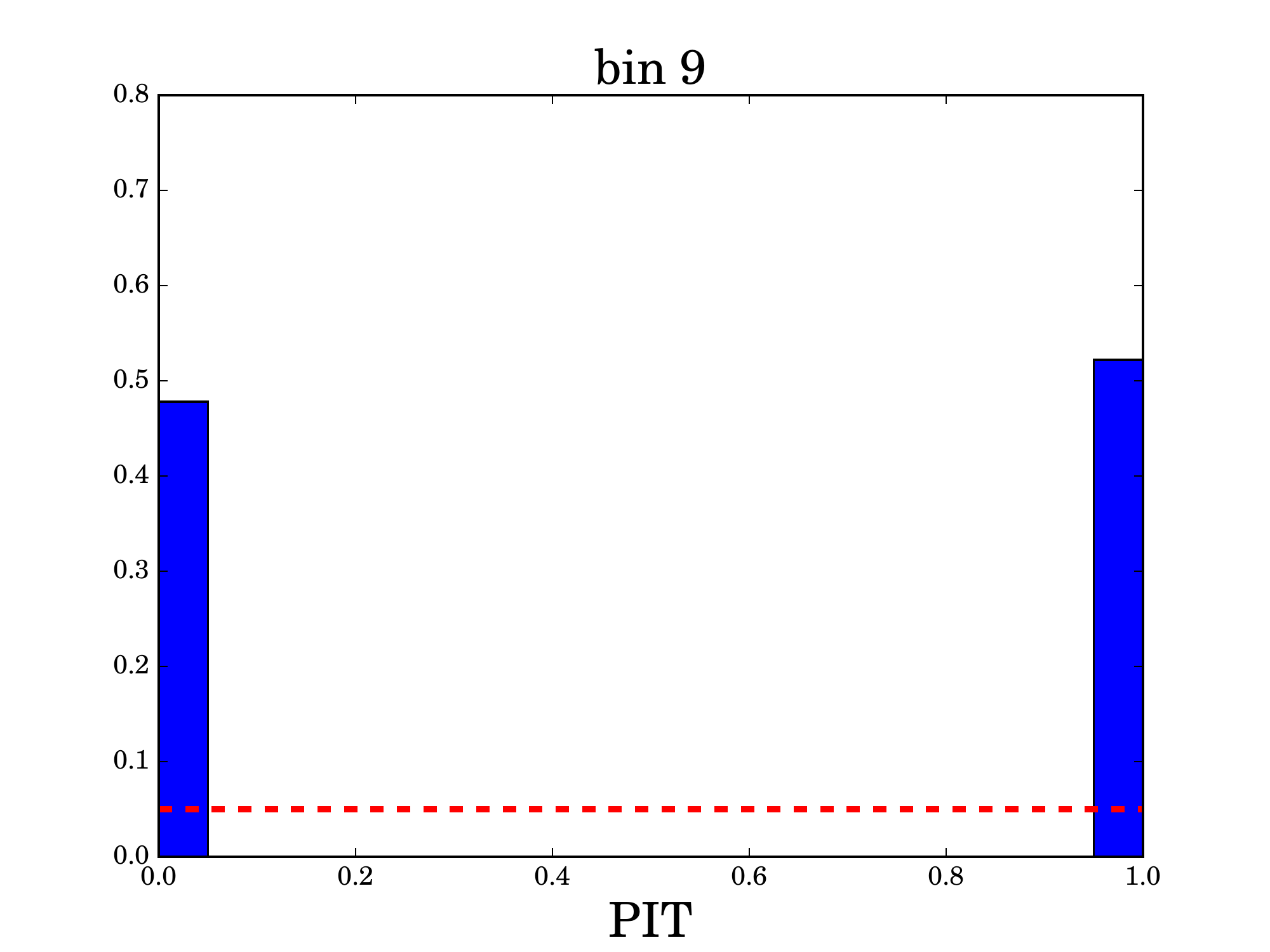}}
   {\includegraphics[width=0.245 \textwidth]{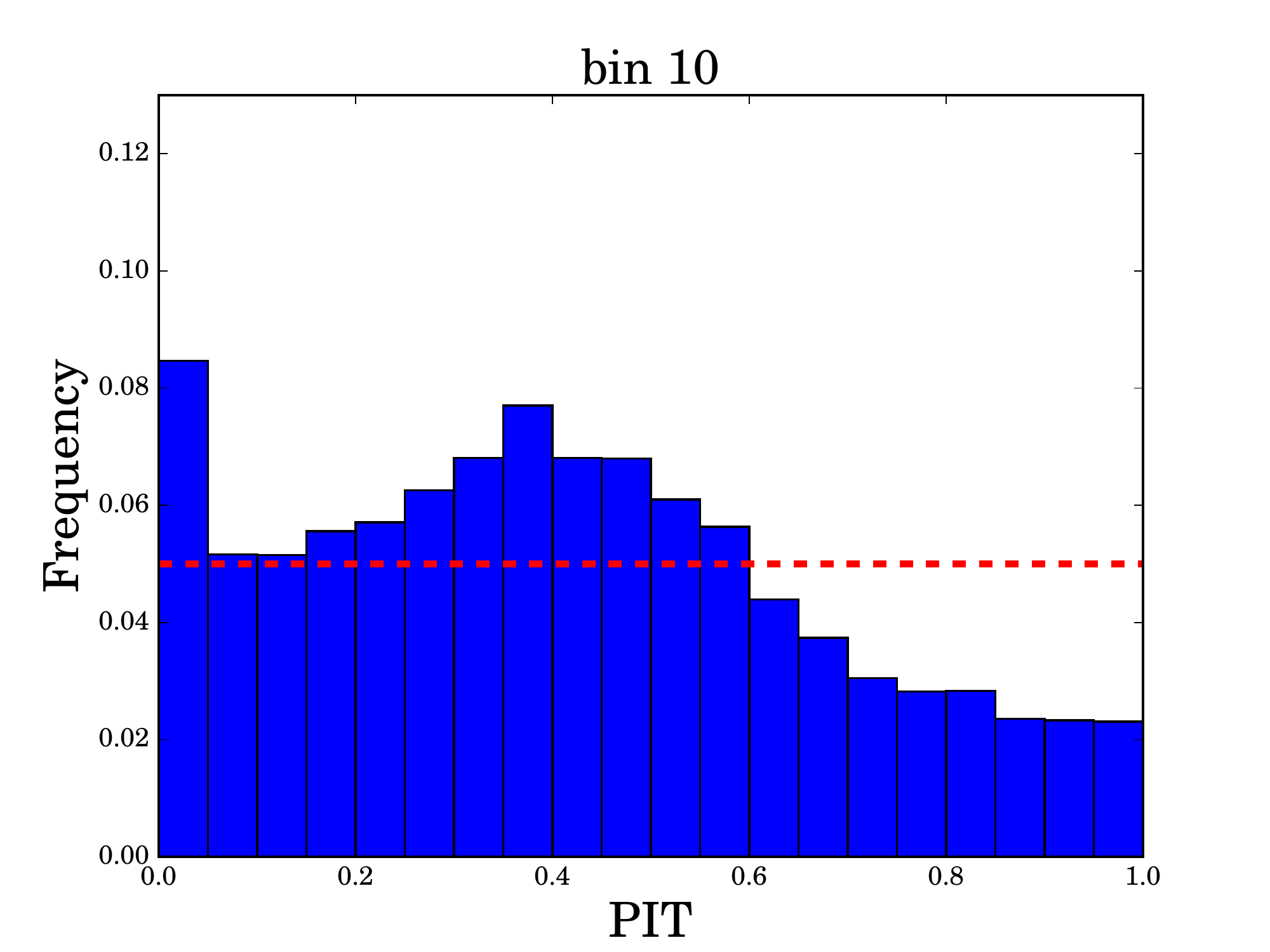}}
  {\includegraphics[width=0.245 \textwidth]{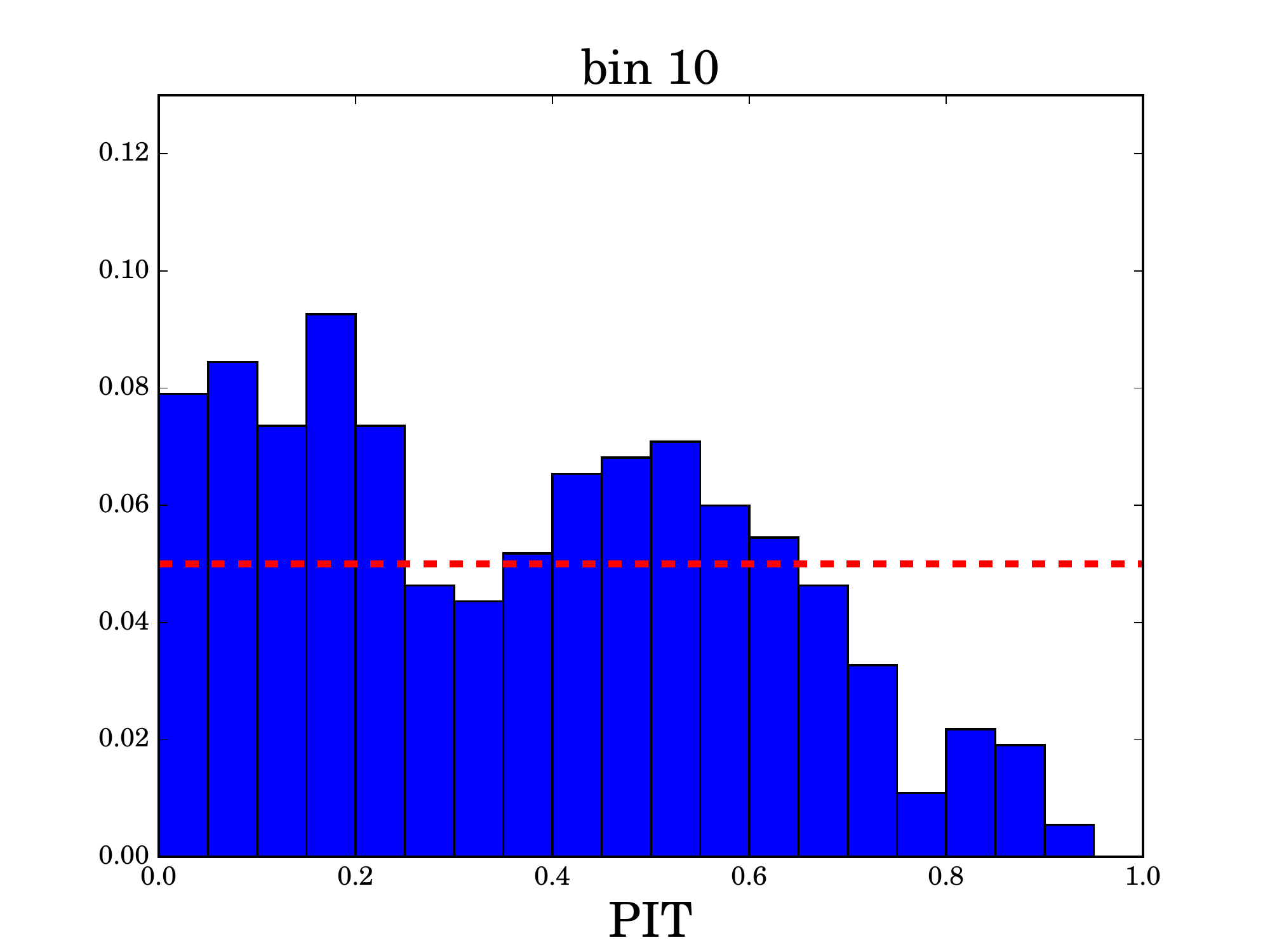}}
  {\includegraphics[width=0.245 \textwidth]{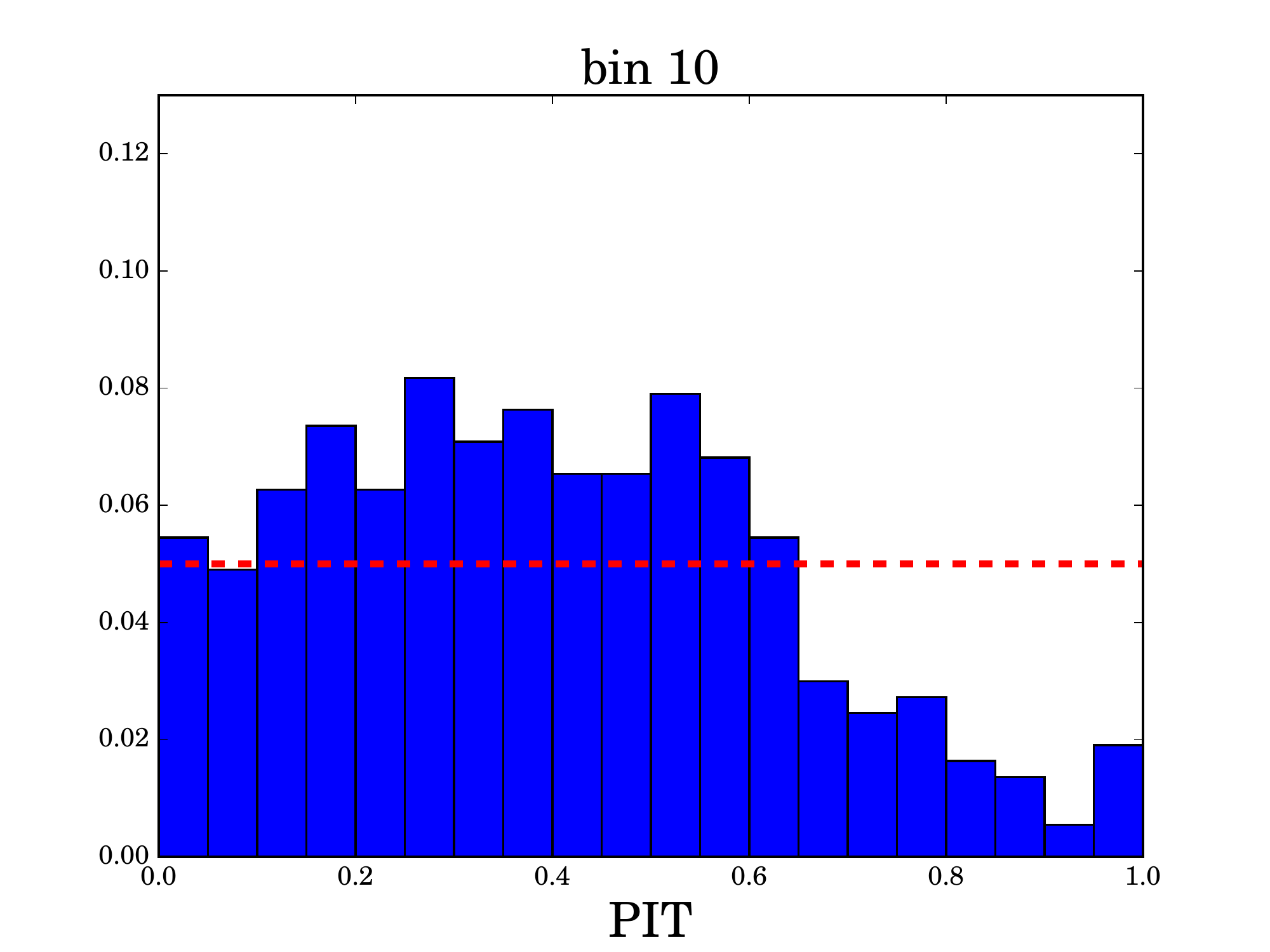}}
  {\includegraphics[width=0.245\textwidth]{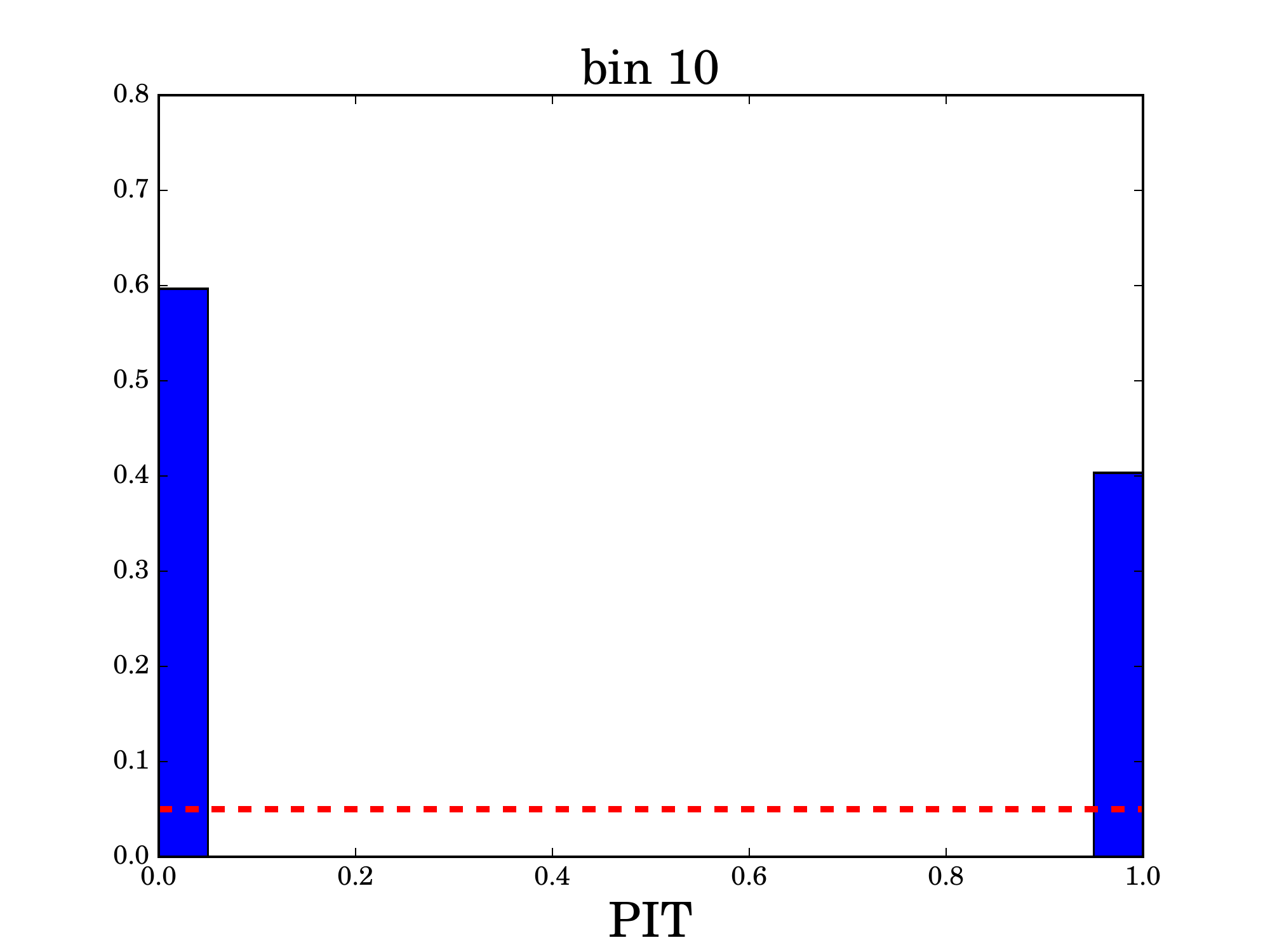}}
\caption{ Probability Integral Transform (PIT) obtained for METAPHOR (first  column panels), ANNz2 (second column panels), BPZ (third column panels), and for the \textit{dummy} PDF, calculated by METAPHOR (fourth column panels) in the second five magnitude tomographic bins from Table~\ref{tab:TOMOG}.}
\label{fig:pitALLtom2}
\end{figure*}

\begin{figure*}
 \centering
  {\includegraphics[width=0.30 \textwidth]{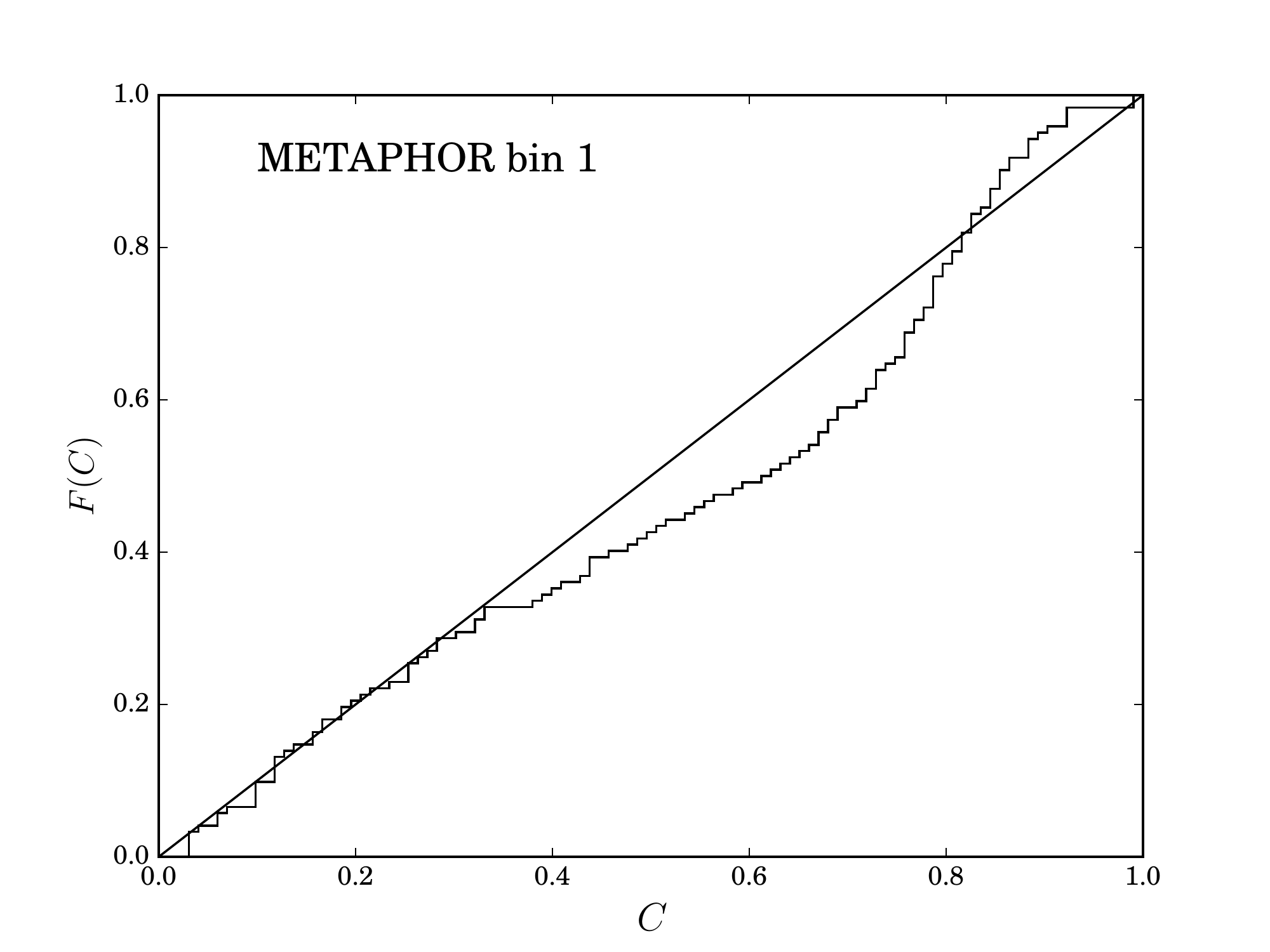}}
  {\includegraphics[width=0.30 \textwidth]{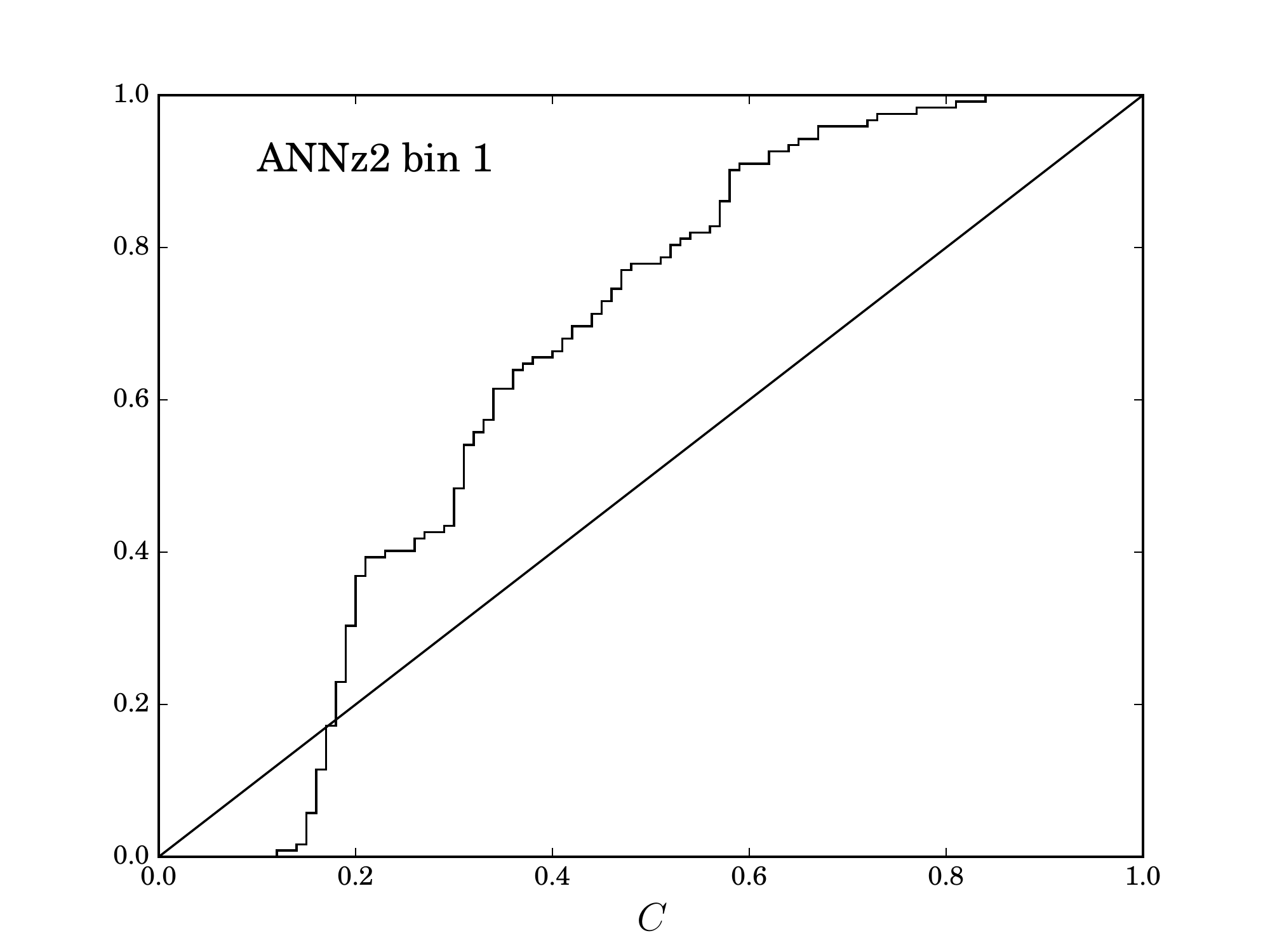}}
  {\includegraphics[width=0.30 \textwidth]{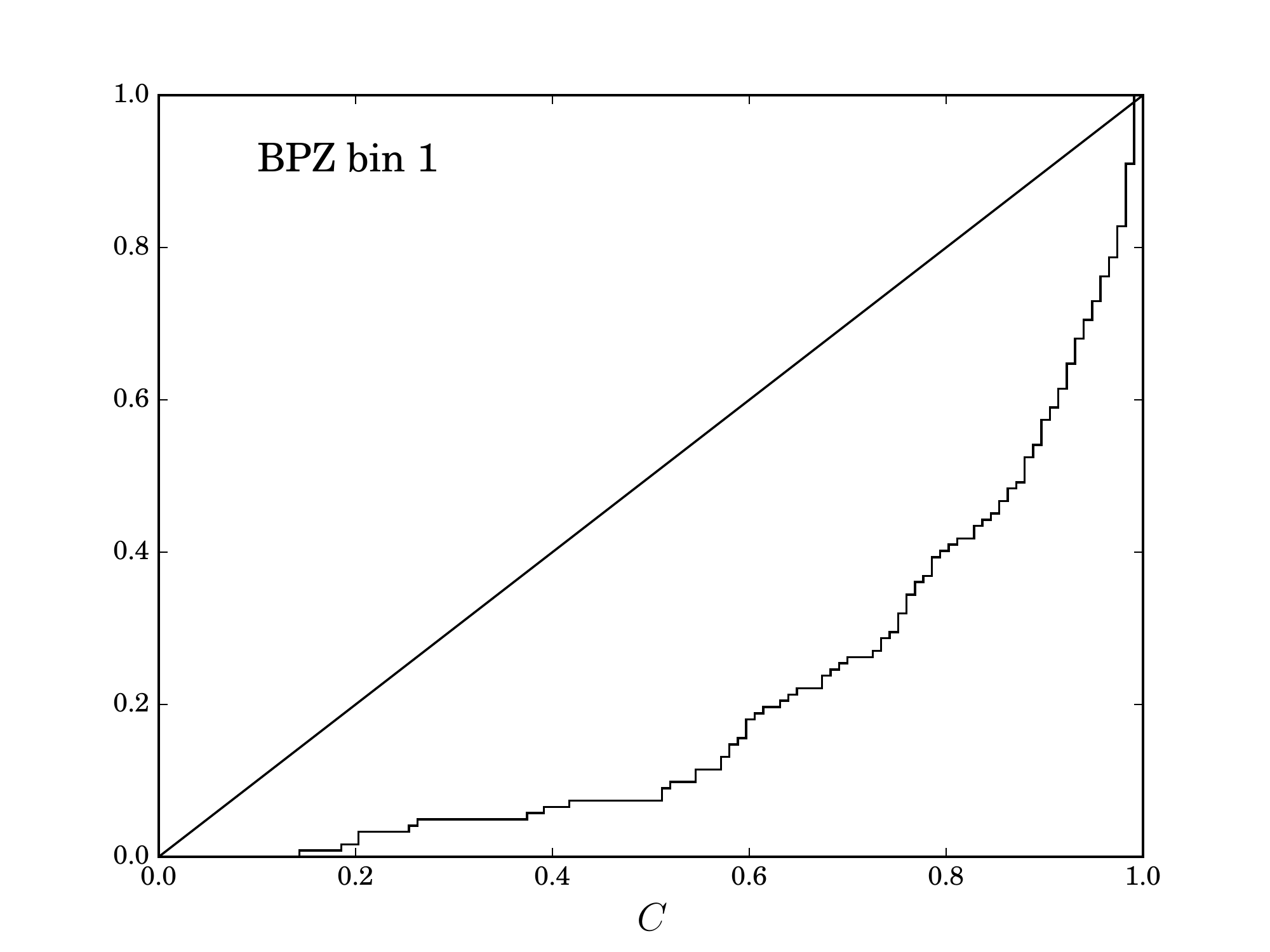}}
  {\includegraphics[width=0.30 \textwidth]{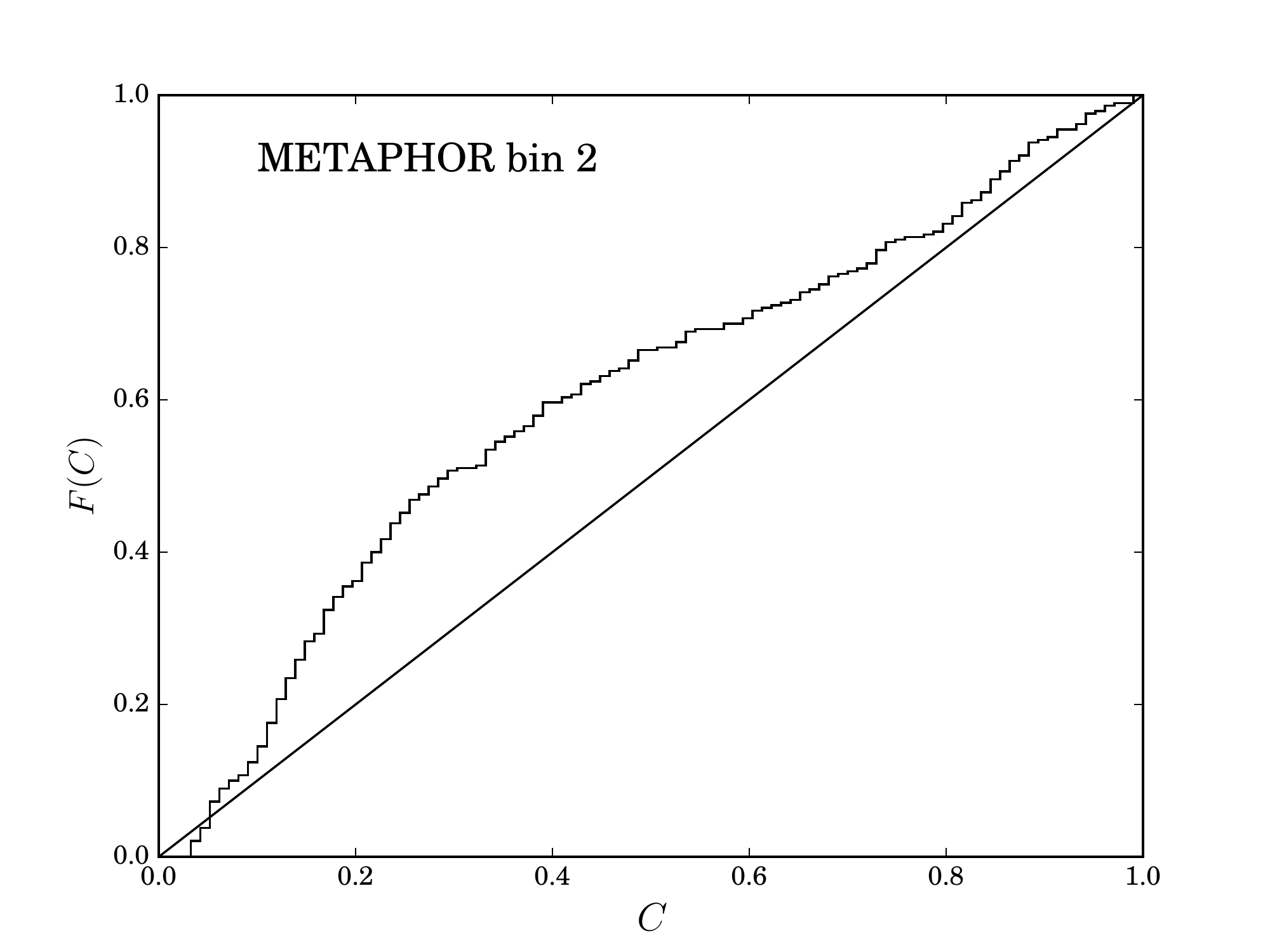}}
  {\includegraphics[width=0.30 \textwidth]{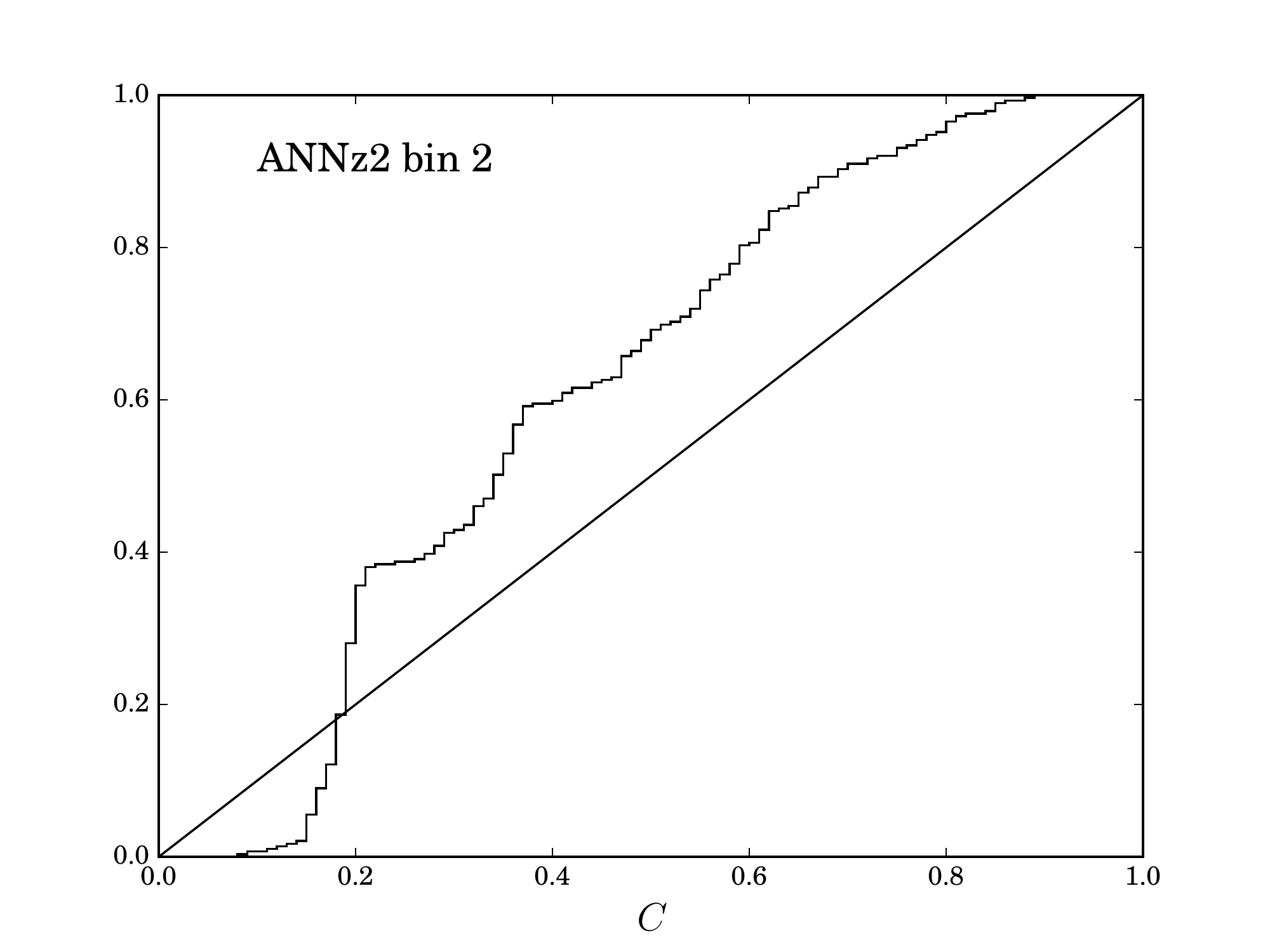}}
  {\includegraphics[width=0.30 \textwidth]{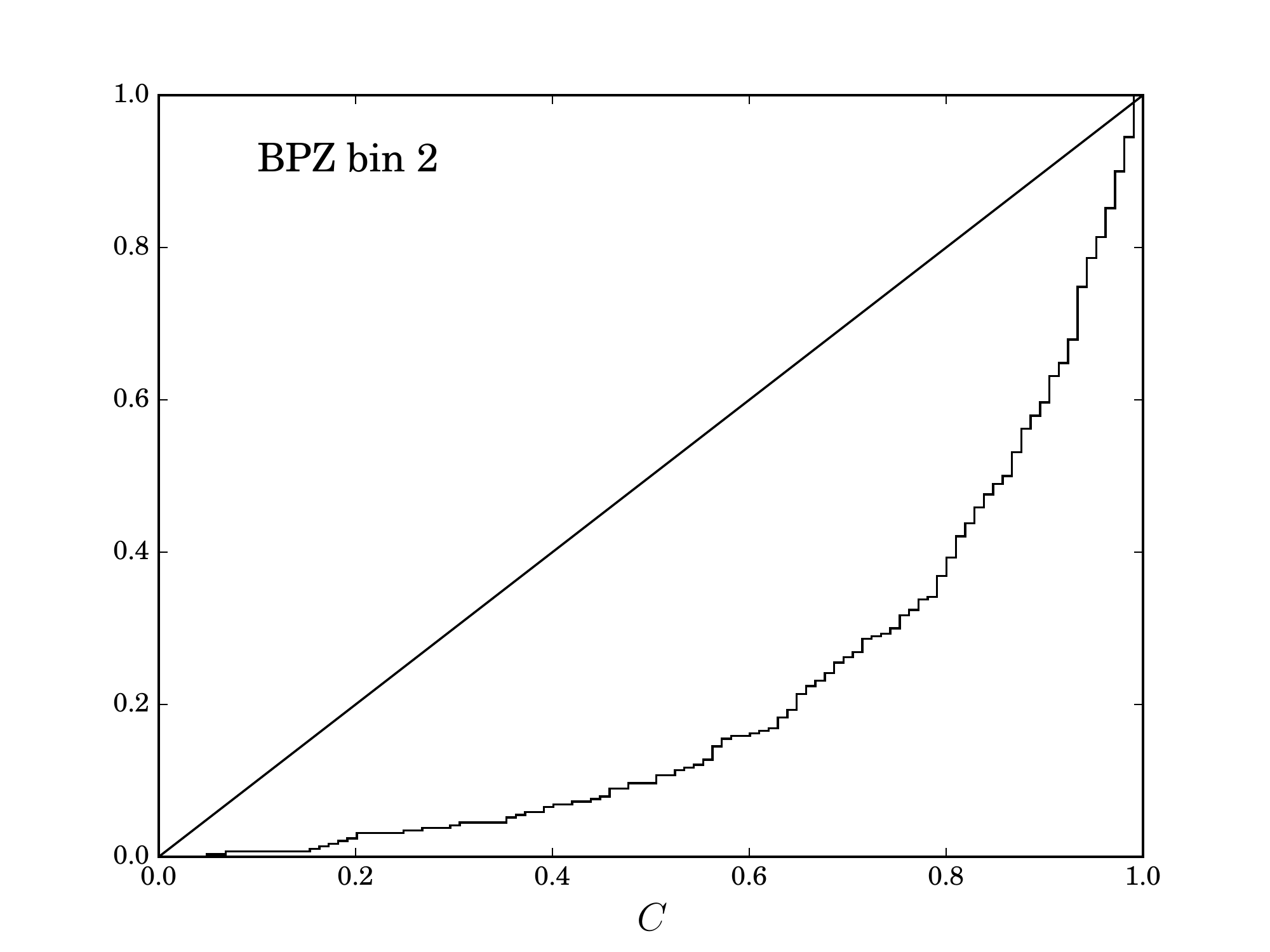}}
  {\includegraphics[width=0.30 \textwidth]{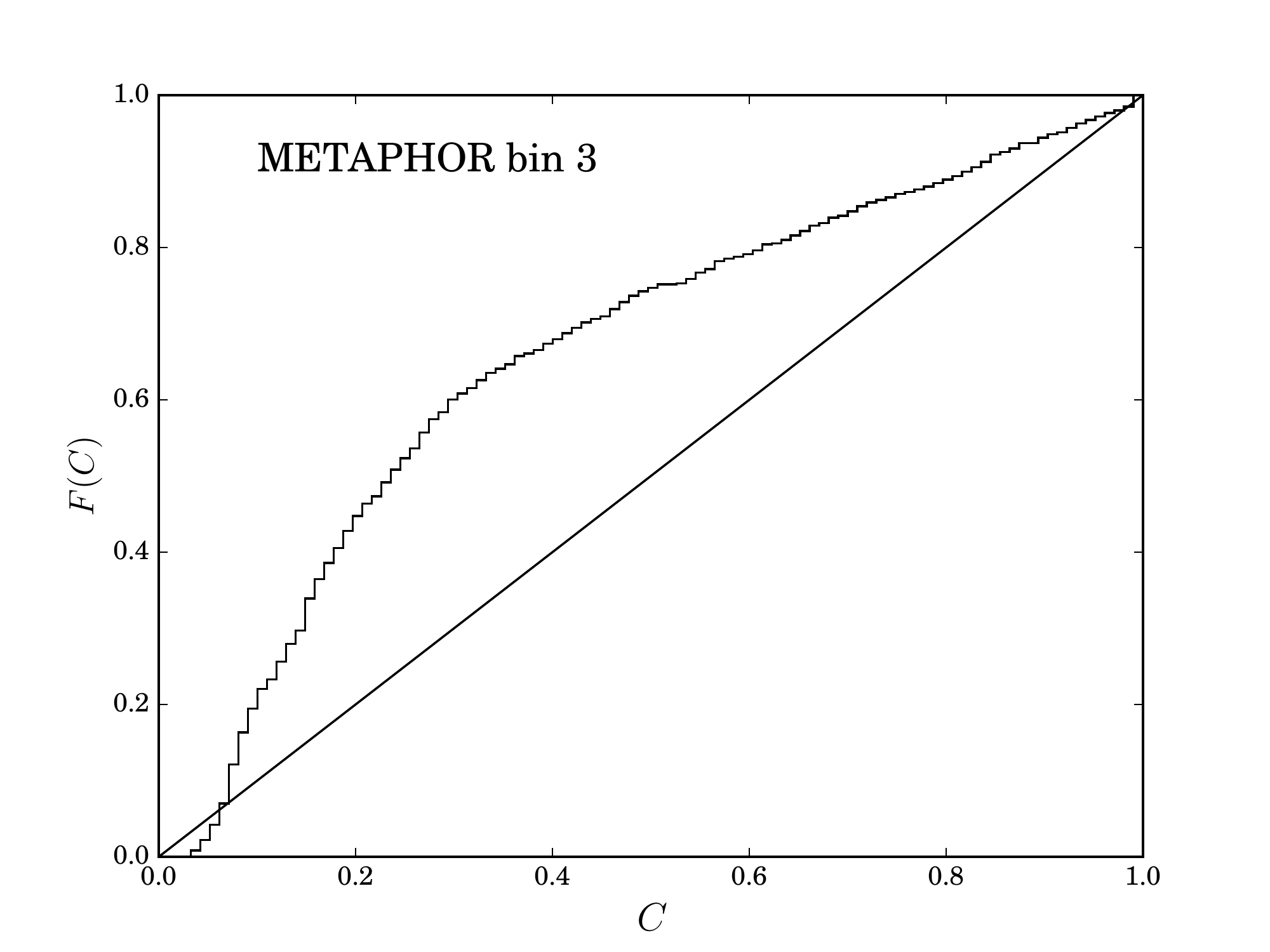}}
  {\includegraphics[width=0.30 \textwidth]{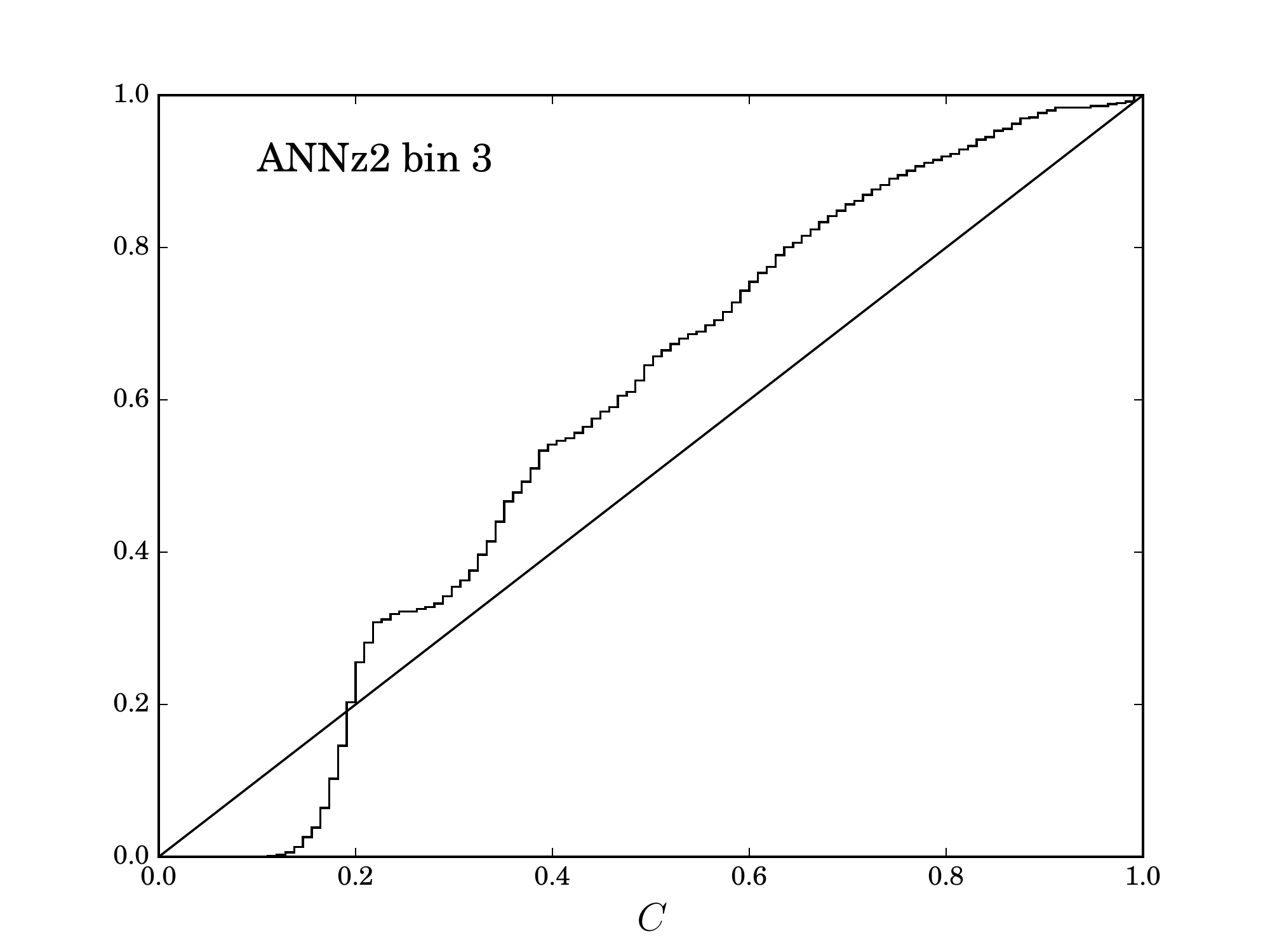}}
  {\includegraphics[width=0.30 \textwidth]{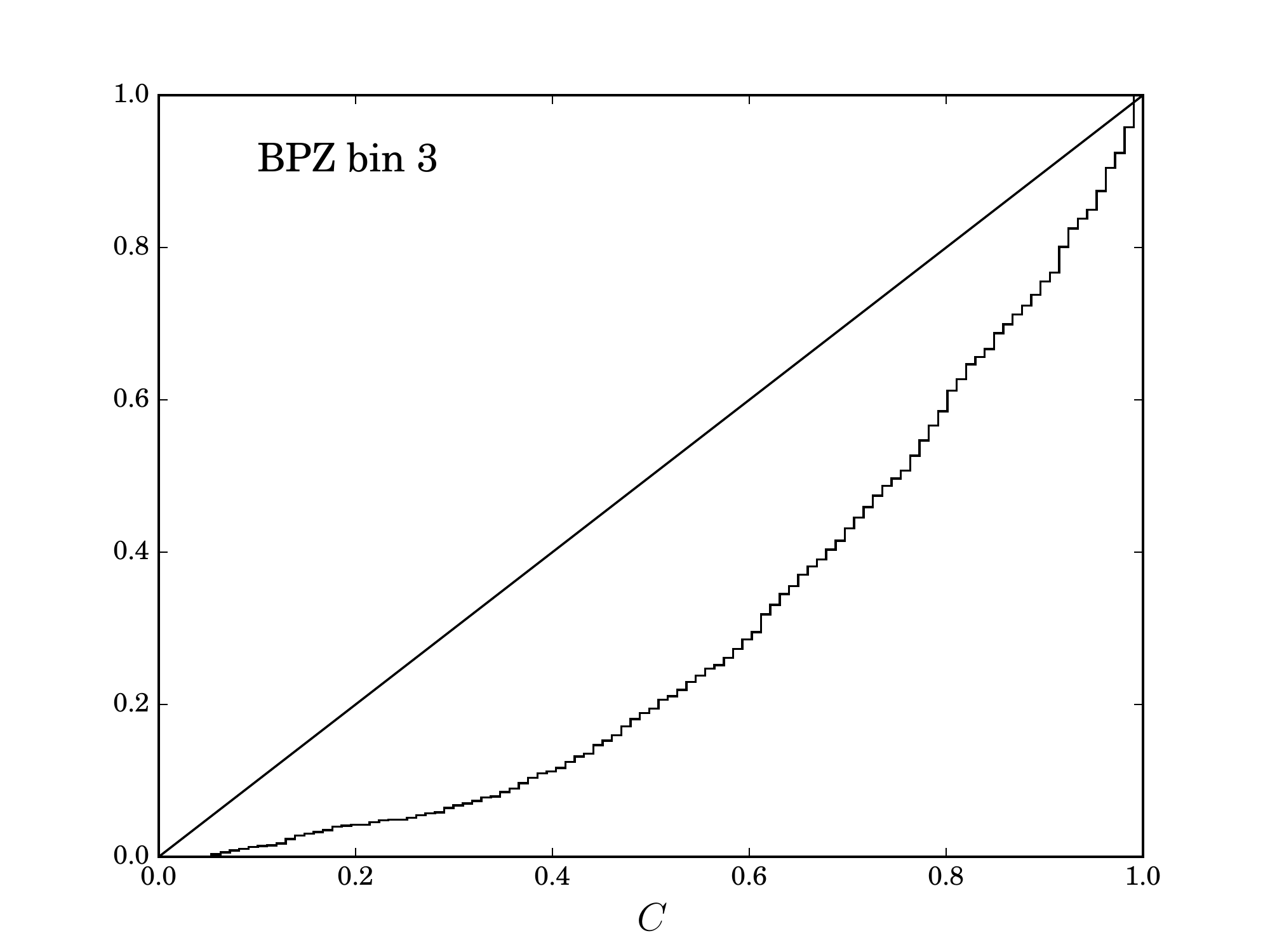}}
  {\includegraphics[width=0.30 \textwidth]{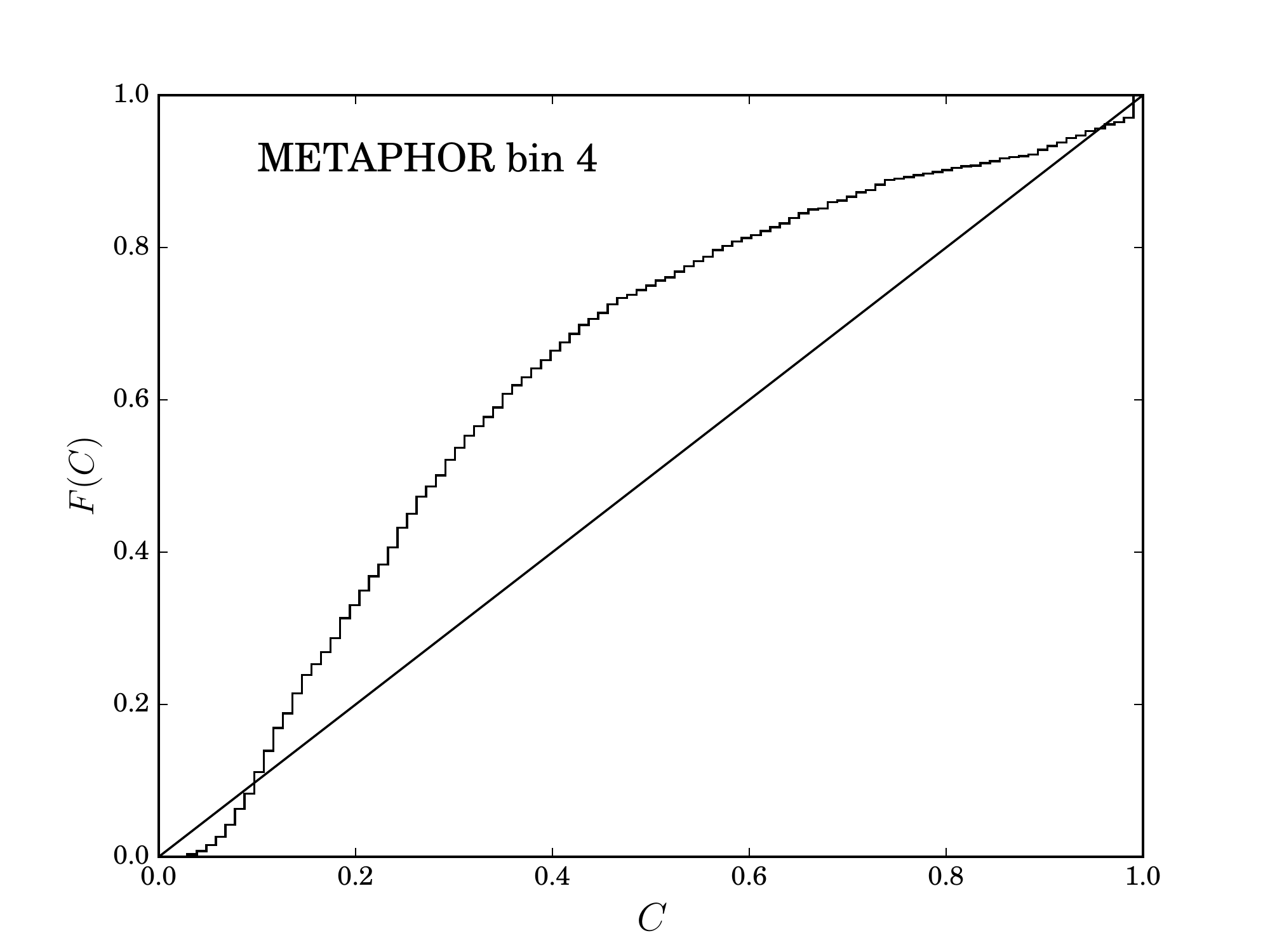}}
  {\includegraphics[width=0.30 \textwidth]{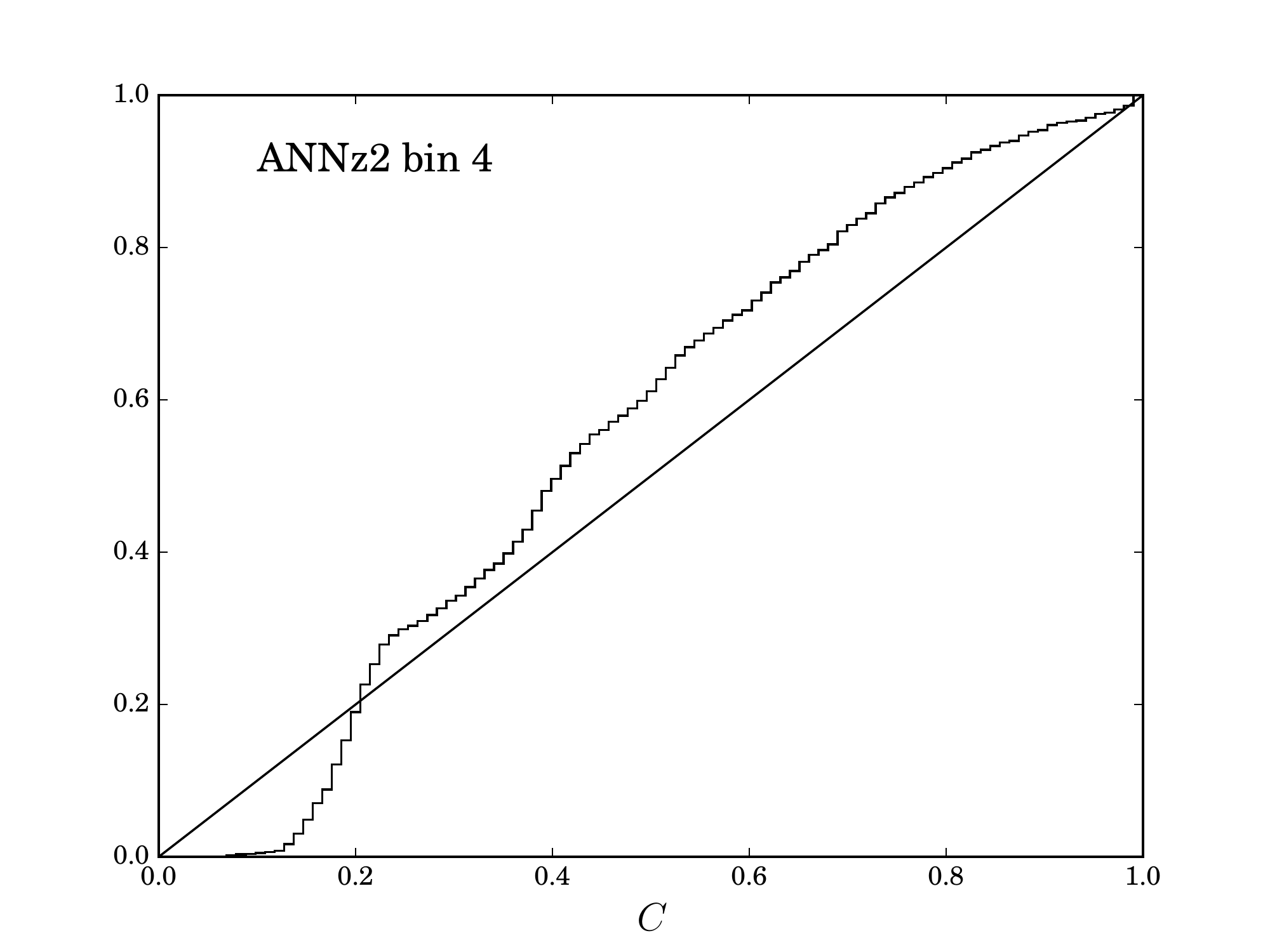}}
  {\includegraphics[width=0.30 \textwidth]{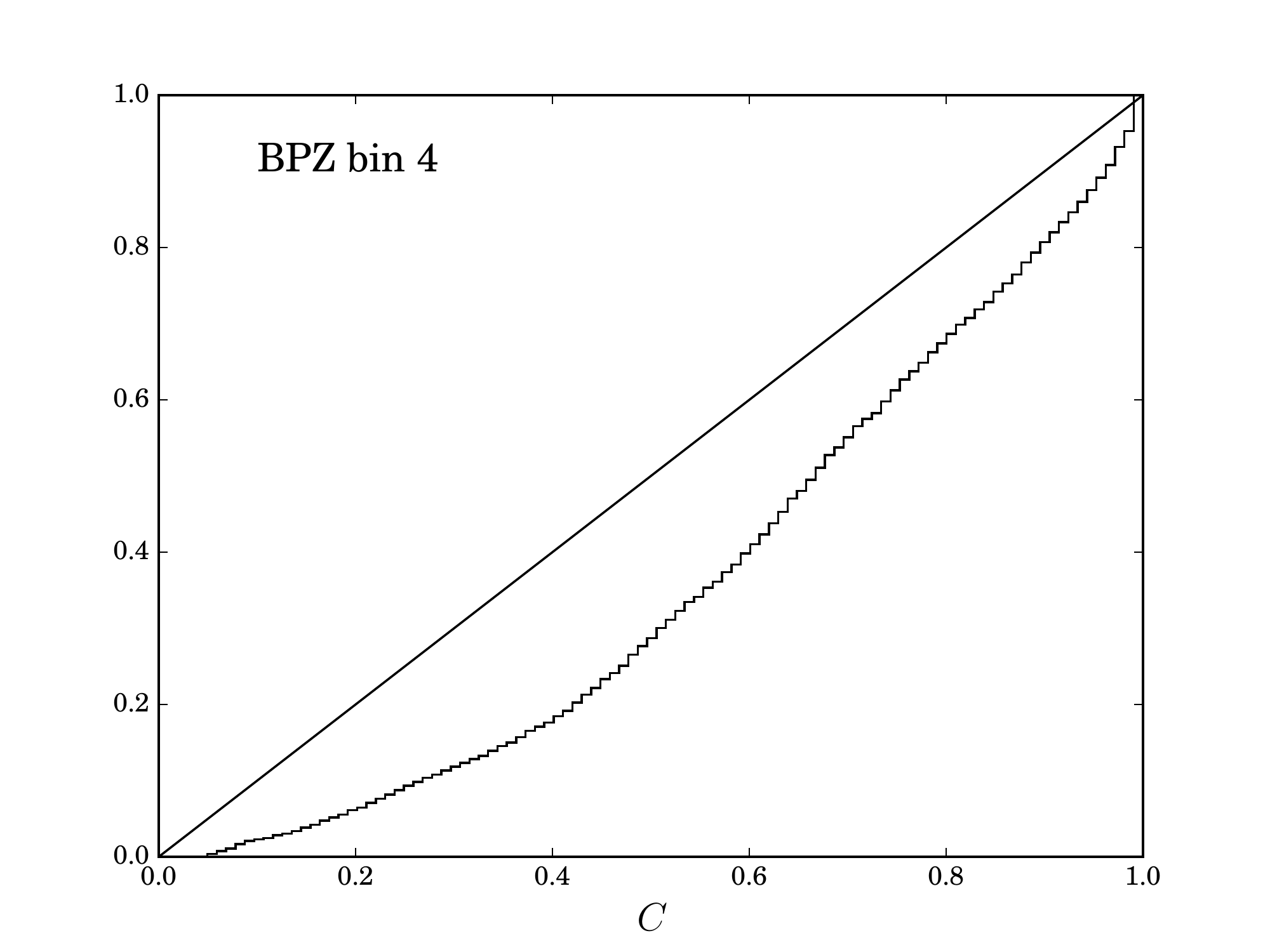}}
  {\includegraphics[width=0.30 \textwidth]{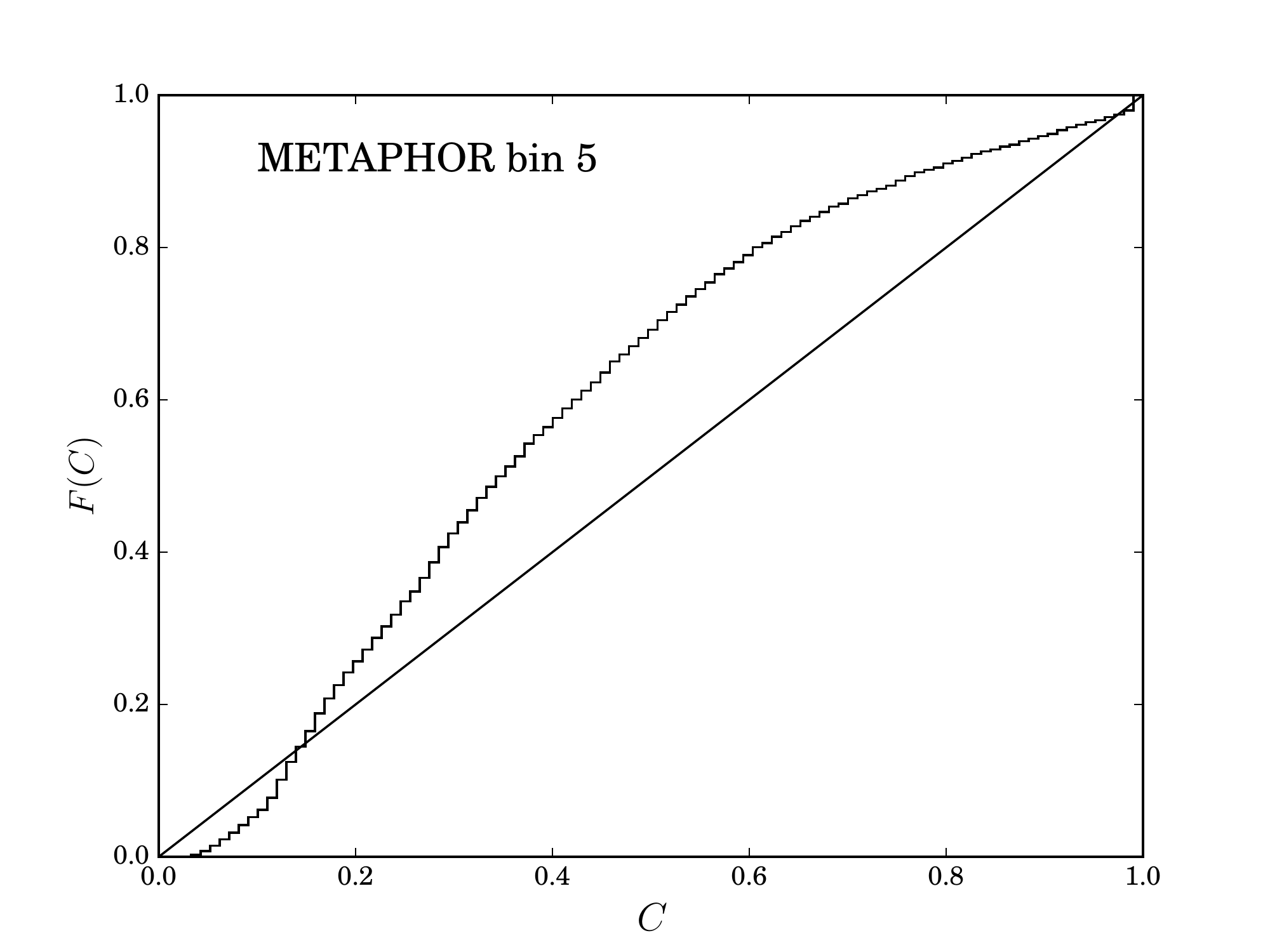}}
  {\includegraphics[width=0.30 \textwidth]{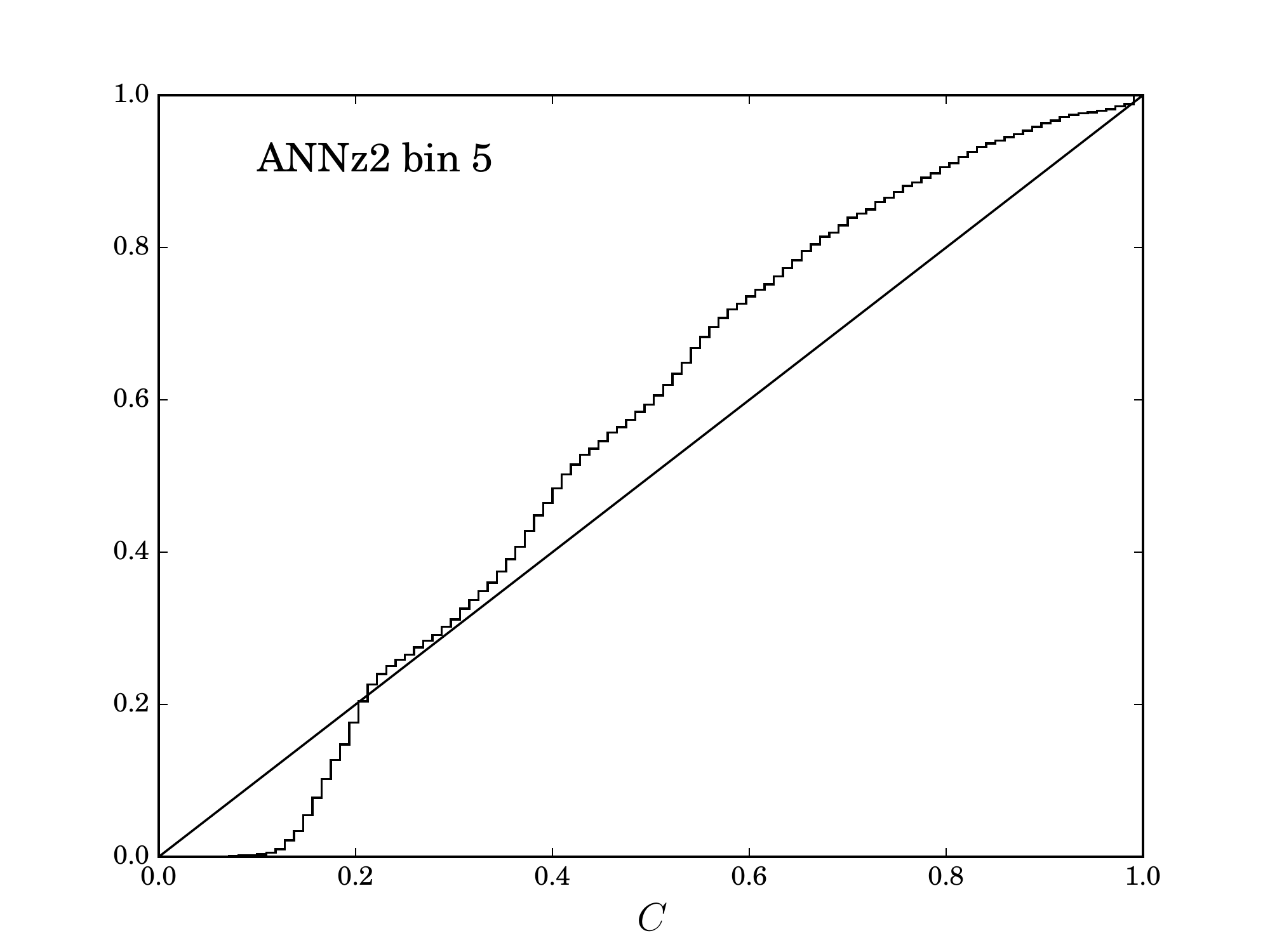}}
  {\includegraphics[width=0.30 \textwidth]{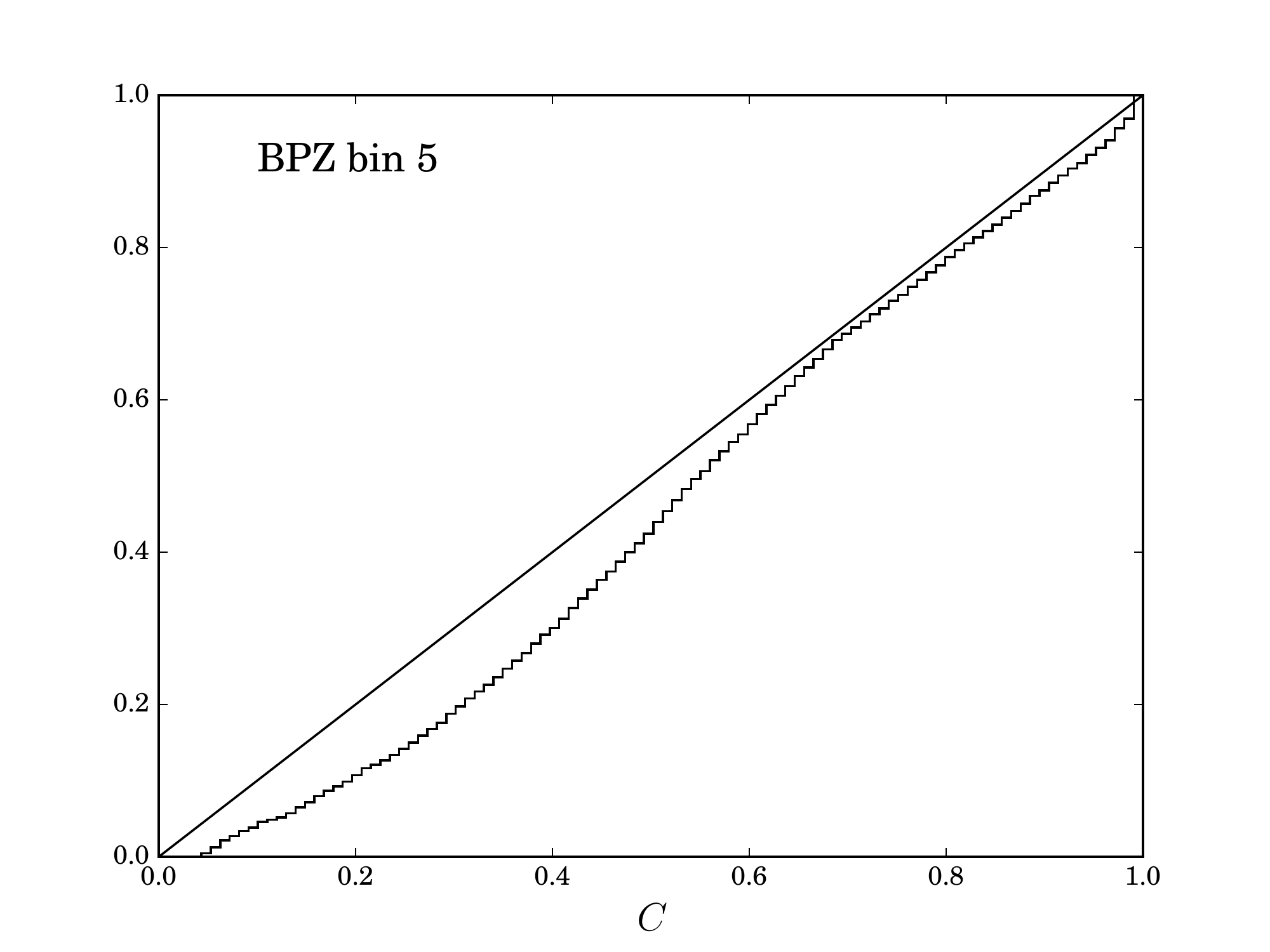}}
\caption{Credibility analysis (see Sec.~\ref{SEC:statindicators}) obtained for METAPHOR, ANNz2 and BPZ for the first five magnitude tomographic bins from Table~\ref{tab:TOMOG}. The credibility plots for the \textit{dummy} PDF are the same as the bottom right panel of Fig.~\ref{fig:WittmanALL} in all the bins.} 
\label{fig:WittmanALLtom1}
\end{figure*}

\begin{figure*}
 \centering
  {\includegraphics[width=0.30 \textwidth]{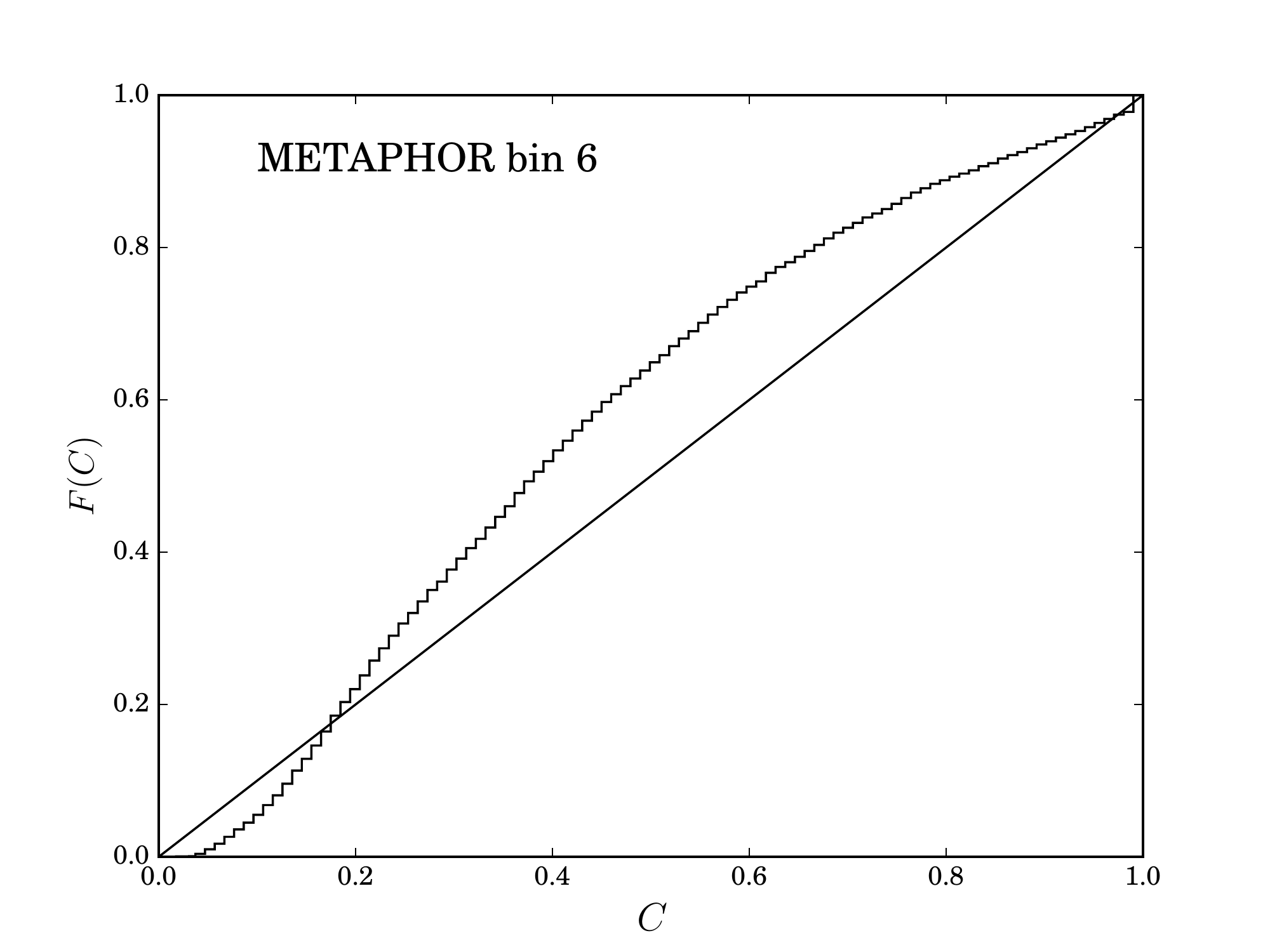}}
  {\includegraphics[width=0.30 \textwidth]{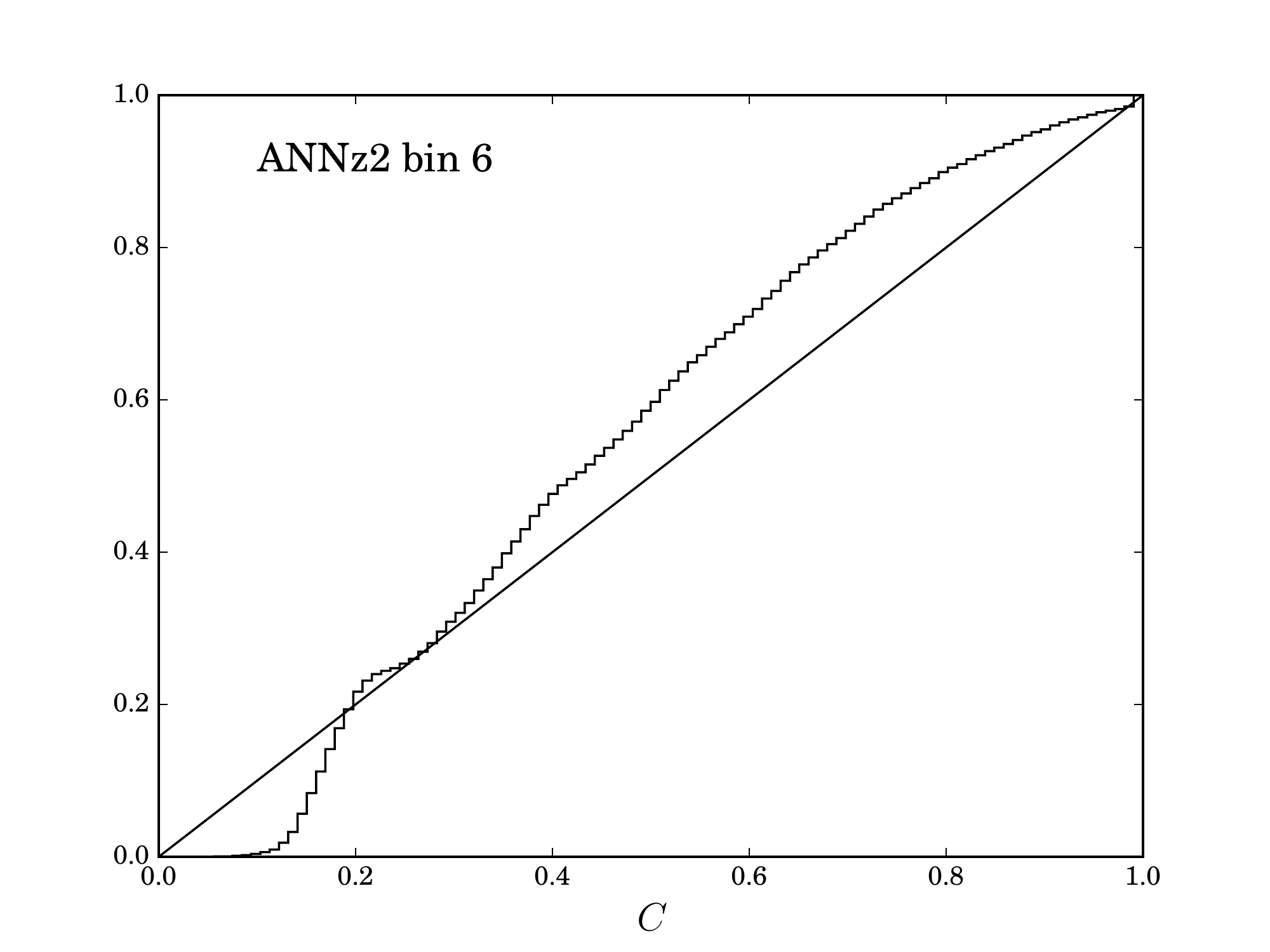}}
  {\includegraphics[width=0.30 \textwidth]{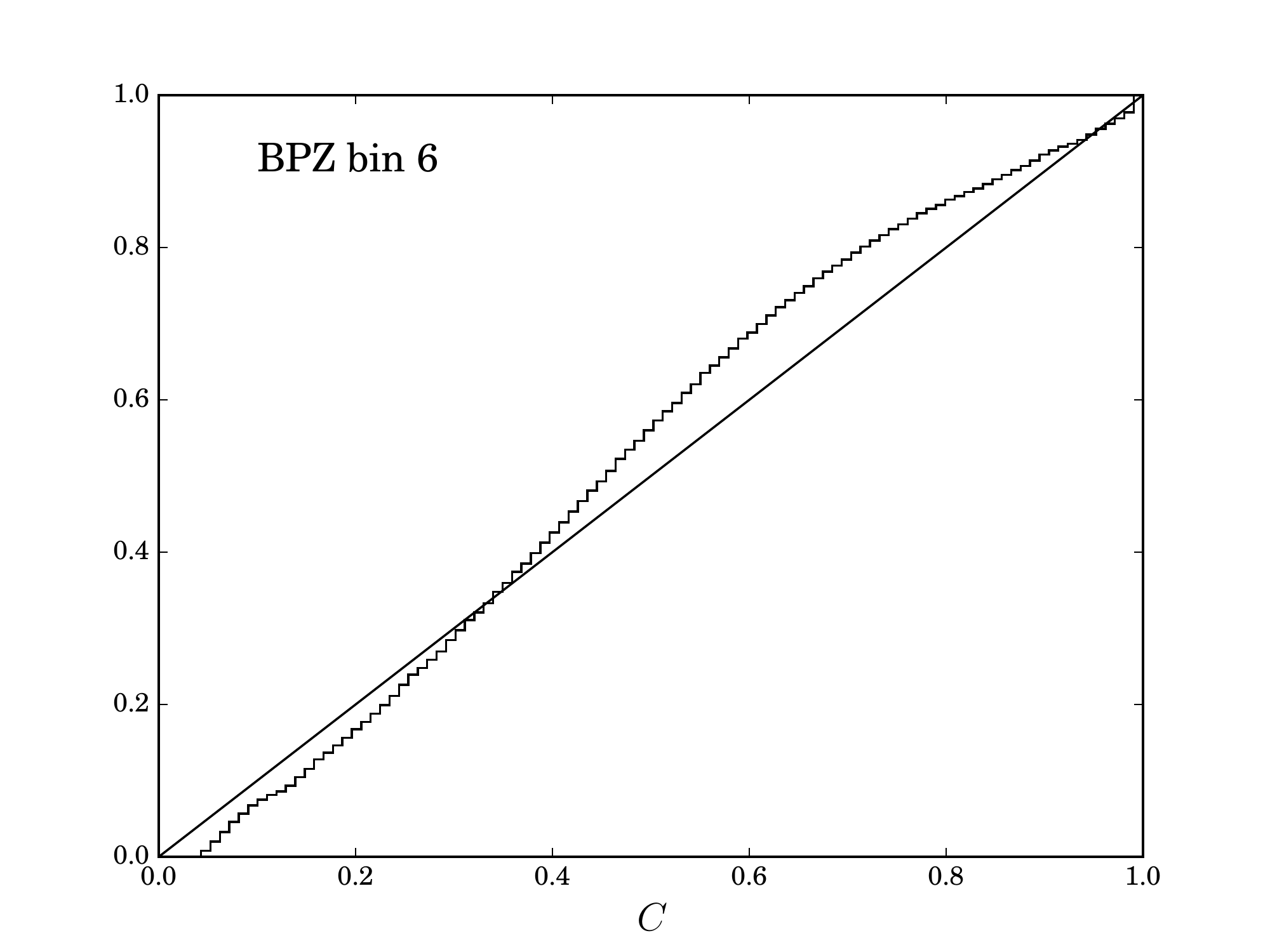}}
  {\includegraphics[width=0.30 \textwidth]{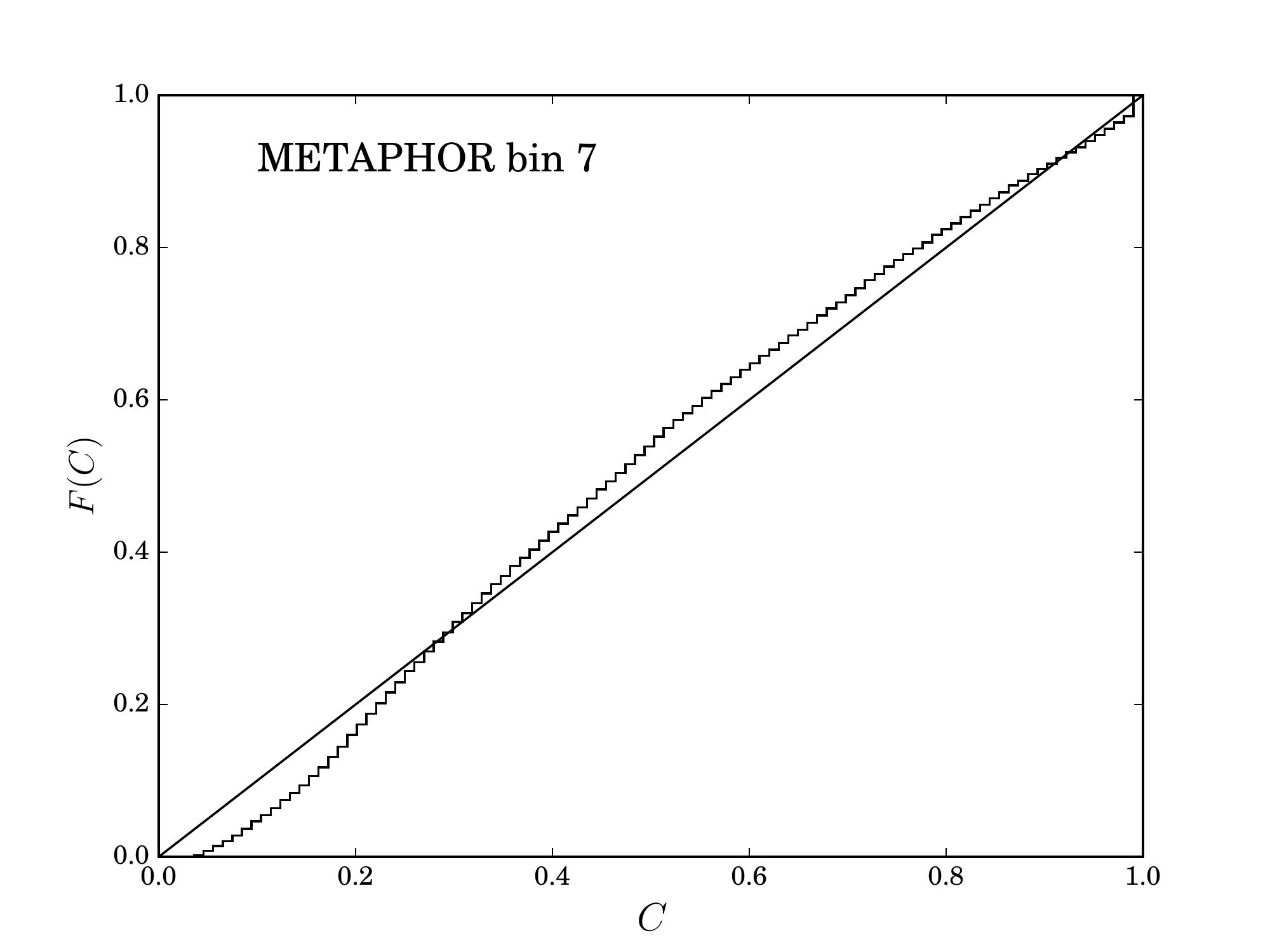}}
  {\includegraphics[width=0.30 \textwidth]{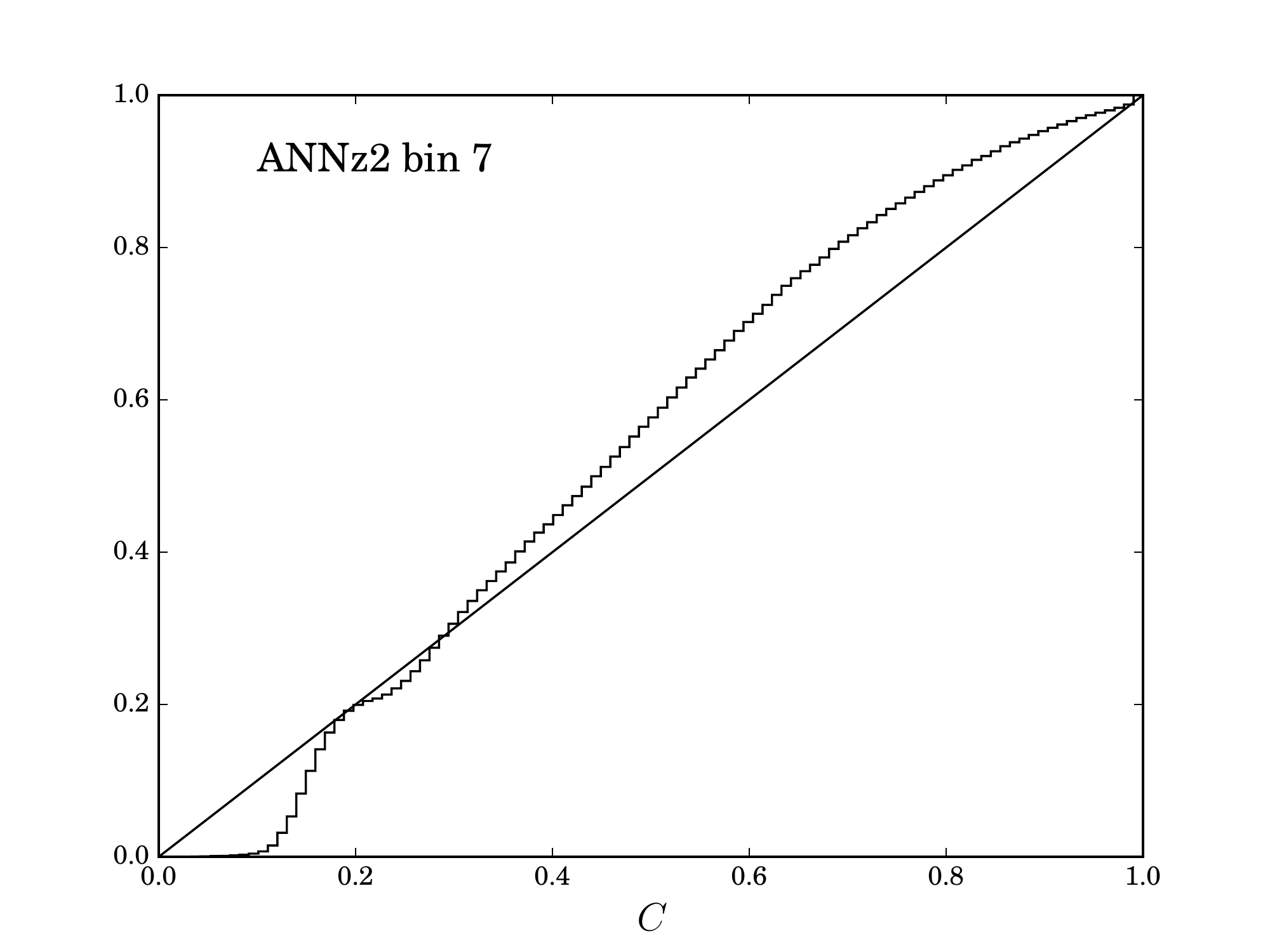}}
  {\includegraphics[width=0.30 \textwidth]{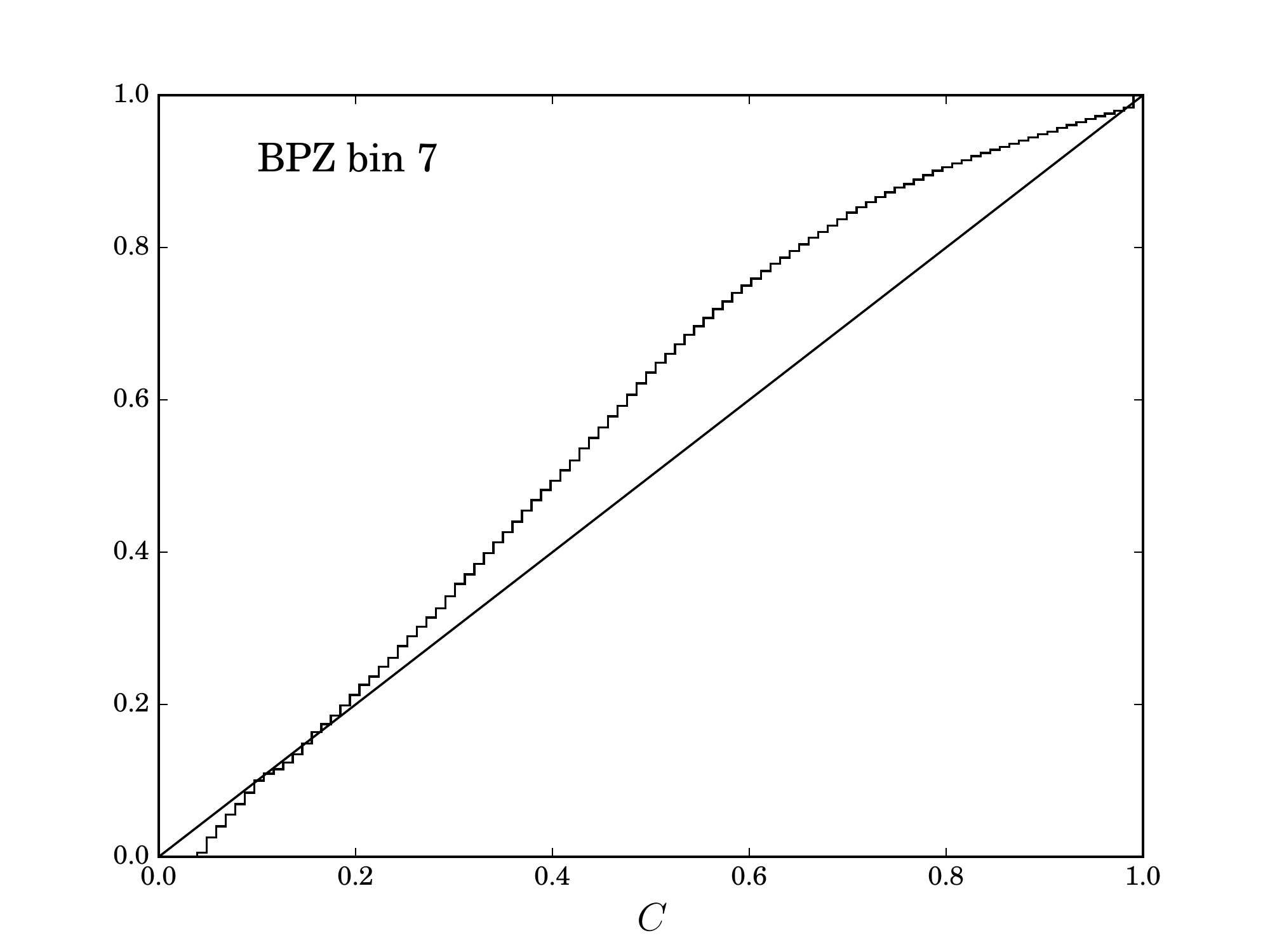}}
  {\includegraphics[width=0.30 \textwidth]{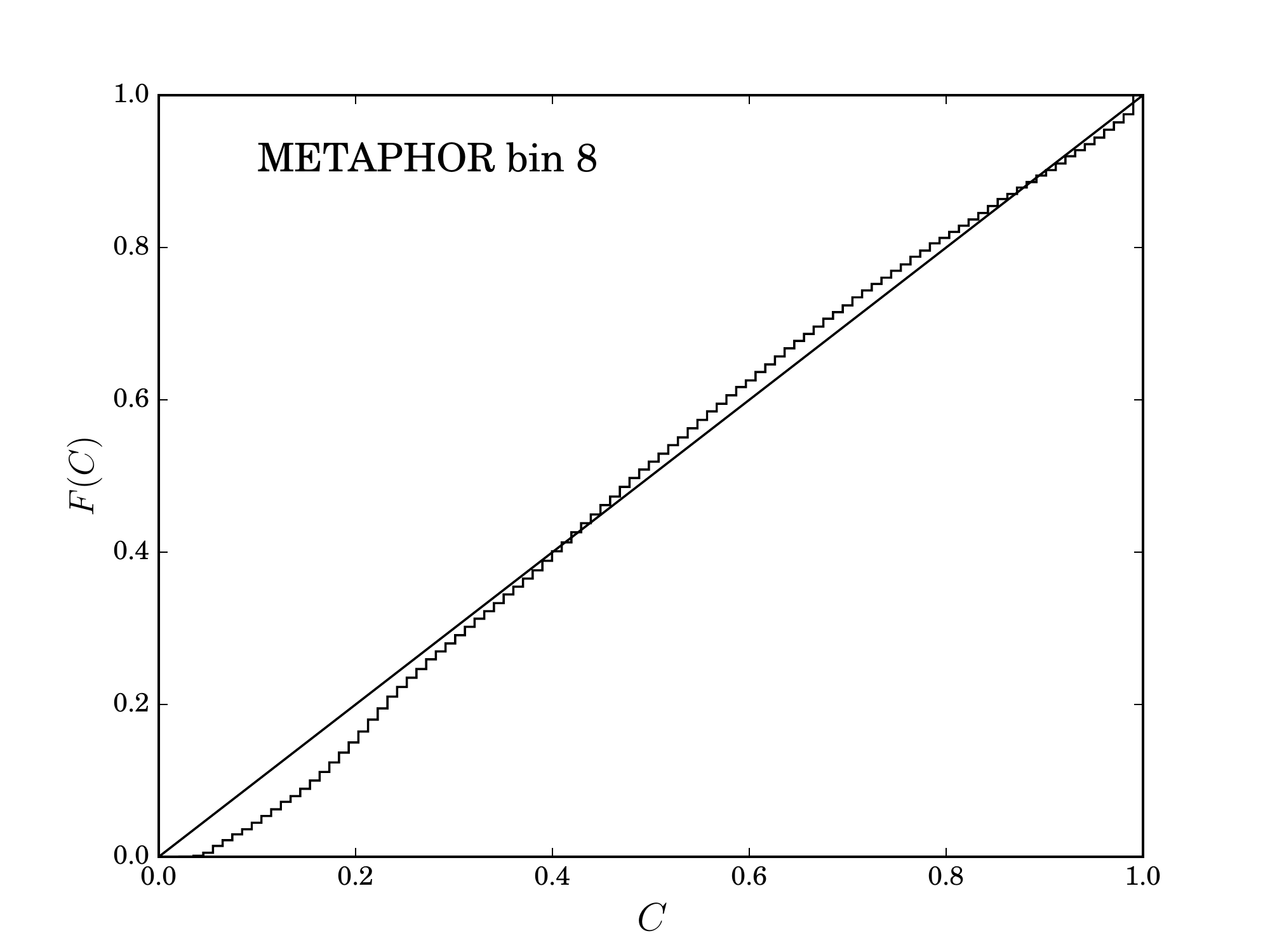}}
  {\includegraphics[width=0.30 \textwidth]{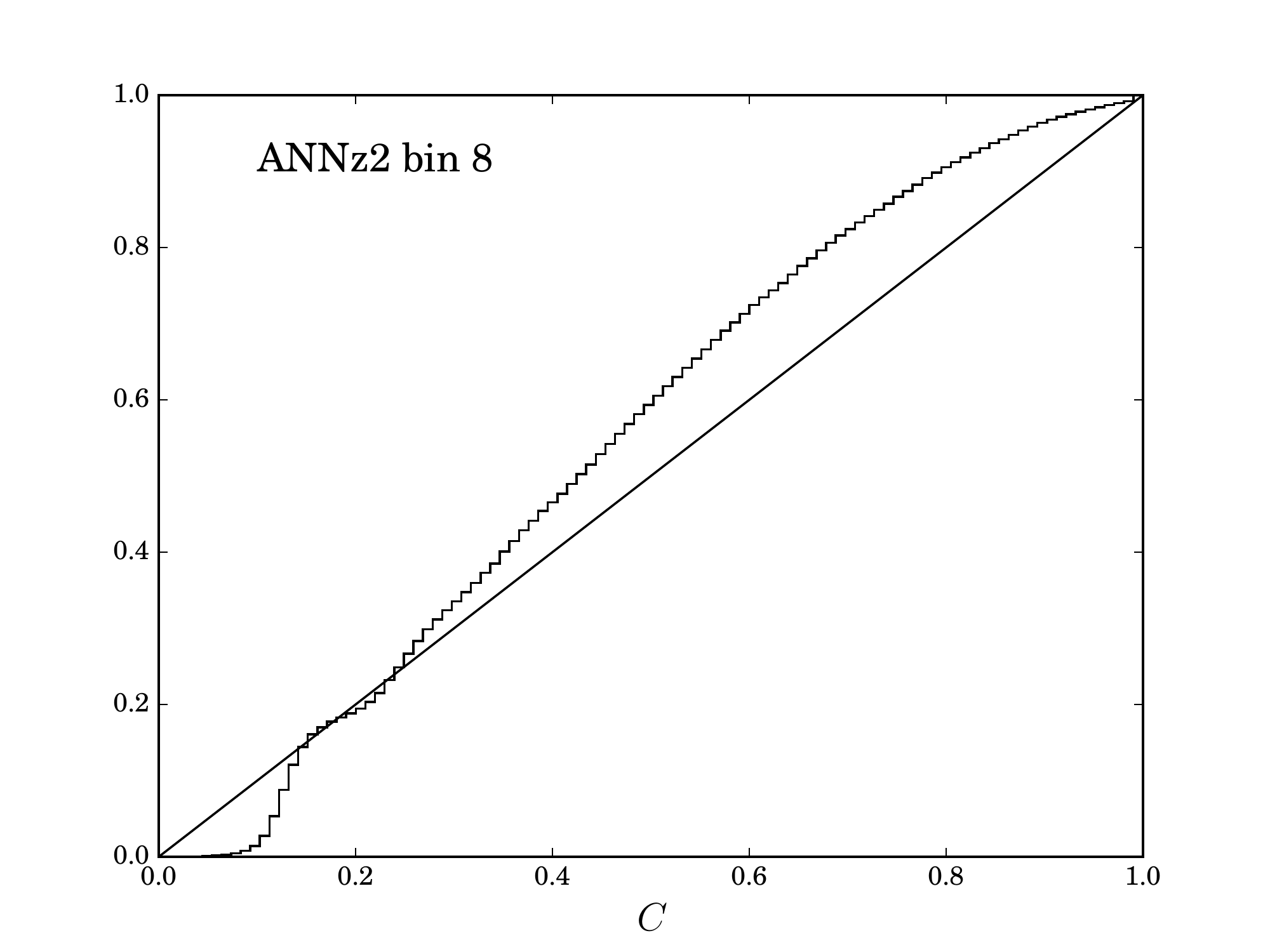}}
  {\includegraphics[width=0.30 \textwidth]{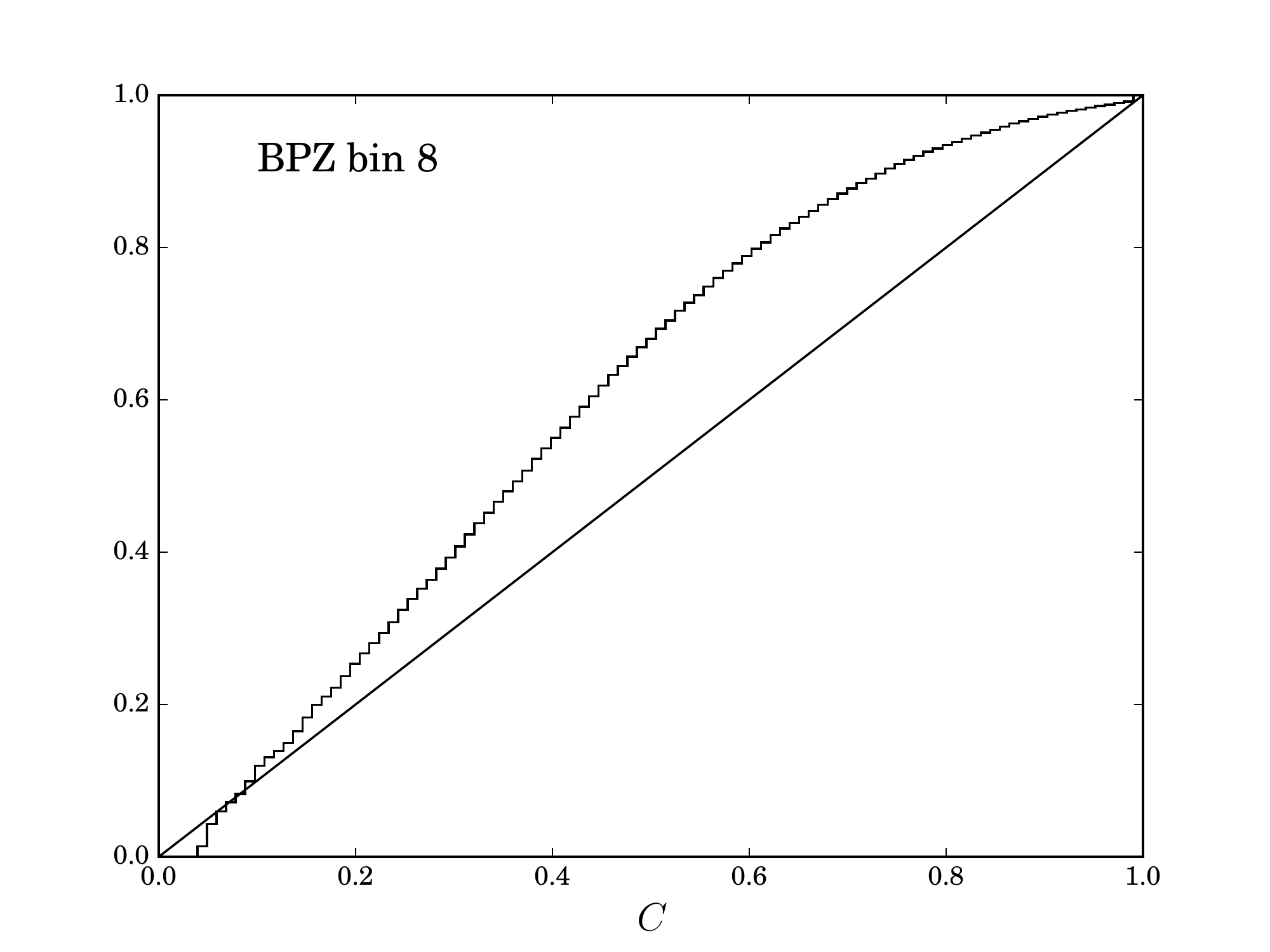}}
  {\includegraphics[width=0.30 \textwidth]{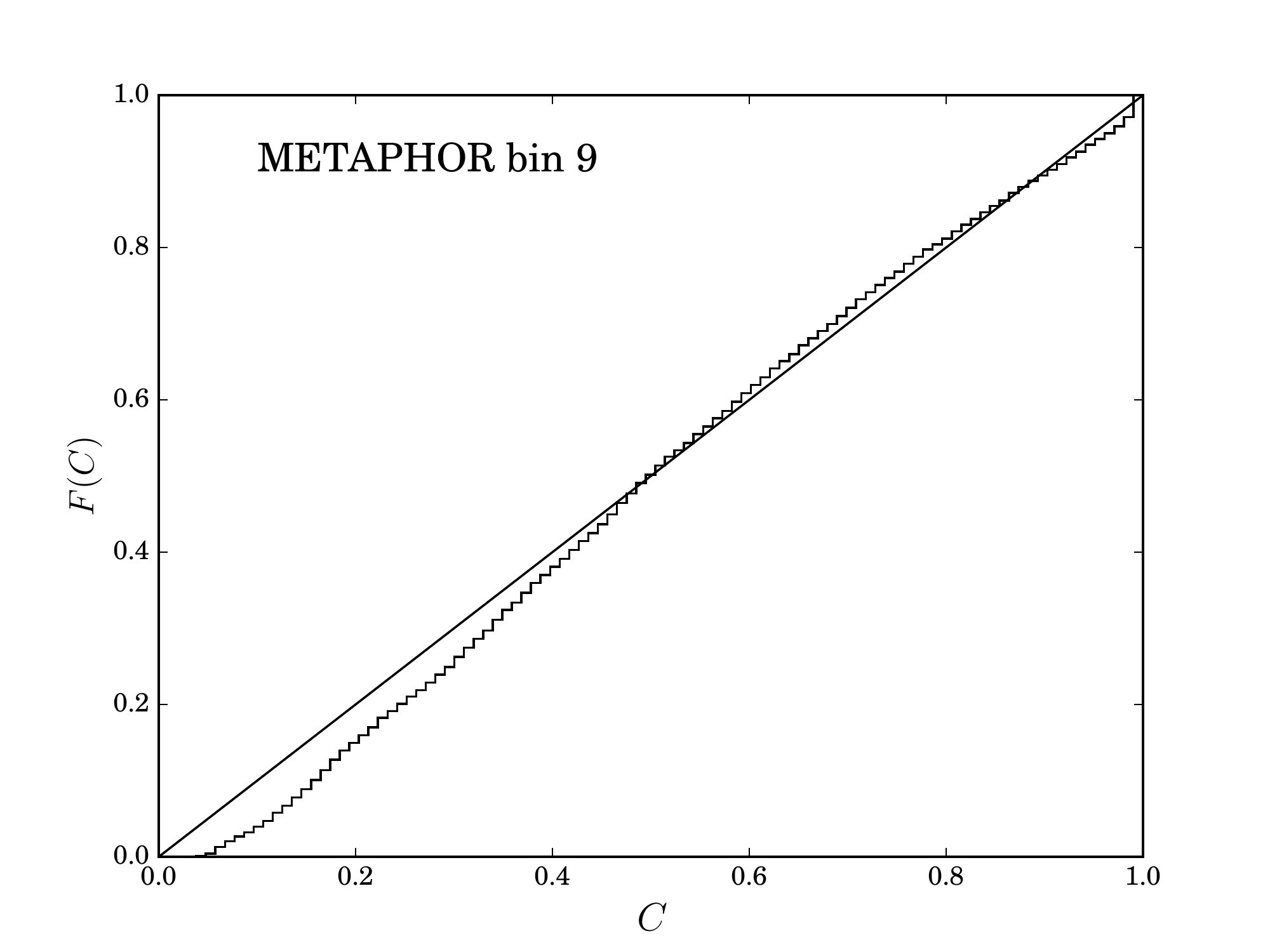}}
  {\includegraphics[width=0.30 \textwidth]{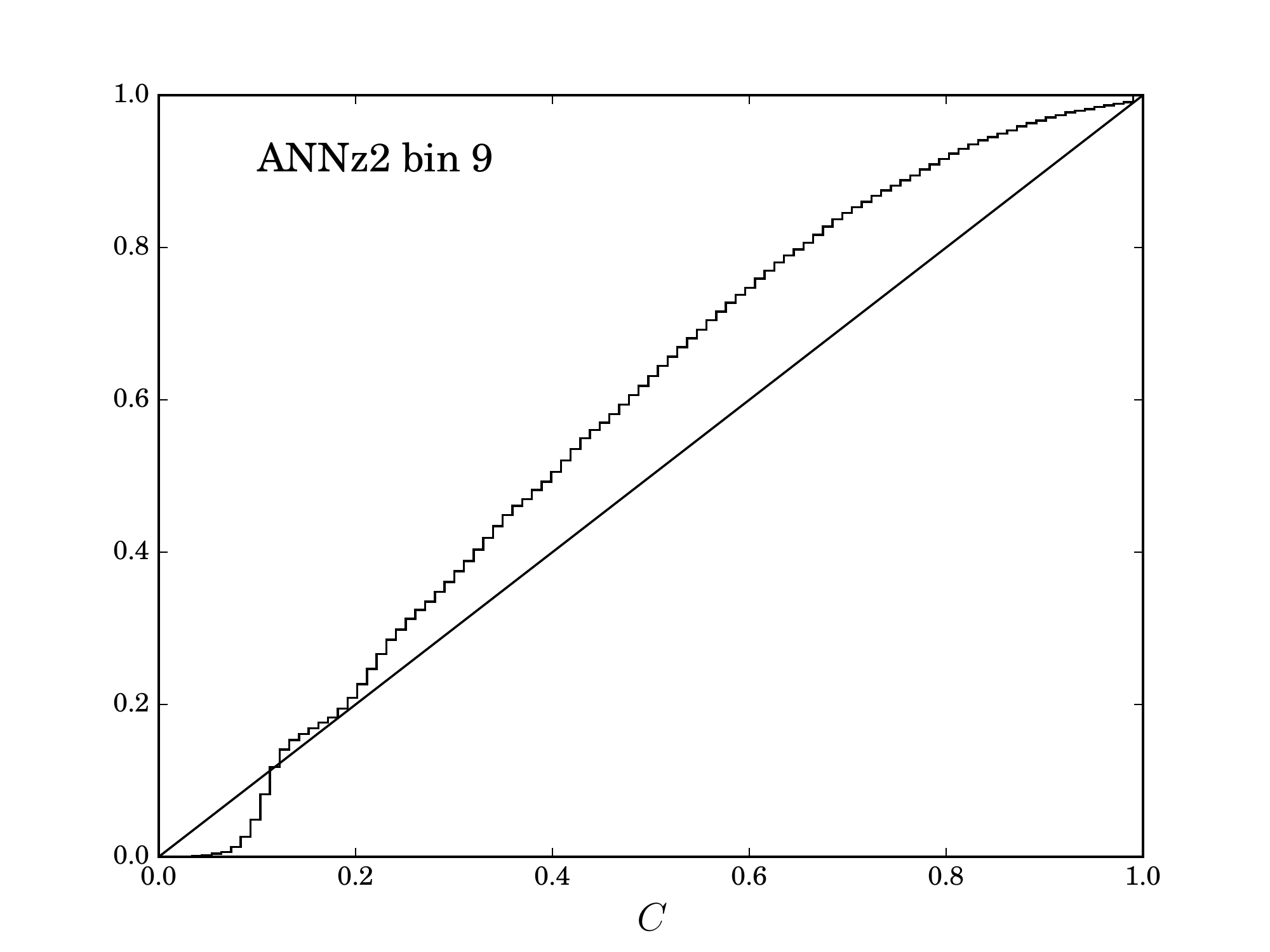}}
  {\includegraphics[width=0.30 \textwidth]{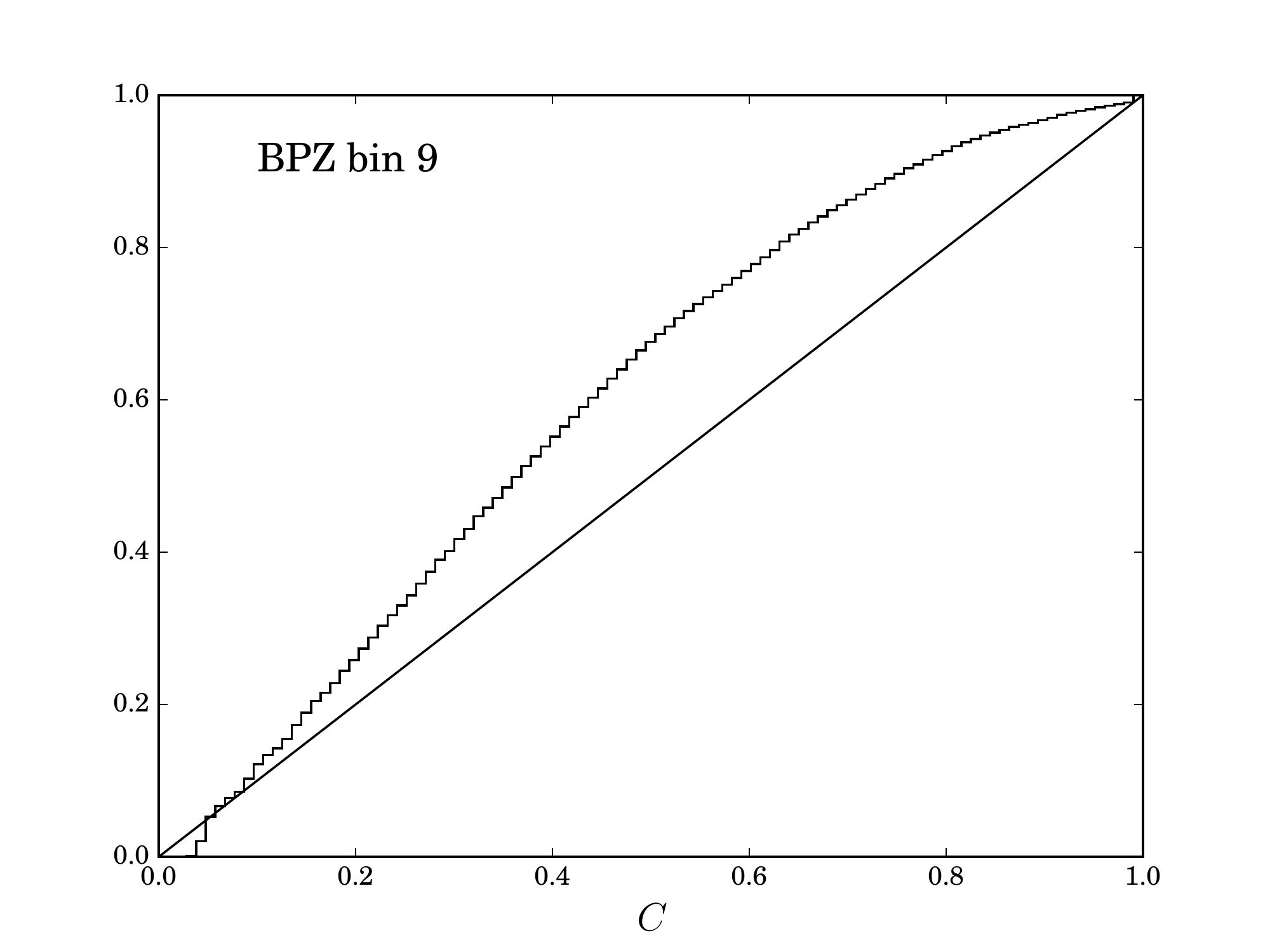}}
  {\includegraphics[width=0.30 \textwidth]{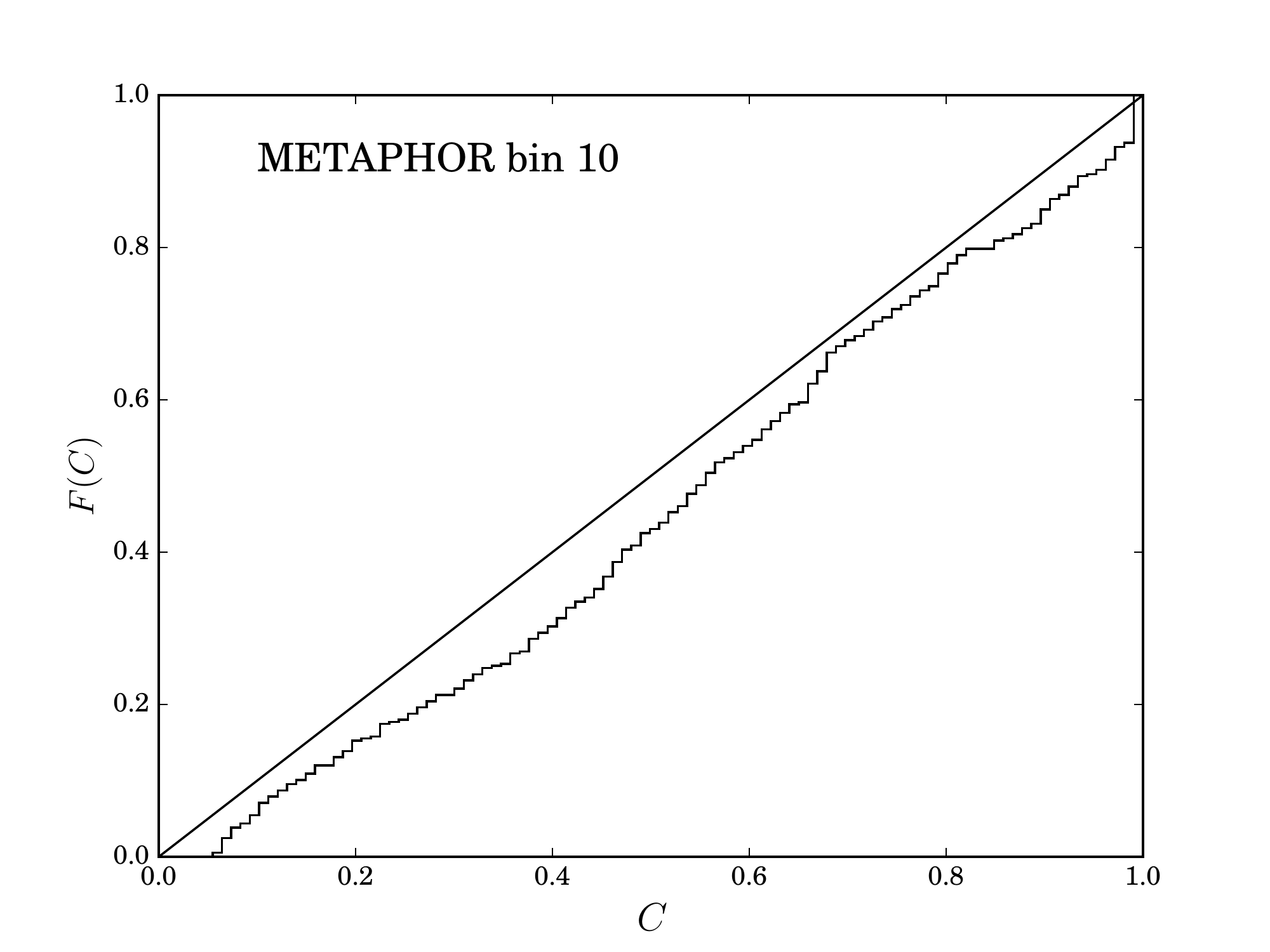}}
  {\includegraphics[width=0.30 \textwidth]{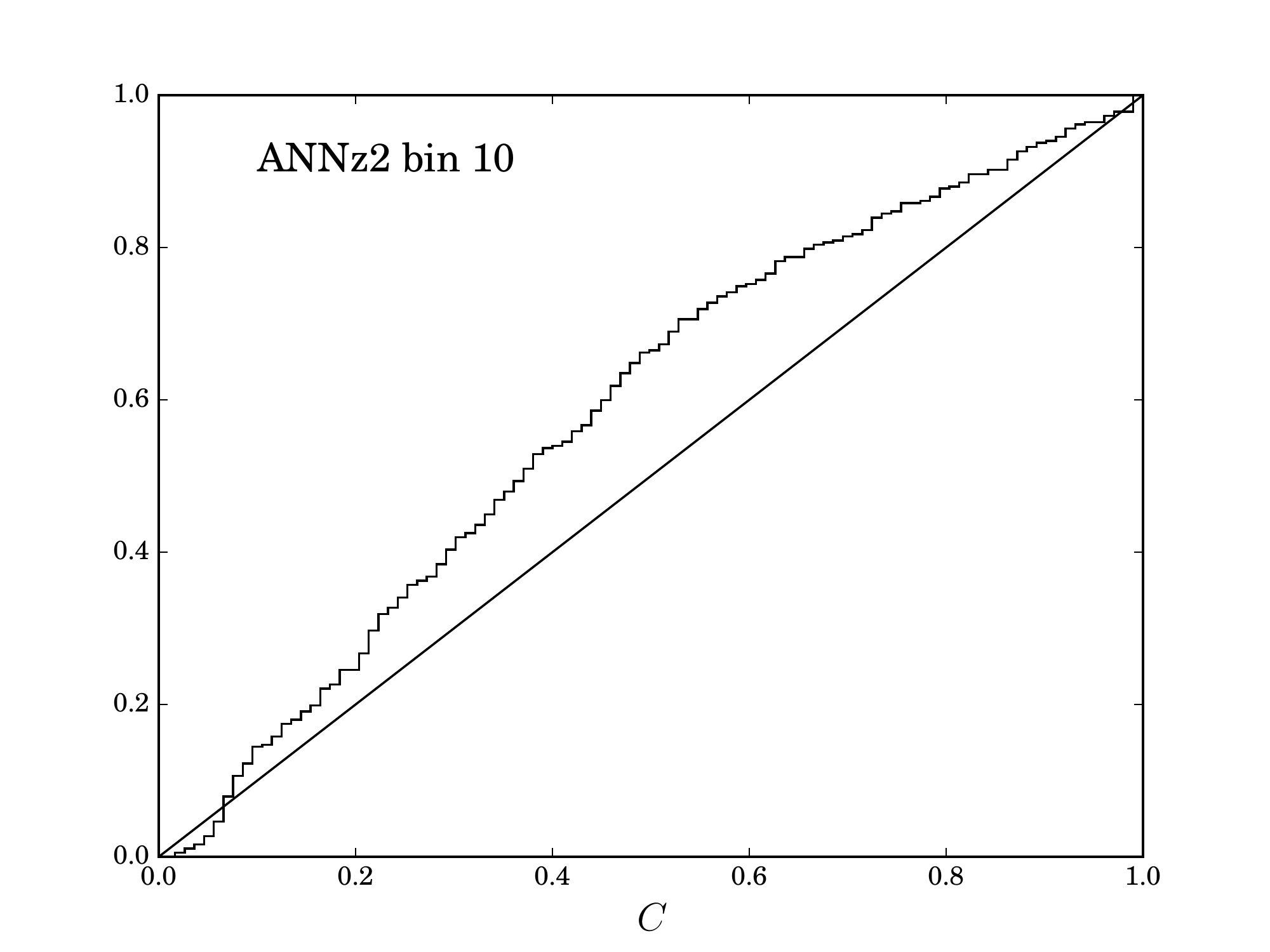}}
  {\includegraphics[width=0.30 \textwidth]{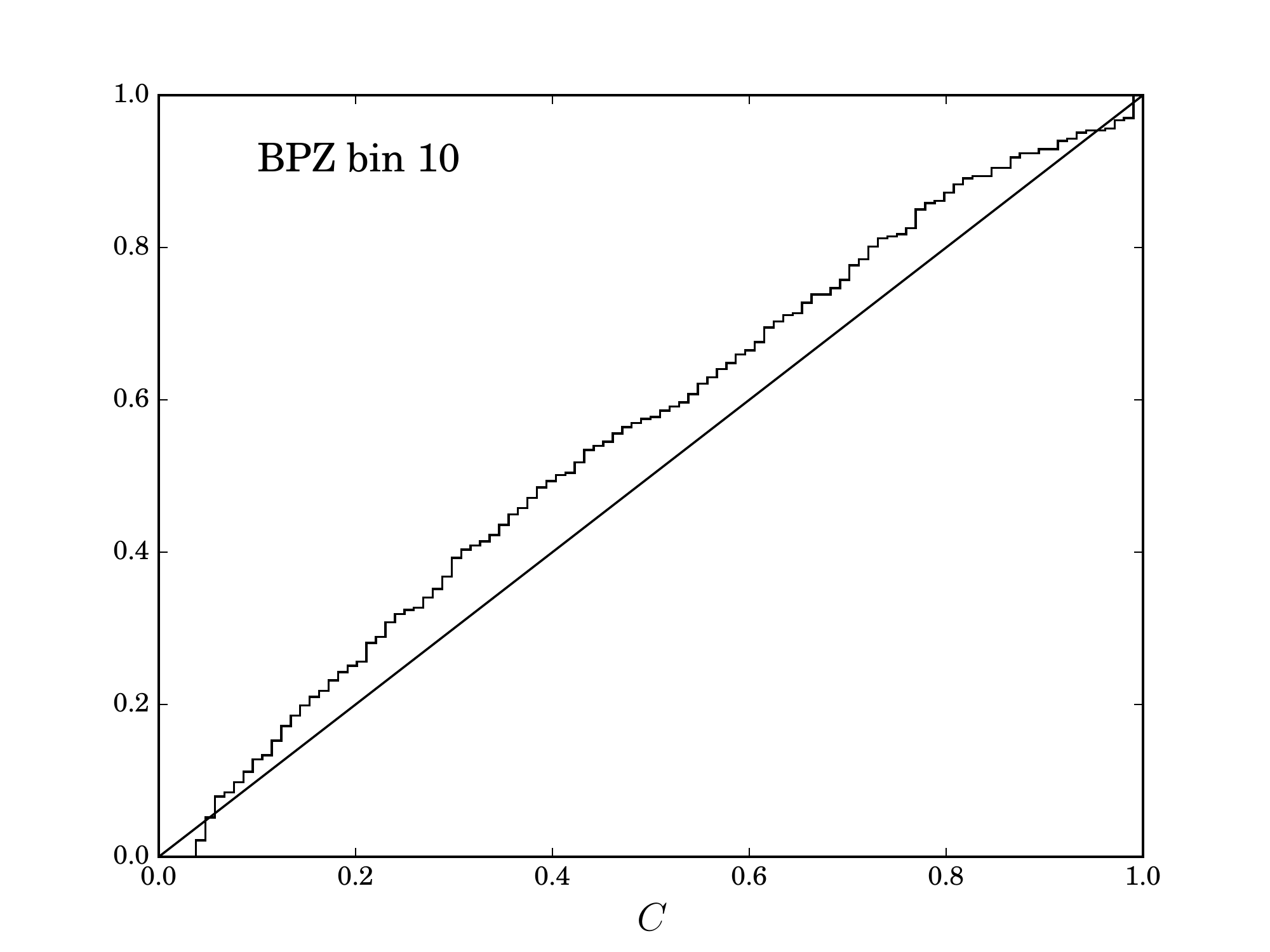}} 
 \caption{Credibility analysis (see Sec.~\ref{SEC:statindicators}) obtained for METAPHOR, ANNz2 and BPZ for the second five magnitude tomographic bins from Table~\ref{tab:TOMOG}. The credibility plots for the \textit{dummy} PDF are the same as the bottom right panel of Fig.~\ref{fig:WittmanALL} in all the bins.} 
\label{fig:WittmanALLtom2}
\end{figure*}

\subsection{PDF Tomography}

Finally, in order to analyze the stacked PDFs obtained by the  four estimation methods in different ranges of magnitude, we performed a binning in $mag\_gaap\_r$  in the range $[16.0, 21.0]$ with a step $\Delta mag=0.5$, resulting in a tomography of 10 bins. The range has been chosen in order to ensure a minimum amount of objects per bin to calculate the statistics. The results in terms of the fraction of residuals and the overall average for the stacked PDFs are reported in Table~\ref{tab:TOMOG}, while the fraction of residuals $f_{0.05}$ is shown as a function of $r$-band magnitude in Fig.~\ref{fig:f005ALL1}.
 
\begin{table*}
\centering
\caption{Tomographic analysis of the stacked PDFs for METAPHOR, ANNz2, BPZ and \textit{dummy} PDF calculated by METAPHOR, respectively, in ten bins of the homogenized magnitude \textit{$mag\_gaap\_r$}.} \label{tab:TOMOG}
 \begin{tabular}{ccccccccccccccc}
 \multirow{2}{*}{\bf Bin}&\multirow{2}{*}{\bf $r$-band }&\multirow{2}{*}{\bf Amount}&\multicolumn{3}{c}{\bf METAPHOR}&\multicolumn{3}{c}{\bf ANNz2}&\multicolumn{3}{c}{\bf BPZ}&\multicolumn{3}{c}{\bf dummy}\\
    &  & &$f_{0.05}$ &$f_{0.15}$ &$\Braket{\Delta z}$ &$f_{0.05}$ &$f_{0.15}$ &$\Braket{\Delta z}$ &$f_{0.05}$ &$f_{0.15}$	&$\Braket{\Delta z}$  &$f_{0.05}$ &$f_{0.15}$	&$\Braket{\Delta z}$ \\ \hline\hline
  $1$   &$]16.0,16.5]$		   & $122$    & $16.3\%$ & $37.2\%$ & $-0.330$ & $80.9\%$ & $99.5\%$ & $-0.016$ & $26.5\%$ & $87.0\%$ & $-0.080$ & $97.5\%$ & $100\%$  & $-0.015$ \\
  $2$   &$]16.5,17.0]$		   & $290$    & $23.9\%$ & $49.0\%$ & $-0.249$ & $81.7\%$ & $99.2\%$ & $-0.015$ & $28.5\%$ & $86.7\%$ & $-0.080$ & $97.9\%$ & $99.3\%$ & $-0.009$ \\
  $3$   &$]17.0,17.5]$		   & $858$    & $34.2\%$ & $62.4\%$ & $-0.185$ & $82.0\%$ & $98.4\%$ & $-0.016$ & $36.4\%$ & $89.7\%$ & $-0.068$ & $95.1\%$ & $98.7\%$ & $-0.006$ \\
  $4$   &$]17.5,18.0]$		   & $1,873$  & $48.0\%$ & $75.7\%$ & $-0.132$ & $81.6\%$ & $97.4\%$ & $-0.017$ & $41.0\%$ & $90.7\%$ & $-0.060$ & $94.2\%$ & $97.8\%$ & $-0.010$ \\
  $5$   &$]18.0,18.5]$		   & $4,427$  & $59.0\%$ & $84.6\%$ & $-0.086$ & $82.2\%$ & $98.2\%$ & $-0.011$ & $45.4\%$ & $92.5\%$ & $-0.050$ & $95.3\%$ & $98.7\%$ & $-0.006$ \\
  $6$   &$]18.5,19.0]$		   & $8,230$  & $64.9\%$ & $89.4\%$ & $-0.067$ & $81.1\%$ & $98.0\%$ & $-0.008$ & $47.6\%$ & $93.1\%$ & $-0.043$ & $94.3\%$ & $98.8\%$ & $-0.008$ \\
  $7$   &$]19.0,19.5]$		   & $15,388$ & $68.9\%$ & $92.6\%$ & $-0.051$ & $79.2\%$ & $97.9\%$ & $-0.008$ & $48.5\%$ & $93.2\%$ & $-0.037$ & $93.7\%$ & $98.9\%$ & $-0.007$ \\
  $8$   &$]19.5,20.0]$		   & $22,952$ & $68.5\%$ & $93.8\%$ & $-0.043$ & $75.9\%$ & $98.0\%$ & $-0.006$ & $47.8\%$ & $92.9\%$ & $-0.033$ & $93.4\%$ & $99.2\%$ & $-0.003$ \\
  $9$   &$]20.0,20.5]$		   & $9,178$  & $65.8\%$ & $94.2\%$ & $-0.040$ & $61.4\%$ & $97.0\%$ & $-0.010$ & $45.4\%$ & $91.6\%$ & $-0.033$ & $89.9\%$ & $98.9\%$ & $-0.007$ \\
  $10$  &$]20.5,21.0]$		   & $367$    & $55.5\%$ & $88.4\%$ & $-0.061$ & $44.5\%$ & $80.4\%$ & $-0.104$ & $43.1\%$ & $88.9\%$ & $-0.033$ & $74.6\%$ & $94.0\%$ & $-0.025$ \\ \hline
 \end{tabular}
\end{table*}

\begin{table*}
\centering
\caption{\textit{zspecClass} fractions for METAPHOR, ANNz2 and BPZ in tomographic bins of the homogenized magnitude \textit{$mag\_gaap\_r$}.} \label{tab:TOMOGzspec}
 \begin{tabular}{cccccccccccccccccc}
 \multirow{2}{*}{\bf Bin }&\multirow{2}{*}{\bf $r$-band }&\multirow{2}{*}{\bf Amount }&\multicolumn{5}{c}{\bf METAPHOR}&\multicolumn{5}{c}{\bf ANNz2}&\multicolumn{5}{c}{\bf BPZ}\\
    &  & &$0$ &$1$ &$2$ &$3$ &$4$  &$0$ &$1$ &$2$ &$3$ &$4$ &$0$ &$1$ &$2$ &$3$ &$4$\\ \hline\hline
  $1$   &$]16.0,16.5]$		   & $122$    & $4.9\%$  & $13.9\%$ & $80.3\%$ & $0.8\%$ & $0.0\%$ & $38.5\%$  & $29.5\%$ & $32.0\%$ & $0.0\%$ & $0.0\%$ & $0.0\%$  & $1.6\%$  & $98.4\%$ & $0.0\%$ & $0.0\%$   \\
  $2$   &$]16.5,17.0]$		   & $290$    & $10.3\%$ & $22.7\%$ & $66.5\%$ & $0.3\%$ & $0.0\%$ & $38.4\%$  & $22.1\%$ & $39.4\%$ & $0.0\%$ & $0.0\%$ & $0.7\%$  & $1.7\%$  & $97.6\%$ & $0.0\%$ & $0.0\%$  \\
  $3$   &$]17.0,17.5]$		   & $858$    & $19.0\%$ & $32.3\%$ & $47.8\%$ & $0.6\%$ & $0.3\%$ & $31.3\%$  & $27.9\%$ & $40.5\%$ & $0.0\%$ & $0.3\%$ & $1.2\%$  & $3.4\%$  & $95.3\%$ & $0.0\%$ & $0.1\%$ \\
  $4$   &$]17.5,18.0]$		   & $1,873$  & $18.7\%$ & $33.4\%$ & $46.1\%$ & $0.6\%$ & $1.2\%$ & $29.6\%$  & $31.0\%$ & $38.9\%$ & $0.0\%$ & $0.5\%$ & $1.9\%$  & $6.0\%$  & $92.0\%$ & $0.0\%$ & $0.05\%$  \\
  $5$   &$]18.0,18.5]$		   & $4,427$  & $16.7\%$ & $31.0\%$ & $50.3\%$ & $0.5\%$ & $0.8\%$ & $24.9\%$  & $35.2\%$ & $39.2\%$ & $0.0\%$ & $0.6\%$ & $4.0\%$  & $10.0\%$ & $86.0\%$ & $0.0\%$ & $0.09\%$   \\
  $6$   &$]18.5,19.0]$		   & $8,230$  & $18.6\%$ & $30.9\%$ & $49.2\%$ & $0.2\%$ & $1.0\%$ & $23.3\%$  & $33.7\%$ & $42.3\%$ & $0.0\%$ & $0.7\%$ & $6.5\%$  & $14.7\%$ & $78.7\%$ & $0.0\%$ & $0.1\%$  \\
  $7$   &$]19.0,19.5]$		   & $15,388$ & $14.7\%$ & $27.6\%$ & $56.7\%$ & $0.2\%$ & $0.8\%$ & $19.5\%$  & $31.3\%$ & $48.6\%$ & $0.02\%$ & $0.6\%$ & $8.4\%$  & $16.3\%$ & $75.2\%$ & $0.0\%$ & $0.07\%$  \\
  $8$   &$]19.5,20.0]$		   & $22,952$ & $12.7\%$ & $24.3\%$ & $66.5\%$ & $0.4\%$ & $0.9\%$ & $17.3\%$  & $28.6\%$ & $53.7\%$ & $0.01\%$ & $0.4\%$ & $9.0\%$  & $16.9\%$ & $74.2\%$ & $0.0\%$ & $0.03\%$ \\
  $9$   &$]20.0,20.5]$		   & $9,178$  & $10.7\%$ & $21.2\%$ & $66.5\%$ & $0.4\%$ & $0.9\%$ & $15.4\%$  & $25.5\%$ & $58.6\%$ & $0.02\%$ & $0.4\%$ & $8.4\%$  & $15.3\%$ & $76.3\%$ & $0.0\%$ & $0.0\%$  \\
  $10$  &$]20.5,21.0]$		   & $367$    & $10.3\%$ & $16.3\%$ & $70.3\%$ & $0.8\%$ & $2.2\%$ & $10.9\%$  & $17.4\%$ & $70.3\%$ & $0.0\%$ & $1.4\%$ & $9.0\%$   & $12.8\%$ & $77.9\%$ & $0.0\%$ & $0.0\%$  \\ \hline
 \end{tabular}
\end{table*}

Given the statistics in Tables \ref{tab:TOMOG} and \ref{tab:TOMOGzspec}, we observe that for BPZ the zspec falls within the PDF in practically all bins and that the highest concentration of PDFs is within $f_{0.15}$. This behavior indicates a broad shape of the PDFs, also confirmed by the \textit{underconfidence} shown in Fig.~\ref{fig:WittmanALL} and by the \textit{overdispersion} in Fig.~\ref{fig:pitALL}. The latter figure also shows the presence of a high bias, visible from the unbalanced trend. Furthermore, the PIT tomography, reported in Figures \ref{fig:pitALLtom1} and \ref{fig:pitALLtom2}, shows a high variability and confirms the general \textit{overdispersion} and bias of the PDFs. Turning to the HPDCI tomography, the overall trend of Fig.~\ref{fig:WittmanALL} indicates a general \textit{underconfidence}, but Fig. \ref{fig:WittmanALLtom1} and \ref{fig:WittmanALLtom2} show an inversion, from a high \textit{underconfidence} to a lower \textit{overconfidence}, compatible with the general variability of BPZ PDFs.\\ 
ANNz2 shows a similar behavior as BPZ for $f_{0.15}$ but has a higher percentage of $f_{0.05}$, which indicates PDFs more centered around zspec. Furthermore, the values of \textit{zspecClass} equal to $3$ and $4$ show that zspec mostly falls within the PDFs, thus indicating that also in the case of ANNz2 the PDFs have a broad shape, albeit to a lesser extent. This is also confirmed by the \textit{overdispersive} trend of the PIT diagram in Fig.~\ref{fig:pitALL} as well as by the \textit{underconfidence} in Fig.~\ref{fig:WittmanALL}. In terms of PIT and HPDCI tomography, ANNz2 shows a more regular behavior than BPZ.\\
METAPHOR shows a stacked PDF with a more pronounced average $\Braket{\Delta z}$ than BPZ and ANNz2 in Table~\ref{tab:TOMOG}, due to the larger tails of its $\Delta z$ distribution. 
Moreover, both Tables \ref{tab:TOMOG} and \ref{tab:TOMOGzspec} indicate that the METAPHOR and especially the BPZ PDFs have a broader shape than those of ANNz2, for example by looking at the percentages for \textit{$zspecClass = 2$}.
This is also reflected by the PIT diagram of Fig.~\ref{fig:pitALL}, which reveals highly biased and \textit{overdispersive} PDFs. In contrast with previous statistics, the HPDCI diagram indicates that METAPHOR is less \textit{underconfident} than BPZ and ANNz2. The tomographic analysis of the PIT diagram reports highly biased PDFs for the bins of Fig.~\ref{fig:pitALLtom1}, while in the other bins of Fig.~\ref{fig:pitALLtom2} METAPHOR shows similar characteristics as ANNz2, albeit with different types of defects. Finally, in terms of HPDCI tomography, Figures \ref{fig:WittmanALLtom1} and \ref{fig:WittmanALLtom2} reveal a general coherence in the behavior of METAPHOR, except in the first and last bin, which are least populated.

\section{Conclusions}\label{SEC:discussion}
Due to the increasing demand for reliable zphot and the intrinsic difficulty to provide reliable error PDF estimation for machine learning methods, a plethora of solutions have been proposed. The derivation of PDFs with machine learning models is in fact conditioned by the mechanism used to infer the hidden flux-redshift relationship. In fact, this mechanism imposes the necessity to disentangle the contributions to the zphot estimation error budget, by distinguishing the intrinsic method error from the photometric uncertainties. Furthermore, due to the large variety of methods proposed, there is also the problem of finding objective and robust statistical estimators of the quality and reliability of the derived PDFs.\\
We believe that it is extremely useful to estimate the zphot error through the intrinsic photometric uncertainties, by considering that the observable photometry cannot be perfectly mapped to the true redshift. Furthermore, the evaluation of a statistically meaningful PDF should consider the effective contribution of the intrinsic error of the method. \\
\indent In \cite{Cavuoti2017}, we presented METAPHOR, a method designed to provide a PDF  of photometric redshifts calculated by machine learning methods. METAPHOR has already been successfully tested on SDSS \citep{Cavuoti2017} and KiDS-DR3  \citep{dJ2017} data, and makes use of the neural network MLPQNA \citep{Brescia2013,Brescia2014a} as the internal zphot estimation engine.\\
\indent Main goal of the present work is a deeper analysis of zphot PDFs obtained by different methods: two machine learning models (METAPHOR and ANNz2) and one based on SED fitting techniques (BPZ), through a direct comparison among such methods.  The investigation was focused on both cumulative (\textit{stacked}) and individual PDF reliability. 
Moreover the methods were subjected to a comparative analysis using different kinds of statistical estimators to evaluate their degree of coherence. Exactly for this reason, by modifying the METAPHOR internal mechanism, we  also derived a \textit{dummy} PDF method (see Sec.~\ref{SEC:thedummypdf}), helpful to obtain a benchmark tool to evaluate the objectivity of the various statistical estimators  applied on the presented methods.\\
\indent Regarding the \textit{dummy} PDF (Table~\ref{tab:stackedstat}), the more the PDF is representative of an almost perfect mapping of the parameter space on the true redshifts, the better are the performances in terms of stacked PDF estimators. However, we have shown that the PIT histogram and the credibility analysis provide important complementary statistical information, the first showing the total \textit{underdispersive} trend of the reconstructed photometric redshift distribution; the second reporting an \textit{overconfidence} of all zphot estimates. Both the \textit{underdispersion} and the \textit{overconfidence} are related to the narrowness of the PDFs: the narrower they are, the more the PIT histogram is \textit{underdispersed} and the results, as determined by the credibility analysis, overconfident.\\
Thus, it appears clear that the statistical estimators used for the stacked PDF (for instance $f_{0.05}$, $f_{0.15}$ and $\Braket{\Delta z}$), are not self-consistent and should be combined with other statistical estimators, such as the PIT diagrams and credibility analysis.\\
\indent Although the credibility analyses of the different methods, based on the Wittman diagram (Fig.~\ref{fig:WittmanALL}) and the Probability Integral Transform diagram (Fig.~\ref{fig:pitALL}), appear comparable in terms of overall results, their tomography (Figures ~\ref{fig:pitALLtom1}, ~\ref{fig:pitALLtom2}, ~\ref{fig:WittmanALLtom1} and ~\ref{fig:WittmanALLtom2}) shows different behaviors at different redshift regimes.\\
Summarizing the results for the three PDF estimation methods analyzed, considering the combination of statistical estimators, ANNz2 is favored by the $f_{0.05}$, $f_{0.15}$, $\Braket{\Delta z}$ and the PIT diagram. However, METAPHOR is more competitive, in particular when considering the confidence analysis. BPZ has the best PDFs in the faintest magnitude bin. Moreover, all three methods show a generally broad shape of their PDFs, albeit to a different extent, with also a bias in the case of BPZ and METAPHOR. However, they show occasional fluctuations in their tomographic analysis. For instance, BPZ reverses its \textit{overconfidence} trend at fainter magnitudes, while METAPHOR and BPZ show a high level of variability along the magnitude bins in terms of \textit{underdispersion} and bias. In the specific case of our method METAPHOR, all mentioned defects require further investigation in terms of the photometric perturbation function.\\
\indent It should be noted that the current comparison is preliminary, since the methods explored in this paper deal with different sources of errors. In fact, ANNz2 takes into account only the internal errors of the method, METAPHOR only those induced by the photometry, BPZ includes both these error sources and, finally, the benchmark (\textit{dummy} PDF) does not include either of these two.\\
\indent All considerations together lead us to affirm that a detailed analysis of the performances, based on a combination of independent statistical estimators, is key to unraveling the nature of the estimated zphot PDFs and to assess the objective validity of the method employed to derive them.  


\section*{acknowledgements}
We would like to thank the anonymous referee for all the comments and suggestions that improved the manuscript. Based on data products from observations made with European Southern Observatory (ESO) telescopes at the La Silla Paranal Observatory under programme IDs 177.A-3016, 177.A-3017 and 177.A-3018, and on data products produced by Target/OmegaCEN, Istituto Nazionale di AstroFisica (INAF)-Osservatorio Astronomico di Capodimonte Napoli (OACN), INAF-Osservatorio Astronomico di Padova (OAPD) and the KiDS production team, on behalf of the KiDS consortium. OmegaCEN and the KiDS production team acknowledge support by NOVA and NWO-M grants.
MBr acknowledges financial contribution from the agreement ASI/INAF I/023/12/1. MBr acknowledges also the \textit{INAF PRIN-SKA 2017 program 1.05.01.88.04} and the funding from \textit{MIUR Premiale 2016: MITIC}.
MBi is supported by the Netherlands Organization for Scientific Research, NWO, through grant number 614.001.451.
GL and MBr acknowledge partial support from the H2020 Marie Curie ITN - SUNDIAL. 
CT is supported through an NWO-VICI grant (project number 639.043.308).
SC acknowledges support from the project ``Quasars at high redshift: physics and Cosmology'' financed by the ASI/INAF agreement 2017-14-H.0.
JTAdJ is supported by the Netherlands Organisation for Scientific Research (NWO) through grant 621.016.402.
DAMEWARE has been used for this work \citep{Brescia2014a}.

\appendix
\section{Analysis of METAPHOR error sources} \label{appendix}
In this appendix we investigated the possibility to quantify the contribution of the method error to the zphot estimation. For instance, such error, in the case of METAPHOR, mostly depends on the random initialization of the neural connection weights in the MLPQNA neural network, used as internal engine to determine the zphot point estimates.\\
Through a test performed on the SDSS DR9 data \citep{Cavuoti2017}, we already showed that $N$ different trainings did not degrade the PDF performance: de facto the error introduced by the method appears negligible. On the other hand, $N$ network trainings are very time consuming. Here we deepened this exercise with METAPHOR pipeline for the KiDS-DR3 data, by performing two different experiments described below.\\ 
We created $100$ training samples, namely $100$ random extractions from the training set used to obtain the KiDS-DR3 PDFs (see Sec.~\ref{SEC:data}). Each of the $100$ training sets contains $10,000$ objects. The experiments are the following:
\begin{itemize}
\item  Experiment \textit{i)}: $100$ training + test executions by keeping unchanged both training and test sets. The single training set has been randomly selected from among the $100$ sets available and the test set corresponds to the sample obtained by cross-matching the KiDS-DR3 photometry with GAMA DR2 spectroscopy (see Sec.~\ref{SEC:data});
\item Experiment  \textit{ii)}: $100$ training + test executions by varying the training set each time and by keeping unchanged the test set (same set as previous experiment).
\end{itemize}

In both experiments, all other setup parameters of the full METAPHOR pipeline have been left unchanged. Therefore, the difference between the two experiments is only the training setup of the internal engine MLPQNA (weights initialization is left random and photometry is fixed at each training of the experiment \textit{(i)}, while weights initialization is left random and training photometry is variable in the experiment \textit{(ii)}). In other words, in the experiment \textit{(i)} we isolated the effect of the random weights initialization, while in the second experiment we kept the sum of the two effects (weights initialization and variable training photometry). Both experiments lasted $\sim11$ days on a $8$-core pentium $i7$.

\begin{table}
\centering
\caption{Statistics of the zphot error stacked PDFs obtained by METAPHOR, for the experiments \textit{(i)} and \textit{(ii)}.}
 \begin{tabular}{|c|c|c}
 {\bf Estimator}		  & {\bf exp \textit{i)}}      & {\bf exp \textit{ii)}}           \\ \hline
 $f_{0.05}$ & $92.2\%$\phantom{AA}	 & $92.1\%$\phantom{A}\\
 $f_{0.15}$ & $98.4\%$\phantom{AA}	 & $98.4\%$\phantom{A}\\
 $\Braket{\Delta z}$ & $-0.008$\phantom{AA}     & $-0.008$\phantom{A}\\ \hline
 \end{tabular}
\label{tab:stackedstatERMET}
\end{table}

\begin{figure}
 \centering
 {\includegraphics[width=0.35 \textwidth]{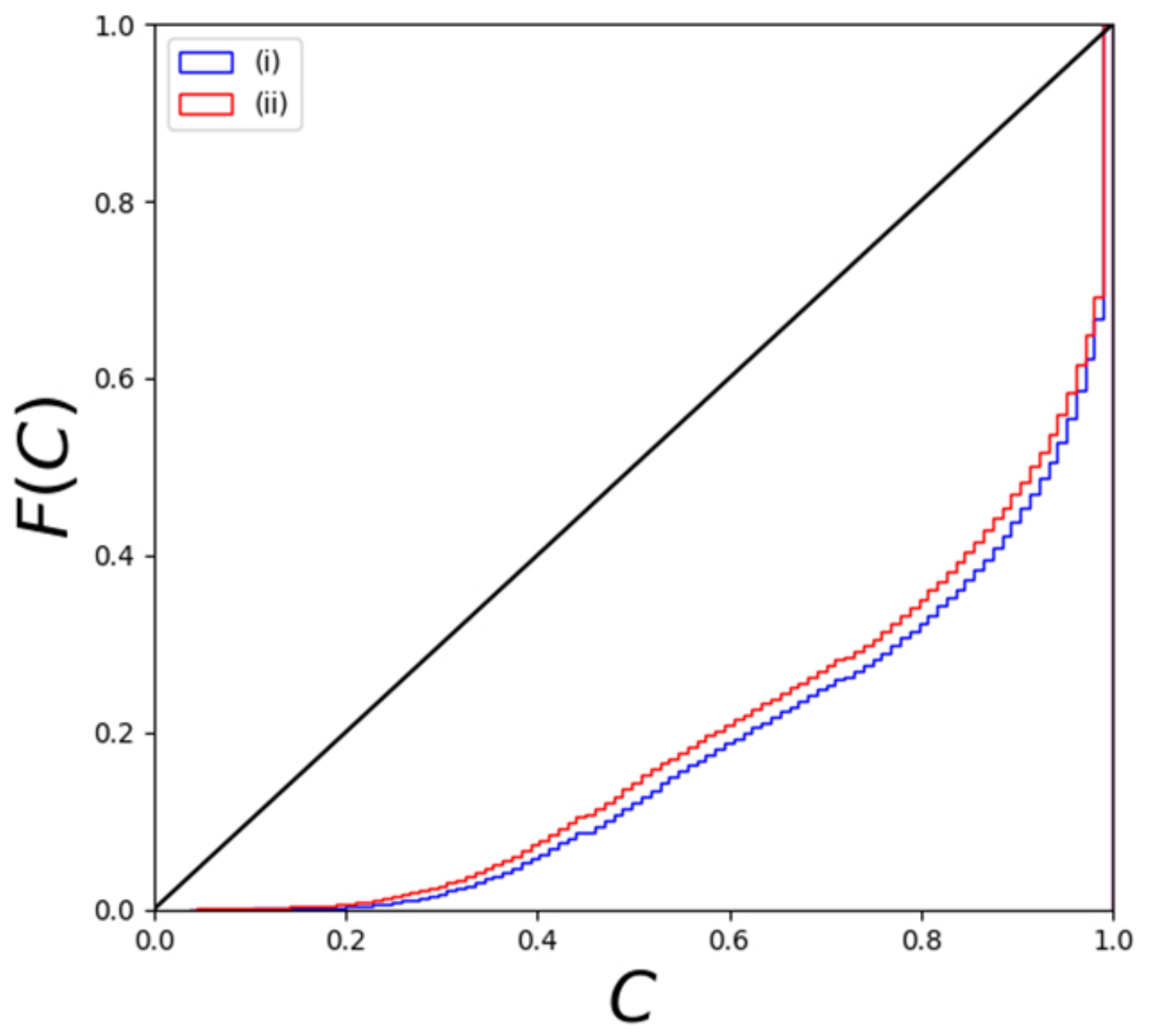}}
\caption{Credibility analysis obtained for the experiments \textit{i} (blue) and \textit{ii} (red).}
\label{fig:CAexp12}
\end{figure}

\begin{figure}
 \centering
 {\includegraphics[width=0.37 \textwidth]{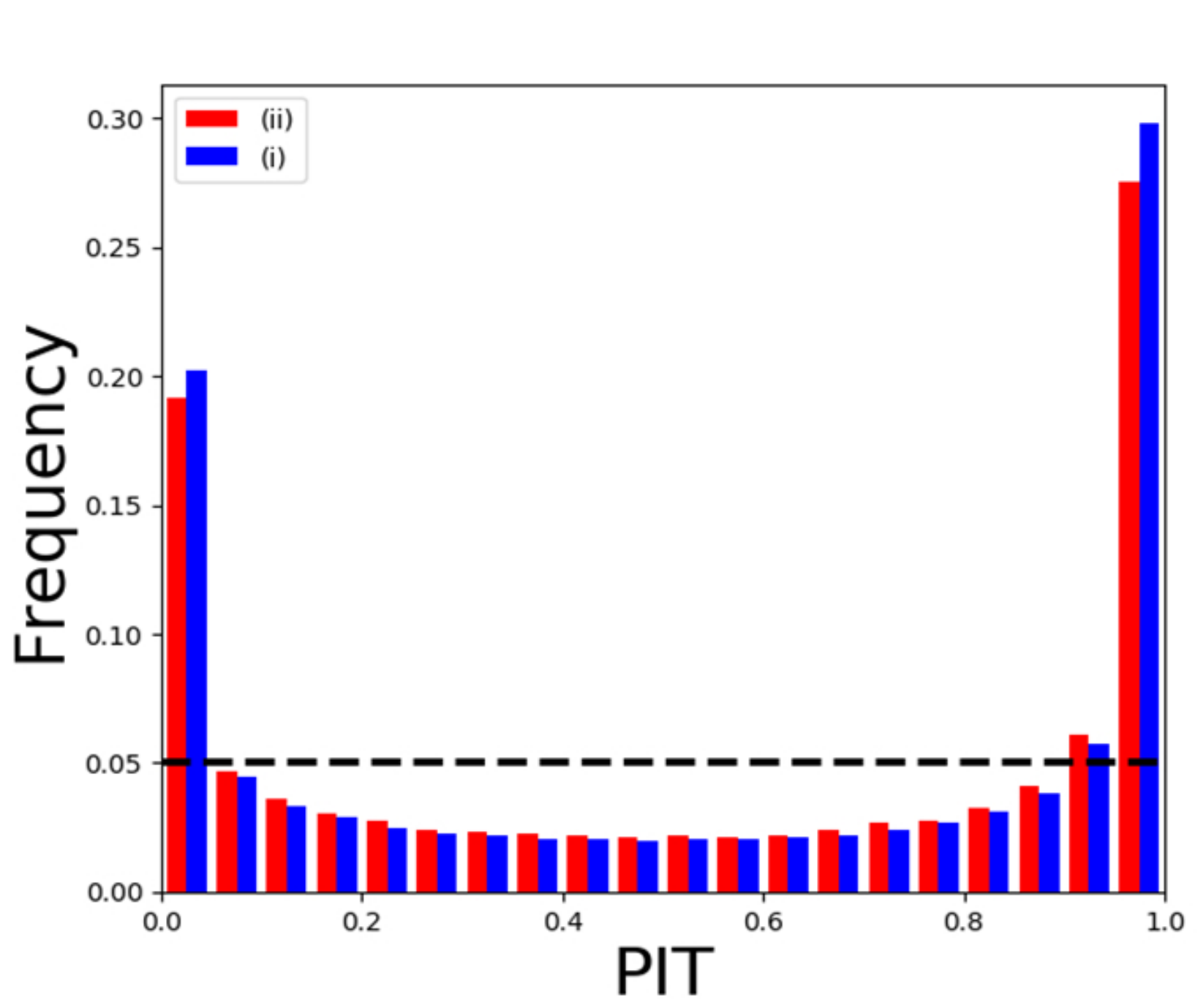}}
\caption{Probability Integral Transform obtained for the experiments \textit{i} (blue) and \textit{ii} (red).}
\label{fig:PITexp12}
\end{figure}

\noindent In Table~\ref{tab:stackedstatERMET} we report the stacked PDF statistics for the two experiments, while the credibility and PIT analyses are shown in figures~\ref{fig:CAexp12} and \ref{fig:PITexp12}, respectively. The results, as expected, are comparable to those obtained for the \textit{dummy} PDF (see Table~\ref{tab:stackedstat}), since in both cases we did not introduce any photometry perturbation. The degradation of the stacked PDF performance (compared to that of the \textit{dummy} PDF) is of the order of $1\%$ and $0.6\%$ for the residual fractions, respectively, $f_{0.05}$ and $f_{0.15}$, while $0.002$ for $\Braket{\Delta z}$. Also the comparison with the credibility and PIT diagrams of the \textit{dummy} PDF (right-bottom diagrams of figures \ref{fig:WittmanALL} and \ref{fig:pitALL}, respectively), reveals only the small differences induced by the $100$ trainings in experiments (\textit{i}) and (\textit{ii}) instead of the single training in the \textit{dummy} case. 
Such statistical variations can be considered negligible if compared to those obtained by the photometry perturbation of the test set (see Sec.~\ref{SEC:thecomparison} and Table~\ref{tab:stackedstat}); where instead the computational cost for experiments \textit{i)} and \textit{ii)} becomes prohibitive (increasing computing time by~$\sim 70\%$).\\
A comparison between experiments \textit{(i)} and \textit{(ii)} in terms of credibility and PIT diagrams, shows very similar results. In particular, by overlapping the two kinds of diagrams (figures~\ref{fig:CAexp12} and \ref{fig:PITexp12}), it appears evident that the sum of contributions of the variable photometry within the training set plus the random weights initialization differ very little from the case in which the photometry is kept unchanged.

\bsp	
\label{lastpage}
\end{document}